# THE SECRET LIVES OF CEPHEIDS

## A MULTI-WAVELENGTH STUDY OF THE ATMOSPHERES AND REAL-TIME EVOLUTION OF CLASSICAL CEPHEIDS

Thesis submitted in March, 2014,
revised thesis submitted in October, 2014, by

Scott Gerard ENGLE BSc (Astronomy & Astrophysics)
Villanova University, PA USA

## STATEMENT OF CONTRIBUTIONS TO THE THESIS

**Leonid Berdnikov** – Astronomer at Sternberg Astronomical Institute, Moscow University

- Provided, in digital form, the O-C data he had gathered/published for β Dor

**Neil Butterworth** – AAVSO observer

- Gathered dSLR photometry of β Dor

**Kenneth Carpenter** – HST Operations Project Scientist at NASA / Goddard Space Flight Center

- Discussion about the stability, performance and analysis of COS UV spectra

**Kyle Conroy** – Graduate Student, Vanderbilt University

- Wrote the Fourier Series fitting routine used in determining Cepheid times of maximum light and light amplitudes

**Nancy Evans** – Astrophysicist at Harvard-Smithsonian Center for Astrophysics (retired)

- PI of the Chandra X-ray proposal for Polaris in 2006 (NASA grant Chandra-GO6-7011A)
- Numerous helpful discussions on various aspects of Cepheids

**Ed Guinan** – Primary Advisor / Professor of Astrophysics & Planetary Science, Villanova University

- Served as PI on all grants used in the thesis
- Many discussions on the direction and methodologies of the study

**Petr Harmanec** – Astronomer, Astronomical Institute of the Academy of Sciences of the Czech Republic

- Discussions on photometric calibrations
- Provided a list of suitable, photoelectric standards

**Graham Harper** – Assistant Professor, Director of Teaching and Learning Undergrad Physics at Trinity College, Dublin

- A fount of astrophysical knowledge, especially UV emission lines and stellar atmospheres

**David Turner** – Professor Emeritus at St. Mary's University, Canada

- Discussion on Cepheid period changes and evolution.
- Provided, in digital form, his updated period change data and also his newest equation used to calculate instability strip crossings and period change rates.

**Rick Wasatonic** – Adjunct Faculty / Research Associate at Villanova University

- Carried out (and continues to carry out) *BV* photometry of Polaris

**Ian Whittingham** – co-Advisor / Professor Emeritus at James Cook University

- Discussions on thesis formatting, organization and presentation

**Grants Associated with the Thesis:**
**NSF grant** AST05-07542
**NASA grants:** HST-GO11726X, HST-GO12302X, HST-GO13019X, XMM-GO050314X, XMM-GO055241X, XMM-GO060374X, XMM-GO06547X, XMM-GO072354X, SST-GO40968X

## ACKNOWLEDGEMENTS

This thesis is dedicated to:

My best buddy Julie, who has been incredibly loving during this rather long road. Her support, through a great many delays, has meant the world to me, and I'll always be lucky to have her.

Mom and Rusty, whose pride has always been felt.

Ms. S and Karen, who always make it feel like I'm part of the family.

Nan, who cut out every (even loosely) Astronomy-related newspaper article, "just in case I was interested…"

Troy, who always had to know where the Moon was.

Ed Guinan, my primary advisor and mentor, who took a chance by allowing me to "wrap up my Polaris study" and then encouraged its expansion into this thesis, which was only possible because of him.

Lena and Ava, who'll grow up to be the best and the smartest of us all.

## ABSTRACT

Classical Cepheids are variable, yellow supergiants that undergo radial pulsations primarily arising from opacity variations in their stellar interiors. Over a century ago, the discovery of a reliable relationship between the period of a Cepheid's pulsations and its luminosity made them "standard candles" and, as such, interest in studying Cepheids boomed. The generally held belief that their pulsations are essentially static over human timescales has sadly led to a narrowing in the field of Cepheid studies. This is in addition to the widespread adoption of high-sensitivity CCD instruments that can quickly saturate when observing nearby Cepheids. The result is that many of the brightest Cepheids, with the longest observational histories, have recently stopped being systematically observed. The primary overall goal of this study is to observe how complex the behaviors of Cepheids can be, and to show how the continued monitoring of Cepheids at multiple wavelengths can begin to reveal their "secret lives."

We aim to achieve this goal through optical photometry, UV spectroscopy and X-ray imaging. Through Villanova University's guaranteed access to ground-based photometric telescopes, we have endeavored to secure well-covered light curves of a 10 Cepheid selection as regularly as possible. Amplitudes and times of maximum brightness were obtained from these lightcurves, and compared to previous literature results. At UV wavelengths, we have been very fortunate to secure numerous high-resolution spectra of two nearby Cepheids – δ Cep and β Dor – with the Cosmic Origins Spectrograph (COS) onboard the Hubble Space Telescope (HST) and additional future spectra have recently been approved. Finally, at X-ray wavelengths, we have (again, very fortunately) thus far obtained images and X-ray (0.3 – 5 keV) fluxes and luminosities of five Cepheids with XMM-Newton and the Chandra X-ray Observatory, and further observations with both satellites have been proposed for (XMM) and approved (Chandra).

Our analysis of optical photometry has shown that 8 of the 10 observed Cepheids have amplitude variability, or hints thereof, and all 10 Cepheids show evidence of period variability (recent, long-term or even possibly periodic). The UV spectra reveal a wealth of emission lines from heated atmospheric plasmas of $10^4$ – $10^5$ K that vary in phase with the Cepheid pulsation periods. The X-ray images have detected the three nearest Cepheids observed (Polaris, δ Cep and β Dor), while the distances of the farther two Cepheids place their fluxes likely at or below detector background levels. The X-ray fluxes for δ Cep show possible phased variability, but possibly anti-correlated with the UV emission line fluxes (i.e. high X-ray flux at phases of low UV flux, and vice versa).

In conclusion, the optical studies have shown that Cepheids may likely undergo period and amplitude variations akin to the Blazhko Effect observed in RR Lyr stars, but on longer timescales. The heating mechanism(s) of their atmospheres appears to be a combination of magnetic/acoustic activity, common in many cool stars, along with pulsation-related effects (shock propagation and possibly convective strength variability). Further data are required to ultimately confirm Blazhko-like cycles in Cepheids, X-ray variability with phase and the particulars of the high-energy variability such as phase-lags between atmospheric plasma emissions of different temperature and the exact contributions of the possible heating mechanisms.

## TABLE OF CONTENTS

## List of Tables

## List of Figures

# CHAPTER 1 – A CEPHEID OVERVIEW

## 1.1 Introduction – The Importance of Cepheids

Type I, or Classical, Cepheids (also known as δ Cep variables and, in this paper, shall hereafter be referred to simply as Cepheids) are a fundamentally important class of pulsating variable stars. Cepheids are among the first classes of variable star discovered, and have played a crucial role in Astronomy and Astrophysics for over two centuries. The study of Cepheids is not only valuable to understand a complex stage of stellar evolution, but to also measure the Universe itself. In Astrophysics and especially Cosmology, the most important aspect of Cepheids is likely their use as extragalactic "standard candles." This is done by way of the Period-Luminosity Law (P-L Law or Leavitt Law hereafter) for Cepheids – first discovered by Henrietta Leavitt in 1912 (Leavitt 1912). The usefulness of the law is bolstered by the fact that the (pulsational) period of a Cepheid can be easily measured through either near-UV – near-IR photometric or spectrophotometric observations, or through radial velocity measurements. A properly calibrated Leavitt Law can then use that period to calculate the Cepheid's luminosity which can, in turn, then be used to calculate the distance to the Cepheid. With the large number of extragalactic Cepheids known, they are at the forefront of studies into the dimensions of our Universe. However, there are some difficulties in calibrating the Leavitt Law. The elimination of these flaws and precise calibration of the Leavitt Law is the aim of several research groups and studies, notably the Hubble Key Project (Freedman et al. 2011).

Cepheids play as important a role in stellar evolution as they do Cosmology. Cepheids exist in an important phase of stellar evolution, as post Hydrogen core-burning stars that have evolved off the main sequence. In the H-R Diagram, Cepheids are aligned into what is called the Classical Cepheid Instability Strip – a sub-section of the general *instability strip* of the H-R Diagram, sometimes called the *classical instability strip*, and marked by the dashed oval shape in Fig. 1. It is within this strip that all Cepheids are found (although non-variable supergiants also exist in this region) and, as with the Leavitt Law, the exact dimensions and extent of the strip are constantly being refined. As they evolve, Cepheids will cross the strip. The "first crossing" of the Instability Strip is from the blue edge of the strip to the red edge – i.e. from higher stellar surface temperatures to lower. Cepheids of sufficient mass will make multiple crossings of the strip as they burn different fuels in different stellar layers. The minimum mass required for a Cepheid to cross the strip more than once is up for debate, and depends on exactly what parameters are used for computing the evolutionary tracks. Aside from the precision of the Instability Strip and the Leavitt Law (and how massive the star must be to carry out multiple crossings), there is no argument that a very specific area of the H-R diagram marks the location of Cepheid variables, that this location is defined by the physical properties (interior structure, chemical composition, density…) necessary to sustain Cepheid pulsations, and that their periods and luminosities are related.

### 1.2 The General Picture of Cepheids

Cepheids are young (50 – 300 Myr), intermediate-mass (typically 4–10 $M_\odot$), luminous (~$10^3$–$10^5$ $L_\odot$), white-yellow (spectral types of approximately F6–K2) Population I supergiants (luminosity classes of Ia, Ib and II) whose radial pulsations produce periodic variations in radius, temperature and, consequently, brightness. The pulsation periods of Cepheids range from around 1.5-days to as long as ~45-days (for SV Vul, but as long as ~60-days or even longer depending on the classification of S Vul and other possible long period Cepheids). The prototype of Cepheids is δ Cep. Although recognized as the prototype of its class, δ Cep was actually the second Cepheid to be discovered, with η Aql owning the distinction of being the first. Both stars were discovered to be variable in 1784. Edward Pigott discovered the variability of η Aql in September of 1784, and John Goodricke (whom Pigott had mentored in Astronomical observing) discovered the variability of δ Cep just a month or so later (Pigott 1785; Goodricke 1786). There are currently more than 800 known Classical Cepheids in the Milky Way and thousands of extragalactic Cepheids (Szabados 2003; 2010). In the visual and photoelectric eras of observational astronomy, Cepheids were often targeted, with numerous studies published per Cepheid. With the widespread adoption of CCD instruments in modern observatories, recent observations of the brightest Cepheids are becoming rarer, as they pose saturation problems for these high-sensitivity instruments. However, the proximity of these bright Cepheids to the Earth makes them attractive targets for time-precious instruments, such as those onboard satellites. Also, the brightest Cepheids usually have the longest observational timelines, crucial for providing insights into long-term behaviors and possibly even evolutionary changes taking place along human timescales.

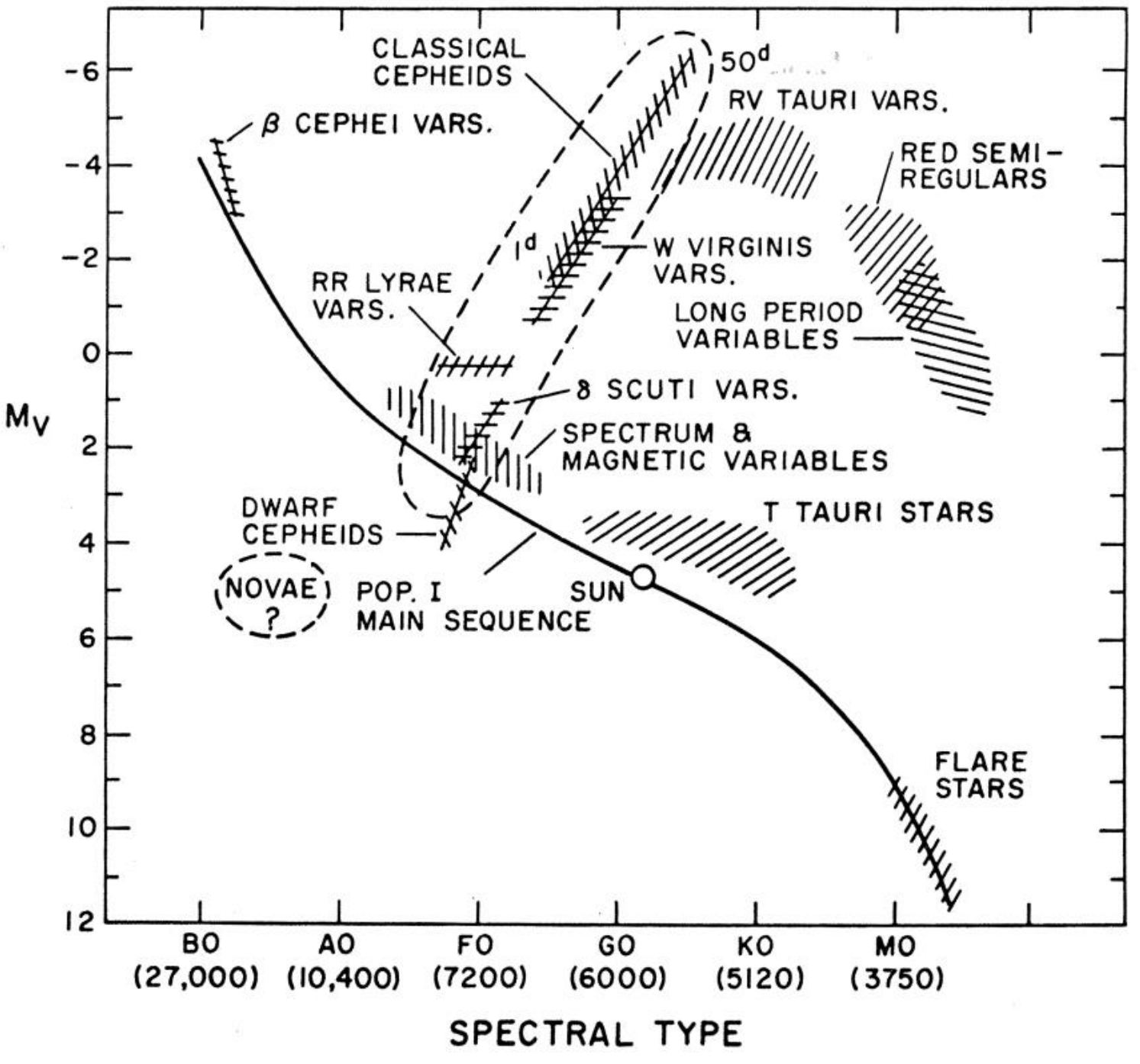

Figure 1 – Hertzsprung-Russell diagram showing the locations of various types of intrinsic variables, including Classical Cepheids at the top-center of the diagram (From Cox 1974).

### 1.3 The Cepheid Pulsation Mechanism

The source of Cepheid light and radial velocity variations posed a long-standing problem for astrophysicists, with a great amount of work devoted to understanding it. Several theories were proposed and studied, with the two most prominent being the pulsations of a single star, or the variability of a binary system. German physicist August Ritter (1879) was the first to propose a simple radial pulsation model to explain variable stars, and even showed that the pulsation period would be proportional to the inverse square-root of the stellar density (this is now known as the *period-density relationship*).

$$P \propto \sqrt{1/\bar{\rho}}$$

where $P$ is the pulsation period and $\bar{\rho}$ is the average density. However, this theory was decades away from acceptance and so Ritter's work unfortunately seemed to gather no interest from astronomers of the time. Aristarkh Apollonovich Belopolsky first showed that Cepheids underwent changes in radial velocity (RV) when he observed and plotted the periodic RV variations in δ Cep (Belopolsky, 1894, 1895). Schwarzschild (1900) then discovered that the color of η Aql varied over its pulsation period, and that the brightness variability had a larger amplitude in photographic (blue) light than in visual (green) light. This showed the brightness changes to be accompanied by temperature changes. Although it was with great difficulty, astronomers labored to include the new data into binary theory, giving rise to more complicated orbital solutions, sometimes involving multiple bodies. In 1913, Brunt (1913) wrote "The problem of the Cepheid Variables" where he gave a rather comprehensive description of the problems associated with the binary theory, yet he never questioned it, instead saying, early in the article,

> *The discovery of the binary nature of δ Cephei was made by Belopolsky in 1894… Since then most of the short-period variables have been shown to be binary stars (Brunt 1913).*

In the same year, Plummer (1913) pointed out the serious problems in interpreting the radial velocity variations of the Cepheid ζ Gem. One year later, Plummer (1914) qualitatively suggested that a radial pulsation mechanism in single stars would avoid the numerous and serious problems in explaining "certain classes" of variable stars. The theory of binarity was dealt its strongest blow by the seminal paper of Shapley (1914), in which he also highlighted its various shortcomings. Shapley made note of the errors associated with fitting binary orbits to Cepheid RV curves. He then combined numerous radial velocity studies of Cepheids with the new stellar classification scheme of Hertzsprung (1909) and Russell (1913), which firmly placed Cepheids in the Supergiant class, and showed that if Cepheids were in fact spectroscopic binaries, the orbital separation of the two stars would place the secondary within the surface of the visible Cepheid. This was the most convincing argument against the binary theory. Despite Shapley's work, some Astronomers continued to research and defend the binary theory into the 1920s and 30s, however the rate of publications was steadily decreasing. One such study was that of the renowned astronomer Jeans (1925) who hypothesized that Cepheids, and long-period variables, were binary stars in the process of fission. One of the latest studies was that of Hoyle & Lyttleton (1943) who revisited Jeans' theory and proposed that Cepheids could be contact systems surrounded by a common envelope that remains independent of the motion of its contained binary system. Despite these later studies, the focus primarily shifted to the dynamics of single stars being responsible for the observed variations. Eddington (1917, 1918) published two papers in which he offered an astrophysical explanation of the stellar pulsations he believed were at work in Cepheids. Eddington proposed that Cepheids act as "heat engines" and, although in his 1917 paper he did not yet understand the *exact* mechanism allowing them to do this, he proposed that:

> *"Possibly during the pulsation, variations of the transparency, which governs the flow of heat, might cause the engine to be fed in the required manner" (Eddington, 1917).*

Though the physics of the process would take some time to develop, the general theory would turn out to be correct. Just over a decade later, Baade (1926) devised a rather important test of the pulsation theory. If the Cepheids were radially pulsating, this would be accompanied by periodic changes in the stellar surface area. The observed luminosity and temperature changes were evidence of this; however, the two variations could be separated to achieve a plot of radius vs. phase. Additionally, the RV data showed the change in radius over time, and could serve as an independent check of the radius variability. His method was improved upon by other astronomers, most notably by Wesselink (1946), and it is now known as the Baade-Wesselink method and is a rather important method of physically calibrating classical pulsators.

Even while the exact cause of their light and velocity variations was still unclear, Cepheids nonetheless rose to a prominent position in astronomy, thanks to the work of a "female computer" – Henrietta Leavitt. Leavitt was hired in 1893 by Edward Pickering, director of the observatory, as one of his "computers" – women whom he hired to measure and catalog the vast amount of photographic plates the observatory possessed and continued to gather. The general view was that women were well-suited to the job, since they earned less than men (so more could be hired for the same overall cost) and were also not allowed to observe on the telescopes, so their time could be focused on plate analysis. Fifteen years after beginning at the observatory, Leavitt remarked in one sentence a behavior that would form one of the most important laws in astronomy: *"It is worthy of notice that in Table VI the brighter variables have the longer periods*" (Leavitt, 1908). At the time the implications were most likely not understood, but this simple sentence announced that ground-breaking research was afoot. Four years later, Leavitt would release a more detailed study confirming the results:

> *"A remarkable relation between the brightness of these variables and the length of their periods will be noticed. In H.A. 60, No. 4, [the 1908 paper] attention was called to the fact that the brighter variables have the longer periods, but at that time it was felt that the number was too small to warrant the drawing of general conclusions. The periods of 8 additional variables which have been determined since that time, however, conform to the same law" (Leavitt & Pickering, 1912).*

In addition to confirming the earlier suspicions of an apparent relationship between period and brightness, Leavitt further noted:

> *"They resemble the variables found in globular clusters, diminishing slowly in brightness, remaining near minimum for the greater part of the time, and increasing very rapidly to a brief maximum. Since the variables are probably at nearly the same distance from the Earth, their periods are apparently associated with their actual emission of light, as determined by their mass, density, and surface brightness" (Leavitt & Pickering, 1912).*

Thus was the "official confirmation" of what would become known as the Cepheid Period-Luminosity Law (P-L Law), also recently redubbed the Leavitt Law. Leavitt did not know it at the time, but she was comparing Population I Cepheids (Classical Cepheids – "they" in the quoted text) and Population II Cepheids (previously called W Virginis Stars before separate sub-classes were created – "the variables found in globular clusters"). The two stellar classes exhibit similar light and velocity changes, but are completely different stars brought to a similar location in the H-R Diagram by ways of unrelated evolutionary tracks. The inclusion of both Cepheid types led to larger uncertainties in the Law, since Type II Cepheids are ~1.5-mag fainter than Type I Cepheids of the same period. This continued until Walter Baade (1944) first discovered two distinct stellar populations of stars in his detailed study of M31, M32 and NGC 205. Later, specifically in regard to problems encountered calibrating the P-L Law, he wrote:

> *"Miss Leavitt's cepheids in the Magellanic Clouds and the classical cepheids in our galaxy are clearly members of population I, while the cluster-type variables and the long-period cepheids of the globular clusters are members of population II. Since the color-magnitude diagrams of the two populations leave no doubt that ... we are dealing with stars in different physical states, there was no a priori reason to expect that two cepheids of the same period, the one a member of population I, the other a member of population II, should have the same luminosity (Baade 1956)."*

The Leavitt Law has played a crucial role in astrophysics since its discovery. The Shapley-Curtis "Great Debate" highlighted the importance of scale in our modern understanding of the Universe. Great astronomers such as Hertzsprung, Shapley and Hubble all used the Law to determine accurate distances to some of the nearest and most prominent extragalactic objects, such as the Magellanic Clouds, globular clusters and the Andromeda Galaxy. The Leavitt Law helped astronomers understand that the Universe was, in fact, *much* larger than originally thought. The tight, linear arrangement of Cepheids in the HR Diagram is illustrative of the Law they have become associated with, but what is responsible for such an arrangement is the pulsation mechanism at work and the internal structure/dynamics of Cepheids.

As mentioned, much work has been devoted to understanding the pulsation mechanisms of Cepheids. The foundation of a modern theory of Cepheid pulsations was laid down in an important paper by Eddington (1941), in which the "valve" mechanism controlling the pulsations was incorrectly ascribed to hydrogen as the crucial element. Further, Eddington considered core nuclear reactions to be the direct driver of the pulsations. Two decades later, Baker & Kippenhahn (1962), Cox (1963) and Zhevakin (1963) would detail a Cepheid pulsation mechanism via the non-adiabatic opacity modulations of ionizing gas regions existing in the internal stellar envelope, specifically partially ionized He II zones in the upper layers of Cepheids. This ionization mechanism causes the layer to absorb heat during compression and then release it during expansion. For Cepheid pulsations to be successfully driven, two basic criteria must be satisfied: a sufficient amount of ionizing material (He II in the case of Cepheids) must exist at the transition within the star from adiabacity to non-adiabacity (Cox 1985). More specifically, a star must contain a concentration of at least 10-15% helium (number fraction), half of which is ionized (Kukarkin 1975). Also, the helium ionization zone can only exist within the temperature range 35000-55000 K (Bohm-Vitense 1958). The stellar depth at which these temperatures exist progresses from nearer to the surface in hotter (earlier-type) stars down to depths nearer the core in cooler (later-type) stars, as shown in Fig. 2. Because of the stringent requirements on ionization zone depth, combined with the dampening effect of strong convection in cooler stars (see Gastine & Dintrans 2011 and references therein), Cepheid pulsations can only be maintained in a specific range of effective temperature. This range is defined as the nearly vertical "classical instability strip" where Type I Cepheids are found on the HR diagram (Fig. 1).

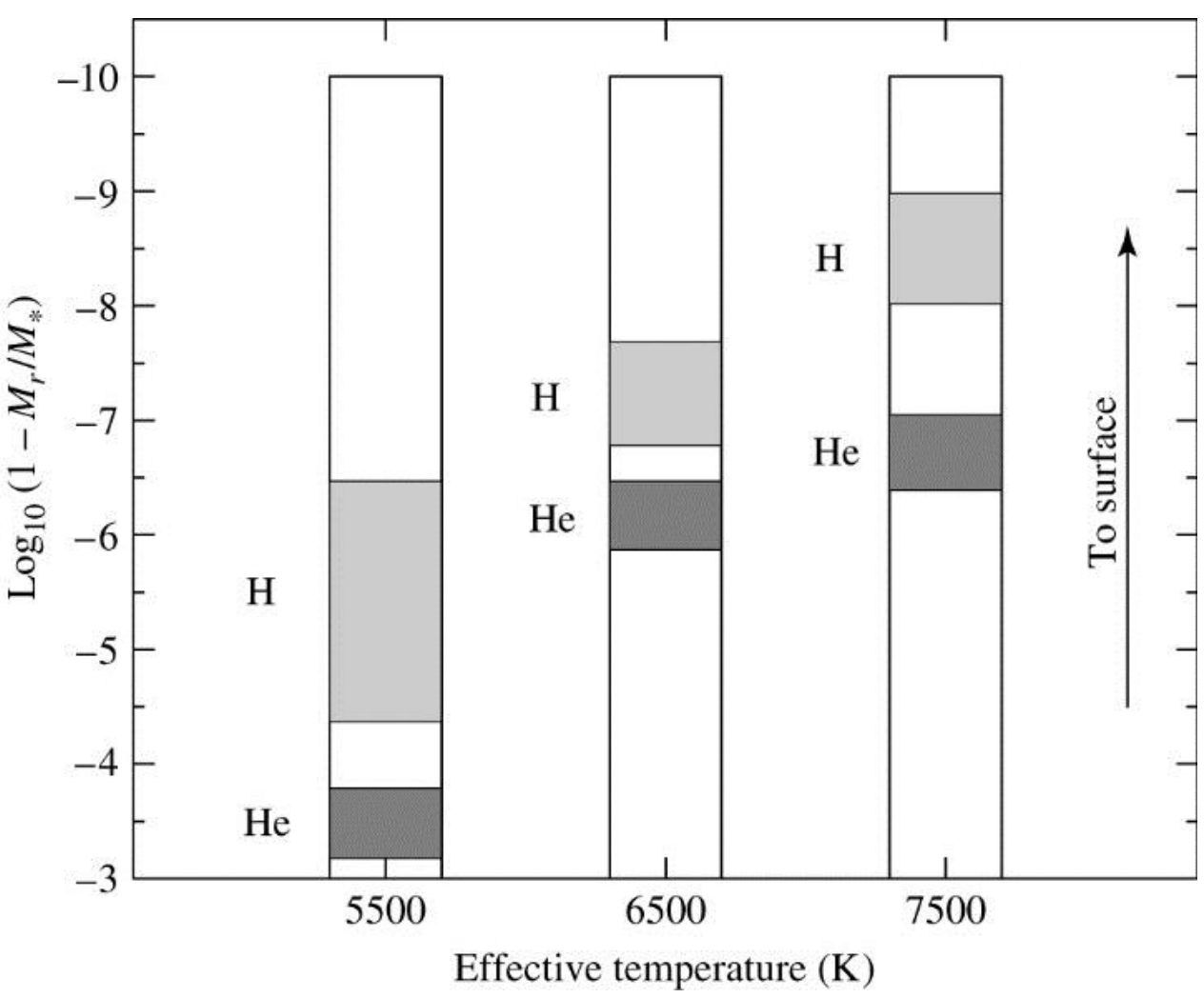

Figure 2 – Hydrogen and Helium ionization zones in stars of different temperatures. For each point in the star, the vertical axis displays the logarithm of the fraction of the star's mass that lies *above* that point. (From Carroll & Ostlie 2006).

Though the He II ionization zone is regarded as the primary driving force behind Cepheid pulsations, it is one of three mechanisms at work. Under adiabatic conditions, the compression of a gas would increase its temperature. In partial ionization zones, however, compression energy (that would normally be converted to thermal energy) is used to further ionize the gas. This prevents the temperature from rising, although the density of the region *is* allowed to rise, and therefore opacity also rises, in accord with *Kramer's opacity law*

$$\bar{\kappa} \propto \rho / T^{3.5}$$

where opacity (or mass absorption coefficient) is indicated with the Greek letter kappa ($\kappa$), $\rho$ is the density and $T$ is the temperature. In this scenario, opacity increases upon compression, and energy is trapped (due to the increased opacity) within the ionization zone during compression. Also, the increased opacity of the region causes pressure to build beneath it, driving stellar expansion. This expansion reduces the opacity, allowing pressure and energy to be released. The consequence is that the energy released upon expansion is larger than it would be in a purely adiabatic case since, upon expansion and cooling of the gas, the ionization energy is radiated back into the region during recombination, and converted into thermal energy. After the release of the stored pressure and energy, gravitational contraction begins the cycle anew. This is referred to as the "kappa mechanism" (Baker & Kippenhahn 1962) and is responsible for "pumping" the pulsations of a Cepheid. In addition, since the temperature of the ionization region remains low compared to the surrounding stellar interior, heat flows from the surroundings into the partial ionization region, further driving its full ionization and increasing opacity. This is the "gamma mechanism" (Cox et al. 1966) and is a secondary driver of the pulsations. The final, third mechanism at play is perhaps the most simple and straightforward: the "r mechanism" or "radius mechanism," so named because, upon compression, the star has a smaller surface area from which it can emit radiation. Consequently, the star stores more radiation when smaller than it would at a larger size and this stored radiation builds pressure, also feeding into the pulsation cycle, aiding in the expansion of the star.

However, as mentioned previously, the instability of the He II region within the star to variations in temperature and pressure is *primarily* responsible for the radial pulsations of Cepheid variables. The *Eddington Valve*, now understood to rely on the ionization of He instead of H, serves to store and release heat at specific times to successfully drive stellar pulsations. Since this valve relies heavily on a specific temperature, composition and location within a star to operate, only stars within a certain area of the H-R diagram can support stable, radial pulsations. This, of course, is why the Cepheid Instability Strip is a prominent feature of the H-R diagram, and it is also why Cepheids obey the Period-Luminosity law for which they have become so well-known and cosmologically important.

### 1.4 The Stellar Evolution of Cepheids

In short, Cepheids are B-type stars (at birth, with masses ranging from ~3–18 $M_{\odot}$, Zombeck 2007) that have evolved off the main sequence and are now passing through the Cepheid Instability Strip. As stellar evolutionary tracks show, a star can enter the instability strip more than once during its post-main sequence lifetime (Fig. 3).

Except for the most massive Cepheids, the first crossing of the instability strip occurs after the exhaustion of hydrogen in the stellar core, when the star has entered the hydrogen shell-burning phase. During this, the star expands, rising above the main sequence and proceeding to the right (towards cooler surface temperatures) on the H-R diagram, towards the red giant region. This initial path through the instability strip occurs as the stellar surface is cooling, and is very rapid, occurring along a Kelvin-Helmholtz (thermal) timescale, generally lasting $10^3$–$10^4$ years (Bono et al. 2000b). After this first crossing of the instability strip, the star exits the red edge of the strip and ascends along the red giant branch of the H-R diagram to the "red giant tip" where core He ignition occurs.

With the onset of core He burning, the star contracts and heats up, descending the red giant branch and moving towards the left on the H-R diagram, towards higher temperatures and bluer colors. This can bring the star into the instability strip for the second time. The star is now undergoing what is called a "blue loop." The exact extent of the blue loop (in color/temperature space) can either cause the star to make two distinct crossings of the instability strip, or it can cause the star to enter the instability trip through the red edge, "turn around" while still within the strip, and evolve back out through the red edge again. In either case, the blueward path of the Cepheid through (or into) the instability strip is referred to as the second crossing, and the redward path through (or back out of) the strip is the third crossing. The specifics of blue loops (and even their very existence for stars of lower masses) are constantly being revised, and generally depend on the abundances and structure of the star as a result of the previous stages of stellar evolution, such as the H core-burning and shell-burning phases, and also factors such as convective core treatment. In general, the extent of the blue loop increases for stars of increasing mass. One consequence of this is that, for a star of low enough mass, the blue loop will not extend far enough to penetrate the red edge of the instability strip. As such, lower mass stars will never undergo more than a first crossing of the strip. This, combined with the short lifetime of the first crossing, can contribute to the lower mass limit observed for Cepheids. For good discussions of blue loops, see Bertelli et al. (2009) and Valle et al. (2009).

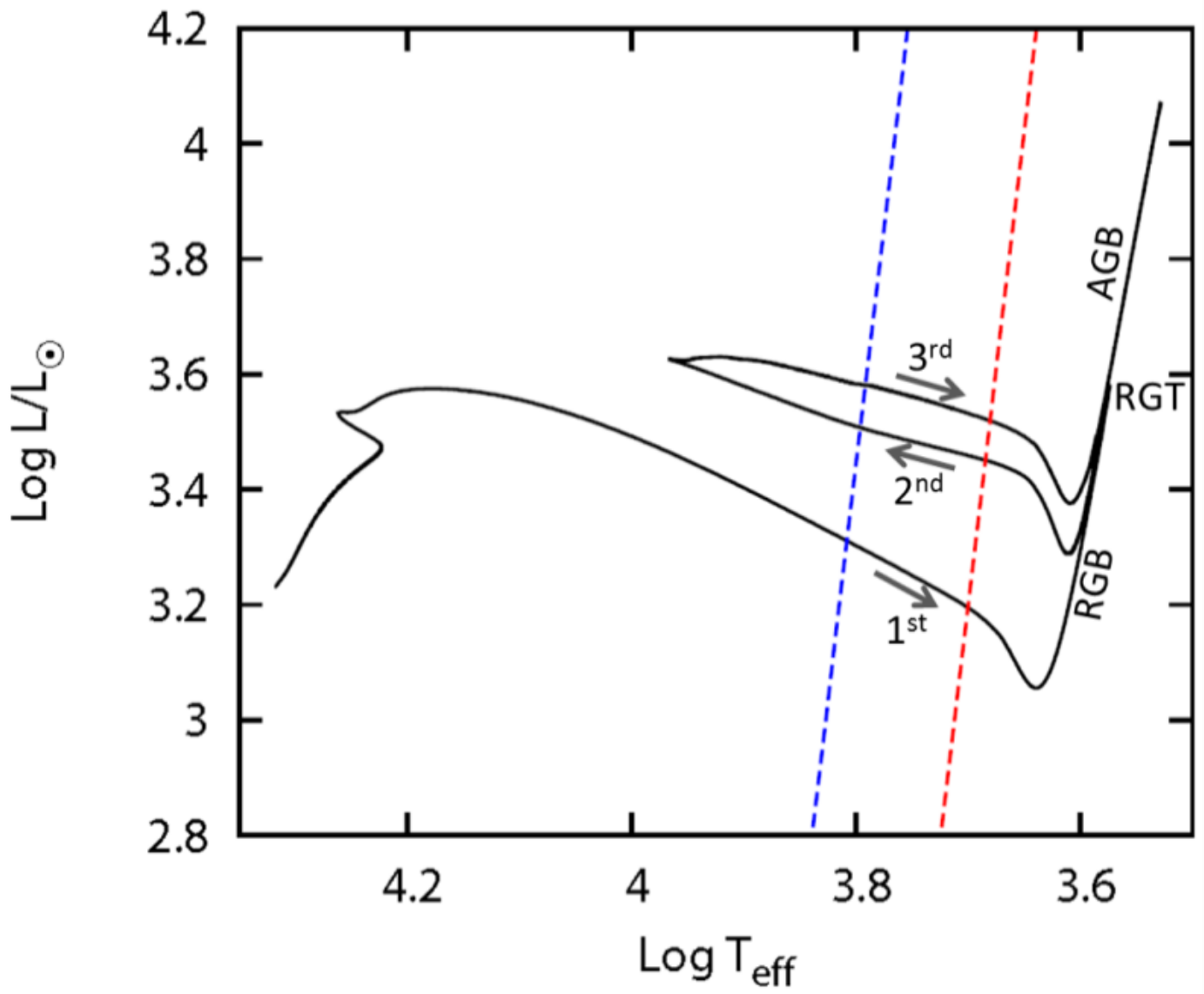

Figure 3 – The evolutionary track of a 7 M$_\odot$ (initial mass) star, computed using the code of Yoon & Langer (2005). The dashed, colored lines show the blue and red edges of the instability strip, based on the study of Bono et al. (2000b). Labels mark the 1$^{st}$, 2$^{nd}$ and 3$^{rd}$ crossings of the instability strip, as well as the direction of the star along the evolutionary path, as well as the red giant branch (RGB), red giant tip (RGT) and the asymptotic giant branch (AGB).

This second crossing of the instability strip is longer than the first crossing (Bono et al. 2000b), as is (in most cases) the third crossing. This is consistent with the findings of Turner (1998), where most Cepheids are found to be in the second or third crossings of the instability strip. The third crossing of the instability strip occurs with He core exhaustion and contraction, or can also be the result of He shell-burning. A second blue loop can occur, resulting in fourth and fifth crossings of the instability strip, but this requires specific conditions that are not often met (Becker 1981).

After evolving along the blue loop(s), stellar evolution proceeds as dictated by the star's mass. The intermediate mass Cepheids evolve onto the asymptotic giant branch, eventually losing their stellar envelopes and leaving a degenerate core. The higher mass Cepheids can proceed along further steps of nuclear burning before ending their lives as supernovae.

### 1.5 Development of the Period–Luminosity Law

As mentioned, Henrietta Leavitt was the first to observe and remark on a relationship between Cepheid periods and luminosities (Leavitt 1908), using Harvard Observatory photographic plates of the Magellanic Clouds. Her 1912 publication was based upon the study of Cepheids in the Small Magellanic Cloud (SMC) alone, which allowed her to comfortably use the apparent magnitudes of the Cepheids in her sample, since they were all assumed to have essentially the same distances (Fig. 4).

This original relationship, having been based on apparent magnitudes, was of limited technical use outside of the SMC. However, it showed that perhaps an absolute law should exist, and the importance of such a law was quickly recognized, motivating further studies. Since Leavitt's original work, many famous Astronomers have contributed to this task: Hubble, Shapley and Baade, to name a few (Hubble 1925; Shapley 1927; Baade 1956). Modern formulations of the Leavitt Law generally take the form:

$$\langle M \rangle = \alpha + \beta \log P$$

where $M$ is the Cepheid's mean absolute magnitude at the particular wavelength of study, $\alpha$ is the zero point of the relationship, $\beta$ is the slope, and $P$ is the pulsation period (usually in days). Even when using the most accurate data possible, there is a luminosity scatter to the relationship, which is a consequence of the appreciable width to the Cepheid instability strip. After much work investigating the scatter found in the relationship, many authors determined that a color term should also be included (see Tammann et al. 2003). The general form of the Period-Luminosity-Color (PLC) relationship is:

$$\langle M_\lambda \rangle = \alpha + \beta \log P + \gamma(CI)$$

where $CI$ now represents the specific color index being used in the relationship, $\lambda$ is the coefficient of the color index, and $M_\lambda$ is the absolute magnitude at a given wavelength bin or photometric band.

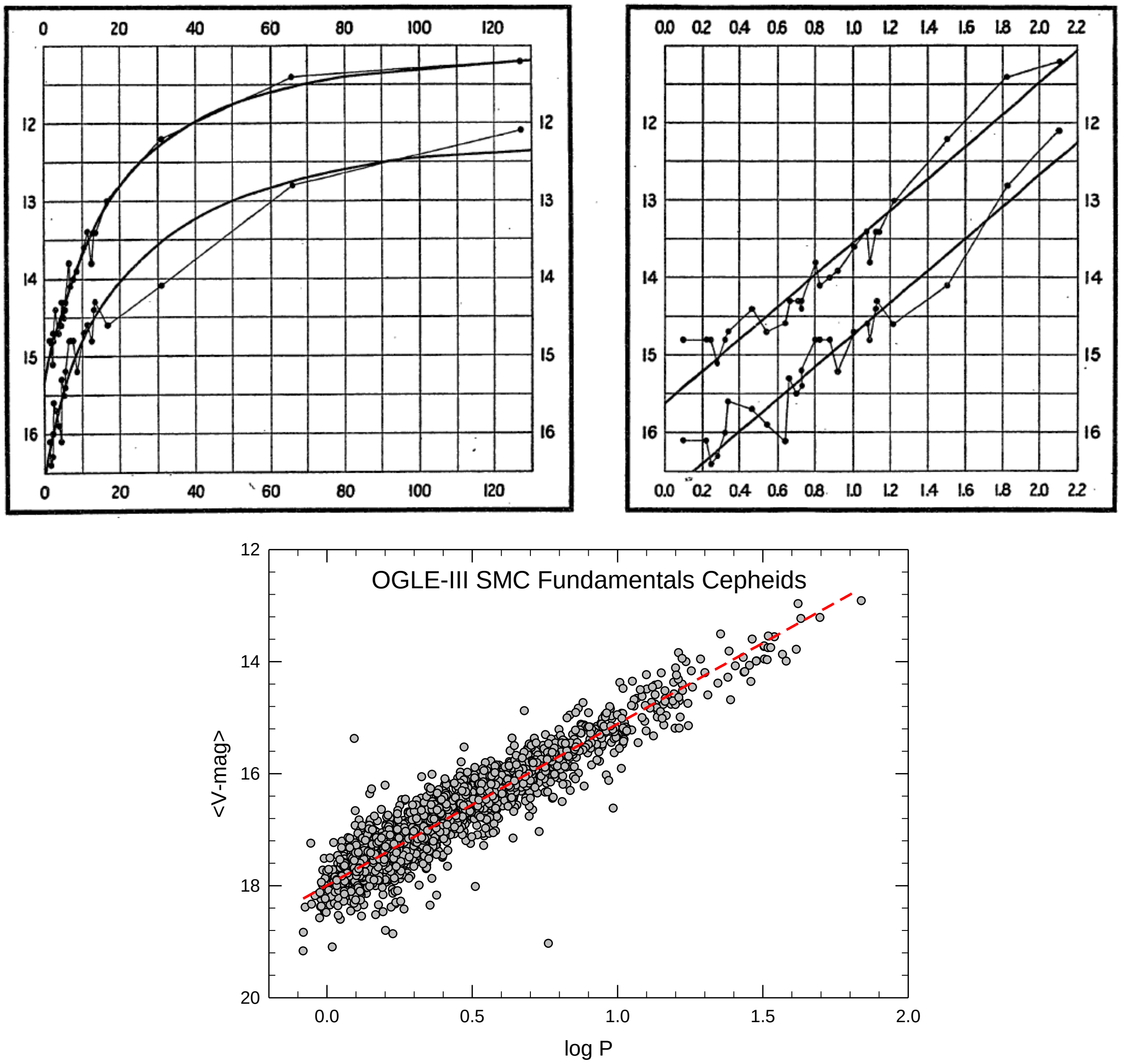

Figure 4 – (Top Plots) The Original "P-L Law" of Leavitt (1912). The y-axes of both graphs gives the apparent magnitudes, and the x-axes give period in days (left) and log period in days (right). (Bottom Plot) The full OGLE-III inventory of SMC fundamental mode Cepheids with available average V-magnitudes and (log) pulsation periods (http://ogledb.astrouw.edu.pl/~ogle/CVS/). The resulting relationship, with 3σ errors, is: $\langle V \rangle = 17.9944(10) - 2.8839(20)\log P$.

In order for the P-L Law to achieve its full potential, there must be a precise zero point. This precision can be most directly obtained through absolute trigonometric parallaxes of Cepheids with a variety of pulsation periods. Benedict et al. (2007) used the very precise Hubble Space Telescope Fine Guidance Sensors (FGS) to determine absolute parallaxes for a sample of 10 galactic Cepheids. In addition, van Leeuwen et al. (2007) analyzed revised *Hipparcos Satellite* parallaxes of a group of Cepheids, combined with the results of Benedict et al. (2007), to return a very accurate PLC relationship of:

$$\langle M_W \rangle = -2.58 - 3.288(\pm 0.151)\log P + 2.45(V - I)$$

where $M_W$ represents the absolute Wesenheit magnitude, an extinction-corrected magnitude, following the prescription of Freedman et al. (2001).

### 1.6 Period Studies of Cepheids

In addition to being one of a Cepheid's most readily determinable properties, and also one that can be determined with the highest precision, the pulsation period of a Cepheid can also provide astrophysically relevant information on the star (e.g. evolutionary status, luminosity). Bezdenezhnyj (2007) conducted a study of 473 Cepheids with known periods found in the 4$^{th}$ edition of the General Catalog of Variable Stars (GCVS), and Fig. 5 below gives the results of his study.

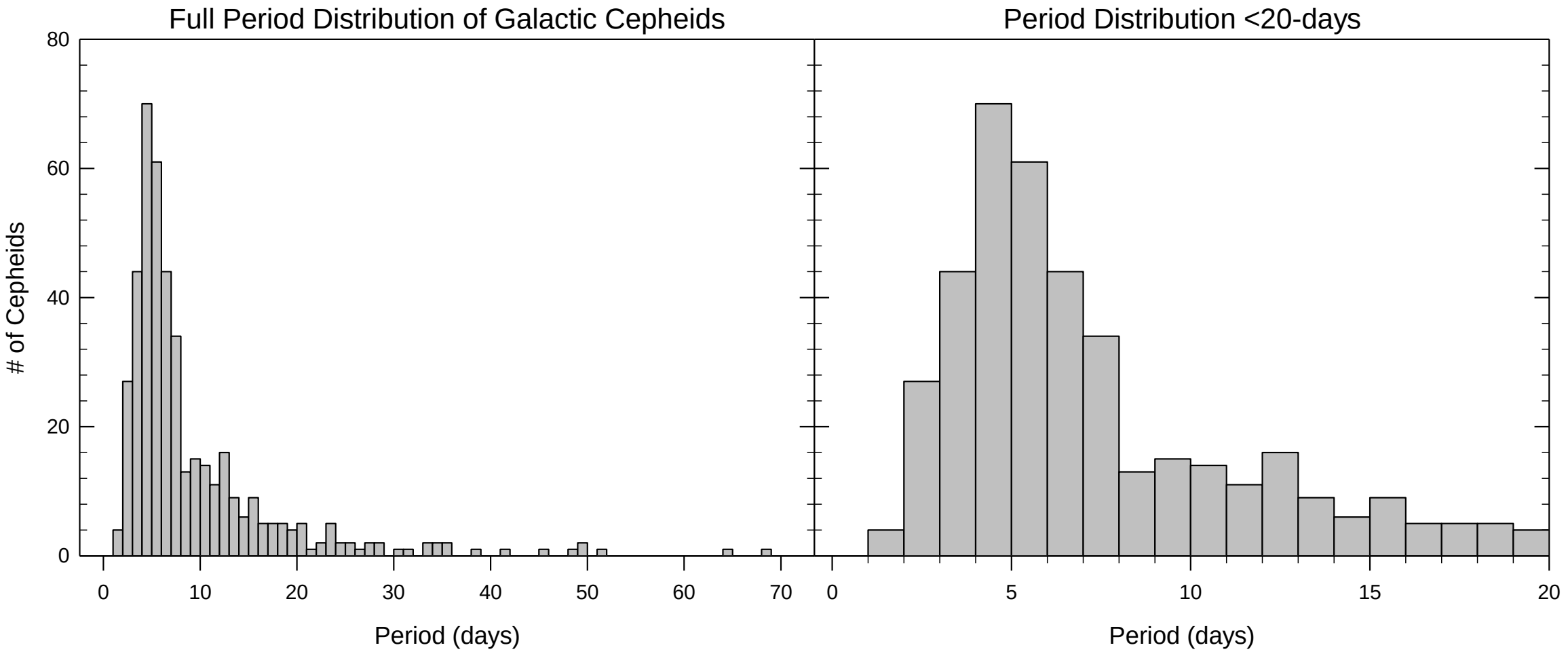

Figure 5 – Results of the galactic Cepheid period distribution study of Bezdenezhnyi (2007). The left-hand plot shows the full distribution, while the right-hand plot shows only the period shorter than 20-days. The strong, primary maximum around periods of 5-days can clearly be seen, along with the possible secondary maximum in periods of around 10-days.

As can be seen, there is a clear maximum in the period distribution for Cepheids with periods around 5-days. There is also a possible secondary maximum at around 10-days. In studying the period distributions of Cepheids in the LMC and SMC, Soszynski et al. (2010) found the most prominent periods to be around 3.2-days in the LMC and 1.6-days in the SMC. This is a metallicity effect, since the LMC has a lower mean metallicity than the Milky Way, and the SMC has an even lower metallicity ($[Fe/H]_{LMC} \approx -0.34$, $[Fe/H]_{SMC} \approx -0.68$ – Storm et al. 2011). Stellar evolutionary tracks show the blue loops to extend out to higher temperatures (and bluer colors) with decreasing metallicity. This extension of the blue loops would allow lower mass stars to undergo second and third crossings of the instability strip. Since stars of lower mass are relatively more numerous, a region (or galaxy) of lower metallicity would naturally have more Cepheids of lower masses and shorter periods.

In addition to statistical studies of Cepheid pulsation periods, there are characteristic light curve behaviors to Cepheids of different periods. Cepheids with "short" and "long" periods display, for the most part, smooth light curves with a shorter-duration rise to maximum light and a longer-duration fall to minimum light (the 'saw tooth' pattern). Hertzsprung (1926) was the first to notice that Cepheids with periods primarily between ~6–16 days (but possibly even longer), however, display a "bump" on their lightcurves. At P ≈ 6 days, a small "bump" (local increase in brightness) appears at an appreciable distance (in phase-space) from maximum light, on the descending branch of the lightcurve, progressing up the descending branch (nearer maximum light) and becoming more prominent as the period of the Cepheid increases. The bump is very close to the phase of maximum light in Cepheids with ~9–12-day periods, proceeding down the ascending branch of the lightcurve as period increases further. The migration of the bump is called *the Hertzsprung progression*, and Cepheids within this period range, although still Classical Cepheids, are sometimes referred to as *Bump Cepheids* or *Bump Resonance Cepheids*.

Two proposed models to explain the Hertzpsrung progression are the echo model and the resonance model (hence the title of Bump Resonance Cepheids). In the echo model, pressure waves are generated within the Cepheid. The inward-traveling wave reflects (or echoes) off the stellar core, and then causes the bump as it reaches the stellar surface. In the resonance model, the bumps are the results of a resonance between second overtone and fundamental pulsations in the Cepheid. For a thorough discussion of the Hertzsprung progression and the proposed models, see Bono et al. (2000a). Fig. 6 shows the lightcurve of 5 selected Bump Cepheids. The gray line in each plot illustrates the approximate location of the bump and the black arrow indicates the direction of progression along the lightcurve (except in the case of VX Cyg where the bump is, in theory, no longer progressing).

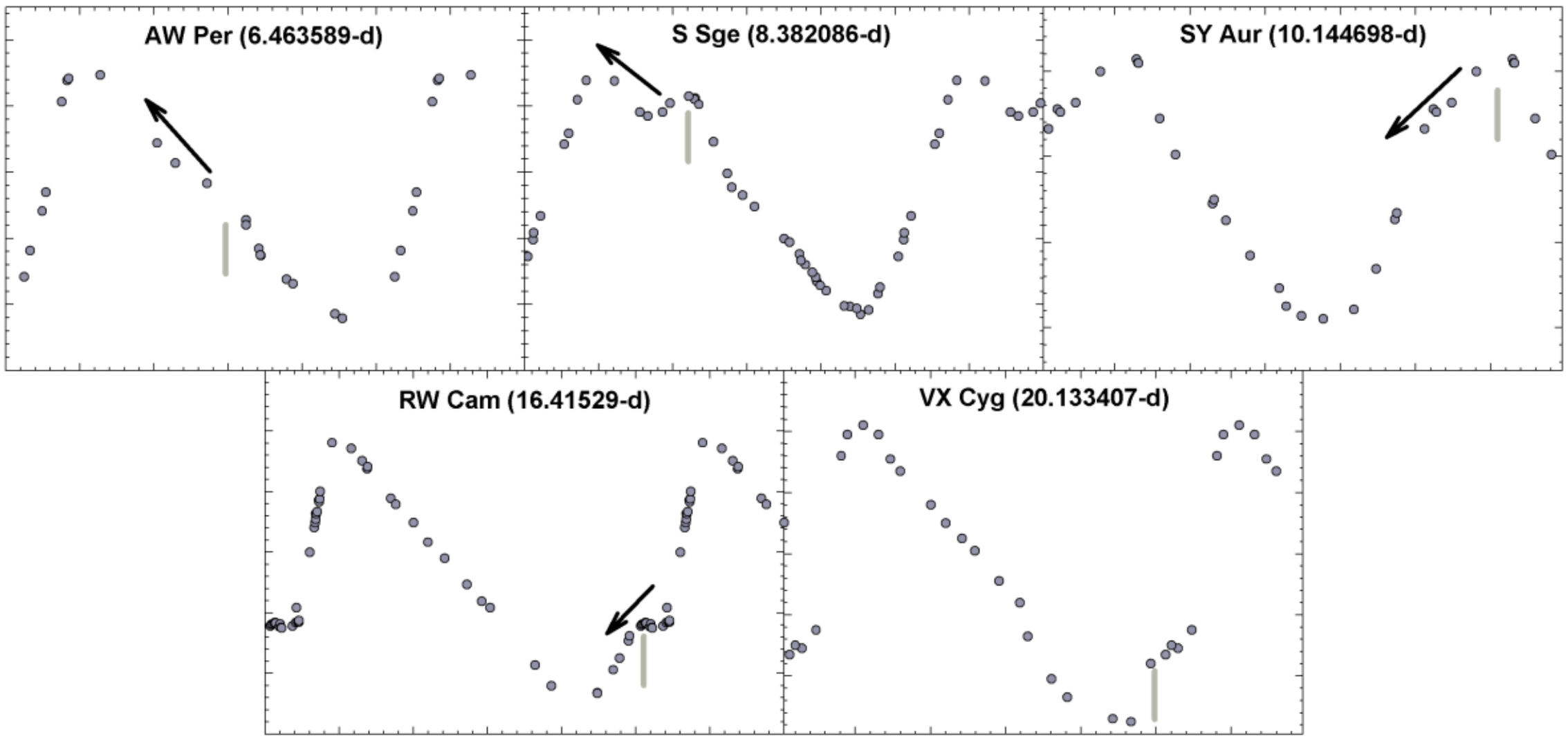

Figure 6 – The lightcurves for five selected Bump Cepheids are shown, (literature data obtained from the **McMaster Cepheid Photometry and Radial Velocity Data Archive** – http://crocus.physics.mcmaster.ca/Cepheid/) covering essentially the full period-span of the Hertzsprung progression. The gray lines under each lightcurve mark the phase of the bump maximum, and the black arrows indicate the direction (in phase-space) in which the bump is progressing.

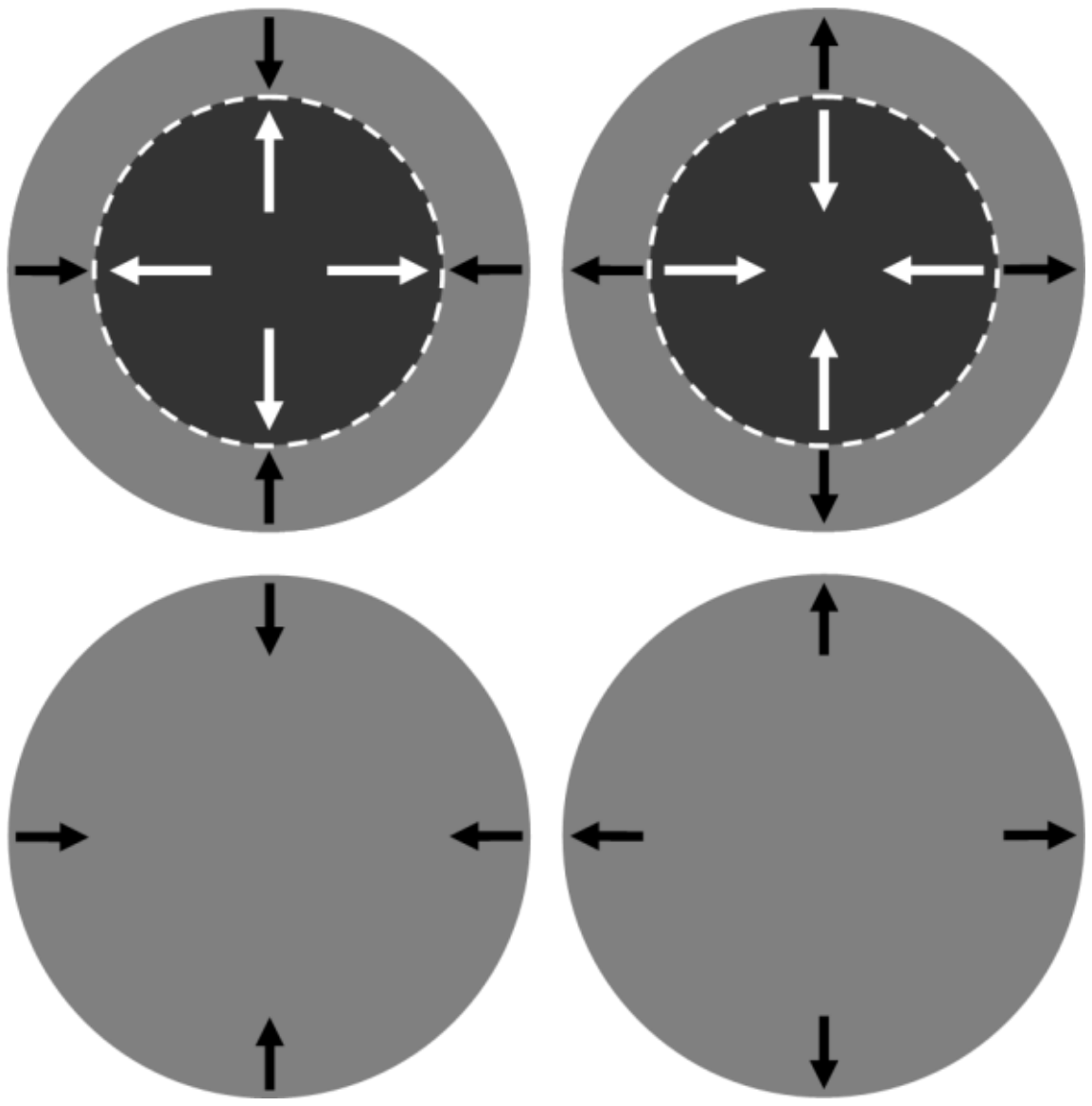

Figure 7 – Illustration of a first overtone pulsator (top row) vs. a fundamental pulsator (bottom row). Arrows indicate direction of pulsation, and the white dashed line shows the pulsation node inside a first overtone pulsator.

Among short-period Classical Cepheids (P < 5-days), there is another subgroup of stars. These Cepheids are generally characterized by smaller amplitudes (< 0.5-mag) and more symmetrical (sinusoidal) light curves (Luck et al. 2008). Their increased symmetry is responsible for their title as *s-Cepheids*. The *Secret Lives of Cepheids* (SLiC) Program (under which this thesis has been carried out – see Section 1.9 for program description) presently contains three s-Cepheids: Polaris (although still debated), SZ Tau and SU Cas. It has been suggested that s-Cepheids pulsate in the first-overtone mode and, where many s-Cepheids have indeed been found to be first-overtone pulsators, there are also some known fundamental mode s-Cepheids (see Luck et al. 2008). When dealing with the Cepheid Period-Luminosity law, s-Cepheids are usually treated as first-overtone pulsators. The 'simplest' pulsations that a Cepheid can undergo are the aptly named fundamental mode pulsations, where the entire star contracts and expands in unison (bottom row of Fig. 7). The first overtone differs from the fundamental mode in that there are two separate "shells" within the Cepheid which are pulsating in the opposite direction of each other – when one shell is expanding, the other is contracting, and vice versa. These shells are separated by what is called a "node" (white dashed line in top row of Fig. 7), with one shell expanding while the other contracts, and then vice versa. It is this opposite behavior between the shells that is responsible for the increased symmetry in first-overtone Cepheid lightcurves, causing them to appear more sinusoidal, and also results in lower luminosity amplitudes.

Although s-Cepheids and other first-overtone Cepheids share many common traits with fundamental mode Cepheids, they do not behave in exactly the same way. It has been discovered that first-overtone Cepheids, while following Period-Luminosity and Period-Radius relationships, do not follow the same relationships as fundamental mode Cepheids. This difference in behavior has played an important role in determining the true status of Polaris, a SLiC program star and argued to be an s-Cepheid pulsating in the first-overtone mode. Nordgren et al. (2000) used the Navy Prototype Optical Interferometer (NPOI) to directly measure the radius of Polaris to be 46 ± 3 $R_{\odot}$, using the original Hipparcos parallax of 7.56 ± 0.48 mas. For Polaris' observed period of ~3.97-d, this radius is too large for it to fit along the fundamental mode Period-Radius relationship. However, when the derived fundamental period of Polaris, 5.59-days, is used (applying a first-overtone/fundamental period ratio of 0.71 – Aerts et al. (2010), p. 12), the Cepheid fits nicely along the period-Radius relationship. The NPOI result would lend credence to the argument that Polaris pulsates in the first-overtone mode.

Cepheids are most often observed pulsating in either the fundamental or first-overtone modes. Higher overtone modes, and even "Beat Cepheids" that simultaneously pulsate in two or more pulsation modes, are also observed but are much rarer.

### 1.7 Photometric Properties of Cepheids

The strict periodicity of most Cepheids is perhaps the main characteristic that made them such intriguing and popular targets for early observers and is also partly responsible for their astrophysical importance as distance indicators. However, many Cepheids exhibit a change in period over time. The first record of this behavior seems to have been published by Eddington (1919) in his paper on δ Cep. In this paper, Eddington mentions earlier studies of a period change in δ Cep made by Chandler, but no publication could be found. There are now many Cepheids for which very rich datasets have been used to display and delineate period changes – some linear, and others possessing random, sometimes even repeating, period variations. The main analytical method of characterizing period changes in Cepheids is by constructing O–C diagrams. After a sufficient number of (most commonly) times of maximum light have been observed for a Cepheid, the behavior of the pulsation period can be understood by plotting the observed (O) times of maximum light minus those that are calculated (C) based on the assumption of a known, static period and essentially stable light curve shape. If an O-C diagram appears entirely linear, then the period used to generate such a plot is incorrect. If the O-C curve appears "concave up," then the period is increasing; and if the curve appears "concave-down," such as in SV Vul (a program Cepheid, which will be shown later), then the period is decreasing. Some different O-C curve "shapes" and their period variability implications are shown in Fig. 8. One small trend that has been noticed is that, on average, Cepheids with longer periods display greater changes in period per year. SV Vul, for example, with a pulsation period of ~45 days, undergoes a long-term period decrease of dP/dt ≈ –231 sec/year. Finally, some O-C plots reveal cyclic changes in the period over time. These periodic changes may be the result of as yet unknown internal changes within the Cepheid, or they may also be the result of the Cepheid being in orbit with a companion star. In the latter case, the orbital motion of a Cepheid would produce a "light-time" effect in which the period would be seen to shorten and lengthen as the star moved towards or away from us.

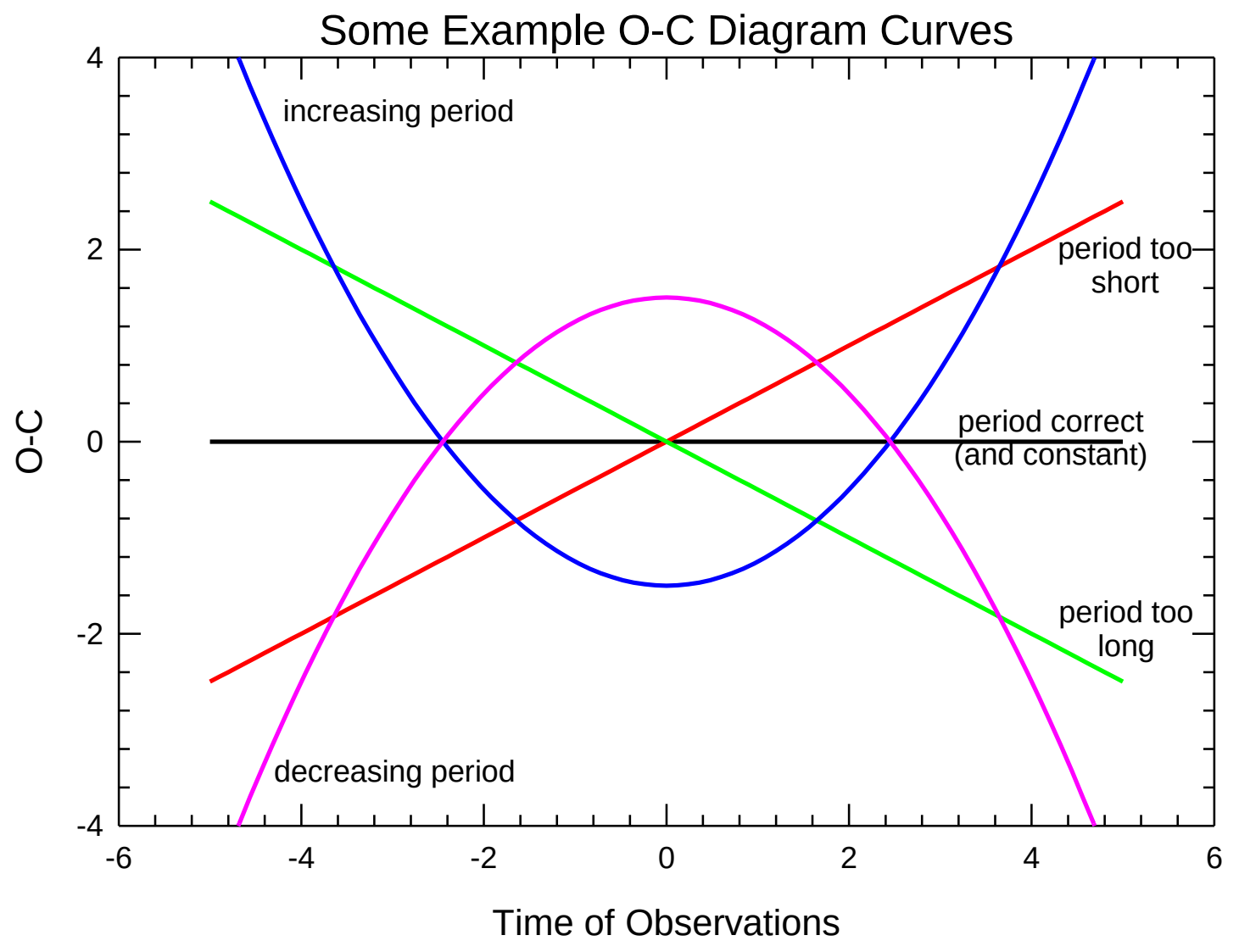

Figure 8 – An illustration of the various shapes that an O-C curve can have, and what basic information the shape can tell us about the period behavior of the variable star.

Cepheid period changes can also give information on the evolutionary state of the Cepheid, as shown in Turner et al. (2006). Specifically, the rate of period change can tell us which crossing of the instability strip the Cepheid is making (see Fig. 9), which is indeed valuable information to have. When looking at Fig. 3, one can see that as Cepheids cross the instability strip they undergo relatively large changes in average temperature as they pass from one edge of the strip to the other. With such temperature changes, one would expect accompanying luminosity changes. As this is not the case, we know that the average radius of a Cepheid also changes while it crosses the instability strip, to compensate for the temperature changes and result in a similar luminosity. When evolving towards the cool edge of the instability strip, the radius of a Cepheid increases, and when evolving towards the hot edge, the radius decreases. The effect of this evolution on pulsation period can be calculated. The basis of this calculation is the well-known period-density relationship for radially pulsating stars, such as Cepheids (Turner et al. 2006):

$$P\rho^{1/2} = \frac{PM^{1/2}}{[(4/3)\pi]^{1/2}R^{3/2}} = Q$$

where $P$ is the pulsation period, $\rho$ is the density, $M$ is the stellar mass, $R$ is the stellar radius and $Q$ is the pulsation constant. The small period dependence $Q \propto P^{1/8}$ (Turner et al. 2006) is taken in to account. Next, the density can be substituted with $\rho = MR^{-3}$ and radius can be substituted using the standard stellar luminosity equation: $L \propto R^2T^4$ where $L$ is the stellar luminosity and $T$ is stellar temperature. It is thus determined that the period should increase as the radius increases (evolving towards cooler temperatures) and the period should decrease as the radius decreases (towards hotter temperatures). The rate of period change, $\Delta P$, can then be predicted through an equation of the form:

$$\frac{\Delta P}{P} = \frac{5}{8}\frac{\Delta L}{L} - \frac{5}{2}\frac{\Delta T}{T}$$

(the specific coefficients used here result from also including a mass-period relation $M \propto P^{0.4}$: Turner, private communication). Using this equation, in combination with stellar evolutionary tracks, the rates of period change for different crossings of the instability strip, as shown in Fig. 9, were calculated. For a thorough discussion of the evolutionary implications of Cepheid period changes, see Turner et al. (2006) and references therein. It is also very important to mention that, following Turner, the calculated period change rates assume that no mass loss is taking place and have also been calculated for the fourth and fifth instability strip crossings. However, as illustrated in Fig. 3 for a 7 $M_{\odot}$ star, modern evolutionary tracks do not display strip crossings beyond the third. Recent studies (Neilson et al. 2012b, 2012c) indicate that Cepheids theorized to be in the fourth or fifth instability strip crossing could in fact be in the second or fourth crossing (respectively), but undergoing enhanced mass loss that would result in larger period change rates. Therefore, period-change studies of Cepheids impact not only our understanding of the targets, but also of stellar evolution and the important topic of mass loss. If confirmed, enhanced rates of mass loss could explain the well-known Cepheid mass discrepancy, where masses calculated from pulsation properties are commonly (and significantly) smaller than those calculated from evolutionary codes (first discussed by Fricke et al. 1972; see Neilson et al. 2011 for a modern discussion). Recent efforts aimed at better understanding (and possibly resolving) this discrepancy concentrate on Cepheids that are members of binary systems, where mass determinations can be made free of either pulsation or evolutionary considerations (e.g. Evans et al. 2008, Pietrzyński et al. 2010).

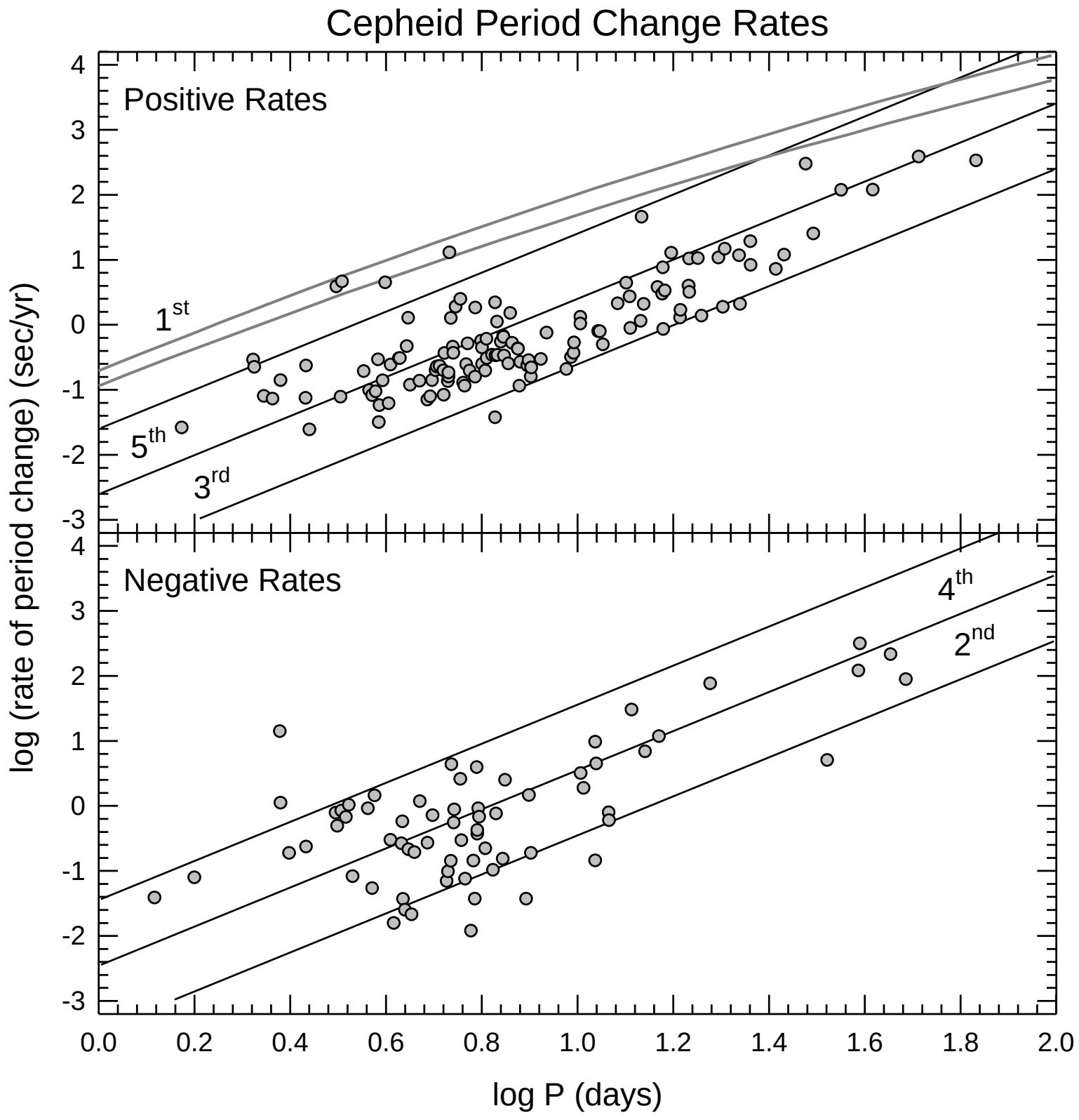

Figure 9 – Galactic Cepheids with known rates of period change, and the possible crossings of the Instability Strip indicated by such rates, as discussed in the text.

In addition to period changes in Cepheids, there are also the very rare instances of amplitude changes. In fact, at the time of this publication, the only two Cepheids for which amplitude changes have been published are Polaris and V473 Lyr. Both Cepheids are seen as atypical, however, so trying to characterize their behaviors as applicable to a broad range of Cepheids would be difficult, at best. Polaris is a low amplitude Cepheid, whose true mode(s) of pulsation is still debated (recent results in favor of a fundamental mode are given by Turner et al. 2013). V473 Lyr has the shortest pulsation period of any known galactic Cepheid (~1.49-days), and also has the only *characterized* amplitude cycle (Stothers 2009). Until recently, Polaris had only been observed as undergoing an amplitude decrease, but newer studies have confirmed that Polaris' amplitude is now increasing again (Engle et al. 2004, Engle & Guinan 2012, Spreckley & Stevens 2008, Bruntt et al. 2008). However, given the length of time already covered by Polaris' amplitude changes to date, it appears that covering a full amplitude cycle (if a repeatable cycle exists) will unfortunately take decades of further data.

The closest possible analog for this behavior, from which a theoretical explanation might be extrapolated, would be the much better (yet still not fully) understood Blazhko effect in RR Lyr stars. RR Lyr stars also occupy the classical instability strip of the H-R diagram, but below and to the left (less luminous, but hotter) than the Cepheids (see. Fig. 1). Many RR Lyr stars are known to undergo variations in their pulsation periods and amplitudes. This behavior is termed the Blazhko effect, as it was first observed by Sergey Blazhko in the RR Lyr variable RW Dra (Blazhko 1907). The mechanism responsible for the Blazhko effect is still debated, but three possible sources have risen to prominence: 1) the resonance model, where resonating pulsation modes within the same star are responsible for the effect; 2) the magnetic model, where the geometric structure or alignment of the stellar magnetic field can augment the pulsations, and; 3) a model where turbulent convection in the ionization zones within variable stars can periodically (or even randomly) vary in strength, and this variation in convective strength would then affect the pulsation amplitude. The third mechanism could also be the work of a stellar magnetic field (Stothers 2006). None of the mechanisms adequately account for every observed aspect of the Blazhko effect. Thus, more than a century after its discovery, the exact cause(s) of the effect is (are) still unknown.

### 1.8 Super-Photospheric Studies of Cepheids

The term "super-photospheric" is used in this paper to refer to the study of plasmas with temperatures well above that of the stellar photosphere. In terms of the Cepheids (and most other cool stars, for that matter), it specifically references activity originating from excited plasmas in the stellar atmospheres. The nearest example of super-photospheric activity in a cool star is, of course, the Sun. The solar atmosphere is broadly comprised of (in order of increasing distance from the photosphere): the chromosphere and transition region, where plasmas are magnetically heated to temperatures on the order of $10^4$ – $10^5$ K, and then the corona, where temperatures extend into the million Kelvin (MK) range. Plasmas in the chromosphere and transition region are the dominant source of emission lines in the UV portion of the solar spectrum, where MK plasmas in the corona are the dominant source of X-ray emission. As such, both regions of the spectrum provide valuable diagnostics into their respective places of origin. Some valuable diagnostic emission lines in the UV region and their approximate formation temperatures (Haisch & Linsky 1976, Doschek et al. 1978 and Doyle et al. 1997) are: the N V $\lambda$1240 doublet (~$2 \times 10^5$ K), C II $\lambda$1335 (~$4 – 5 \times 10^4$ K), O I $\lambda$1358 (~$1 – 2 \times 10^4$ K), the Si IV $\lambda$1400 doublet (~$6 – 8 \times 10^4$ K) and the C IV $\lambda$1550 doublet (~$1 \times 10^5$ K). These emission lines represent a very wide range of temperatures and can offer a detailed look into a stellar atmosphere. In the Sun (and, again, most other cool stars), magnetic activity is the dominant heating mechanism of the outer atmospheres. However, other mechanisms also contribute to the overall heating, such as energetic particles and shocks released from microflare and nanoflare events near the solar photosphere (which are also a result of the magnetic field). The magnetic fields of the Sun and other cool stars are believed to be generated by a solar/stellar magnetic dynamo. The current dynamo model for the Sun is the *shear-interface model* (the $\alpha – \omega$ model), where convective motion and differential rotation combine to twist, break and reconnect the once orderly solar magnetic field (caused simply by the large amount of conductive materials in motion throughout the solar interior). However, in certain cases (such as stars lacking strong enough differential rotation, but possessing robust inward-outward motions) a *turbulent dynamo* (the $\alpha^2$ model) can also produce similar magnetic fields and activity. For more information about these concepts see Noyes et al. (1984), Canfield (2003), Chabrier & Küker (2006) and Parker (2009). Note that this is simply a small selection of papers, as *numerous* references exist for the dynamo mechanism in cool stars. For variable stars undergoing robust radial pulsations, such as Cepheids, one must also consider the effects of such pulsations, not only on the stellar interiors and the possible dynamo mechanisms at work, but also on the possibility of the pulsations generating shocks which could propagate through the stellar atmospheres and contribute to the heating of plasmas.

The history of Cepheid super-photospheric studies is rather sparse (with most studies being ~30 years old or older) and, as a result, can sometimes go overlooked. The earliest of such studies appears to be that of Kraft (1957), who observed and analyzed the Ca II *H* (3968.5 Å) and *K* (3933.7 Å) emission lines in a large number (20+) of brighter Cepheids. These emission lines originate in plasmas with temperatures in the $8 - 15 \times 10^3$ K range; similar to temperatures found in the lower chromospheres of solar-type stars. Kraft noted that the Ca II *HK* emissions peaked in Cepheids around $\phi \approx 0.8 - 1.0$ (just after the Cepheid has begun to expand from minimum radius). Due to the star's expansion, a shock is expected to pass through the Cepheid photosphere at this phase, which is sometimes referred to as the "piston phase" of Cepheids. From this, Kraft concluded that "the transitory development of Ca II *H&K* emission in Classical Cepheids is associated with the appearance of hot material low in the atmosphere. These hot gases are invariably linked with the onset of a new impulse." Kraft also noted that the Ca II emissions were stronger in Cepheids with longer pulsation periods when compared to those with shorter periods. This important study laid the groundwork for Cepheid atmospheric studies, but the Ca II lines can only probe relatively cool atmospheric plasmas. More than two decades would pass before an investigation into the higher temperature plasmas of Cepheids was conducted.

Schmidt & Parsons (1982, 1984a,b) made the most thorough and revealing study of Cepheid atmospheres using spectra from the International Ultraviolet Explorer (IUE) satellite. The wavelength range of IUE (~1200 – 3200 Å) covers a number of important emission lines with temperatures of $10^4 - 10^5$ K, equivalent to those found in solar-type chromospheres and transition regions. In accord with the results of Kraft (1957), Schmidt & Parsons found these emissions to vary systematically with the pulsation phase, peaking shortly before maximum light (which is also shortly after minimum radius). Again, this is the phase where the stellar photosphere begins to expand. Lines normally associated with hotter, transition region temperatures were also found, but were not as strong as chromospheric emissions and were more easily contaminated by (what appeared to be, as discussed later) the Cepheid photospheric continuum in all but the longest exposed (and ideally phase-space located) spectra. Still, the detections of such lines served as evidence that Cepheid atmospheres could be more complex (and possibly hotter), and that perhaps a more modern, and higher resolution, instrument could reveal such lines in a more concrete way.

A decade later, Böhm-Vitense & Love (1994) studied IUE archival spectra of the 35-day Cepheid *ℓ* Car with the purpose of trying to distinguish between line emissions from a solar-like magnetically heated chromosphere and transition region vs. those from an outward-moving shock. Böhm-Vitense and Love measured several emission lines present in the IUE spectra finding that, as with the Schmidt and Parsons studies, the emission line fluxes were variable in phase with the Cepheid pulsations. The line emissions also peaked at or near the phase of minimum radius, as found in other studies. Böhm-Vitense and Love concluded that an outward-moving shock must be at least partially responsible for the UV line fluxes and variability observed, because of the tight phasing of enhanced emissions just before maximum light, along with the excitation of the highly ionized C IV doublet prior to the excitation of the lower ionization C II. This is expected from a shock, as the higher temperatures required for C IV emission would be rapidly achieved. Then, after shock passage, the plasmas would cool down, allowing C II emissions to be observed. Unfortunately, the IUE observations did not cover the 0.8 – 0.9 phase-space, which prevented a detailed measure of the phase of maximum C IV emission, but it was still observed to peak earlier in phase than C II.

Motivated by these results, an analysis of FUSE (~920 – 1190 Å) spectra was carried out, and new observations were successfully proposed for. Sadly, although multiple FUSE observations of additional Cepheids were approved, just one observation (for β Dor) was carried out before the fatal failure of the mission's guidance system. The "FUSE Cepheid database" therefore includes only Polaris and β Dor (and only β Dor has multiple observations, but of somewhat poor phase coverage). However, both the C III 977Å and O VI 1032/1038Å line emissions were not only found in the spectra of both Cepheids, but were found to increase in strength at the piston phase of β Dor ($\phi \approx 0.8$, Engle et al. 2009). This result is in agreement with the phase of peak N V 1240 Å emission (N V forming at comparable, albeit slightly cooler, temperatures to O VI), found from our HST/COS observations (discussed later). The C III emission and variability was somewhat expected, given that it forms at similar temperatures (~ $5 – 10 \times 10^4$ K) to emission lines studied in the archival IUE spectra. However, the discovery of the O VI doublet was "a pleasant surprise," given that it forms in plasmas with temperatures of ~ $3 – 4 \times 10^5$ K (Redfield et al. 2002). The O VI lines were, at that time, the hottest plasma emissions observed from a Cepheid. This study raised the questions: just how hot is a Cepheid's outer atmosphere, and how is it being heated?

Given the phase-timing of the enhanced emissions, again the most plausible explanation is the formation of a shock that excites the atmospheric plasmas surrounding the photosphere. However, a pulsation-driven $\alpha^2$ equivalent dynamo mechanism (Chabrier & Küker 2006) where the inward-outward convective motions are enhanced, or even replaced, by pulsational motions is also a viable and interesting alternative.

UV line emissions from $10^6$ K (MK) plasmas are rare, typically weak, and only appear in the most active of stars (e.g. see Linsky et al. 1995 for a discussion of the coronal *Fe XXI* $\lambda$1354 Å emission line for Capella). Thus, to detect and study Cepheids at these high temperatures, X-ray observations are needed. The only pointed X-ray observations of Cepheids, prior to those carried out as part of this study, were those of the *Einstein Observatory* (*High Energy Astrophysical Observatory 2 – HEAO-2*) and the *Röntgen Satellite* (*ROSAT*). *Einstein* observations were approved for δ Cep, β Dor and Polaris, although no successful pointings on Polaris were ever achieved. Fig. 10 shows the Einstein Imaging Proportional Counter (IPC) images (taken in 1980/81) for the δ Cep and β Dor fields, with the Cepheids' approximate locations encircled. As the figures show, the Einstein satellite was unable to detect either Cepheid. However, the Einstein IPC did not have the sensitivity of later missions, and the exposures of δ Cep and β Dor are a mere 3563 and 838 seconds, respectively. Given the short exposures and low sensitivity, it was quite possible that both Cepheids could in fact be weak, soft X-ray sources and yet still go undetected in the observations. The online Einstein catalog quotes upper limits on the log $L_X$ values of δ Cep and β Dor as 29.1 and 29.4, respectively. However, X-ray activity in Cepheids was considered possible in papers such as Bejgman & Stepanov (1981) and Sasselov & Lester (1994). Their suggested mechanism was that shocks forming within the atmospheres of Cepheids and RR Lyrae stars could produce sufficient temperatures to excite and ionize heavy elements present in Cepheid atmospheres, producing soft X-ray activity. Another possible cause for X-ray activity in Cepheids is also linked to their pulsations. As a Cepheid pulsates, the elements within the atmosphere will undergo turbulent mixing and substantial motions. This could generate a "faux dynamo" (as previously mentioned) which would then generate soft X-ray activity. The "faux" quality of the dynamo is that the dynamo is not generated through convective transport, but rather through "pulsational transport." Despite such studies and theories, the failure of early efforts to detect X-rays with pointed observations reinforced the idea that Cepheids are not significant X-ray sources. Even with *log* $L_X \approx 29$ erg/s, the problem with detecting X-rays from Cepheids is that these stars (except for Polaris at ~133 pc – van Leeuwen 2007) are far away (d > 250 pc). Thus, they are expected to have relatively weak X-ray fluxes ($f_X < 10^{-14}$ ergs/s/cm$^{2)}$. Though Polaris was in fact detected on the 3σ level in a ROSAT High Resolution Imager (HRI) archival image (several years after the observation was carried out), a definitive detection of X-rays from Polaris and other "nearby" Cepheids had to wait for the arrival of powerful X-ray observatories like the X-ray Multi-Mirror Mission (XMM-Newton) and the Chandra X-ray Observatory, as discussed later.

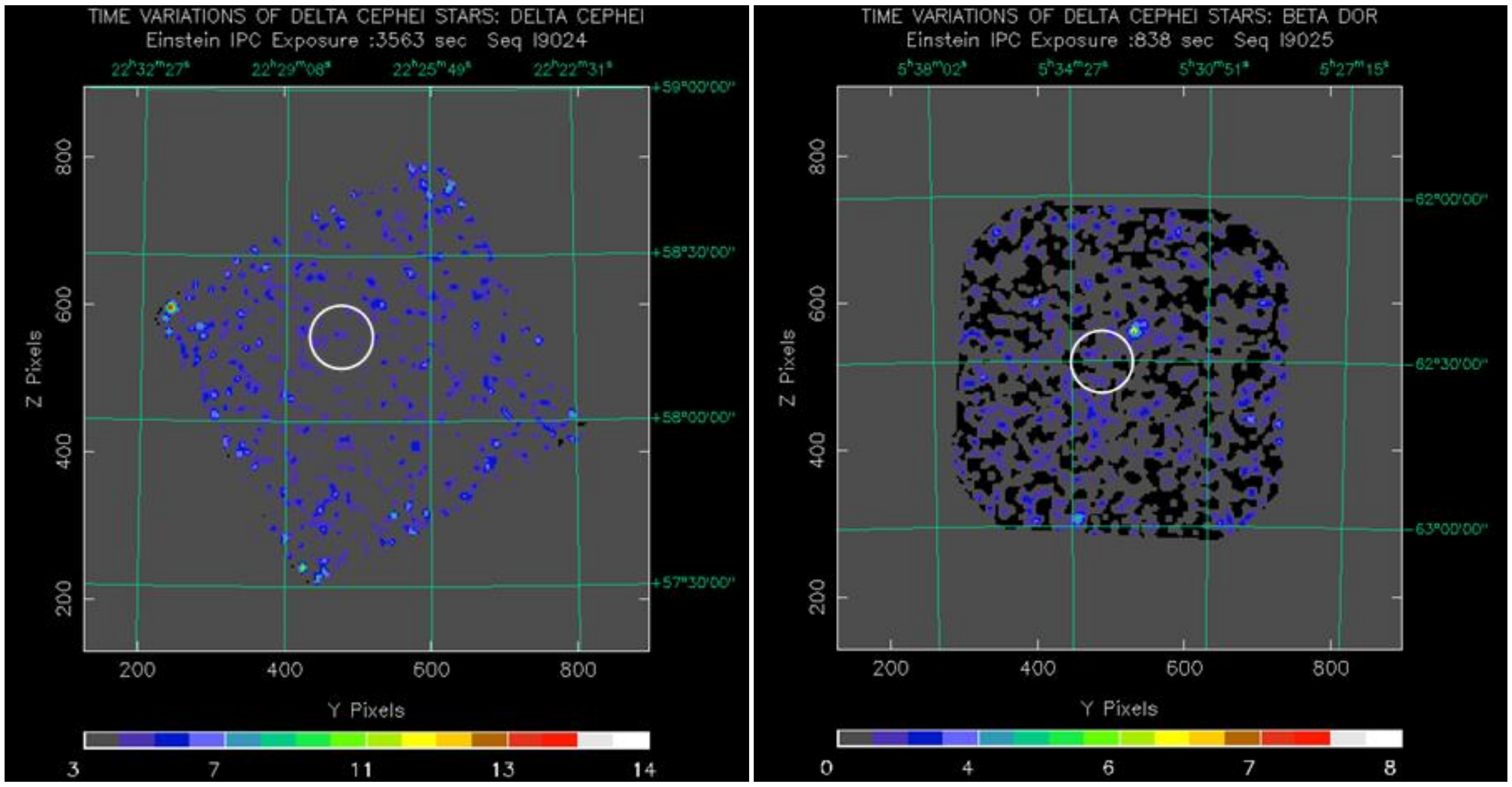

Figure 10 – Einstein IPC observations of δ Cep (left) and β Dor (right) are shown. The white circle in each image marks the location of the respective Cepheids. As can be seen, there are no detections of X-rays from either Cepheid above the noise level of the observations. This is not surprising, however, given the Cepheid distances and the short duration of the exposures.

## 1.9 Thesis: The *Secret Lives of Cepheids* (SLiC) program

The overall aim of this thesis is to expand our understanding of Cepheid variability. We look to do this by documenting previously unconfirmed or unknown behaviors in one of Astronomy's oldest and most important classes of variable stars. The thesis can be divided into two main studies: the optical study and the high-energy study.

For the optical study, ground-based photometry has been carried out for a selection of 10 bright Cepheids (δ Cep, η Aql, EU Tau, Polaris, SU Cas, SV Vul, SZ Cas, SZ Tau, VY Cyg and ζ Gem), selected primarily to represent a range of pulsation periods. However, target selection could also be influenced by: a long timeline of observations, which would aid in the study, or perhaps pre-existing evidence for the period/amplitude variability we are looking for. We combine our own photometry with that found in the literature for each Cepheid:

- Providing well-covered, modern, multi-band (either Johnson/Cousins *UBVRI* or Strömgren *uvby*) lightcurves for each Cepheid, from which new ephemerides have been calculated [note: *BV* photometry only has been carried out for Polaris, as discussed later]
- Extending O-C diagrams for each Cepheid and calculating a new rate of period change, if one is found,
- Studying the V-band (or equivalent) amplitudes of each Cepheid over time, to search for possible amplitude changes

In the high-energy (X-ray–UV) study new, high-quality UV spectroscopy has been carried out for Polaris, δ Cep and β Dor with the Cosmic Origins Spectrograph (COS) onboard the Hubble Space Telescope (HST), along with pointed X-ray observations from either the Chandra or XMM-Newton satellites. For this aspect of the study, the stars chosen represent some of the nearest Cepheids, important for efficiently achieving quality data from satellites where observing time is rather precious. In this study:

- The UV data will be analyzed to improve upon the studies of Schmidt and Parsons and Böhm-Vitense and Love and give further insights into the mechanism(s) responsible for the super-photospheric activity
- The X-ray data will be analyzed to confirm X-ray activity in Cepheids other than Polaris, to establish the temperature of the X-ray emitting plasmas and also to search for possible X-ray variability in addition to that of the UV emission lines

To conclude, we will summarize the work done and the results obtained. We will discuss the implications of the work, and possible explanations for the results. We will also outline what future work we have planned, and would like to undertake, in order to achieve the fullest understanding possible of the Secret Lives of Cepheids.

# CHAPTER 2 – THE OPTICAL STUDY

## 2.1 Instrument and Observing Method

All of the Cepheids studied here are bright enough to be well-suited for photoelectric photometry. Many would quickly saturate CCD detectors, making the selection and accurate measure of nearby comparison stars very difficult. Also, every target has been the subject of multiple other photoelectric studies in recent decades. Choosing to continue with photoelectric photometry for this program allows the issue of target brightness to be easily handled, and also allows a more direct comparison to previous studies.

The overwhelming majority of the Cepheid photometry obtained in this program comes from the *Four College Automatic Photoelectric Telescope* (FCAPT) housed at *Fairborn Observatory* (Observatory Director – Lou Boyd). The observatory is located roughly 15-km SSE of Patagonia, AZ (observatory coordinates: 31º23’11.8” N 110º41’40.5” W), and at an elevation of ~1723-m. The FCAPT is a classical Cassegrain design with a 0.75-m (30-inch) f/2 primary mirror and a 0.2-m (8-inch) f/8 secondary. The FCAPT mount is a custom-made horseshoe symmetrical design. The telescope is controlled by a 2.8 GHz Pentium 4 computer running Red Hat Linux 5.2, which uses ATIS software for telescope operation. The photometer head consists of (in order of light path): fused silica entrance window; diaphragm wheel; ND filter wheel; flip mirror; fused silica fabry lens; two 10 position bandpass filter wheels; Hamamatsu R943-02 photomultiplier operating at 1750 volts with a divide by 4 prescaler and a 16-bit counter read every 0.1-seconds. "Geneva" statistics (Hayes et al. 1988) are provided for each integration to help detect APT problems. Total integration time is selected in ATIS. The flip mirror turns 180 degrees to select a 25-mm f/0.95 relay lens and a Panasonic GP-MF602 integrating video camera for acquisition and centering, using an Imagenation CX-100 frame grabber. Integrating time and filter are calculated from *V* and *B–V* data for each target.

Table 1 gives the available filters and selectable settings for each of the five wheels. One position from each of the five motors (wheels) must be selected for every observation. These settings are determined by the operating system from information in the ATIS input file. There is a sixth stepper in the photometer head which focuses the CCD camera. The entire interior of the photometer head including the filters, PMT, and CCD camera is cooled using a shared 4º C recirculating water chiller. The photometer head is also flooded with -30º C dew point air from a site air drier to prevent condensation.

The FCAPT requires no on-site observer, making it an excellent facility for the long-term monitoring of variable stars. It automatically observes targets from a list maintained and prioritized by astronomers at institutions with dedicated access to the telescope. Observations were carried out in a very usual fashion for photoelectric set-ups, with variable star measures being bracketed by comparison, check and sky measures (all 10-second integrations): *sky – comp – check – var – comp – sky – var – comp – sky – var – check – comp – sky* being the general order when a night included 3 measures of the variable star. Between 3 and 5 measures of the variable star would be carried out in a single run, depending on the available telescope time in a given night. Also, anywhere from 1 to 3 separate observing sequences would be carried out per Cepheid, per night, again depending on available telescope time, but also depending on the length of the Cepheid's period. More than one run per night would often be carried out on Cepheids of shorter periods, to more efficiently fill in the lightcurves. The photometric filters used in this program were designed to match, as closely as possible, the Johnson/Cousins *UBVRI* and Strömgren *uvby* systems. Cepheids were observed in either system based on brightness and/or photometric history (i.e. which system featured more prominently in the literature for each Cepheid). In the case of brighter Cepheids, either the Strömgren system was used or, if the Johnson/Cousins system was preferred, attenuating (neutral density) filters were also employed to mitigate saturation effects.

**Table 1 – Wheel Settings for the 0.8-m *Four College Automatic Photoelectric Telescope* (FCAPT)**

| **Wheel #1 – Diaphragm** | | **Wheel #4 – Filter Wheel 1a** | |
|---|---|---|---|
| Position 1 | 15 arcmin opening | Position 0 | Clear |
| Position 2 | 90 arcsec opening | Position 1 | *U* |
| Position 3 | 60 arcsec opening | Position 2 | *B* |
| Position 4 | 45 arcsec opening | Position 3 | *V* |
| **Wheel #2 – Flip Mirror** | | Position 4 | *R* |
| Position 0 | CCD | Position 5 | *I* |
| Position 1 | PMT | Position 6 | *u* |
| **Wheel #3 Neutral Filter Wheel** | | Position 7 | *v* |
| Position 1 | Clear | Position 8 | *b* |
| Position 2 | 1.25-mag | Position 9 | *y* |
| Position 3 | 2.5-mag | **Wheel #4 – Filter Wheel 1b** | |
| Position 4 | 3.75-mag | Position 1 | Clear |
| Position 5 | 5.0-mag | Position 2 | *Hβ-wide* |
| | | Position 3 | *Hβ-narrow* |
| | | Position 4 | *Hα-wide* |
| | | Position 5 | *Hα-narrow* |
| | | Position 6 | Clear |
| | | Position 7 | Opaque |
| | | Position 8 | *Wing1 A* |
| | | Position 9 | *Wing2 B* |
| | | Position 10 | Clear |

A photometric reduction program written by George McCook (Villanova U.) has been used on FCAPT data since the mid-1990s. The program automatically converts from local time to Heliocentric Julian Date, and also includes atmospheric extinction corrections (coefficients given in Table 2). *UBVRI* and *uvby* standard stars were also observed with the FCAPT, covering a wide range of colors (spectral types). Using the standard star data, instrumental magnitudes were then transformed into standard magnitudes for each Cepheid. A final transformation was then applied to convert Stromgren *y*-mag into Johnson *V*-mag. The transformation equations used in this study are given in Table 3.

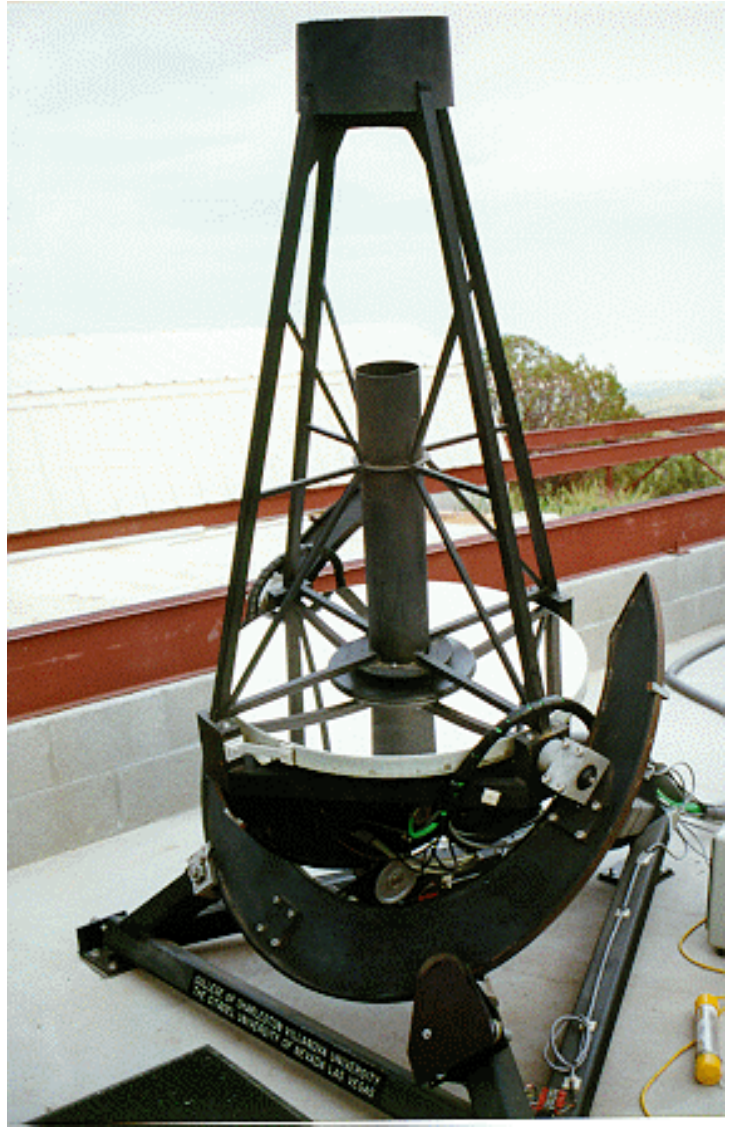

Figure 11 – The FCAPT installed at the Fairborn Observatory.

**Table 2 – Average Extinction Coefficients for Filters Used in This Study**

| | *U* | *B* | *V* | *R* | *I* |
|---|---|---|---|---|---|
| Johnson/Cousins | 0.659 | 0.385 | 0.237 | 0.19 | 0.143 |
| | *u* | *v* | *b* | *y* | |
| Strömgren | 0.503 | 0.286 | 0.176 | 0.129 | |

**Table 3 – Photometric Transformation Equations for the FCAPT**

| ***UBVRI*** | ***uvby*** |
|---|---|
| | |
| $U = u + 0.0015 + 0.0491(U–B)$ | $u = u' – 0.1538 + 0.1192(u–v)$ |
| $B = b + 0.0125 – 0.0179(B–V)$ | $v = v' + 0.0091 + 0.0008(v–b)$ |
| $V = v + 0.0106 – 0.0262(B–V)$ | $b = b' – 0.0947 + 0.2148(b–y)$ |
| $R = r + 0.1382 – 0.4611(V–R)$ | $y = y' – 0.0233 + 0.0530(b–y)$ |
| $I = i + 0.1537 – 0.5444(R–I)$ | |
| | |
| $(U–B) = -0.0111 + 1.1402(u–b)$ | $(u–v) = -0.02632 + 1.004(u'–v')$ |
| $(B–V) = 0.0115 + 0.9875(b–v)$ | $(v–b) = 0.1131 + 0.8448(v'–b')$ |
| $(V–R) = -0.0473 + 1.3340(v–r)$ | $(b–y) = -0.07307 + 1.194(b'–y')$ |
| $(R–I) = -0.0037 + 1.0495(r–i)$ | |
| | |
| **Strömgren *y* to Johnson *V*** | |
| | |
| $V = y – 0.12[(b–y) – 0.55]^2$ (Budding & Demircan 2007, p. 94) | |

The FCAPT has a limitation, however, in that the telescope mount does not provide sufficient space for the instrument cluster to allow the telescope to slew above a declination of approximately 76º. Accordingly, Polaris could not be observed with the FCAPT. Originally, Polaris was observed with a photoelectric telescope on the campus of Villanova University, through a filter designed to match the Strömgren *y*-band. Data output from this telescope was also designed so that it could be run through the reduction program of the FCAPT. However, this telescope suffered a hardware crash in 2006, and was retired. In the second half of 2009, AAVSO observer David Williams carried out photoelectric V-band photometry. Finally, collaborator Richard Wasatonic (Villanova U.) kindly agreed to carry out photoelectric *BV* photometry of Polaris beginning in late 2010, and continues to do so.

## 2.2 Data Analysis for the Optical Study

Each Cepheid in the Optical Study was observed with the goal of obtaining a fully-covered (i.e. to be considered for this study, the phases of minimum and maximum light must be well-detailed) light curve in each observing season. Therefore, high-priority was given to a Cepheid during phases near maximum and minimum light. The minimum criteria were set to allow both products of the light curve to be realized: the time of maximum light and the V-band light amplitude (to facilitate comparison with the oldest, visual observations). Both products were determined by running a Fourier series fit to the standardized, phase-folded photometry. The Fourier series used was of the form:

$$m(\varphi) = A_0 + \sum_{n=1}^{N} A_n \cos[2\pi n(\varphi) + B_n]$$

where $m(\varphi)$ is the calculated magnitude at a given phase, $\varphi$. $A_0$ is the mean magnitude of the light curve, and *N* is the final order of the fit. Finally, $A_n$ and $B_n$ are the amplitude and phase of the $n^{\text{th}}$ order, respectively. For each observed light curve, a synthetic light curve was built from the fitted Fourier coefficients. The synthetic maximum and minimum magnitudes were used to calculate the overall light curve amplitude, and the offset of the nearest observation to the synthetic phase of maximum light was used to calculate the observed time (HJD) of maximum light. Due to the various observational hindrances, primarily stretches of inclement weather or mechanical difficulties during the prime observing season of specific Cepheids, a fully-covered light curve could not be obtained for every target, every year. For almost all Cepheids, multiple fully-covered light curves we obtained.

Literature searches were conducted for all observed Cepheids to obtain these products from archival, fully-covered light curves. Depending on the source quality (e.g. phase-coverage of the light curve, level of details given in the study [standardizations, observations techniques], etc.), amplitudes and times of maxima were either quoted directly from the source study, or re-determined by Fourier series fit. For each target's O-C diagram, archival data was obtained from the reference given in the stellar properties table (after "Ephemeris for O-C diagram"). For consistency with each source of O-C data, the diagrams presented here either calculate individual errors for the data points, or use a weighting scheme. The plots with error bars indicate studies where individual errors were calculated. For the plots without error bars, a weighting scheme was employed in the following fashion: early visual or photographic measures were weighted 0.5 – 1, and modern photoelectric measures were weighted 2 – 3, with the specific weight depending primarily on the phase-coverage at and around maximum light. For amplitude vs. time plots, errors were determined for recent measures in standard photometric systems (i.e. after the introduction and wide adoption of the Johnson photometric system) and for earlier data sets where individual observations were made accessible. This study aimed to employ a more objective method of error estimation that would account for both the observational scatter and the phase coverage near times of maximum and minimum light. To do this, an appropriate Fourier series order was first determined for each target. Then, Fourier series fits of neighboring orders were run. The standard deviation of these fits was calculated, and assigned as the error for each amplitude value.

**2.3 δ Cep**

The prototype of all Cepheids, δ Cep is also the 14$^{th}$ variable star ever discovered (http://spider.seds.org/spider/Vars/vars.html), and the 2$^{nd}$ Cepheid for which light variations were observed. It is also currently the 2$^{nd}$ nearest Cepheid (only Polaris is nearer), and the member of a wide binary system with HD 213307 at a projected distance of 40" away from the Cepheid. The companion is an A0-type star ($T_{eff} \approx$ 10,048 K) according to Prugniel et al. (2007), or a B7 – B8 III – V star with its own F0-type companion, according to the HST parallax study of Benedict et al. (2002). Further, δ Cep is a member of the Cep OB6 star cluster (de Zeeuw et al. 1999), along with the cooler and more luminous K1.5Ib star ζ Cep. Majaess et al. (2012) found the cluster-derived distance of 277 ± 15-pc to agree well with distances derived from other methods, including that from the HST parallax given in Table 4, along with selected other relevant properties of δ Cep (Table adapted from Matthews et al. (2012) and references therein). δ Cep displays the typical "saw tooth" lightcurve, with a quicker rise to maximum brightness, and a much slower fall to minimum brightness, as shown in Fig. 12. The overall form of this variability is mirrored by the color indices.

**Table 4 – Relevant Stellar Properties of δ Cep**

| | |
|---|---|
| Spectral Type | F5Ib – G1Ib[1] |
| $T_{eff}$ (K) | 5500 – 6600[1] |
| Mass (pulsational) ($M_{\odot}$) | 4.5 ± 0.3[2] |
| Mass (evolutionary) ($M_{\odot}$) | 5.7 ± 0.5[2] |
| Mean Luminosity ($L_{\odot}$) | ~2000[3] |
| Mean Radius ($R_{\odot}$) | 44.5[3] |
| Distance (pc) | 273 ± 11[4] |
| Ephemeris (this study) 2455479.905 + 5.366208(14) × E | |
| Ephemeris for O-C diagram (Berdnikov et al. 2000) 2412028.956 + 5.3663671 × E | |

[1]Andrievsky et al. (2005); [2]Caputo et al. (2005); [3]Matthews et al. (2012); [4]Benedict et al. (2007)

Photometry gathered as part of this program started in December, 2007, and provided well-covered light curves at two separate epochs. Fourier analysis of the resulting light curves yielded two times of maximum light, which were added to those found in the literature (Berdnikov et al. 2000), and are presented as the red data points in Fig. 13. As can be seen from the "concave down" parabolic shape of the O-C curve, the period of δ Cep is decreasing over time. From a quadratic fit to the O-C data, the rate of period change is calculated to be –0.100565 ± 0.000172 sec/yr. According to Turner et al. (2007), this rate would mean that δ Cep is currently making its second crossing of the instability strip. The parameters of the quadratic fit to the O-C data are shown in Fig. 13.

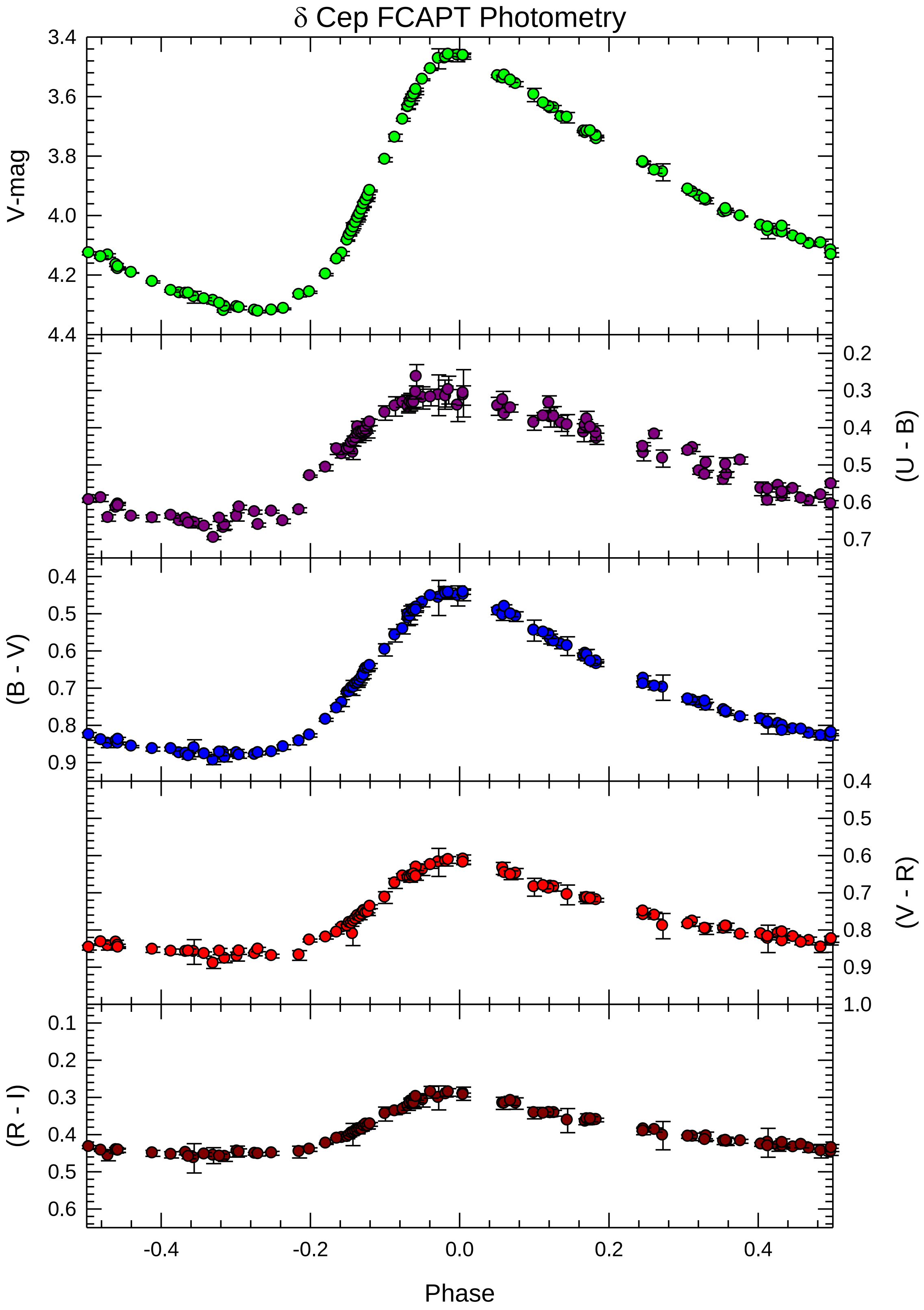

Figure 12 – The *UBVRI* data obtained for δ Cep, phased to the new ephemeris determined in this study (given in Table 4).

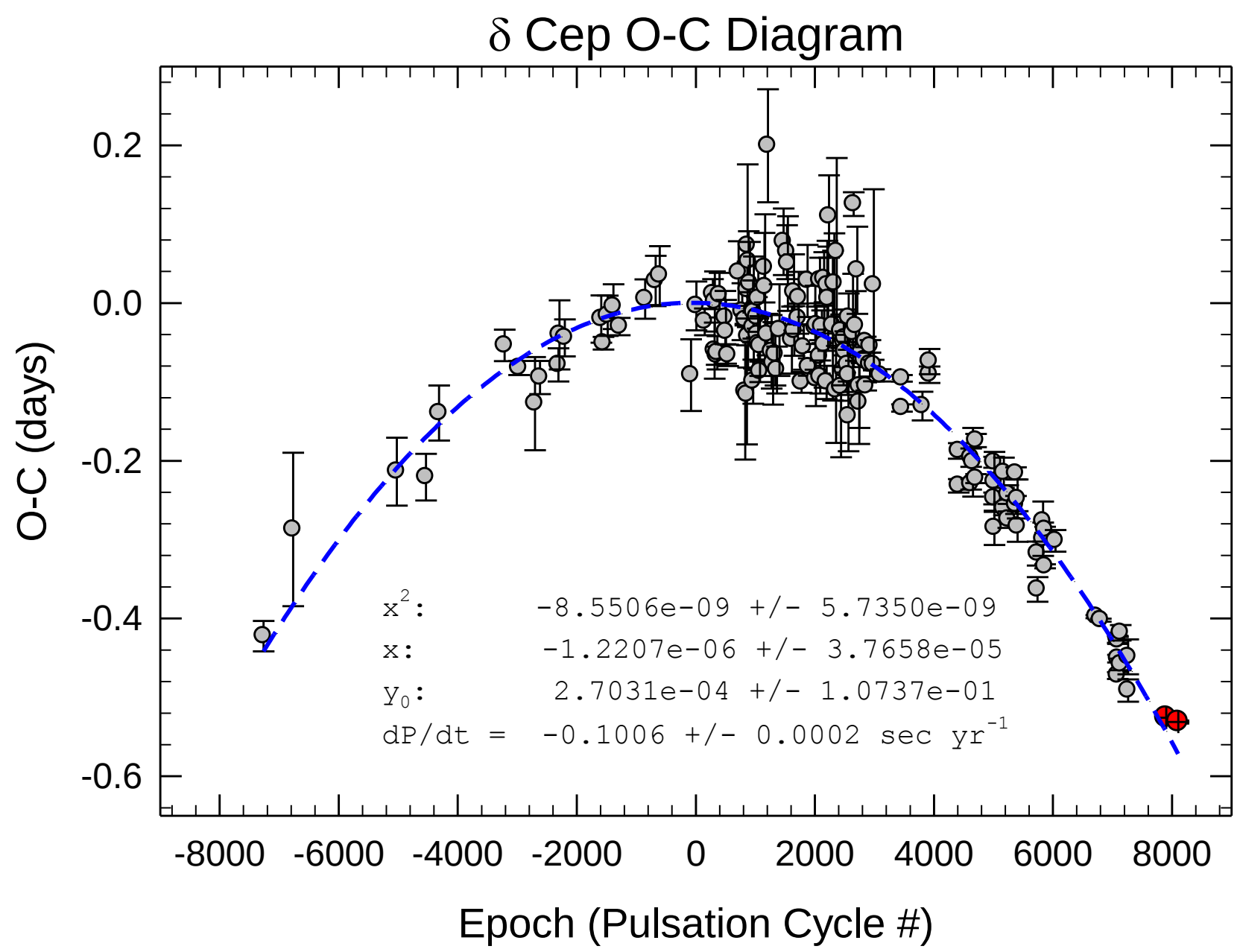

Figure 13 – The O-C diagram for δ Cep, showing the decreasing trend in its pulsation period. Coefficients of the quadratic fit are given in the plot, along with the rate of period change (dP/dt = -0.1006 sec/yr). The O-C points determined from light curves obtained as a part of this program are plotted as the red filled and crossed circles.

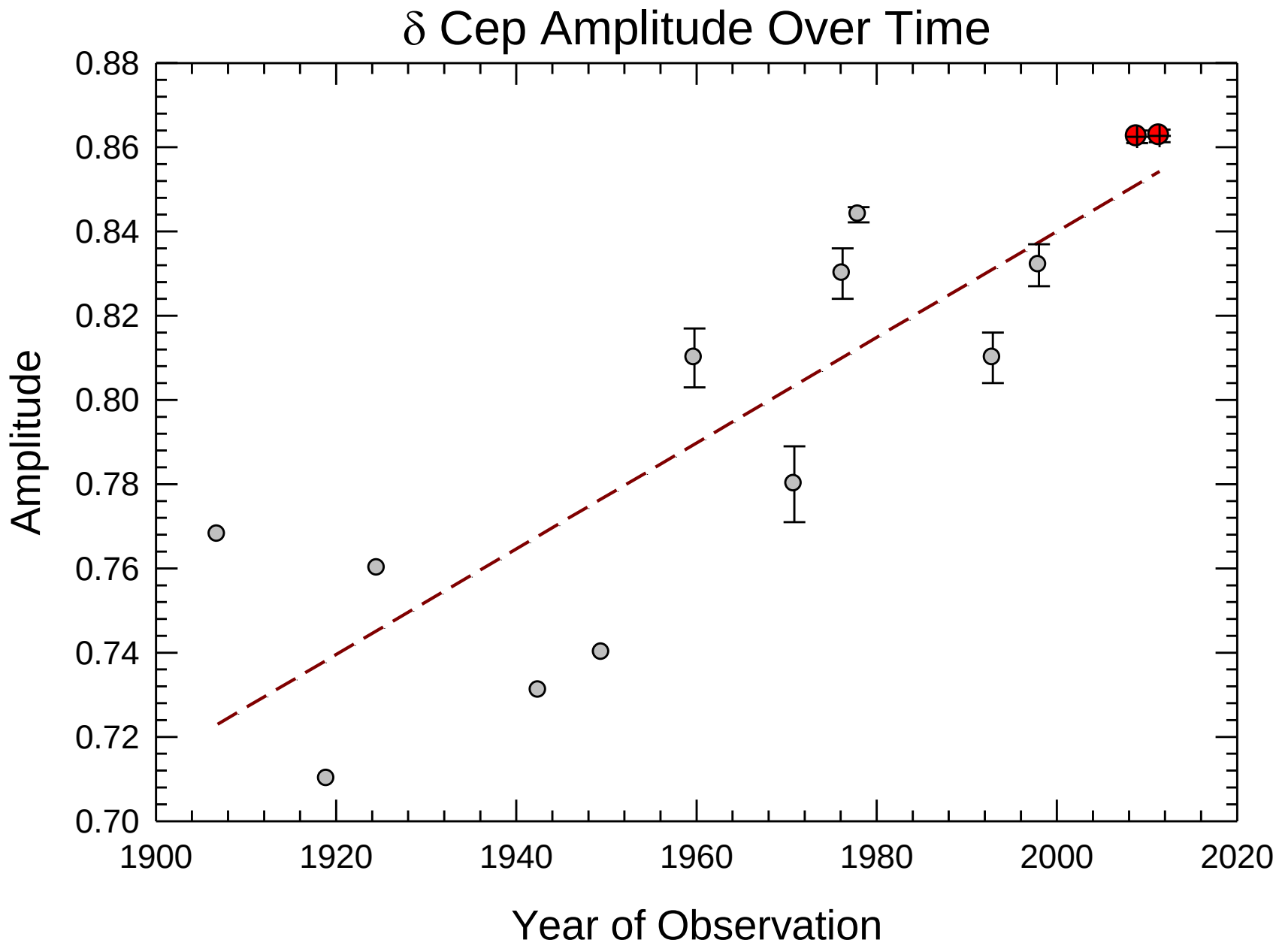

Figure 14 – The observed light amplitudes of δ Cep are plotted vs. the mid-time of the observation set. Points measured as part of this program are indicated by red crosses.

Brightness variability amplitudes were also obtained for the two light curves, and compared to those from archival observations deemed to be of high reliability. The results are shown in Fig. 14, where it can be seen that the light amplitude of δ Cep appears to be increasing over time. If a simple linear trend is assumed, as is plotted in the figure, then the light amplitude of δ Cep appears to be increasing at dA/dt = 1.3-mmag per year. The sparseness of amplitude measures prevents a detailed investigation of possible short-term amplitude variations, although there is a slight hint of amplitude periodicity on a timescale of ~20 – 25 years. It bears repeating, though, that this is only speculation at this point, thanks to the sparse dataset.

### 2.4 η Aql

η Aql is the 12$^{th}$ variable star discovered, and the 1$^{st}$ Cepheid for which light variations were observed. With a pulsation period of ~7.2-days, η Aql is an excellent example of a "bump Cepheid," displaying a prominent Hertzsprung bump in the middle of the descending phase of its light curve (around $\phi \approx 0.33$), as shown in Fig. 15. As with δ Cep, η Aql is known to have a hot companion of spectral type B9.8 V (Evans 1991), which was recently resolved with HST to lie 0.7" from the Cepheid (Evans 2011).

**Table 5 – Relevant Stellar Properties of η Aql**

| | |
|---|---|
| Spectral Type | F7Ib – G2Ib[1] |
| $T_{eff}$ (K) | 5300 – 6400[2] |
| Mass (pulsational) ($M_\odot$) | 4.1 ± 0.4[3] |
| Mass (evolutionary) ($M_\odot$) | 5.7 ± 0.6[3] |
| Mean Luminosity ($L_\odot$) | ~2500 |
| Mean Radius ($R_\odot$) | ~49 |
| Distance (pc) | $424^{+330}_{-130}$ |
| Ephemeris (this study) 2455856.689 + 7.177025(86) × E | |
| Ephemeris for O-C diagram (Berdnikov et al. 2000) 2411999.693 + 7.1765468 × E | |

[1]Kraft (1960); [2]Luck & Andrievsky (2004); [3]Caputo et al. (2005); Neilson et al. (2012a); van Leeuwen (2007)

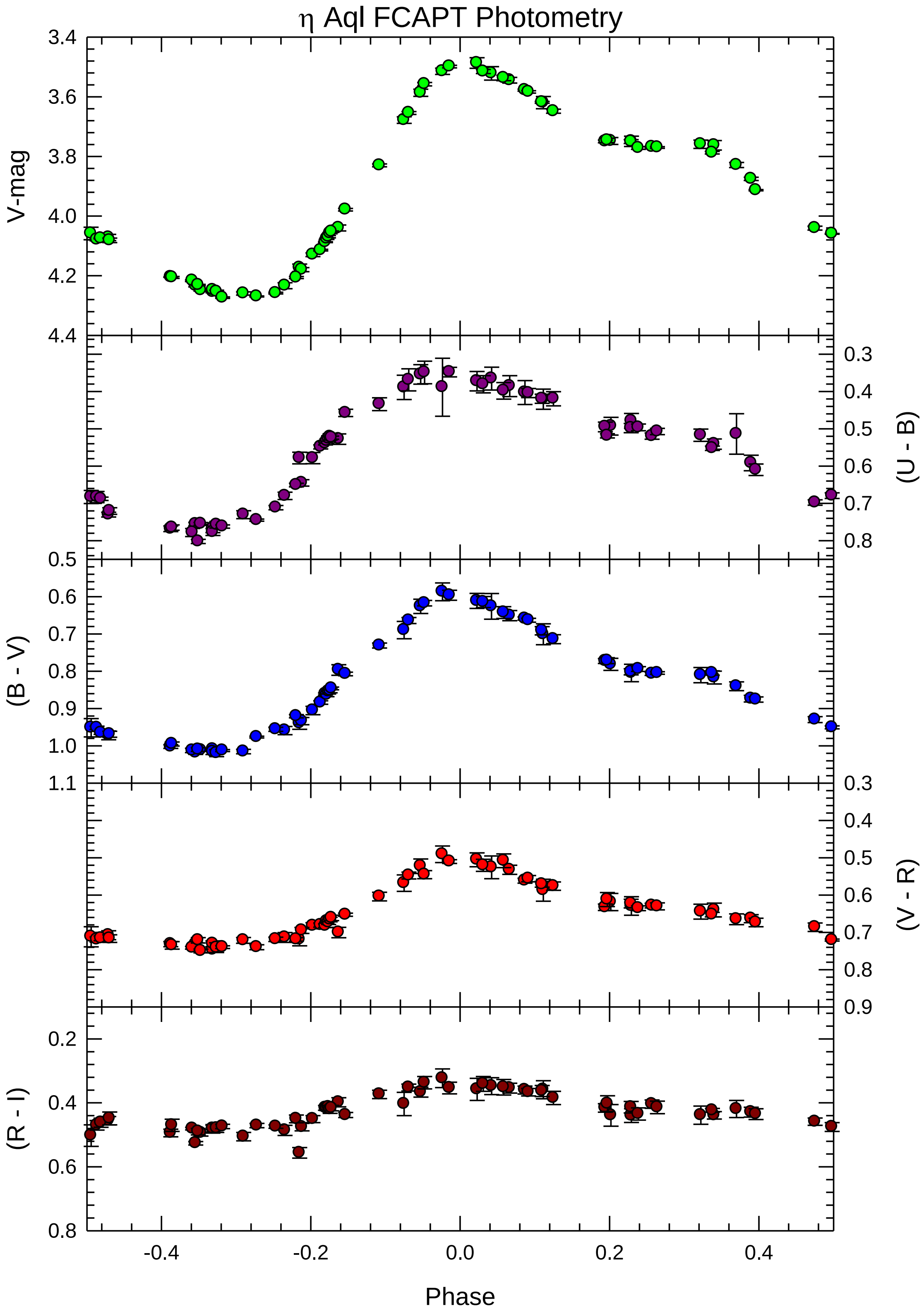

Figure 15 – The *UBVRI* data obtained for η Aql obtained with the FCAPT, phased to the new ephemeris determined in this study (given in Table 5).

A more recent addition to the program, with photometry beginning in June of 2008, one fully-covered light curve of η Aql has been obtained thus far. The Fourier-obtained time of maximum light is plotted in Fig. 16, along with times of maximum light found in the literature (Berdnikov et al. 2000). As shown by the O-C curve, the pulsation period is steadily increasing over time. The rate of period change is calculated to be dP/dt = 0.255 ± 0.001 sec/yr, indicative of a Cepheid undergoing its third crossing of the instability strip (Turner et al. 2007).

The light amplitude of η Aql, when plotted with those found in the literature (Fig. 17), appears to show no coherent variability patterns, either short- or long-term. It’s possible that the amplitude of η Aql has remained essentially constant since the time of its discovery.

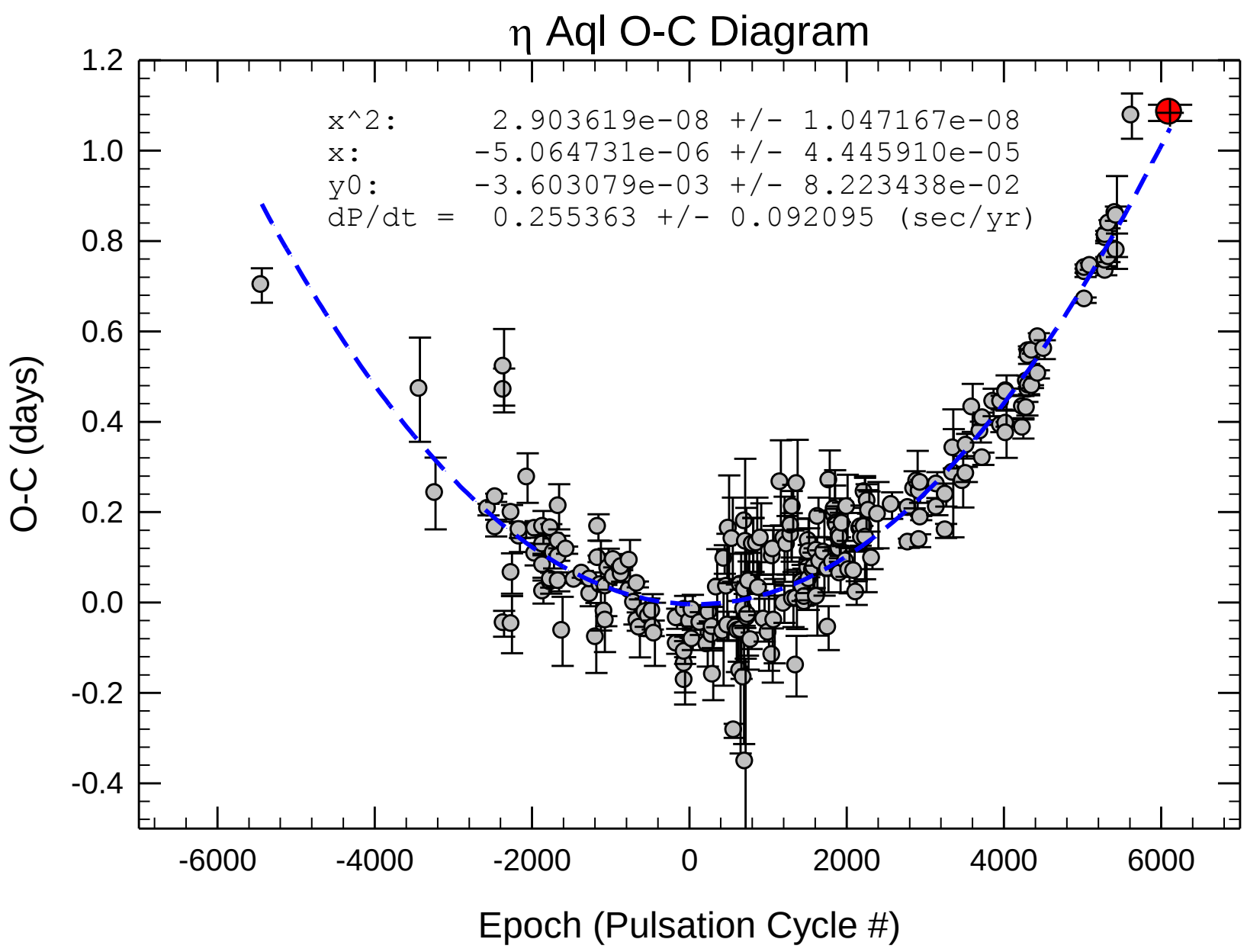

Figure 16 – The O-C diagram for η Aql, showing the increasing trend in its pulsation period. Coefficients of the quadratic fit are given in the plot, along with the rate of period change (dP/dt = 0.255 ± 0.001 sec/yr). The point determined from data obtained as a part of this program is plotted as the red filled and crossed circle.

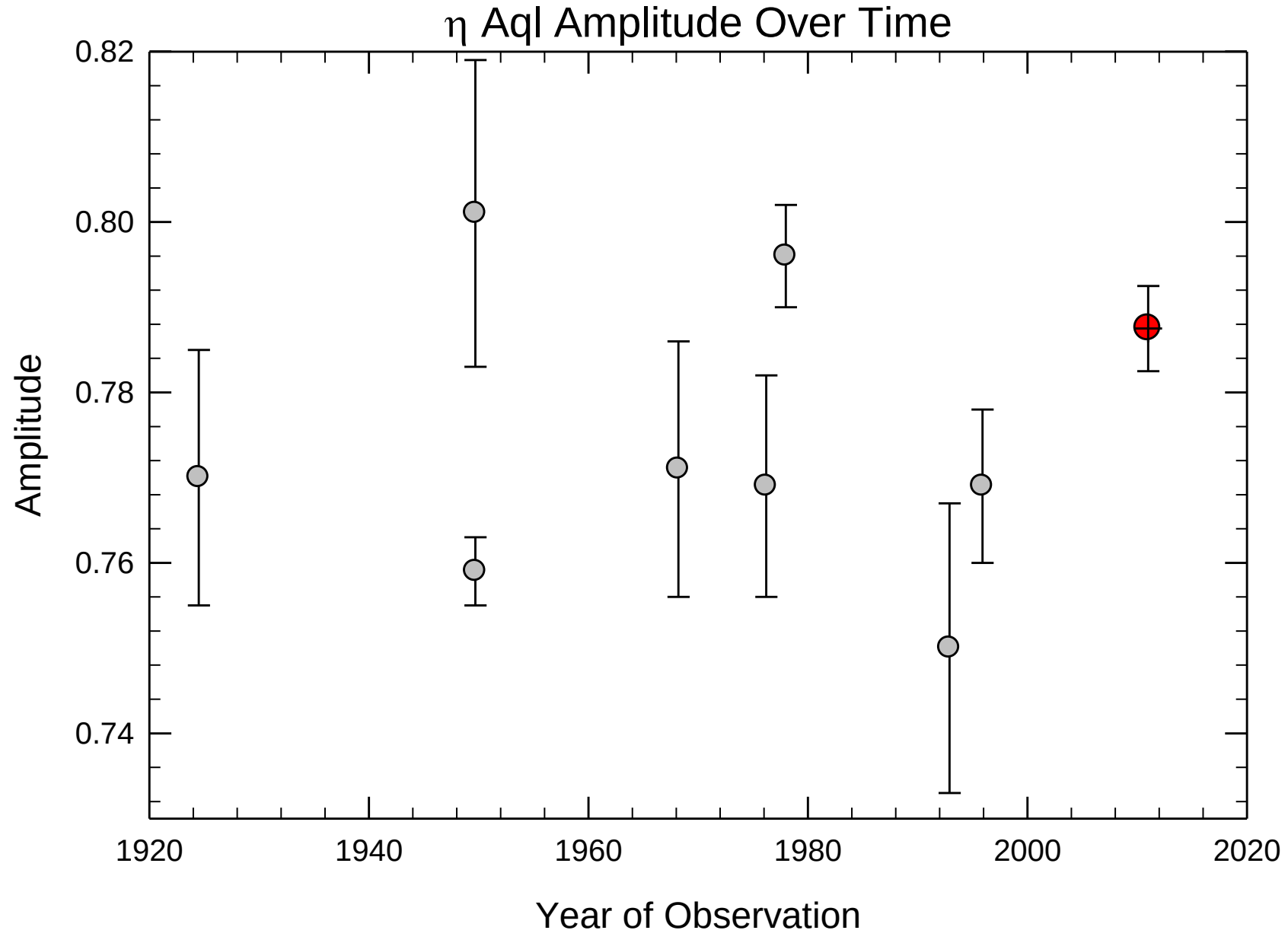

Figure 17 – The observed light amplitudes of η Aql are plotted vs. the mid-time of the observation set. Points measured as part of this program are indicated by red crosses.

### 2.5 EU Tau

EU Tau is a somewhat “newer” Cepheid, especially in terms of those in this program. This is because earlier studies concluded EU Tau was a W UMa-type variable, until Guinan (1966, 1972) discovered that it was a Cepheid. With its modest V-band amplitude of ~0.32-mag, and more symmetrical light curve (Fig. 18), EU Tau is classified as an s-Cepheid by studies such as Andrievksy et al. (1996), who also suggested that EU Tau could not be in its first crossing of the instability strip, given that it is carbon-deficient. The revised *Hipparcos* parallax for EU Tau is 1.17 ± 1.04 mas. This is an extremely large error, but understood given the distance of the Cepheid. Therefore, the distance calculated by Gieren et al. (1990) is given in Table 6. EU Tau has an unresolved, hot companion of spectral type A1 – A2 V (Kovtyukh et al. 1996).

**Table 6 – Relevant Stellar Properties of EU Tau**

| | |
|---|---|
| Spectral Type | F5II – G5II[1] |
| $T_{eff}$ (K) | ~6200 – 6600[2] |
| Mass (pulsational) ($M_{\odot}$) | 4.55 ± 0.54[3] |
| Mass (evolutionary) ($M_{\odot}$) | 4.95 ± 0.2[3] |
| Mean Luminosity ($L_{\odot}$) | ~1040[3] |
| Mean Radius ($R_{\odot}$) | ~30[3] |
| Distance (calculated – pc) | 1191 ± 57[3] |
| Ephemeris (this study)<br>2455618.816 + 2.102299(15) × E | |
| Ephemeris for O-C diagram (Fernie 1987)<br>2442583.623 + 2.1025112 × E | |

[1]Buscombe & Foster (2001); [2]Bersier et al. (1997); [3]Gieren et al. (1990)

One fully-covered light curve was obtained in this program for EU Tau, and the time of maximum light was added to those found in, or determined from, the literature. As shown in Fig. 19, the period of EU Tau is not constant. A quadratic fit has been run through the data, which assumes a steadily decreasing period, but it is important to note that, due to the large data gap before the observations of Peña et al. (2010), this assumption may not hold true. It is entirely possible that, between the observations of Berdnikov (2008) and Peña et al. (2010), an abrupt change in the period took place. The period could have been constant before the observations of Berndnikov, then shifted, and will now continue for some time at its current, slightly shorter value. If the star does have a steadily decreasing period, the quadratic fit returns a period change rate of dP/dt = –0.336897 ± 0.001724 sec/yr. This value is one possibility due to the sparseness of recent data but, if true, would indicate that EU Tau is undergoing its fourth crossing of the instability strip (albeit with a rather quick period change rate for short-period, fourth crossing Cepheids – Turner et al. 2007). This would agree with the finding of Andrievksy et al. (1996).

A single light amplitude was also determined for EU Tau and plotted against those found in, or determined from, the literature (Fig. 20). A linear trend of dA/dt = -0.2-mmag per year has been run through the data; however, we note that the overall spread in the amplitudes is very small, and within the measurement errors. Thus, it is also likely that EU Tau has undergone no significant change in amplitude.

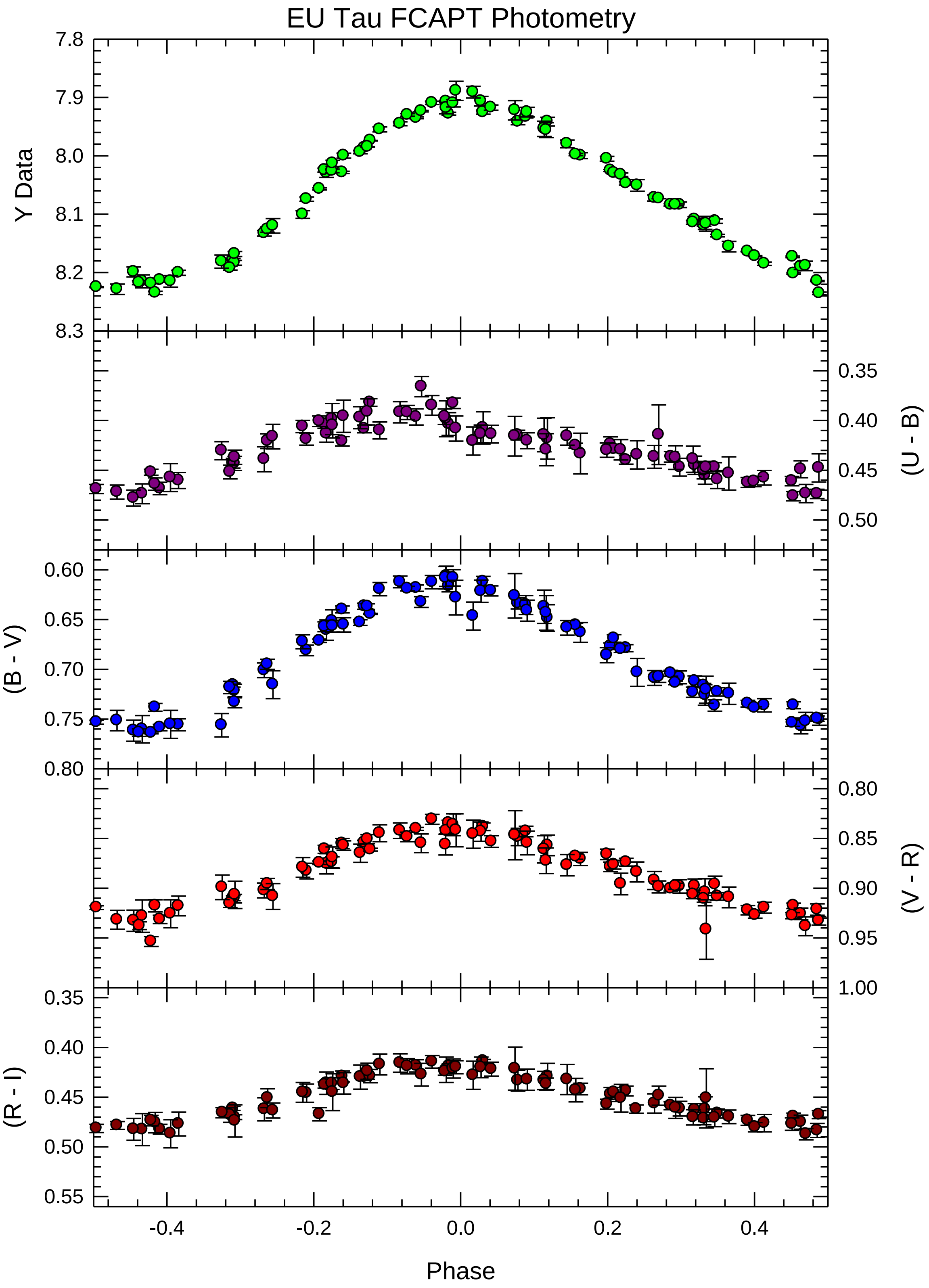

Figure 18 – The *UBVRI* data obtained for EU Tau, phased to the new ephemeris determined in this study (given in Table 6).

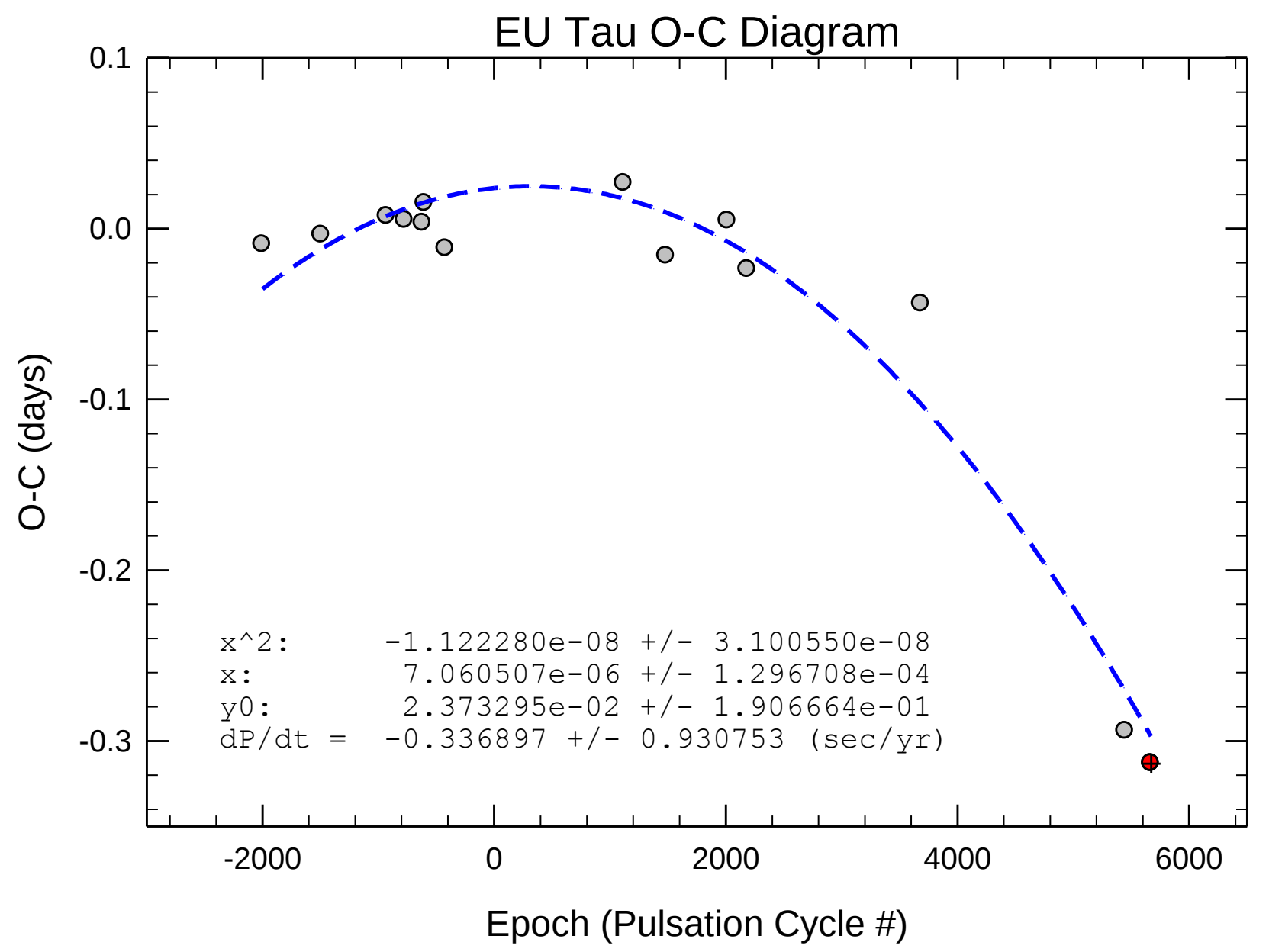

Figure 19 – The O-C diagram for EU Tau. Given the large, recent gap in the data, an unambiguous conclusion cannot be determined. The period of the Cepheid could be smoothly decreasing, as with δ Cep (which is the behavior assumed by the fit), or the Cepheid could have undergone a sudden shift to a shorter period. Coefficients of the quadratic fit are given in the plot, along with the rate of period change (dP/dt). The point determined from data obtained as a part of this program is plotted as the red filled and crossed circle.

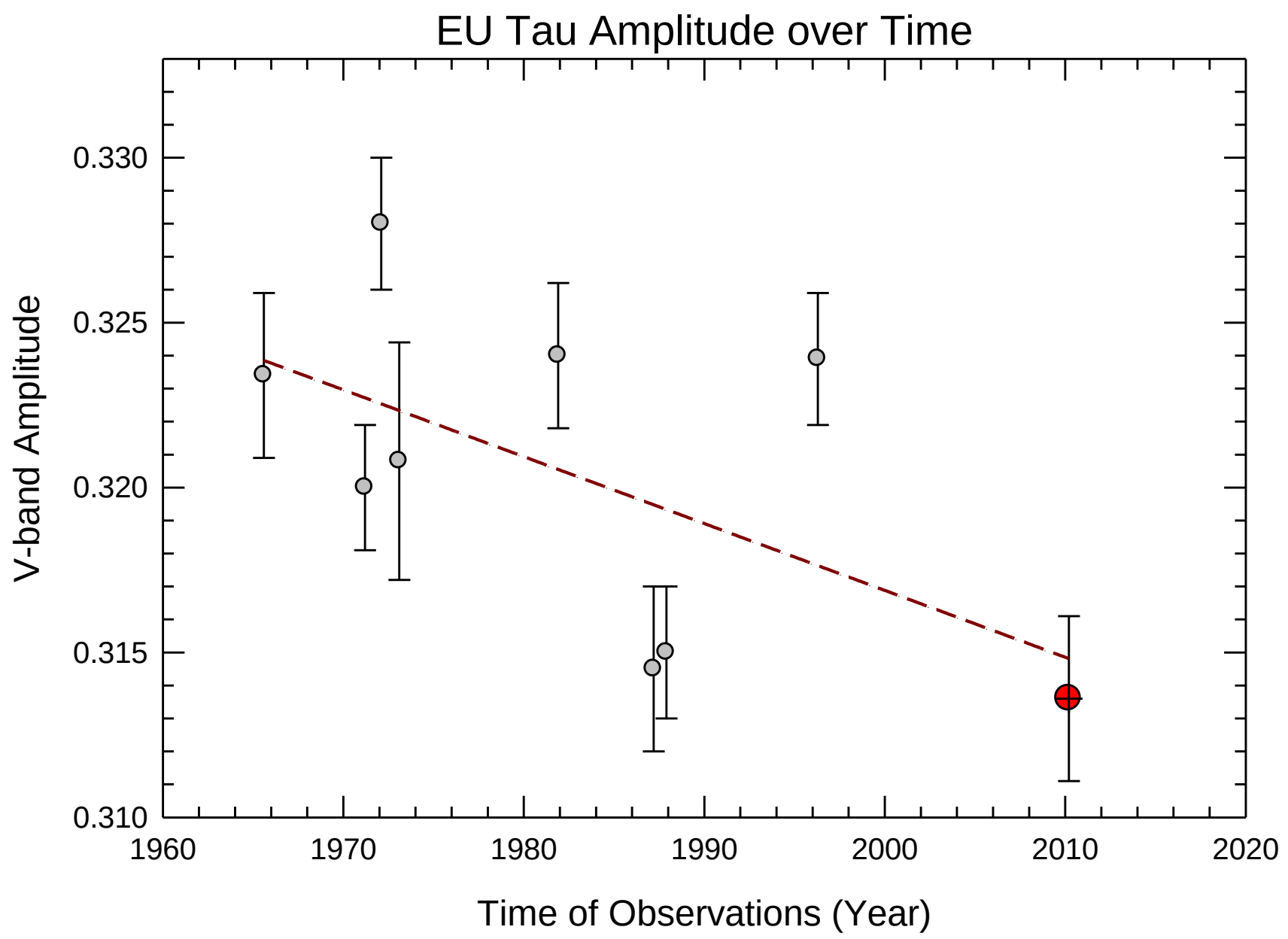

Figure 20 – The observed light amplitudes of EU Tau are plotted vs. the mid-time of the observation set. Points measured as part of this program are indicated by red crosses.

### 2.6 Polaris

Polaris is best known as the North Star, since it is easily visible to the naked eye and lies within 1° of the north celestial pole. Numerous literary references to Polaris have been made throughout history, with many remarking on its constant, steady nature (referring to its essentially static location) as a quality that can be aspired to in humans. Perhaps the most famous quote is that of Julius Caesar who, in Shakespeare's 1599 play *The Tragedy of Julius Caesar*, boasts:

*But I am constant as the northern star,*
*Of whose true-fix'd and resting quality*
*There is no fellow in the firmament.*
*The Tragedy of Julius Caesar (III, i, 60-62)*

However, in the latter half of the 19th century the "constant" nature of Polaris was becoming suspect. Hertzpsrung (1911) then confirmed the periodic variability of Polaris, and compared several aspects of it to the Cepheids. Since then, numerous photometric studies have been conducted of Polaris. By far the nearest Cepheid (δ Cep is second nearest, and lies at twice the distance), Polaris is also considered somewhat special among Cepheids because the amplitude of its brightness variation is very low (currently ~0.06-mag – Fig. 21). In fact, the original motivation for the *Secret Lives of Cepheids* program was the study of Arellano Ferro (1983), where the (surprising) declining light amplitude of Polaris was first reported. Since then, numerous studies have confirmed and expanded on this behavior, including a very thorough analysis by Turner et al. (2005). Polaris has what appears to be a very accurate revised *Hipparcos* parallax of $\pi_{\mathrm{Hipp}} = 7.54 \pm 0.11$-mas, but doubts still exist as to Polaris' true distance, and whether it is a fundamental mode or an overtone pulsator. One reason for the doubts is the difference between the spectroscopically derived absolute magnitude of $M_{\mathrm{V}} \approx -3.0$ for Polaris (Kovtyuhk et al. (2010), and the value of around $M_{\mathrm{V}} \approx -3.6$ that is calculated using parallax values (thus the spectroscopic magnitude implies that Polaris is closer than the parallax value indicates).

In addition to the changes in Polaris' amplitude, and questions about its precise distance and pulsation mode, the period of Polaris' pulsations is increasing. Fig. 22 shows the O-C data for Polaris, with our own times of maximum light (red filled, and crossed, circles) added to those of Turner et al. 2005 and Spreckley & Stevens (2008). The newly calculated rate of period change is currently dP/dt = 4.47 ± 0.08 sec/yr (Fig. 22), which is quicker than most other Cepheids of similar period, suggesting that Polaris is currently evolving through the short-lived and rarely-observed first crossing of the instability strip.

The declining nature of Polaris' light amplitude prompted further questions. Arellano Ferro (1983) theorized that Polaris might have been evolving out of the instability strip and, therefore, on its way toward becoming a non-variable supergiant. Subsequent studies, such as that of Fernie et al. (1993), entitled “Goodbye to Polaris the Cepheid”, support the theory. However, observations taken as part of this program beginning in the early 2000’s indicated that Polaris’ light amplitude had not only stopped decreasing, but appeared to begin increasing again (Davis et al. 2002; Engle et al. 2004; Engle & Guinan 2012). Additional studies (Turner et al. 2005 and Spreckley & Stevens 2008) have since supported the original claim. Fig. 23 shows the amplitudes of Polaris going back over a century (from Turner et al. 2005) with the amplitudes observed as part of this program plotted as red filled and crossed circles. The “return” of Polaris’ amplitude hints at the possibility of a cycle, as opposed to a simple decline, but more years of observation will be required before a full understanding of Polaris’ variable amplitude can be achieved

**Table 7 – Relevant Stellar Properties of Polaris**

| | |
|---|---|
| Spectral Type | F7 – F8 Ib–II[1] |
| $T_{eff}$ (K) | ~6000 – 6050[2] |
| Mass (pulsational) ($M_\odot$) | 4.5 ± 2.0[3] |
| Mass (evolutionary) ($M_\odot$) | 5.8 ± 0.5[3] |
| Mean Luminosity ($L_\odot$) | ~2200[4] |
| Mean Radius ($R_\odot$) | ~46[3] |
| Distance (pc) | 133 ± 2[5] |
| Ephemeris (this study) 2455909.910 + 3.972433(520) × E | |
| Ephemeris for O-C diagram (Turner et al. 2005) 2428260.727 + 3.969251 × E | |

[1]Wielen et al. (2000); [2]Turner et al. (2013); [3]Nordgren et al. (2000); [4]Spreckley & Stevens (2008); [5]van Leeuwen (2013),

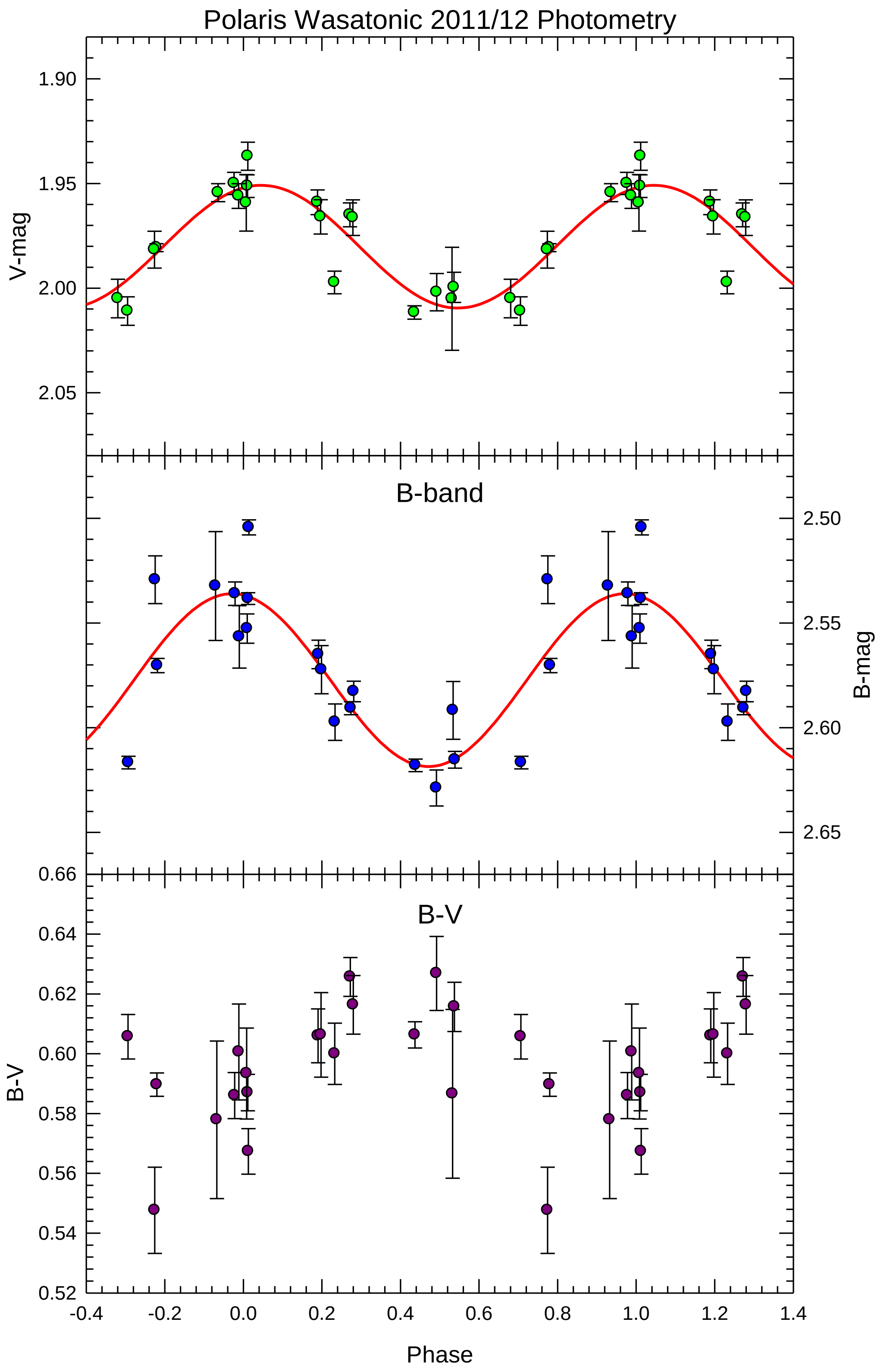

Figure 21 – The most recent *BV* data obtained for Polaris, by observer Rick Wasatonic, phased to the new ephemeris determined in this study (given in Table 7).

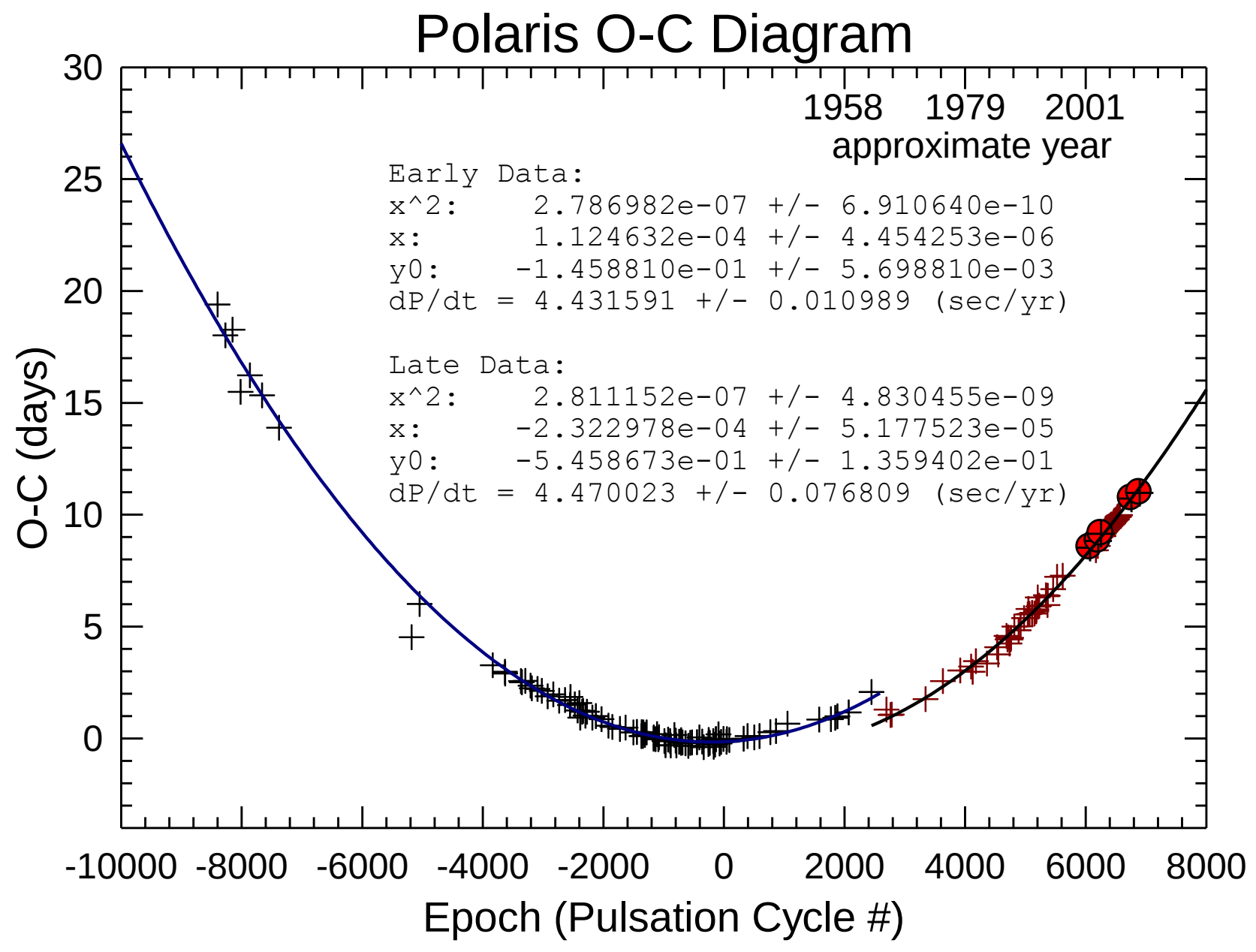

Figure 22 – The O-C diagram for Polaris, which shows the increasing period of Polaris over time. The data has been divided into two epochs: before the "period glitch" in 1963 – 64, and after. Coefficients of the quadratic fit to each epoch are given in the plot, along with the rates of period change (dP/dt = 4.47 ± 0.08 sec/yr). Points determined from this program are plotted as red filled and crossed circles.

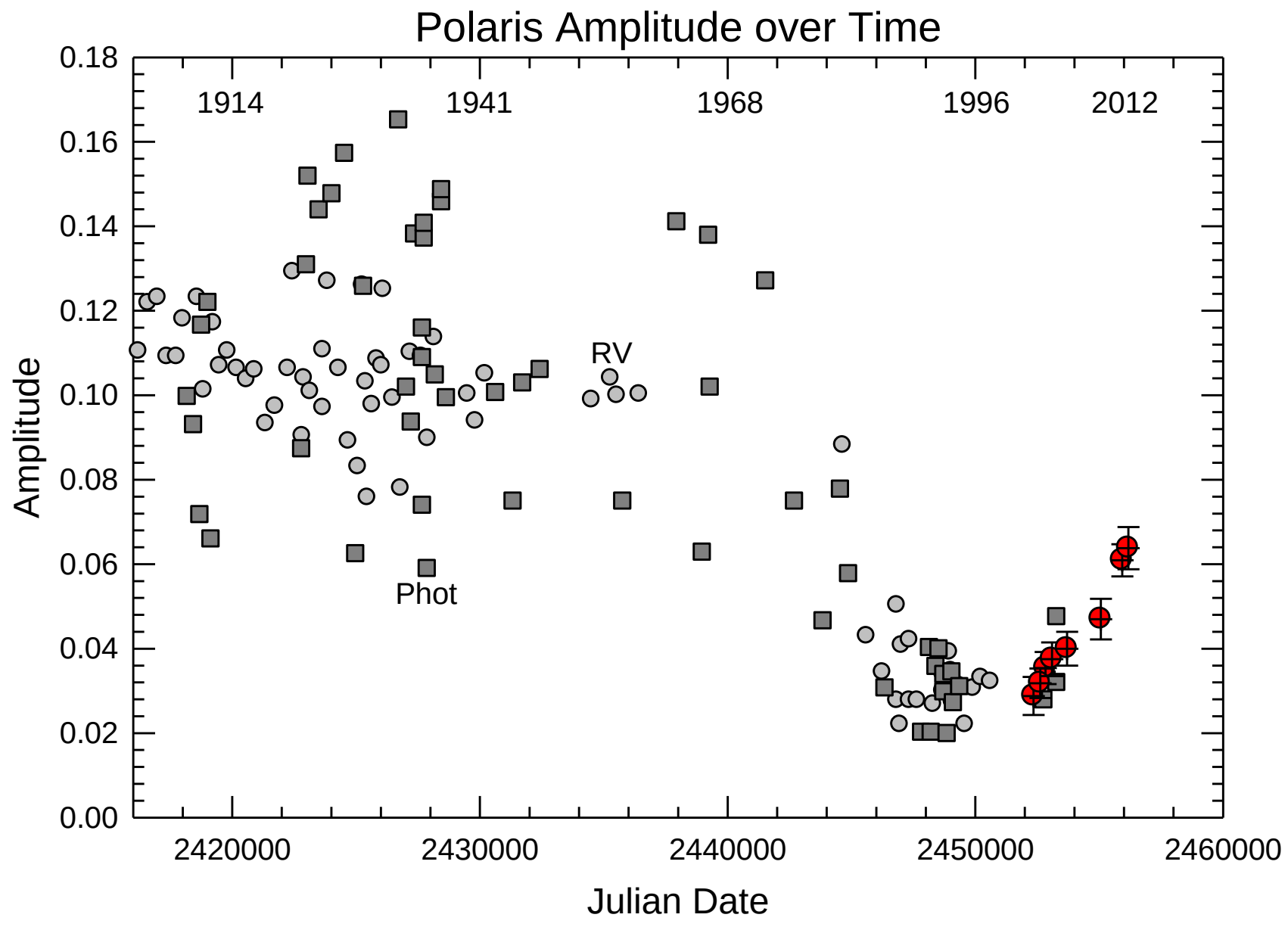

Figure 23 – The observed light amplitudes of Polaris are plotted vs. the mid-time of the observation set (in JD – data obtained from David Turner [private communication]). Points measured as part of this program are indicated by red crosses.

**2.7 SU Cas**

SU Cas is currently the shortest period Cepheid of the program (~2-days, see Table 8), and a probable member of the open cluster Alessi 95 (Turner et al. 2012). Based on factors such as rate of period change, amplitude of variability (presently ~0.4-mag – see Fig. 24) and luminosity, SU Cas is classified as an s-Cepheid, pulsating in the first overtone, with a fundamental period of ~2.75-days (Milone et al. 1999; Turner et al. 2012). In addition to open cluster membership, IUE observations have revealed SU Cas to have a B9.5 V companion (Evans 1991). SU Cas offers calibrator possibilities, with a reliable trigonometric parallax, along with cluster/binary companions, from which distances and luminosities can be calculated and compared.

**Table 8 – Relevant Stellar Properties of SU Cas**

| | |
|---|---|
| Spectral Type | F5 – F7 Ib – II[1,2] |
| $T_{eff}$ (K) | ~6100 – 6600[1] |
| Mass (pulsational) ($M_{\odot}$) | 6.5 ± 0.6[2] |
| Mass (evolutionary) ($M_{\odot}$) | 5.5 ± 0.3[2] |
| Mean Luminosity ($L_{\odot}$) | ~1500[2] |
| Mean Radius ($R_{\odot}$) | ~33[2] |
| Distance (pc) | 395 ± 30[3] |
| Ephemeris (this study)<br>2455199.614 + 1.949330(3) | |
| Ephemeris for O-C diagram (Szabados 1991)<br>2441645.913 + 1.949325 | |

[1]Luck et al. (2008); [2]Milone et al. (1999); [3]van Leeuwen (2007)

In this program, three well-covered light curves were obtained for SU Cas, from which times of maximum light and amplitudes were derived. Fig. 25 shows the O-C diagram for SU Cas, with the times of maximum light from this program (red filled and crossed circles) being added to the literature values from Szabados 1991 and Berdnikov et al. 2003. There is a slight gap in recent observational data for SU Cas. One can see that a period increase has taken place, and a quadratic fit has been run through the data, assuming a steady period increase over time. From this fit, a period change rate of dP/dt = 0.0204 ± 0.0002 sec/yr has been calculated. This rate places SU Cas in the third crossing of the instability strip. It is, of course, very important to note that, just as with EU Tau, SU Cas may not be undergoing a constant period increase, but rather may have undergone an abrupt shift to a slightly longer period. Further data will show us whether the period is constant at its current, slightly longer value, or if it is still increasing.

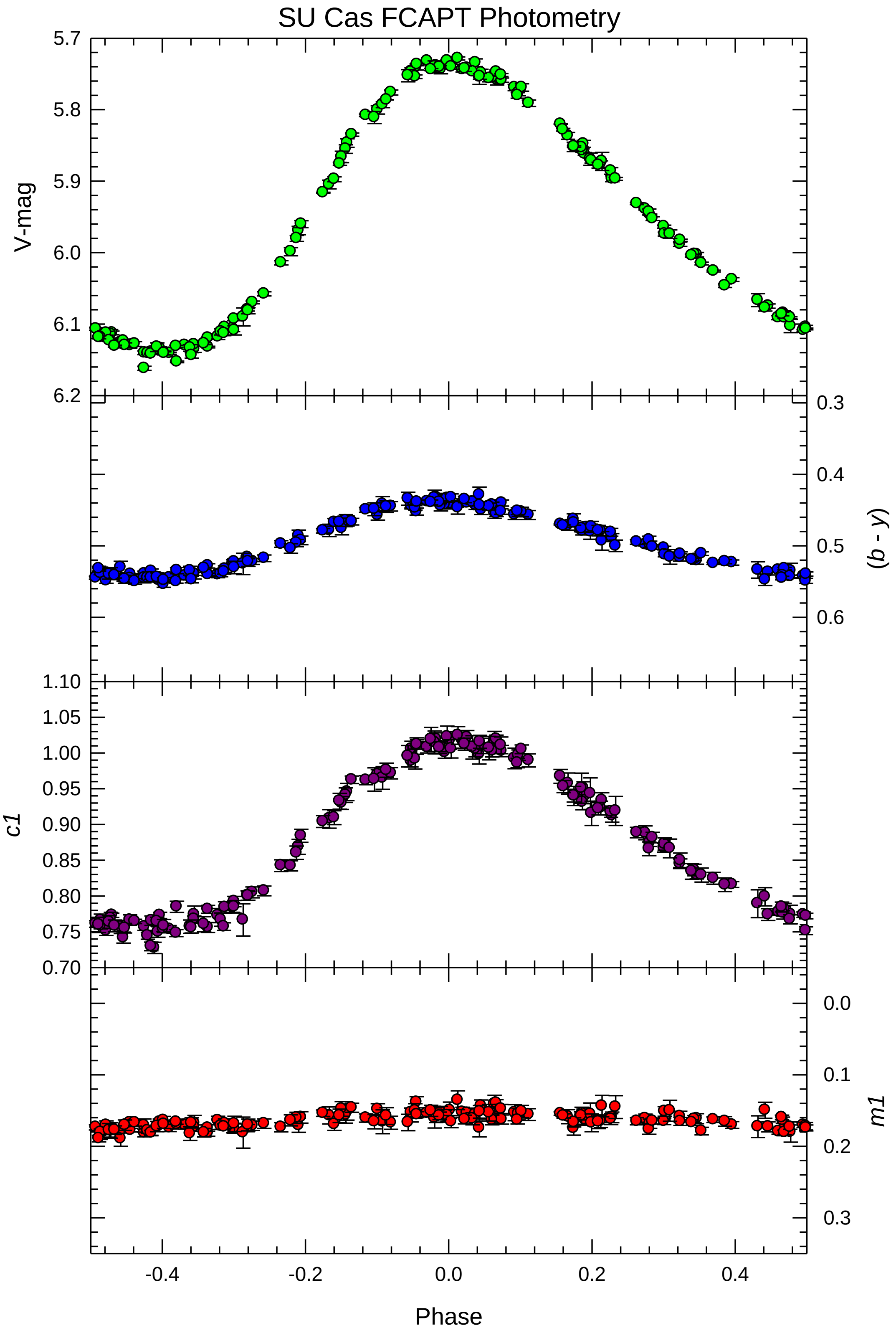

Figure 24 – The *uvby* data obtained for SU Cas. The *y*-band data have been transformed to standard *V*-band magnitudes.

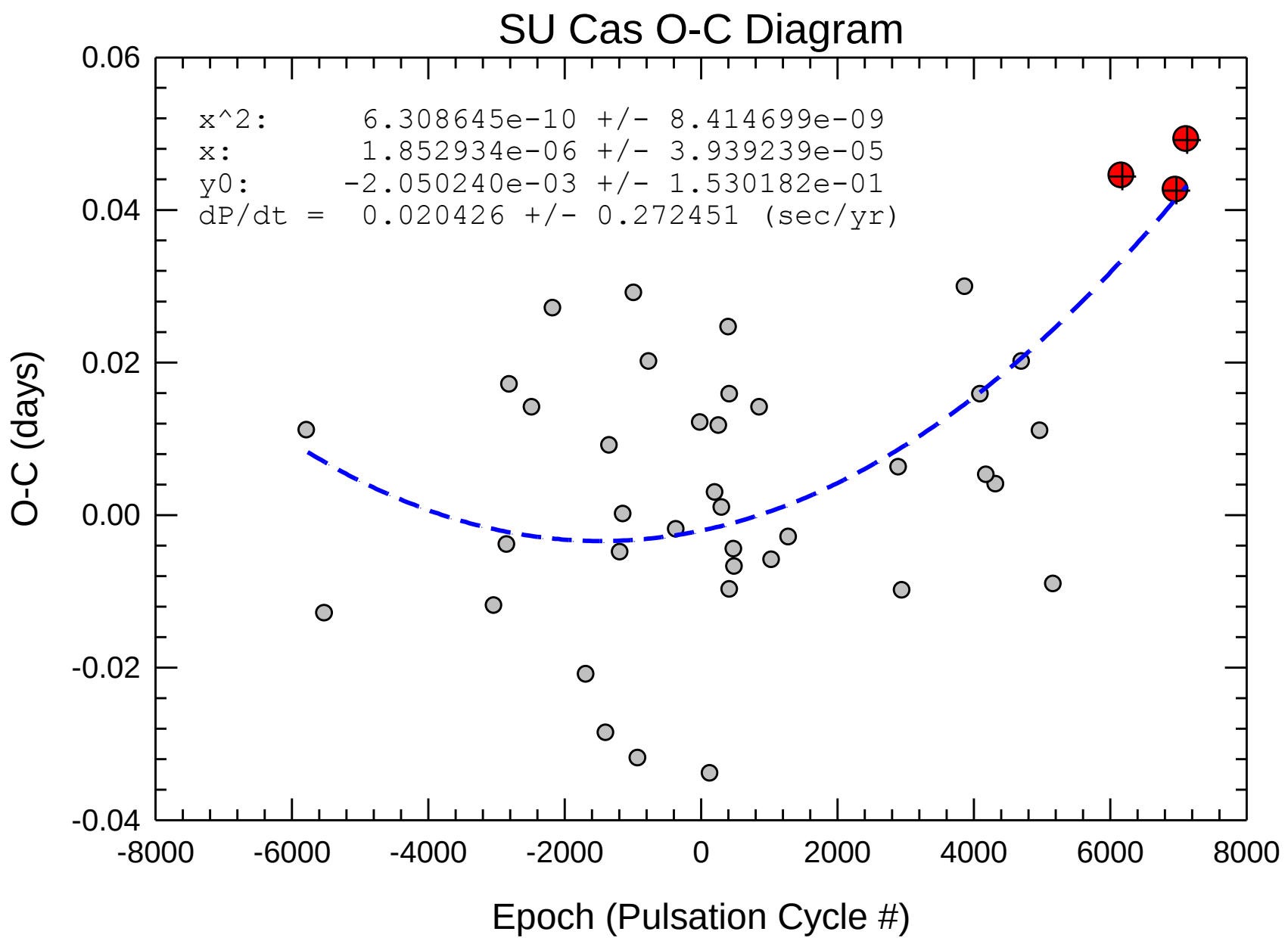

Figure 25 – The O-C diagram for SU Cas. The recent data show an increasing period trend, which before was hinted at but with ambiguity. Coefficients of the quadratic fit are given in the plot, along with the rates of period change (dP/dt = 0.0204 ± 0.0002). Points determined from this program are plotted as red filled and crossed circles.

Fig. 26 below gives the observed amplitudes of SU Cas over time. A linear fit run through the data shows a possible increase of 0.80-mmag over time. However, as can be seen in the graph and the very small rate of increase found, the trend is not concrete and could be covered by the observational errors. As with most Cepheids, the historic data is simply too sparse to *definitively* reveal long- or short-term variations in the amplitude. The variations found between the most modern lightcurves, ours included, as well as between the three amplitudes we have observed as part of this program are enticing but (as always) additional regularly obtained data are necessary to truly understand the amplitude behavior.

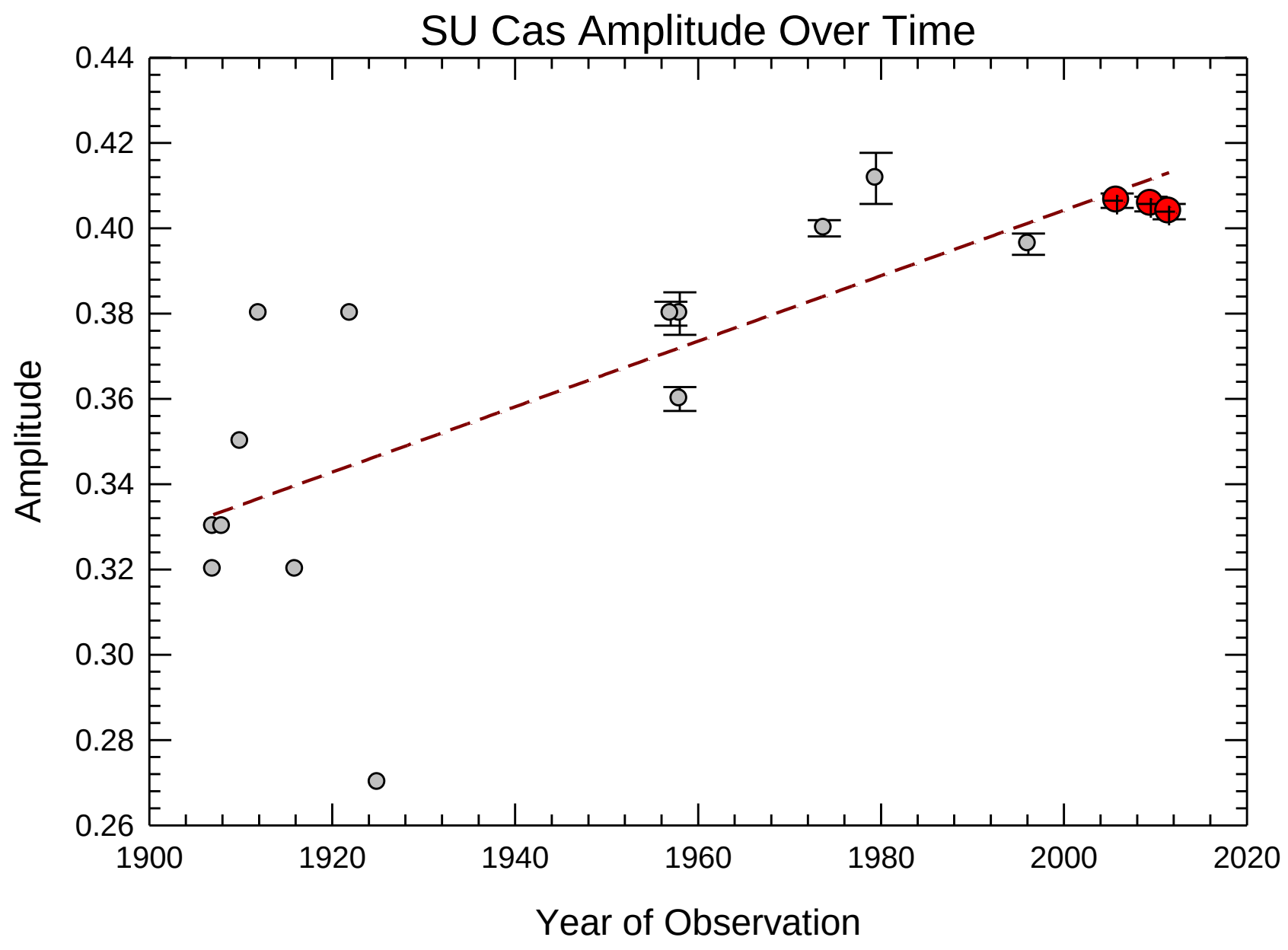

Figure 26 – The observed light amplitudes of SU Cas are plotted vs. the mid-time of the observation set. Points measured as part of this program are indicated by red crosses. A possible linear trend of increasing amplitude over time is hinted at, but relies somewhat on the older, less accurate observations.

## 2.8 SV Vul

SV Vul is the longest period Cepheid in this study (~45-days, see Table 9), and among the longest period Cepheids known in the galaxy. The lightcurve has a full amplitude (~1-mag, see Fig. 27) and prominently displays the asymmetric, saw-tooth shape. The revised Hipparcos parallax for SV Vul is 0.80 ± 0.81 mas. As with EU Tau, this is an unacceptably large error in distance, from a scientifically relevant standpoint, but understandable given the instrumental limitations at such distances. As such, the distance given for SV Vul in Table 9 is that determined by Turner & Burke (2002) based upon the Cepheid's membership in the NGC 6834/Vul OB1 cluster.

**Table 9 – Relevant Stellar Properties of SV Vul**

| | |
|---|---|
| Spectral Type | F7 Iab – K0 Iab[1] |
| $T_{eff}$ (K) | ~4900 – 6100[2] |
| Mass (pulsational) ($M_\odot$) | 17.5 ± 3.5[3] |
| Mass (evolutionary) ($M_\odot$) | 15.1 ± 0.6[3] |
| Mean Luminosity ($L_\odot$) | ~20000[4] |
| Mean Radius ($R_\odot$) | ~200[4] |
| Distance (calculated – pc) | ~2200[4] |
| Ephemeris (this study)<br>2453564.224 + 44.993007(1705) | |
| Ephemeris for O-C diagram (Berdnikov 1994)<br>2448894.50 + 45.02397 | |

[1]Kraft (1960); [2]Luck et al. (2001); [3]Caputo et al. (2005); [4]Turner & Burke (2002)

As part of this program, five times of maximum light and three light amplitudes were measured for SV Vul; this difference is because, in two observing seasons, the phases around maximum light were well-covered, but the phases around minimum light were not. In looking at the lightcurves presented in Fig. 27, scatter in both amplitude and phase can already be seen. As with other Cepheids in the program, an O-C diagram and amplitude over time plot were constructed to better understand exactly the long-term behavior/stability of this Cepheid's pulsations.

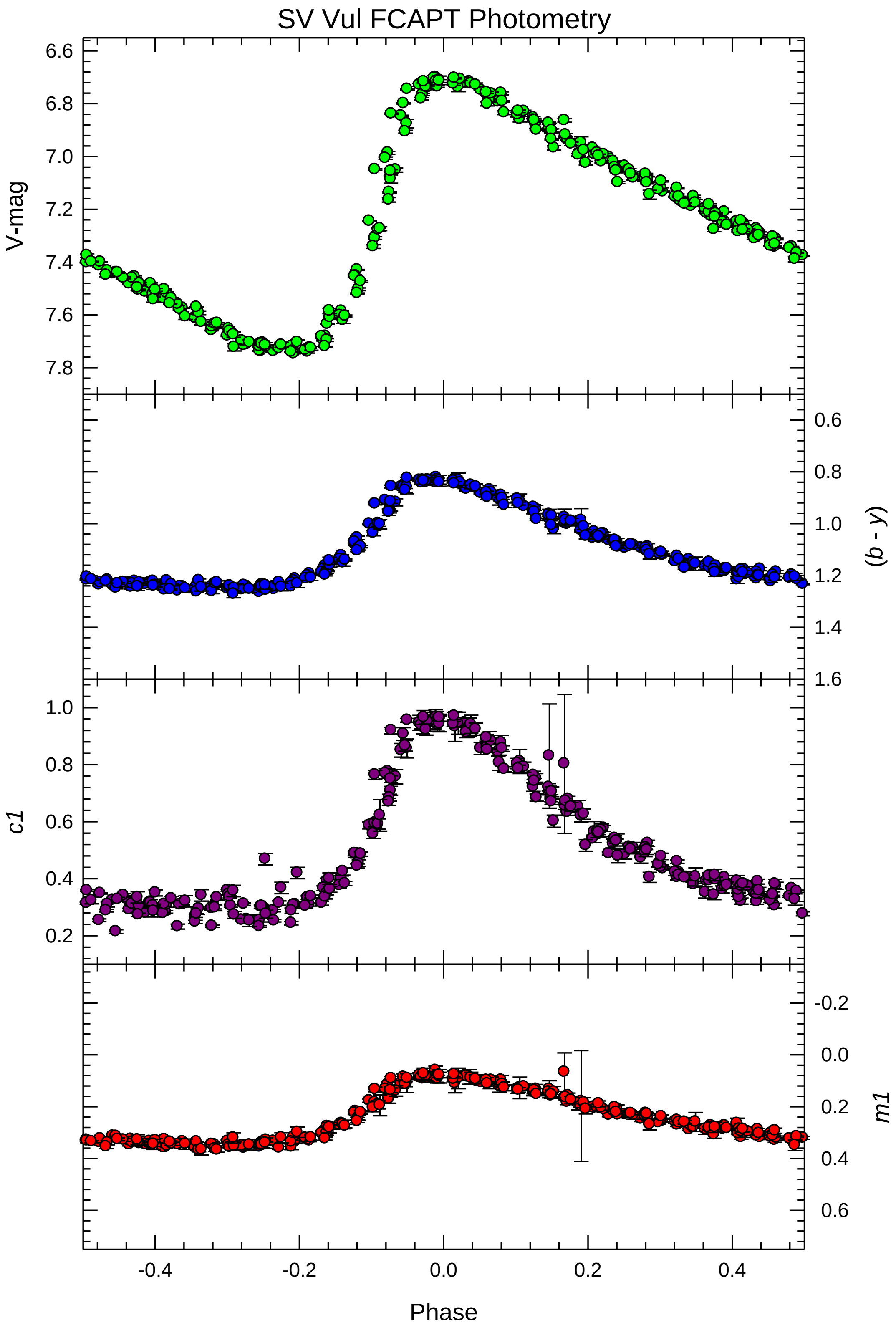

Figure 27 – The *uvby* data obtained for SV Vul. The *y*-band data have been transformed to standard *V*-band magnitudes, and phased to the newly calculated ephemeris given in Table 9.

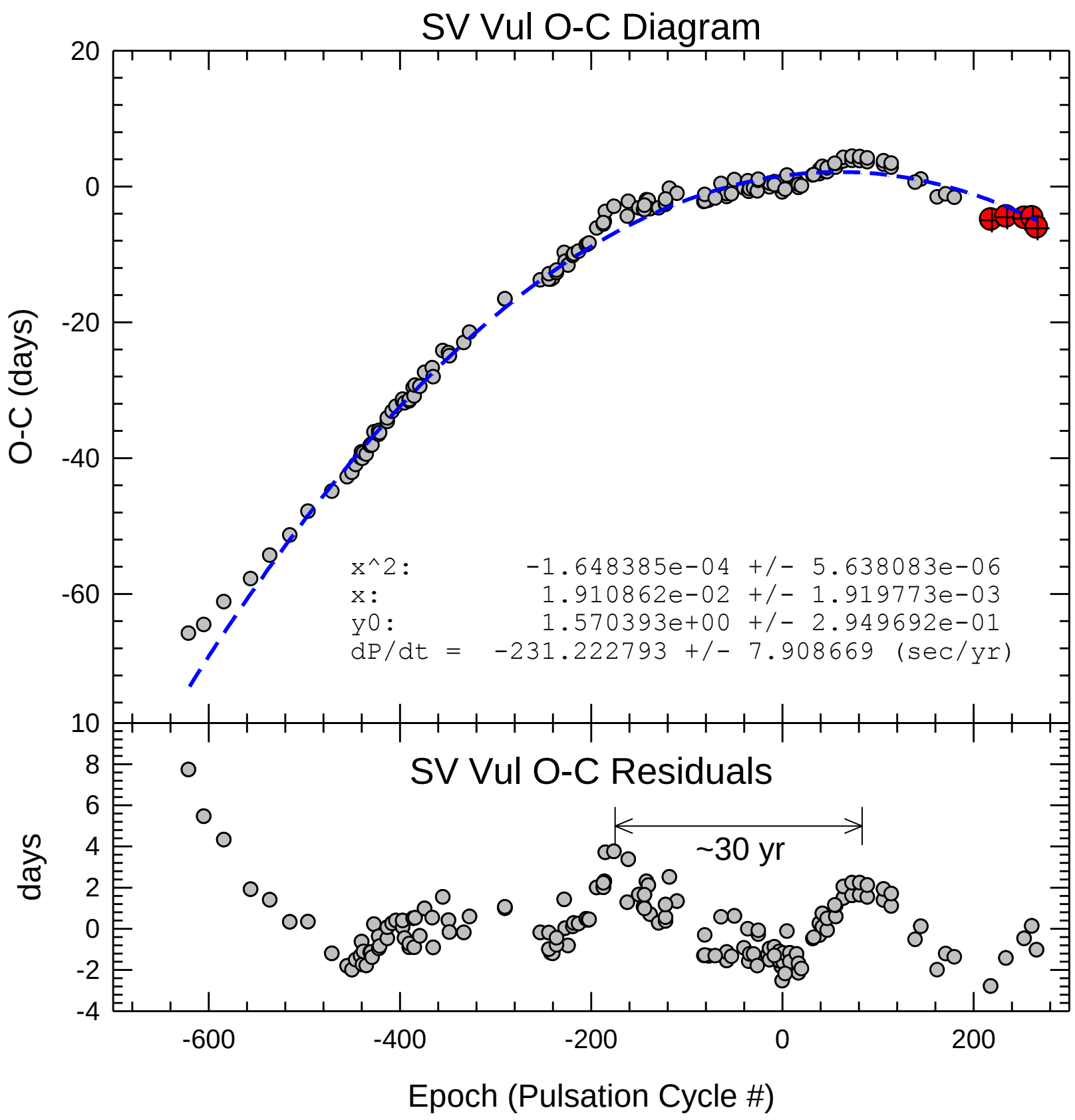

Figure 28 – The O-C diagram for SV Vul, showing the long-term decreasing period trend. Coefficients of the quadratic fit are given in the plot, along with the rates of period change (dP/dt = -231.223 ± 7.909). Points determined from this program are plotted as red filled and crossed circles. On top of the overall trend, there is a very interesting cyclic (~30-yr) behavior, as shown in the residuals plotted in the lower panel. The amplitude of the O-C residuals is too large to be the result of an unseen companion star.

Fig. 28 above shows the O-C diagram for SV Vul. In the top plot, the long-term decreasing behavior of the period is easily visible, and the quadratic fit to the data reveals a period change rate of dP/dt = –231.223 ± 7.909 sec/yr, indicating that SV Vul is currently undergoing its second crossing of the instability strip. However, there is more to the period variability, as shown in the bottom plot where the residuals to the quadratic fit are shown. There seems to be a cyclic variability to the period, with a cycle length of ~30-yr. O-C cycles such as these are not unheard of in a Cepheid, although they are rare. As mentioned earlier, the specific cause of such cyclic variations in period is not yet known.

The light amplitude of SV Vul is also variable, as shown in Fig. 29. For the past 30+ years, SV Vul has been observed more regularly than most Cepheids in the program, and the amplitude over time plot shows that. A linearly increasing trend in the amplitude has been run through the data, but such a trend is almost entirely dependent on the oldest (visual) datasets, and their large spread in amplitudes. Most striking about the graph, however, is the recent 30+ years of data, where the possibility of an amplitude cycle is seen. The amplitude of SV Vul, in this timespan, is seen to vary from a minimum of just under 1.0-mag, to a maximum of as much as ~1.15-mag. Such a span of amplitudes would lie outside what could be assigned to observational error, and points to a real variability in the light amplitude. It is important to note that Epoch 0 for SV Vul in the O-C diagram (Fig. 28) occurred in 1978–1979, corresponding to the previous minimum in light amplitude in Fig. 29. The amplitude then appears to have roughly followed a ~30-yr cycle, as the period has done. However, within the timespan of this program, the amplitude and period behaviors have not mirrored one another, with the amplitude falling while the period has apparently begun to lengthen again. It is important to note that photometry of all Cepheids continues, and the most recent light curve being gathered for SV Vul (not yet ready for inclusion here) indicates a preliminary amplitude equal to that of the earliest observed by the FCAPT. Therefore, the amplitude may have begun to increase again. What must also be taken into account is that a fully-covered lightcurve of SV Vul almost always consists of more than one ~45-day pulsation, making the possible effects of cycle-to-cycle variations in both period and amplitude (as recently found in *Kepler* satellite photometry of V1154 Cyg – Derekas et al. 2012) difficult to account for. As such, a point-to-point comparison of the O-C and amplitude data is most likely not as appropriate as a comparison of the overall trends would be. In summary, the amplitude variability of SV Vul is an exciting find that, taken together with the cyclic behavior of the period change, could offer strong evidence for a Blazhko-type behavior from this long-period Cepheid.

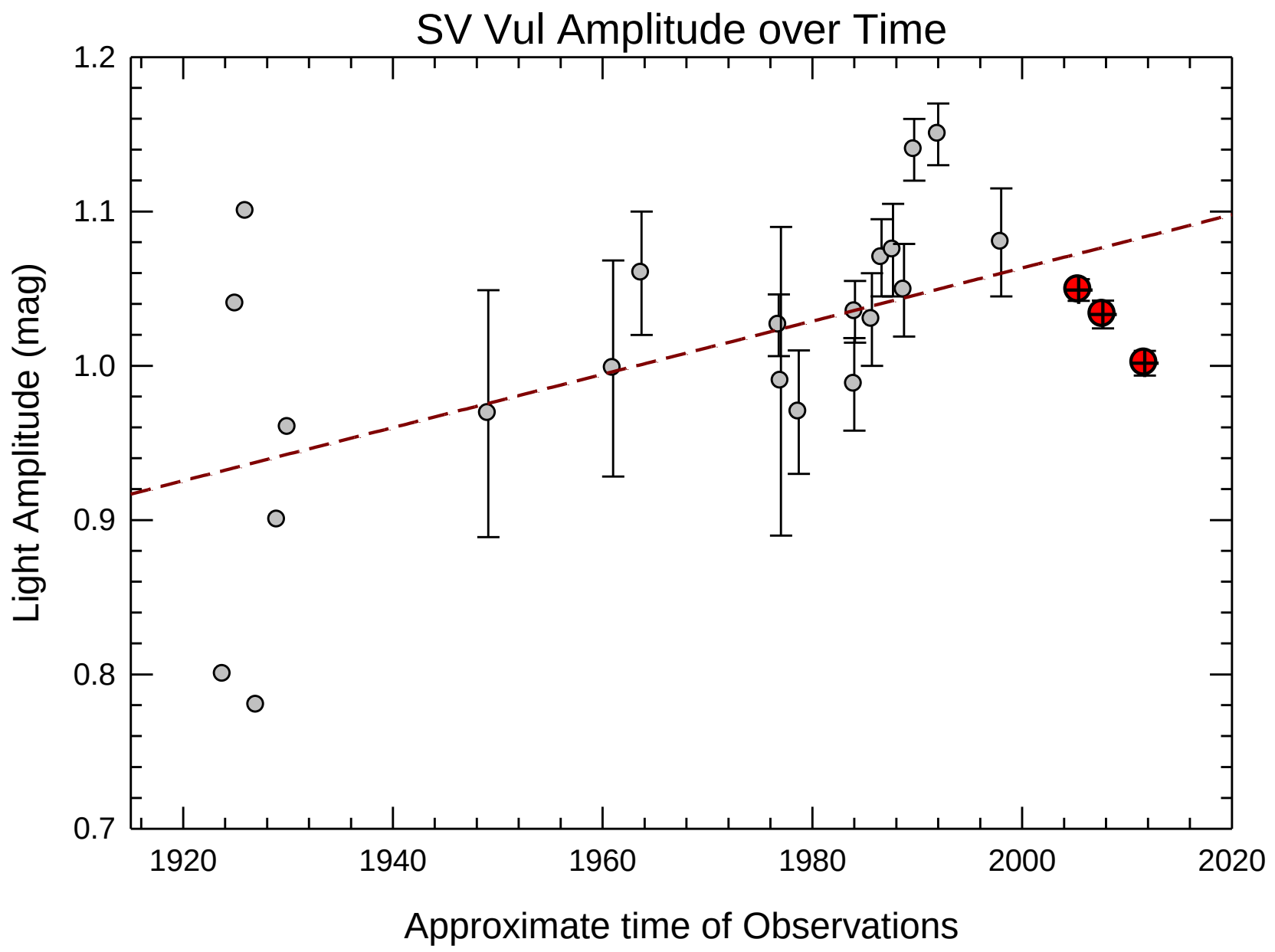

Figure 29 – The observed light amplitudes of SV Vul are plotted vs. the mid-time of the observation set. Points measured as part of this program are indicated by red filled, crossed circles. A possible linear trend of increasing amplitude over time is hinted at, but relies somewhat on the older, less accurate observations.

### 2.9 SZ Cas

Few studies have been conducted on the stellar parameters of SZ Cas, which lies near the *h* and $\chi$ Per double cluster, and is a possible member of the Per C1 star complex (de la Fuente Marcos & de la Fuente Marcos 2009). SZ Cas pulsates with a low amplitude (~0.38-mag in the V-band – see Fig. 30) for a Cepheid with its period, and the lightcurve has greater symmetry. Both are indicative of first overtone pulsations. Pop & Codreanu (2001) studied amplitude changes in SZ Cas, finding a ~11-year cycle of amplitude variability. Our study differs from theirs in that we combined the photometry of Berdnikov (1986) with Berdnikov (1992), to get a fully covered light curve, and we also believed it best to disregard the later Berdnikov (1995) dataset on the basis that a large data gap exists around maximum light, making an amplitude estimate unreliable at best.

**Table 10 – Relevant Stellar Properties of SZ Cas**

| | |
|---|---|
| Spectral Type | F6 – G4 Ib[1] |
| $T_{eff}$ (K) | ~5800 – 6400[1] |
| Mass (pulsational) ($M_{\odot}$) | ~6.5[2] |
| Mass (evolutionary) ($M_{\odot}$) | ~9[2] |
| Mean Luminosity ($L_{\odot}$) | ~8500[2] |
| Mean Radius ($R_{\odot}$) | ~70[3] |
| Distance (calculated – pc) | ~2100[1] |
| Ephemeris (this study)<br>2454142.258 + 13.637772(257) | |
| Ephemeris for O-C diagram (from Szabados 1981)<br>2443818.142 + 13.637747 | |

[1]Luck et al. (2006); Cox (1979); [3]Ivanov (1984)

As part of this program, two fully covered lightcurves of SZ Cas have been obtained, from which times of maximum light and amplitudes were calculated. Fig. 31 gives the O-C diagram for SZ Cas, which shows the historic steady increase in period, and the residuals to the quadratic fit. As with many Cepheids, however, observations have become sparse in recent decades. The given quadratic function yields a good fit for the data, showing a steady period increase of dP/dt = 39.187 ± 0.685 sec/yr, indicating a possible $5^{th}$ crossing of the instability strip (or perhaps a $3^{rd}$ crossing with larger period change, possibly due to enhanced mass loss) but there are two additional and interesting behaviors at work. The first is what could be a period change cycle, as seen in the residuals, similar to SV Vul but of a much longer period (almost the full length of the dataset gathered to date). The second is that the most recent data hint that the period has perhaps stopped increasing. Both behaviors will require further observations to confirm.

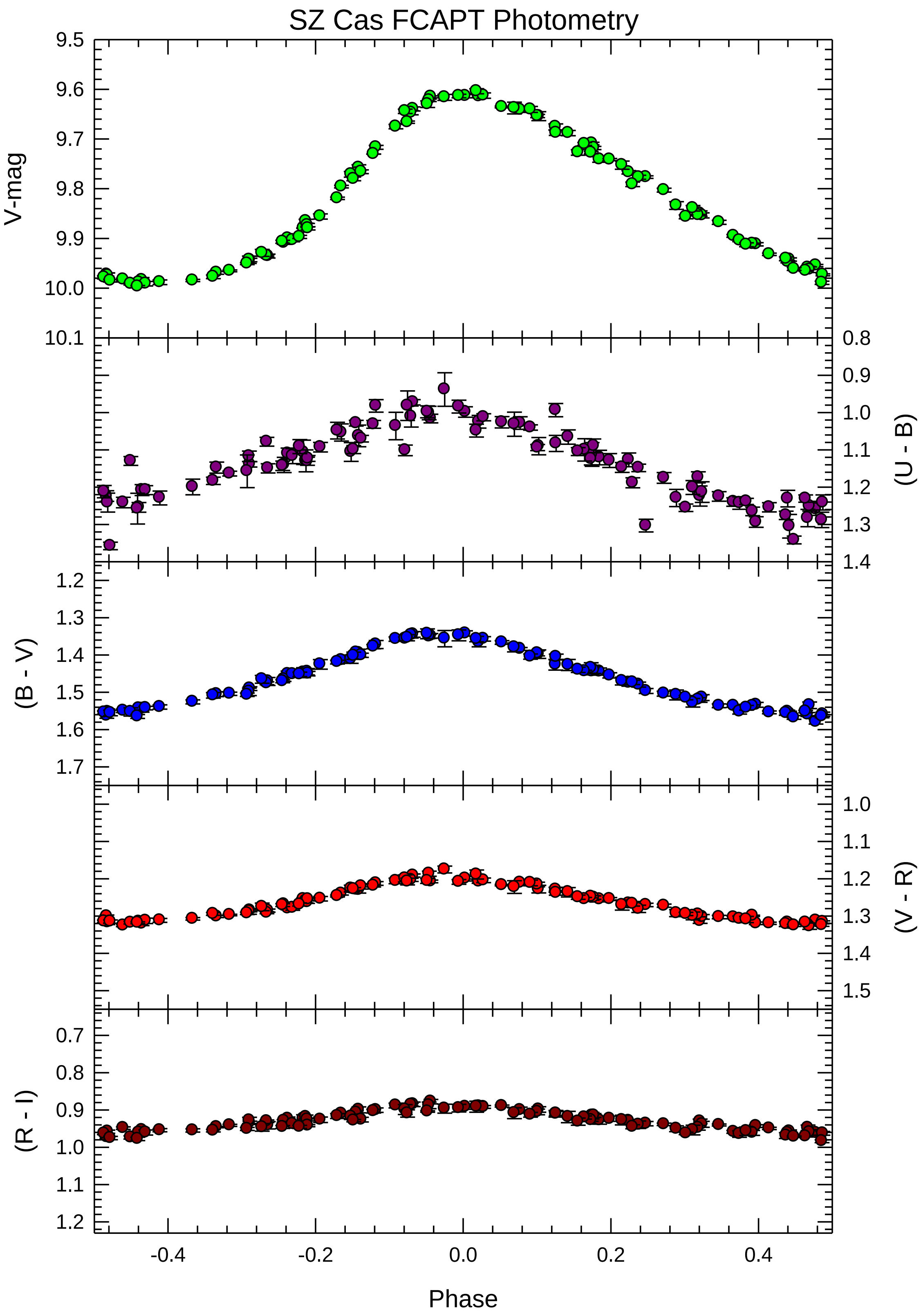

Figure 30 – The UBVRI data obtained for SZ Cas, phased to the new ephemeris calculated (given in Table 10).

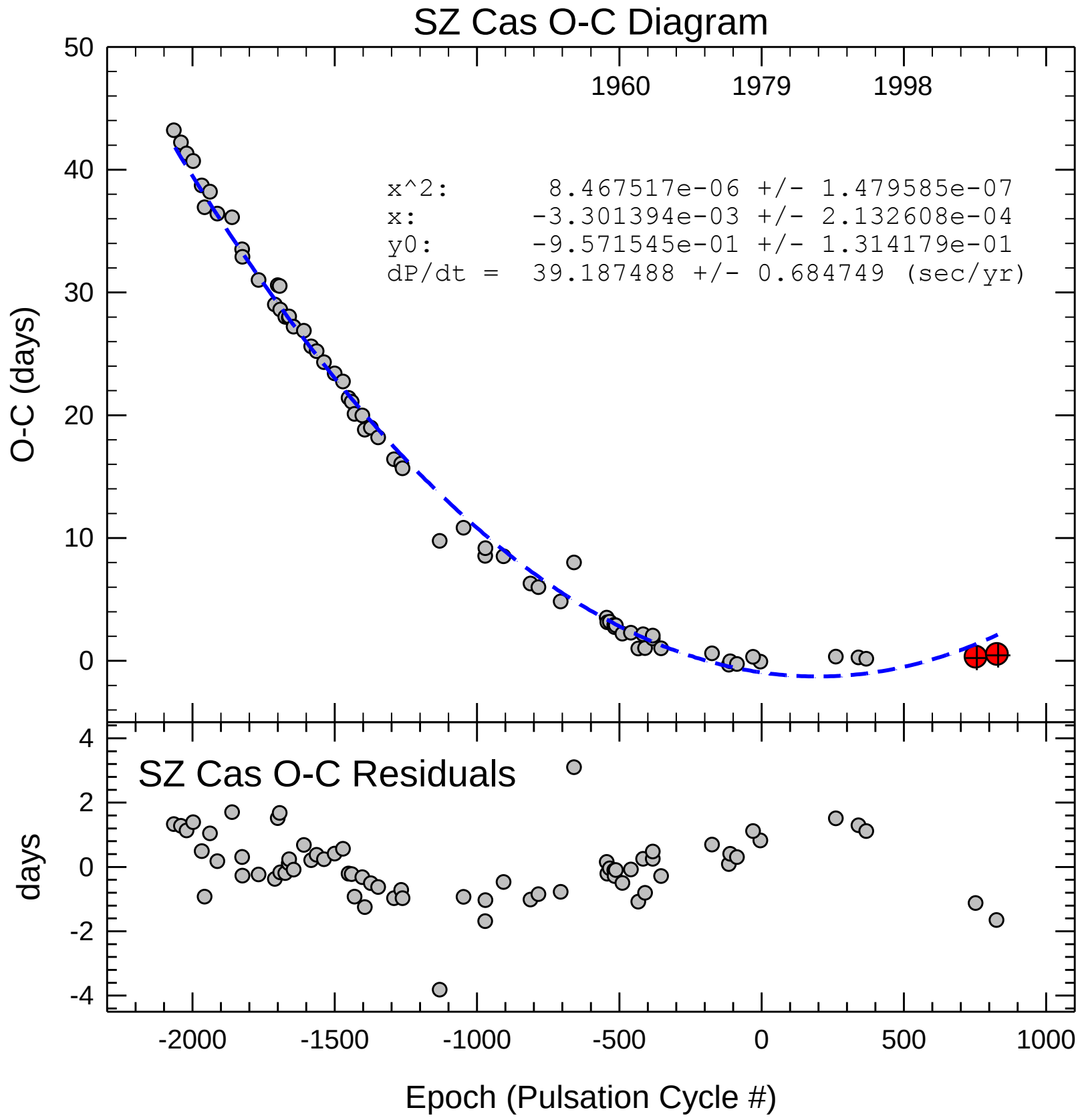

Figure 31 – The O-C diagram for SZ Cas, showing the long-term increasing period trend. Coefficients of the quadratic fit are given in the plot, along with the rate of period change (dP/dt). Points determined from this program are plotted as red filled and crossed circles. On top of the overall trend, there is evidence for a long-term possibly cyclic behavior, as shown in the residuals plotted in the lower panel, but the recent data of this program seem to break the cycle. It is also possible that the long-increasing period of SZ Cas has recently stabilized, but only further data will tell for sure.

Fig. 32 shows the amplitude values for SZ Cas over time. The amplitude appears to have undergone a noticeable decline over the past few decades. The amplitude variability does not seem to correlate to that of the period, as may be the case with SV Vul. However, the sparseness of the amplitude measures must yet again be taken into account. It does seem that amplitude changes have occurred in the past and hopefully future datasets will begin to provide a much clearer picture of the amplitude behavior of SZ Cas. If amplitudes going back to ~1960 are taken into account, a linear fit returns an amplitude change of –0.96 mmag/yr (data from ~1980 to present gives –1.4 mmag/yr), but the true amplitude behavior appears much more complex than a simple linear decrease.

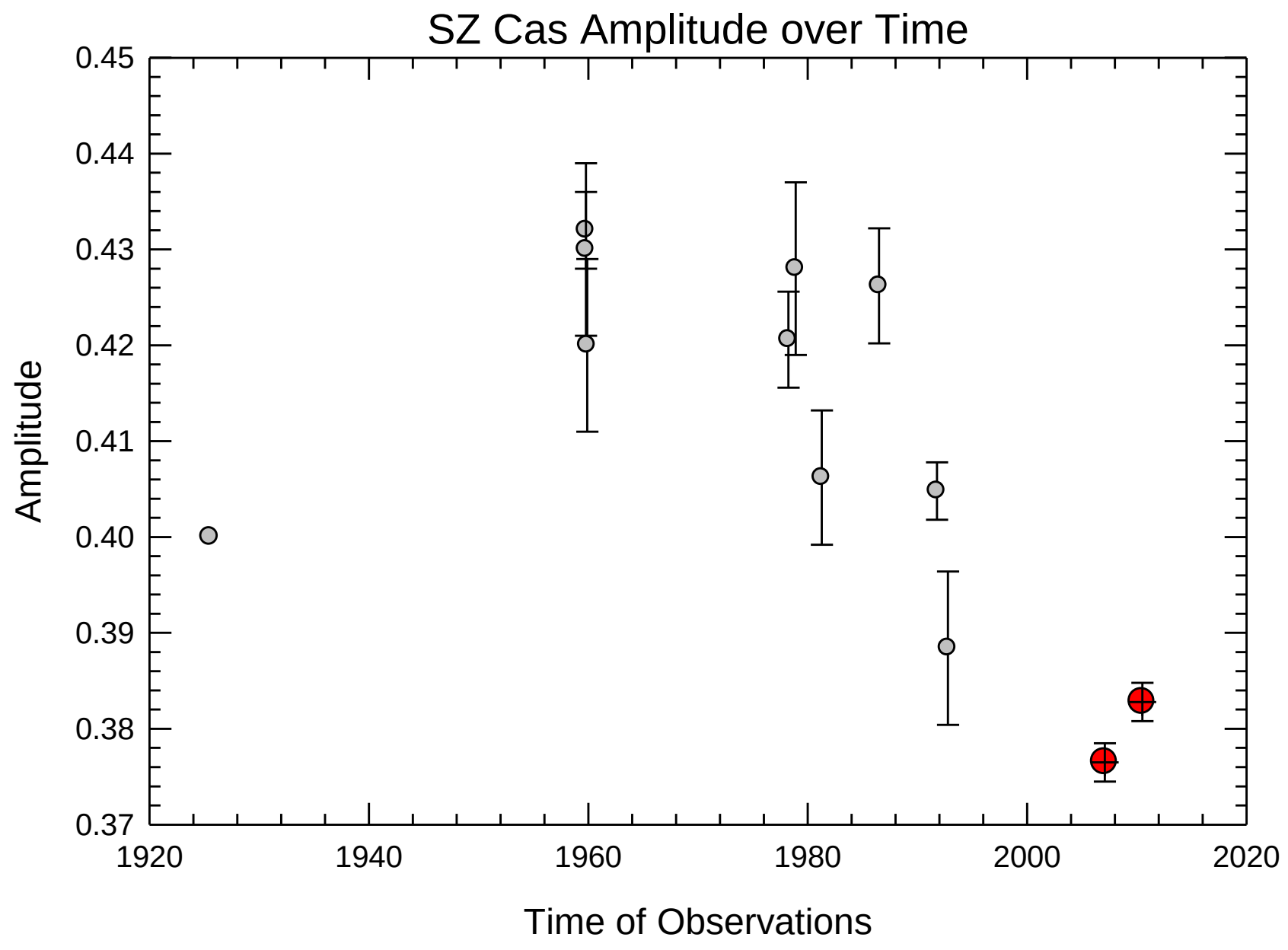

Figure 32 – The observed light amplitudes of SZ Cas are plotted vs. the mid-time of the observation set. Points measured as part of this program are indicated by red filled, crossed circles. There seems to be a very real variability, characterized mainly by a sharp decrease in amplitude in the 1980s and 90s, with a possible resurgence of the amplitude in the 2000s.

## 2.10 SZ Tau

The ~3.15-day Cepheid SZ Tau is suggested as a possible member member of the young open cluster NGC 1647 (Turner 1992; Rastorguev & Dambis 2011). However, it has also been ruled out as a possible member by Gieren et al. (1997) due to the difference between the cluster distance and the distance indicated for the Cepheid from the Leavitt Law. The lightcurve of SZ Tau displays a high symmetry, a lower V-amplitude of ~0.36 – 0.37-mag (see Fig. 33), and as such it is theorized to be a first overtone s-Cepheid (Postma 2008). Some relevant properties of SZ Tau are given in Table 11.

As part of this program, three fully covered lightcurves were obtained for SZ Tau, from which times of maximum light and amplitudes were calculated. The period variability of SZ Tau has been reported in several other studies, such as that of Berdnikov & Pastukhova (1995) whose data the O-C diagram shown in Fig. 34 is based on. Clearly the period of SZ Tau displays a rich and complex variability beyond what a simple quadratic fit could capture. However, it appears there is indeed a steady period change of dP/dt = –0.353 ± 0.156 sec/yr at work, indicating a possible $4^{th}$ crossing of the instability strip, but the overtone nature of this Cepheid lends some ambiguity to its exact crossing. The residuals plotted at the bottom of Fig. 34 reveal the complex additional variability present in the period. The residuals do not seem to show an obvious cycle to them, but a hand-drawn curve has been overplotted to show a possible cycle of ~59-yrs. It is unknown exactly why the period of SZ Tau exhibits such complex variability.

The amplitude behavior of SZ Tau over time seems to be a much simpler case, as seen in Fig. 35. A linear fit to the data reveals that SZ Tau's amplitude is currently increasing at a rate of ~0.8-mmag per year (0.08-mag per decade). There is a hint of cyclic variability in addition to the linear increase, but it is not concrete. It is interesting to note that the amplitude of SZ Tau has been increasing for essentially the entire ~100-yrs that have elapsed since its discovery as a Cepheid (Schwarzschild 1910). In fact, the light amplitude of SZ Tau is now almost 25% stronger than it was in the early 1900s.

**Table 11 – Relevant Stellar Properties of SZ Tau**

| | |
|---|---|
| Spectral Type | F6 – F9 Ib[1] |
| $T_{eff}$ (K) | ~5750 – 6300[1] |
| Mass (pulsational) ($M_\odot$) | ~4.9[2] |
| Mass (evolutionary) ($M_\odot$) | ~5.3[2] |
| Mean Luminosity ($L_\odot$) | ~2140[1] |
| Mean Radius ($R_\odot$) | ~42.5[1] |
| Distance (pc) | $417^{+185}_{-95}$[3] |
| Ephemeris (this study) 2454409.139 + 3.148841(11) | |
| Ephemeris for O-C diagram (Berdnikov & Pastukhova 1995) 2430600.957 + 3.148946 | |

[1]Postma (2008); [2]Gieren (1989); [3]van Leeuwen (2007)

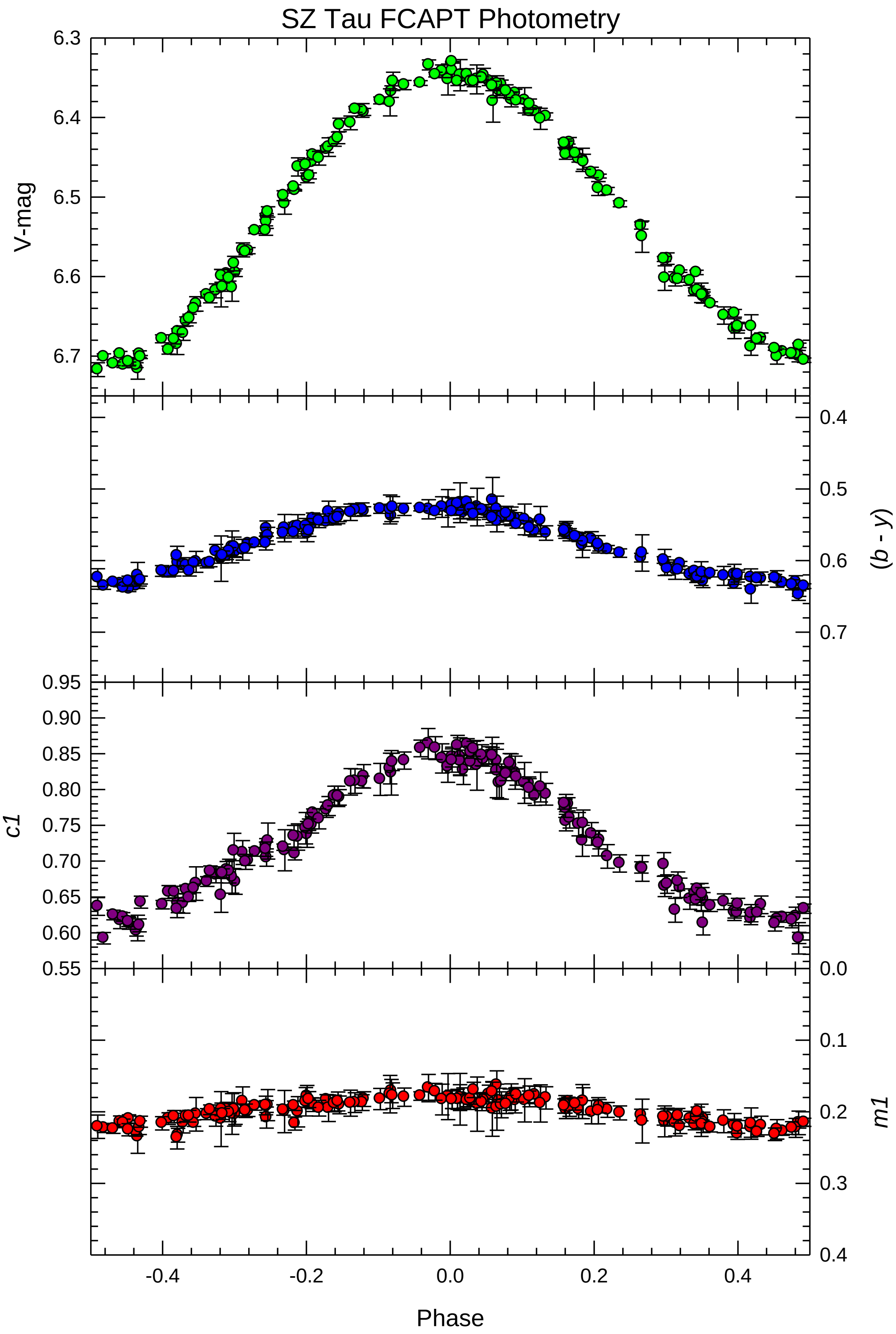

Figure 33 – The *uvby* data collected for SZ Tau. The *y*-band data have been transformed to standard *V*-band magnitudes, and phased to the newly calculated ephemeris given in Table 11.

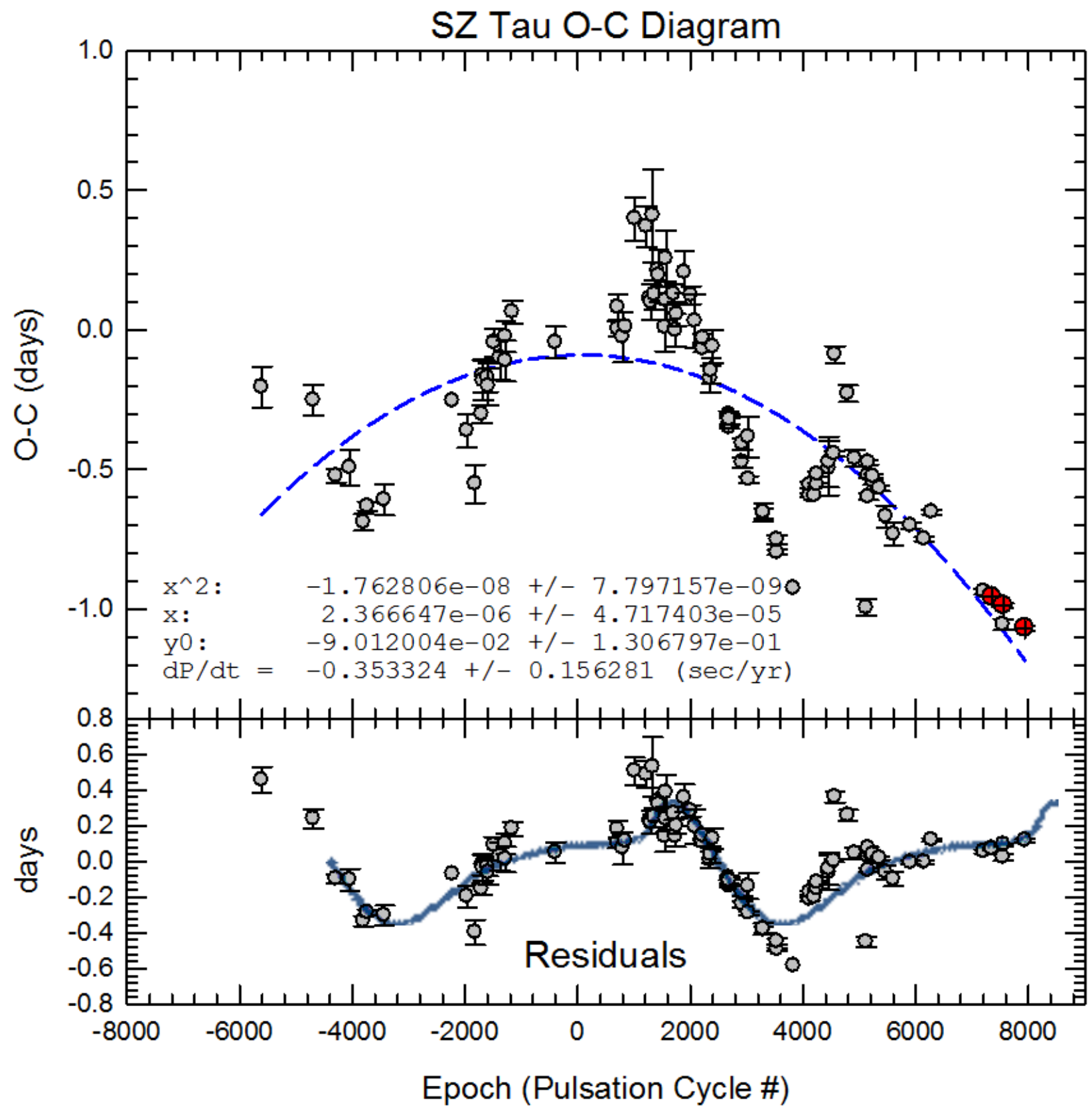

Figure 34 – The O-C diagram for SZ Tau, showing some very complex period behavior. There seems to be an overall trend of decreasing period (indicated by the fit), but with a possibly cyclic variability overlaid, as with SZ Cas and SV Vul. Further data is required to confirm the ~59-year cyclic behavior. Coefficients of the quadratic fit are given in the plot, along with the rate of period change (dP/dt). Points determined from this program are plotted as red filled and crossed circles.

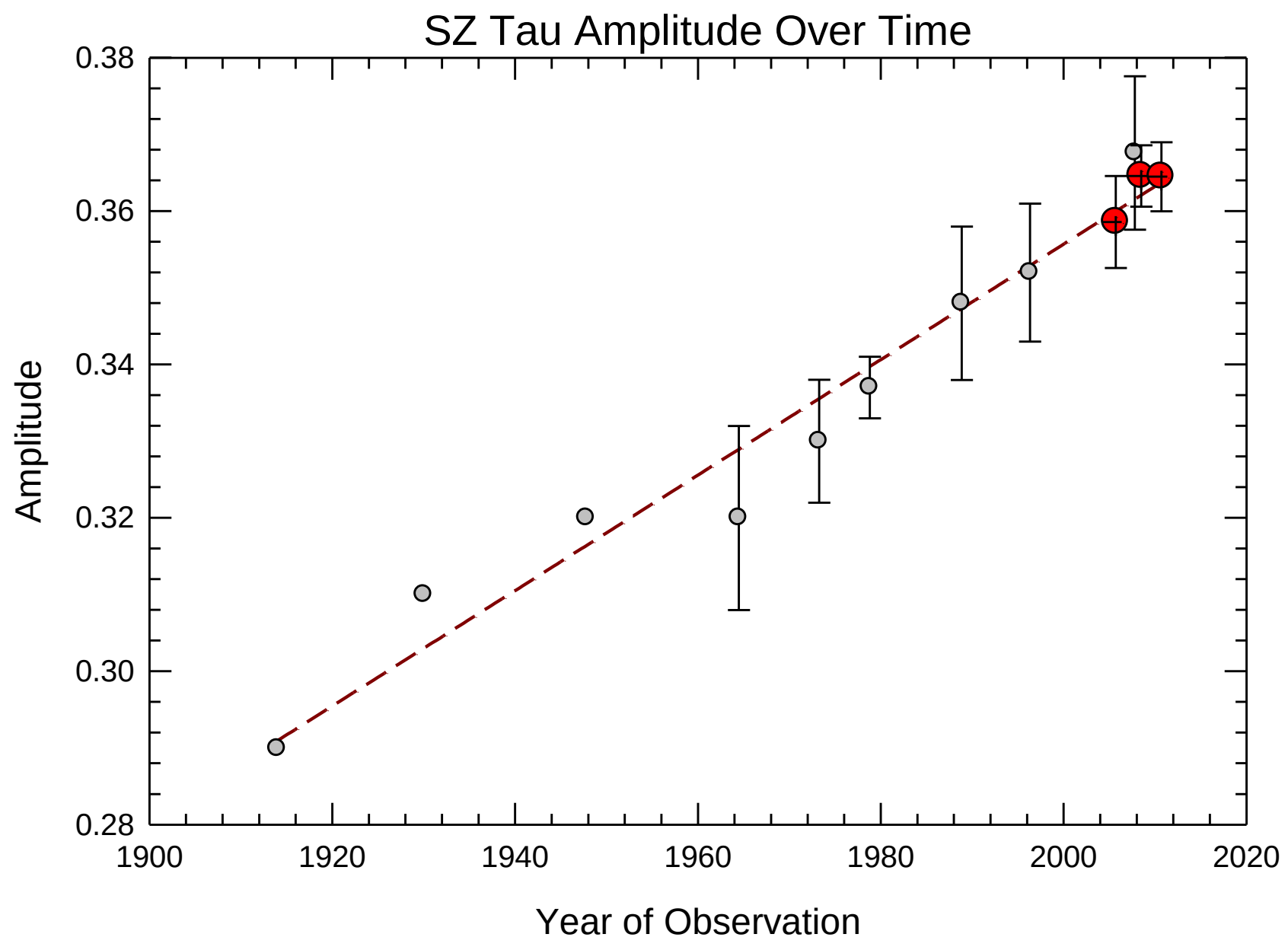

Figure 35 – The observed light amplitudes of SZ Tau are plotted vs. the mid-time of the observation set. Points measured as part of this program are indicated by red filled, crossed circles. A constant increasing trend is very obvious in the data. Hints of additional amplitude variability can be seen, but the sparseness of the data prevents firm conclusions.

### 2.11 VY Cyg

Along with SZ Cas, VY Cyg is one of the fainter Cepheids in our study and, as such, does not possess a rich observational history, like some of the brighter Cepheids. Klagyivik & Szabados (2009) infer that VY Cyg has an as yet unseen blue companion, based on its light amplitudes across different photometric bands when compared to other Cepheids. This would confirm the earlier results of Madore (1977) who suggested that VY Cyg has a B7 main sequence companion. VY Cyg has a revised *Hipparcos* parallax of $\pi_{Hipp}$ = 0.05 ± 1.12 mas (van Leeuwen 2007). As with EU Tau and SV Vul, the distance of this Cepheid is far too large to allow for a usable trigonometric parallax measurement. Given in Table 12 is the calculated distance of VY Cyg, taking into account the absolute magnitude determined by Kovtyukh et al. (2010) and the dereddened (Kovtyukh et al. 2008) apparent magnitude observed as part of this program, adjusted for the presence of the hot companion (Madore 1977).

**Table 12 – Relevant Stellar Properties of VY Cyg**

| | |
|---|---|
| Spectral Type | F6 – G1 Ib[1] |
| $T_{eff}$ (K) | ~5500 – 6300[2] |
| Mass (pulsational) ($M_\odot$) | ~4.5[3] |
| Mass (evolutionary) ($M_\odot$) | ~5[3] |
| Mean Luminosity ($L_\odot$) | ~2000[3] |
| Mean Radius ($R_\odot$) | ~49[4] |
| Distance (calculated – pc) | ~1600 (see text) |
| Ephemeris (this study)<br>2455852.511 + 7.857125(33) × E | |
| Ephemeris for O-C diagram (Szabados 1980)<br>2443045.282 + 7.856982 × E | |

[1]Luck et al. (2006); [2]Kovtyukh et al. (2010); [3]Gieren (1989);
[4]Cogan (1978)

As part of this study, three fully-covered lightcurves of VY Cyg have been obtained (Fig. 36), from which times of maximum light and amplitudes were determined. Fig. 37 shows the O-C diagram for VY Cyg. A quadratic fit has been run through the data, assuming a linear increase in period, which is calculated to be dP/dt = +0.249 ± 0.003 sec/yr, placing VY Cyg in its third crossing of the instability strip, very similar to η Aql which is of similar period, amplitude and stellar properties

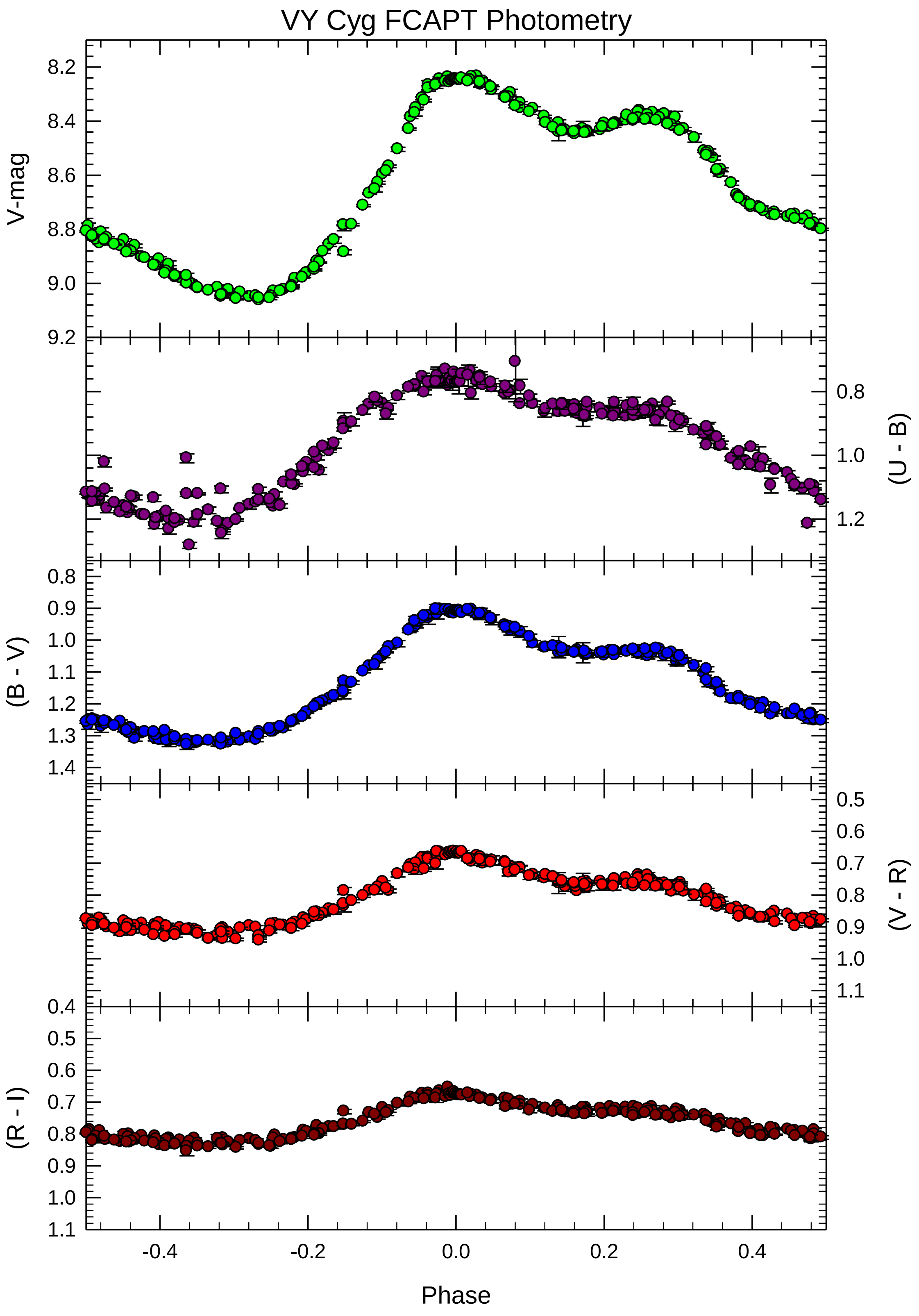

Figure 36 – The *UBVRI* data obtained for VY Cyg, phased to the new ephemeris determined in this study (given in Table 12).

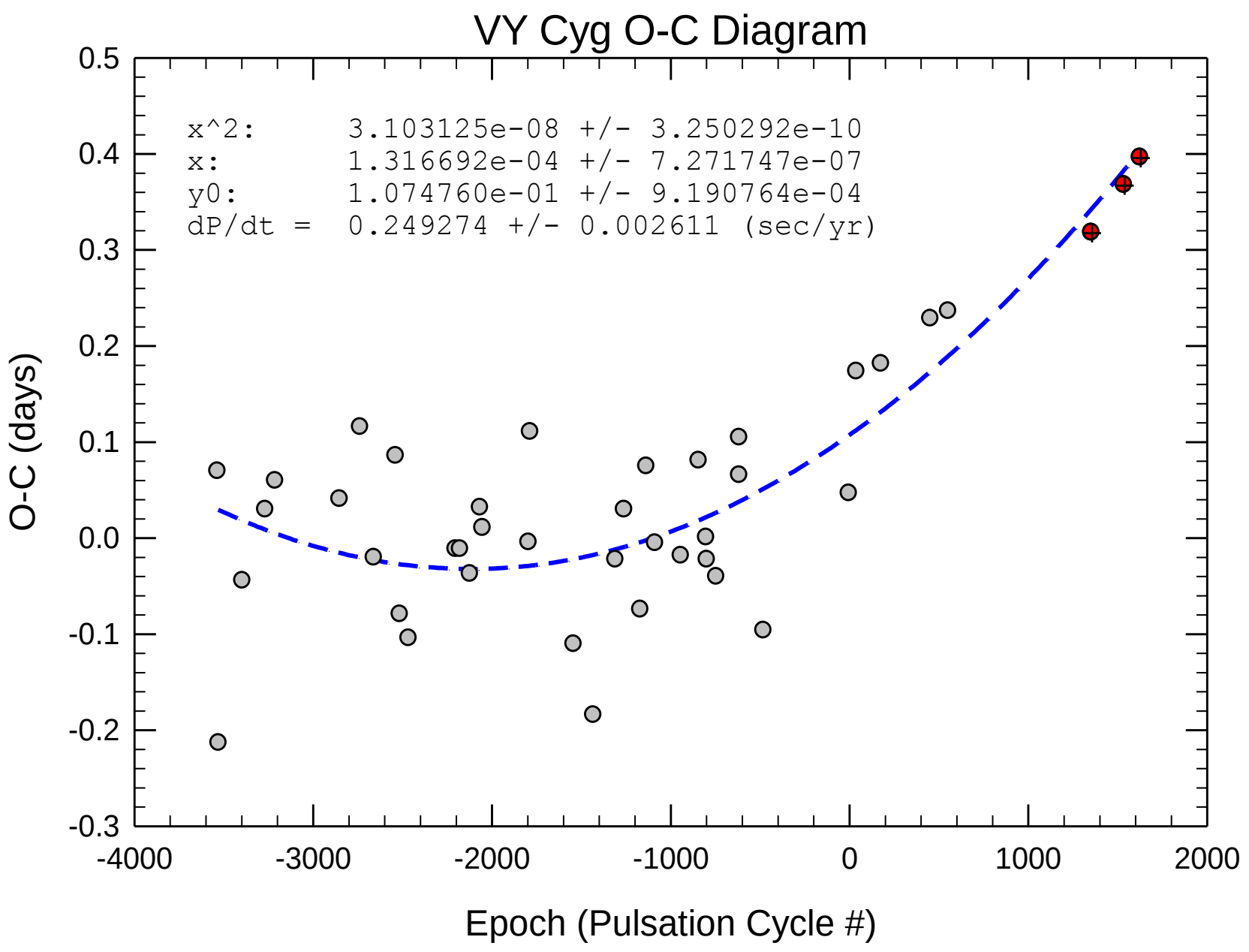

Figure 37 – The O-C diagram for VY Cyg. Our data provides further confirmation and characterization of the recent increasing period trend. Coefficients of the quadratic fit are given in the plot, along with the rate of period change (dP/dt). Points determined from this program are plotted as red filled and crossed circles.

The amplitude behavior of VY Cyg over time is shown in Fig. 38. A linear fit of the amplitudes is given, which indicates a steady period increase of ~0.6-mmag per year, but this is substantially influenced by lower amplitude values reported over a century ago. It is rational to assume that no linear trend in VY Cyg is confidently displayed in the data (coincidentally, as is also the case with η Aql). There is an appreciable spread in amplitudes reported in the past few decades, however, as with other Cepheids, the sparseness of the data precludes a deeper investigation of possible coherent variability in the amplitude over time.

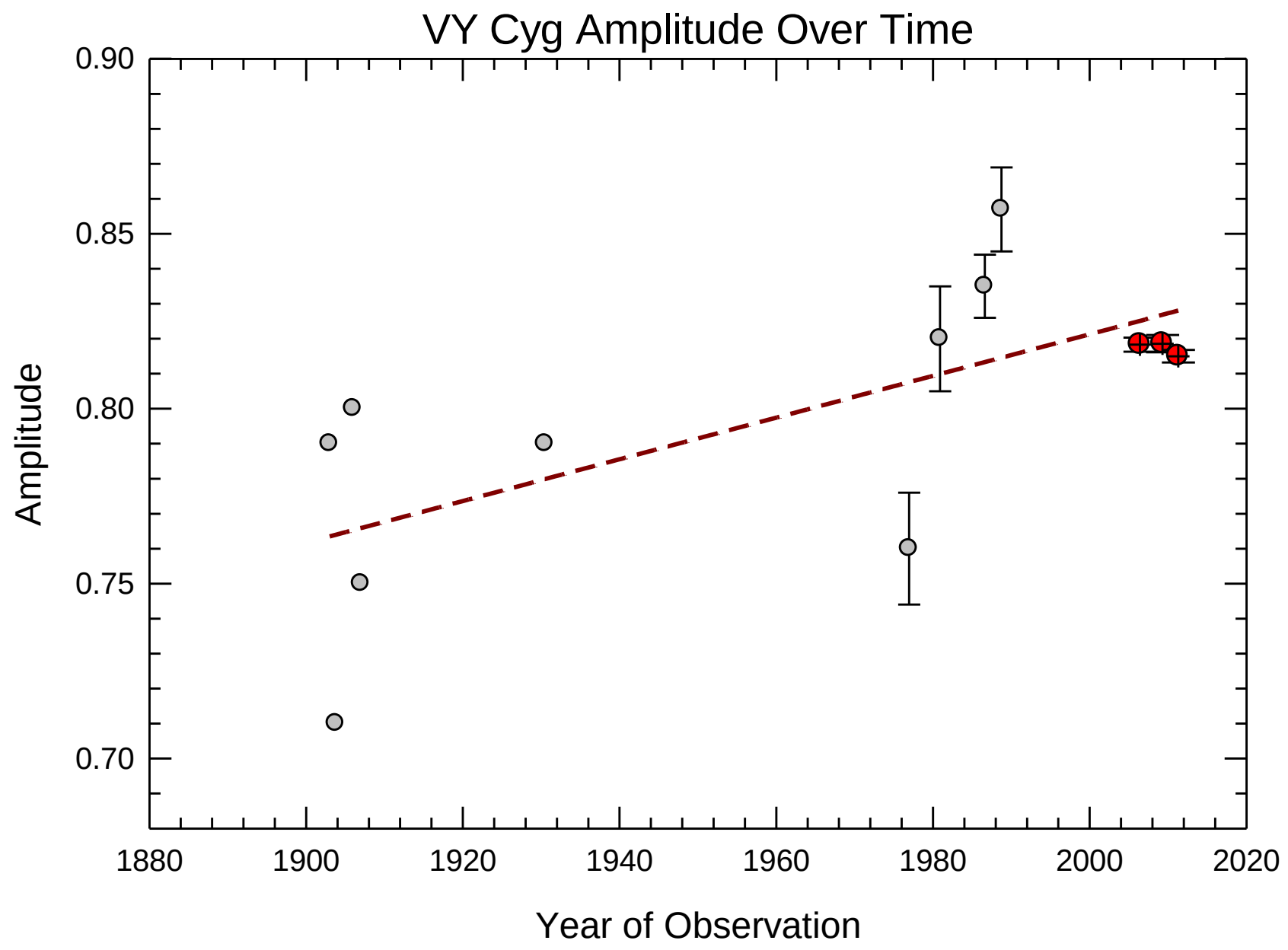

Figure 38 – The observed light amplitudes of VY Cyg are plotted vs. the mid-time of the observation set. Points measured as part of this program are indicated by red crosses. A constant increasing trend is assumed by the fit, but is not concrete since it mainly relies on the oldest, least accurate data. The amplitude measurements of the 1970s and 80s show a spread in amplitudes beyond the realm of observational error, but the lack of observations in the 1990s prevents a further investigation.

### 2.12 ζ Gem

The discovery of light variability in ζ Gem is attributed to Julius Schmidt in either 1844 or 1847, depending on the source (Allen (1899), p. 235; Henroteau (1925)). However, suspicions of variability were originally raised as early as 1790. Radial velocity variability was discovered independently by Belopolsky (1899) and Campbell (1899), and comparisons to δ Cep started shortly thereafter. In fact, for some time after, δ Cep and ζ Gem both enjoyed elevated status as the prototypes of two separate but related classes of variable stars – the δ Cep-type stars, or Cepheids, and the ζ Gem-type stars, or Geminids, to which other more symmetric Cepheids like SZ Tau and Polaris were assigned (Ludendorff (1916) and Henroteau (1925)). To date, no unresolvable companions have been detected, but ζ Gem does have a resolved F3.5 – 4 V companion, ζ Gem B, which lies ~1.4-arcmin from the Cepheid (Majaess et al. 2012). Majaess et al. have also concluded that ζ Gem is the member of a newly discovered cluster, making it an important calibrator Cepheid, for which fundamental stellar parameters, and distance, can be obtained in a variety of ways and then compared.

**Table 13 – Relevant Stellar Properties of ζ Gem**

| | |
|---|---|
| Spectral Type | F7 – G3 Ib[1] |
| $T_{eff}$ (K) | ~5200 – 5800[2] |
| Mass (pulsational) ($M_{\odot}$) | ~5.5[3] |
| Mass (evolutionary) ($M_{\odot}$) | ~6[3] |
| Mean Luminosity ($L_{\odot}$) | ~2900[4] |
| Mean Radius ($R_{\odot}$) | ~65[3] |
| Distance (pc) | $360^{+25}_{-22}$ [5] |
| Ephemeris (this study)<br>2451909.770 + 10.149401(101) | |
| Ephemeris for O-C diagram (Berdnikov et al. 2000)<br>2423838.469 + 10.1522686(135) | |

[1]Kervella et al. (2001); [2]Luck et al. (2008); [3]Gieren (1989); [4]Mallik (1999); [5]Benedict et al. (2007)

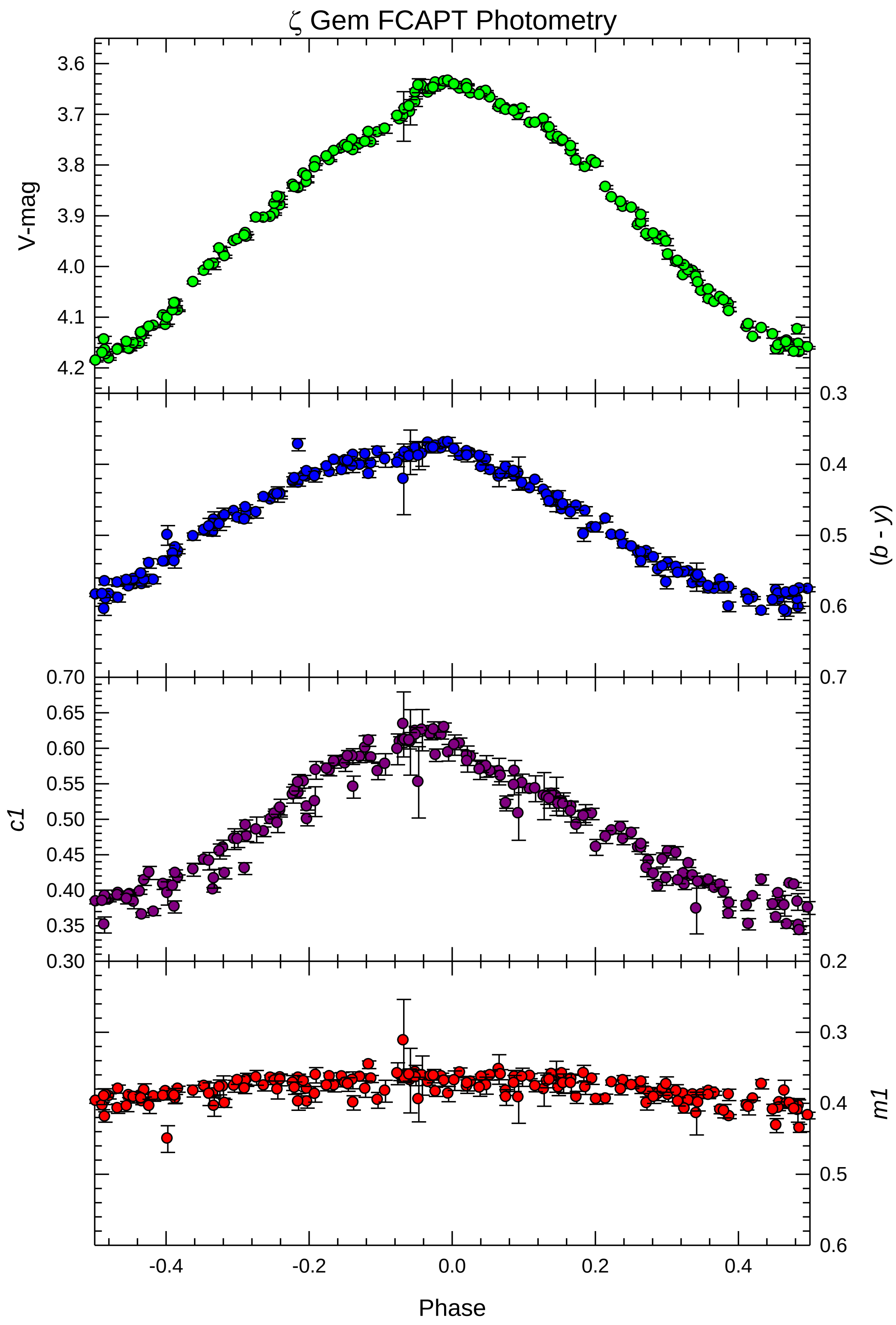

Figure 39 – The *uvby* data collected for ζ Gem. The *y*-band data have been transformed to standard *V*-band magnitudes, and phased to the newly calculated ephemeris given in Table 13.

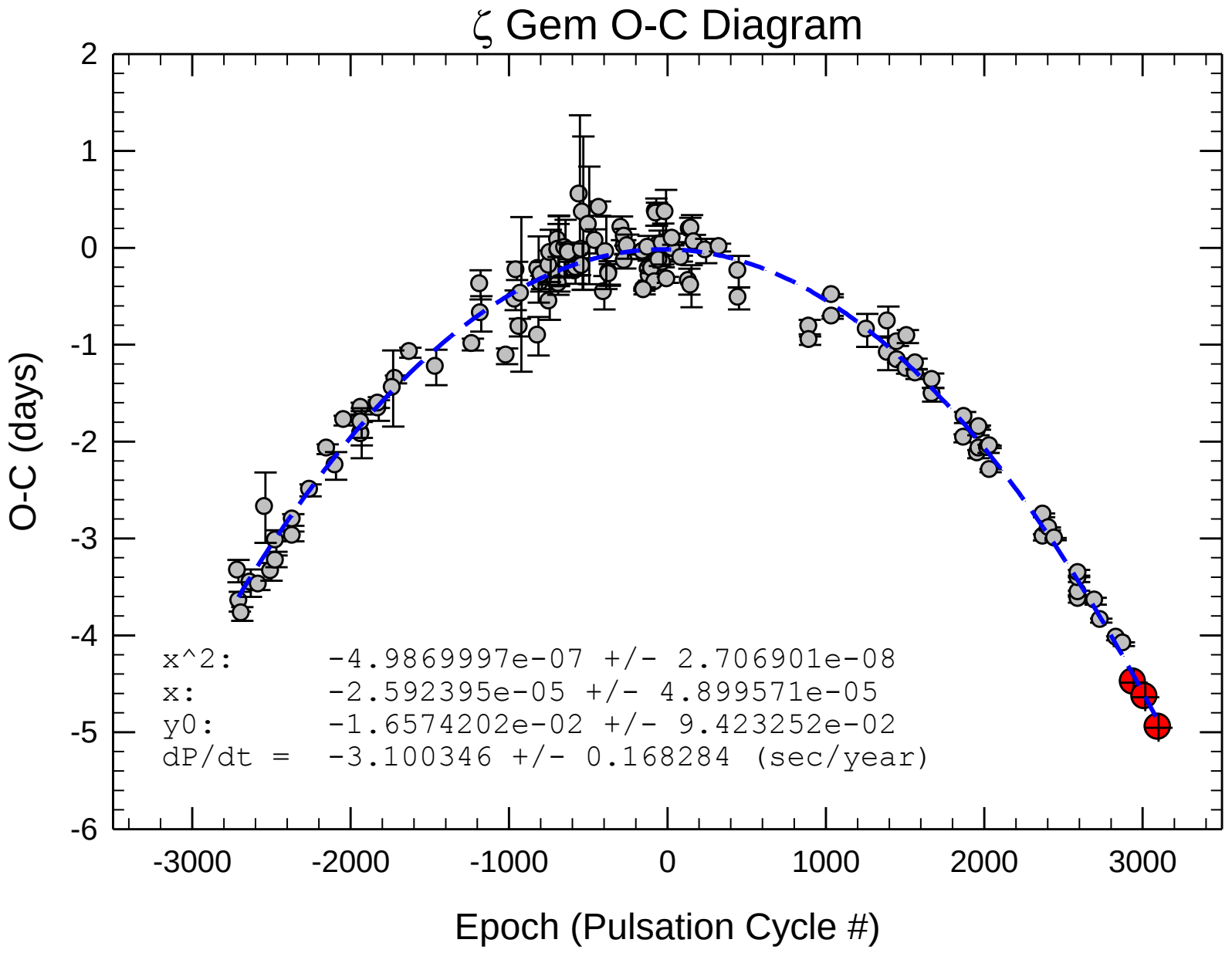

Figure 40 – The O-C diagram for ζ Gem, with the long-term, steadily decreasing period trend easily visible. No hints of further variability are evident. Coefficients of the quadratic fit are given in the plot, along with the rate of period change (dP/dt). Points determined from this program are plotted as red filled and crossed circles.

Three fully covered lightcurves have been obtained for ζ Gem in this program (Fig. 39). Fig. 40 shows the O-C diagram for ζ Gem, where the period is seen to have been undergoing a smooth, steady decline for the full ~160-yr dataset. No coherent variability shows up in the residuals to the quadratic fit, which gives a period change rate of dP/dt = –3.100 ± 0.011 sec/yr. This rate alone could place ζ Gem in either the second or fourth crossing of the instability strip. However, Turner (2010) has classified it as a fourth crossing Cepheid, since its light amplitude is below that expected for a second crossing Cepheid and its color is better matched to a fourth crossing Cepheid. The amplitude behavior of ζ Gem over time shows a probable linear increase, with the best fit giving ~1-mmag per year (0.01-mag per decade – see Fig. 41). On top of the linear trend, however, the more recent data shows a possible ~50-yr cycle to the amplitude. If this cycle is true, it will naturally take a couple decades of regular observation to definitively confirm.

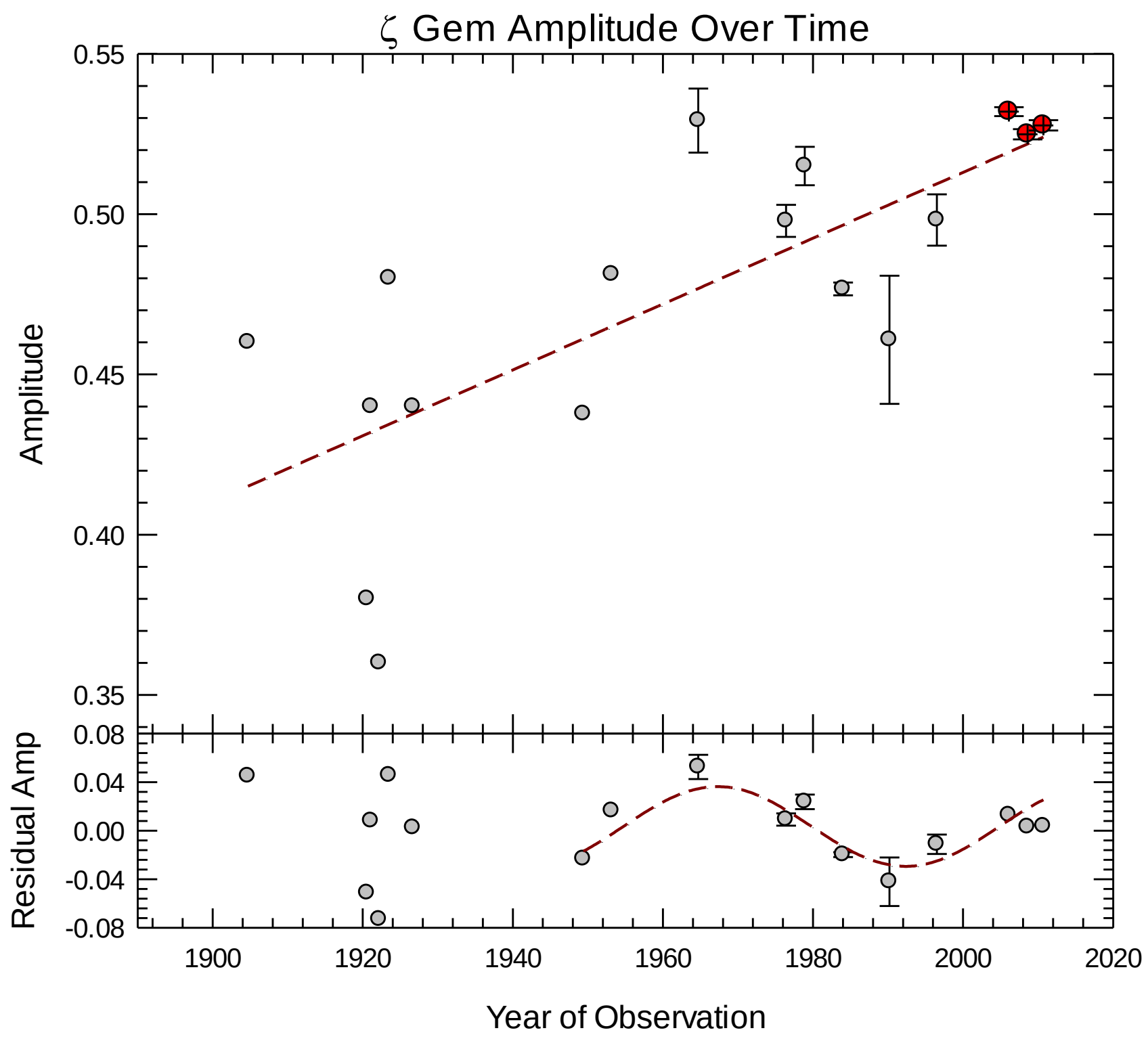

Figure 41 – The observed light amplitudes of ζ Gem are plotted vs. the mid-time of the observation set. Points measured as part of this program are indicated by red crosses. A constant increasing trend is assumed by the fit, but is not concrete since it is strongly influenced by the two low amplitudes around 1920. The amplitude measurements from the 1940s until recently show a possible cyclic behavior of ~50-years (see residuals), but further data is necessary to confirm.

### 2.13 β Dor

The radial velocity variability of β Dor was discovered by Palmer through observations taken in 1930/1904 (Applegate 1927). Wilson was the first to note, at an *American Astronomical Society* meeting, the similarities between β Dor and other Cepheids, suggesting that it might be a member of the class ([*the paper is listed in the ADS as*] N/A 1921). The first confirmation of brightness variability for β Dor is that of Shapley & Walton (1927). Unfortunately, with a declination below -62º, β Dor is a Cepheid that most ground-based telescopes in the northern hemisphere will be unable to observe. As such, no photometric observations have been carried out for β Dor as a part of this program. Nevertheless, it is one of the nearest and brightest Cepheids (see Table 14) which is a large advantage for spectroscopic studies, as a sufficient signal-to-noise value can be quickly obtained. Table 14 gives some important stellar properties for β Dor.

**Table 14 – Relevant Stellar Properties of β Dor**

| | |
|---|---|
| Spectral Type | F4 – G4 Ia – II[1] |
| $T_{eff}$ (K) | ~5500[1] |
| Mass (pulsational) ($M_{\odot}$) | ~6[3] |
| Mass (evolutionary) ($M_{\odot}$) | ~7[3] |
| Mean Luminosity ($L_{\odot}$) | ~3200[4] |
| Mean Radius ($R_{\odot}$) | ~68[2] |
| Distance (calculated – pc) | $318^{+18}_{-15}$ [5] |
| Ephemeris (Berdnikov et al. 2003)<br>2440905.3 + 9.8426 × E | |
| Ephemeris used for O-C Diagram<br>2425659.1126 + 9.8426 × E | |

[1]Kervella et al. (2004); [2]Taylor & Booth (1998); [4]Turner (2010); [3]Gieren (1989); [5]Benedict et al. (2007)

Although no new photometry was carried out for β Dor as part of this study, recent AAVSO observations exist. We acknowledge with thanks the 2012-2013 observations of β Dor, carried out by Neil Butterworth, and accessed through the *AAVSO International Database*. A single time of maximum light was obtained from the V-band data, and added to the O-C data presented by Berdnikov et al. (2003 – *data obtained by private communication*). Fig. 42 shows the O-C diagram for β Dor. A parabola is fit to the data, showing a slow, steady increase in the pulsation period of dP/dt = 0.468 ± 0.016 sec/yr. This rate of period increase places β Dor well within the third crossing of the instability strip, and represents a good fit to the dataset, although a second interpretation of the data is also possible, where the period underwent a sudden lengthening around Epoch 1800 (calendar year 1977), and has since held steady at 9.8426-days.

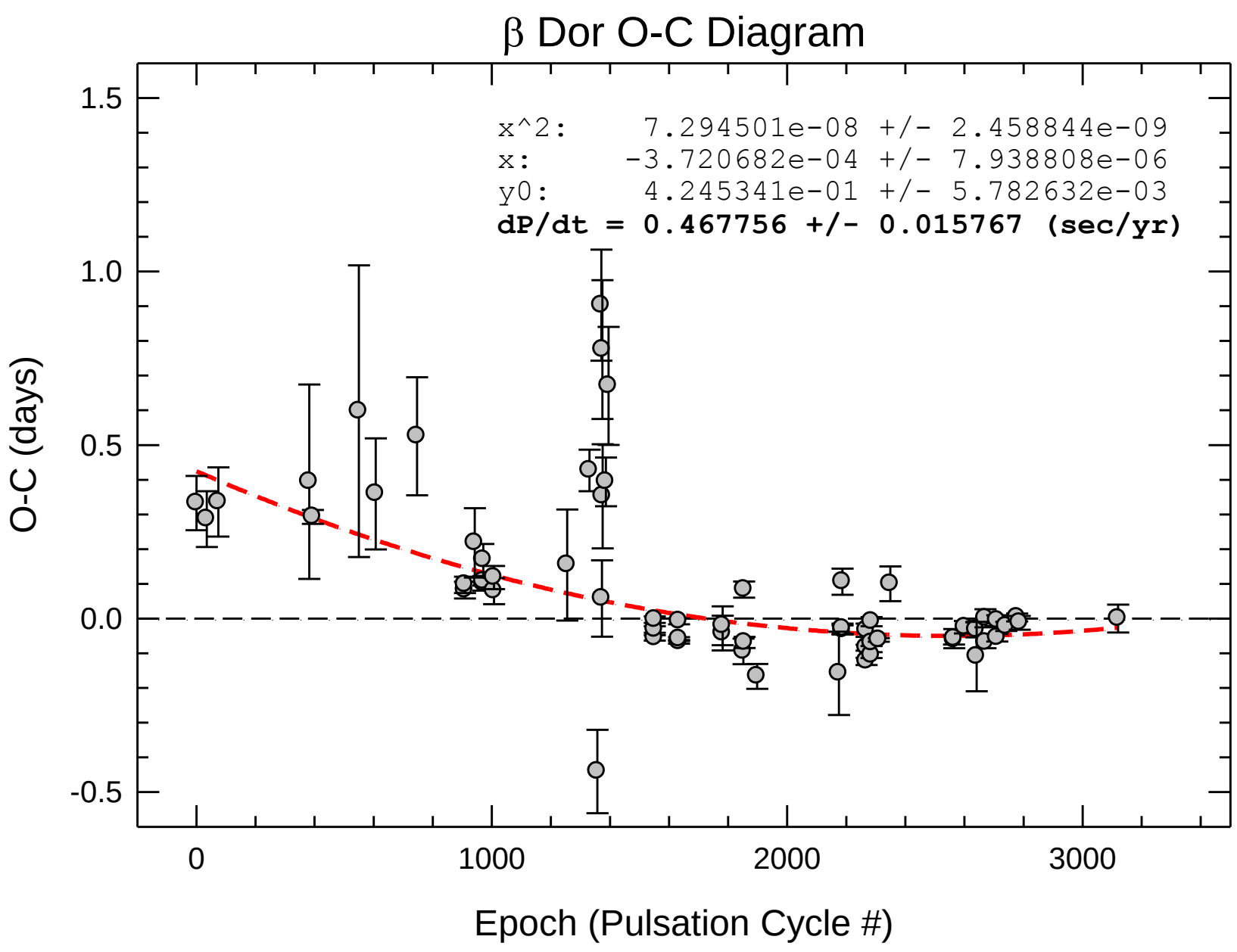

Figure 42 – The O-C diagram for β Dor, with the long-term, steadily increasing period trend shown. No hints of further variability are evident. Coefficients of the quadratic fit are given in the plot, along with the rate of period change (dP/dt). The latest data point is that determined from the 2012 – 2013 AAVSO data (Observer: Neil Butterworth).

## 2.14 A Brief Summary of Chapter 2

New *UBVRI* or *uvby* photometry has been gathered for ten Cepheids, to determine changes in pulsation period and light amplitude. Although no new photometry of the Cepheid β Dor could be gathered in this study, since it is a southern hemisphere target, an updated O-C diagram was built using available literature data and recent photometry gathered by AAVSO observer Neil Butterworth. Thus, the optical study currently contains period studies for eleven Cepheids, and amplitude studies for ten. Table 15 briefly describes the results. All eleven Cepheids display some form of period variability. Six of the eleven Cepheids (δ Cep, η Aql, Polaris, VY Cyg, ζ Gem and β Dor) show steady, persistent increases/decreases in period, and an additional two (EU Tau and SU Cas) display either a slow, steady period change or a sudden shift in period. The remaining three Cepheids (SV Vul, SZ Cas and SZ Tau) display a combination of steady period change and a cyclic variation in the rate of period change, as shown in the residual plots below their O-C diagrams.

The amplitude datasets are much sparser than the O-C diagrams, mainly because the phase-ranges around both maximum *and* minimum light must be well-covered. Nevertheless, there is only one Cepheid, η Aql, which shows no evidence of coherent amplitude variability. EU Tau displays a possible, albeit very small, decrease in amplitude. However, given the margins of error, a steadily decreasing amplitude is unlikely. VY Cyg shows a large spread in recent amplitude values beyond the margin of errors, but a steadily changing amplitude is unlikely. For SV Vul, a long-term amplitude change is merely a possibility due to the large spread of early amplitude measures. However, the Cepheid does display a recent amplitude cycle of similar length (~30-yr) to variations also observed in its rate of period change. This is an interesting behavior, but one that will likely take years of additional data to completely understand. The amplitude of Polaris was once thought to be steadily declining, but is now increasing again and it is unknown if the amplitude will simply attain and hold some previous, larger value or if the Cepheid is undergoing an amplitude cycle. SZ Cas also shows an amplitude that recently declined but now appears to be growing. However, the data is too sparse to present a firm conclusion. SU Cas shows a slowly increasing amplitude, with a small possibility of cyclic behavior. Similarly, SZ Tau shows a reliable, long-term amplitude increase with only a small possibility of additional variations. Both δ Cep and ζ Gem show long-term increases in its amplitude, but with stronger evidence recent additional, cyclic behaviors.

**Table 15 – Period / Amplitude Behaviors of Program Cepheids**

| **Cepheid** | **Period (days)** | **Period Change (sec/yr)** | **Crossing** | **Long- and Short-term (right column, if present) Amplitude Change (mmag/yr)** | |
|---|---|---|---|---|---|
| SU Cas | 1.949 | +0.0204 (or abrupt shift) | $3^{rd}$ | +0.8 | Possible |
| EU Tau | 2.102 | –0.337 (or abrupt shift) | $4^{th}$ | –0.2 (tentative result; likely within errors) | |
| SZ Tau | 3.149 | –0.353 (with complex additional variability) | $4^{th}$ ? | +0.8 | Hinted at, but recent data would not fit the expectations. |
| Polaris | 3.972 | +4.470 | $1^{st}$ | +2.96 (current) | Previously declining, currently increasing – possible cycle. |
| δ Cep | 5.366 | –0.1006 | $2^{nd}$ | +1.4 | Possible |
| η Aql | 7.177 | +0.255 | $3^{rd}$ | No coherent amp. variability | |
| VY Cyg | 7.857 | +0.249 | $3^{rd}$ | +0.6 (unlikely) | Large spread in amplitude, but sparse data limit conclusions. |
| β Dor | 9.843 | +0.468 | $3^{rd}$ | -- | -- |
| ζ Gem | 10.149 | –3.1003 | $4^{th}$ | +1.03 | Possible ~50-yr cycle. |
| SZ Cas | 13.638 | +39.187 (with possible additional cycle) | $3^{rd}$ ($5^{th}$ ?) | -0.96 | Complex: possible recent increase after previous decline |
| SV Vul | 44.993 | –231.223 (with likely ~30-yr cycle) | $2^{nd}$ | +1.7 (possible) | A ~30-yr cycle, similar to the period behavior. |

# CHAPTER 3 – THE HIGH-ENERGY (UV – X-RAY) STUDIES

Several high-energy (specifically UV and X-ray, in this program) studies of Cepheids have been carried out in previous decades, using earlier-generation instruments, most notably *IUE*, *Einstein* and *ROSAT*. The results of these various studies lead us to question just what Cepheid atmospheres would look like when viewed through modern instruments, such as HST, Chandra and XMM. The current Cepheid sample for the high-energy study is unfortunately more limited than the optical study. Two main factors are primarily responsible for this.

First, although Cepheids are optically luminous, allowing the photometric observations to be efficiently carried out even with small telescopes, they are *far* less luminous in the UV and X-ray regimes. Achieving a minimally sufficient signal-to-noise (S/N) ratio at UV and X-ray wavelengths ordinarily requires kiloseconds of exposure time. Fortunately for this study (and numerous others), COS is one of the most sensitive UV spectrographs ever flown. Using roughly one half of an HST orbit per observation, after target acquisition (usually ~900–1100-sec), we were able to achieve S/N ratios of ~10 – 20 for the emission lines studied. For the X-ray images, however, exposure lengths of ~30 – 70-ksec (almost 20 hours) are necessary to secure useful data.

The second factor is the availability of UV and X-ray instruments. Villanova has guaranteed access to the ground-based telescopes used for the optical studies. When dealing with satellite instruments, the competition is very strong and the observing time is extremely precious. The exposure times required, the need for multiple observations of each target to achieve phase coverage, and the willingness of time allocation committees, all combine to dictate the pace at which the high-energy study can proceed. Nevertheless, we are extremely fortunate to have gathered as much data as we have over the past few years, and we are continually attempting to expand the Cepheid high-energy database.

### 3.1 Ultraviolet Studies with HST-COS

During 2009-13, we were awarded 29 HST orbits for COS spectroscopy of three Cepheids: two visits (2 orbits/visit) for Polaris, and 19 visits (1 orbit/visit) for δ Cep & β Dor and 6 visits (1 orbit/visit) for the ~35-day Cepheid ℓ Car. Table 15 summarizes the important information on the COS Cepheid observations carried out in time for inclusion in the current study. Our targets were well-acquired, and the spectra required no special processing. Thus, the data used in this study are the CALCOS pipeline-processed data available from the MAST archive (http://archive.stsci.edu). The original aim of the COS observations was to improve upon previous studies using the IUE satellite, since what appeared to be photospheric continuum flux added to uncertainties in emission line identification and flux measures. The hope was that COS, with its finer spectral resolution, high-sensitivity and low noise, would return superior spectra for which much more precise measurements could be carried out. In addition, closely-spaced emission lines that were blended together in IUE spectra could be individually resolved in COS data and perhaps even less-prominent emission lines would be detected. Since the only reliable data in the ~1200 – 1600 Å region available for Cepheids (at the beginning of this project) was low-resolution (~1–2 Å) IUE spectra, it was quite frankly unknown exactly what the COS observations would show.

**Table 16 – The HST-COS "Cepheid Inventory" To Date**

| Target Name | Dataset | Start Time (UT) | Stop Time (UT) | Exp Time | COS Grating | Central Wavelength |
|---|---|---|---|---|---|---|
| POLARIS | LB5001010 | 12/25/2009 8:56 | 12/25/2009 9:39 | 2364.384 | G130M | 1281.623 |
| POLARIS | LB5001020 | 12/25/2009 10:18 | 12/25/2009 10:49 | 1400.064 | G160M | 1581.854 |
| POLARIS | LB5001030 | 12/25/2009 10:54 | 12/25/2009 10:58 | 243 | G185M | 1914.848 |
| POLARIS | LB5001040 | 12/25/2009 11:03 | 12/25/2009 11:07 | 244 | G185M | 1941.725 |
| POLARIS | LB5001050 | 12/25/2009 11:11 | 12/25/2009 11:15 | 243 | G185M | 1883.428 |
| POLARIS | LB5002010 | 12/27/2009 8:51 | 12/27/2009 9:33 | 2364.384 | G130M | 1281.646 |
| POLARIS | LB5002020 | 12/27/2009 10:12 | 12/27/2009 10:43 | 1400.096 | G160M | 1581.87 |
| POLARIS | LB5002030 | 12/27/2009 10:49 | 12/27/2009 10:53 | 243 | G185M | 1914.413 |
| POLARIS | LB5002040 | 12/27/2009 10:57 | 12/27/2009 11:01 | 244 | G185M | 1941.38 |
| POLARIS | LB5002050 | 12/27/2009 11:05 | 12/27/2009 11:09 | 243 | G185M | 1883.207 |
| V-DEL-CEP | LBK809010 | 10/19/2010 0:12 | 10/19/2010 0:30 | 924.992 | G130M | 1291 |
| V-DEL-CEP | LBK809020 | 10/19/2010 1:30 | 10/19/2010 1:48 | 923.968 | G160M | 1589 |
| V-DEL-CEP | LBK817010 | 12/12/2010 6:12 | 12/12/2010 6:29 | 924.992 | G130M | 1291 |
| V-DEL-CEP | LBK817020 | 12/12/2010 6:34 | 12/12/2010 6:52 | 923.904 | G160M | 1589 |
| V-DEL-CEP | LBK818010 | 10/30/2010 16:24 | 10/30/2010 16:41 | 924.96 | G130M | 1291 |
| V-DEL-CEP | LBK818020 | 10/30/2010 16:54 | 10/30/2010 17:11 | 923.968 | G160M | 1589 |
| V-DEL-CEP | LBK819010 | 10/31/2010 4:04 | 10/31/2010 4:22 | 925.024 | G130M | 1291 |
| V-DEL-CEP | LBK819020 | 10/31/2010 4:26 | 10/31/2010 4:44 | 924.032 | G160M | 1589 |
| V-DEL-CEP | LBK820010 | 12/13/2010 23:42 | 12/14/2010 0:00 | 925.024 | G130M | 1291 |
| V-DEL-CEP | LBK820020 | 12/14/2010 0:04 | 12/14/2010 1:13 | 924.032 | G160M | 1589 |
| V-DEL-CEP | LBK821010 | 10/22/2010 1:41 | 10/22/2010 1:59 | 925.024 | G130M | 1291 |
| V-DEL-CEP | LBK821020 | 10/22/2010 2:56 | 10/22/2010 3:14 | 924.032 | G160M | 1589 |
| V-DEL-CEP | LBK822010 | 10/31/2010 2:28 | 10/31/2010 2:46 | 925.024 | G130M | 1291 |
| V-DEL-CEP | LBK822020 | 10/31/2010 2:50 | 10/31/2010 3:08 | 924.032 | G160M | 1589 |
| V-DEL-CEP | LBK823010 | 10/29/2010 22:59 | 10/29/2010 23:28 | 924.928 | G130M | 1291 |
| V-DEL-CEP | LBK823020 | 10/29/2010 23:32 | 10/29/2010 23:50 | 923.936 | G160M | 1589 |
| V-DEL-CEP-1 | LBK815010 | 6/13/2011 16:50 | 6/13/2011 17:12 | 1152.032 | G130M | 1291 |
| V-DEL-CEP-1 | LBK815020 | 6/13/2011 17:17 | 6/13/2011 17:37 | 1024.032 | G160M | 1589 |
| V-DEL-CEP | LC2307010 | 1/18/2013 18:29 | 1/18/2013 18:49 | 767.008 | G130M | 1291 |
| V-DEL-CEP | LC2307020 | 1/18/2013 18:52 | 1/18/2013 20:12 | 763.072 | G160M | 1589 |
| V-BET-DOR | LBK801010 | 3/17/2011 8:19 | 3/17/2011 8:37 | 927.04 | G130M | 1291 |
| V-BET-DOR | LBK801020 | 3/17/2011 8:41 | 3/17/2011 8:59 | 927.008 | G160M | 1589 |
| V-BET-DOR | LBK811010 | 3/17/2011 9:55 | 3/17/2011 10:13 | 927.04 | G130M | 1291 |
| V-BET-DOR | LBK811020 | 3/17/2011 10:17 | 3/17/2011 10:35 | 927.04 | G160M | 1589 |
| V-BET-DOR | LBK812010 | 2/16/2011 10:38 | 2/16/2011 10:56 | 927.04 | G130M | 1291 |
| V-BET-DOR | LBK812020 | 2/16/2011 11:00 | 2/16/2011 11:18 | 927.008 | G160M | 1589 |
| V-BET-DOR | LBK813010 | 2/27/2011 22:45 | 2/27/2011 23:03 | 927.04 | G130M | 1291 |
| V-BET-DOR | LBK813020 | 2/27/2011 23:07 | 2/27/2011 23:25 | 927.04 | G160M | 1589 |
| V-BET-DOR | LBK814010 | 11/14/2010 1:53 | 11/14/2010 2:10 | 927.04 | G130M | 1291 |
| V-BET-DOR | LBK814020 | 11/14/2010 2:15 | 11/14/2010 2:33 | 926.944 | G160M | 1589 |
| V-BET-DOR-1 | LBK810010 | 6/23/2011 5:03 | 6/23/2011 6:08 | 1127.008 | G130M | 1291 |
| V-BET-DOR-1 | LBK810020 | 6/23/2011 6:12 | 6/23/2011 6:32 | 1036 | G160M | 1589 |
| V-BET-DOR-1 | LBK816010 | 8/3/2011 15:00 | 8/3/2011 15:21 | 1126.976 | G130M | 1291 |
| V-BET-DOR-1 | LBK816020 | 8/3/2011 15:26 | 8/3/2011 15:46 | 1036.032 | G160M | 1589 |

It was *well-known*, however, that IUE suffered from scattered optical light contamination, though it turned out to be much more of an issue than we had anticipated. Fig. 42 shows comparisons of representative IUE and COS spectra for each of the three Cepheids observed with COS at the time of writing. The differences are dramatic. The much finer resolution was expected, but as the figure shows, a great deal of what was originally considered continuum flux from the photosphere turned out to be scattered light. This was the reason that IUE spectra of the program Cepheids could unambiguously show *only* the strongest emission lines (if any). In the case of Polaris, as is shown in the top panel of Fig. 43, there was uncertainty as to whether *any* emission lines were present in the spectra; there was only the possible detection of the strong, but blended, oxygen/sulfur lines near ~1300 Å. The scattered light is not present in the COS spectra of Cepheids, however, which display a wealth of emission lines not detected in the archival IUE spectra (Figs. 43 and 44). These lines define rich and complex Cepheid atmospheres and, as is well known, offer excellent atmospheric diagnostic potential (Linsky et al. 1995 and references therein), since different line species originate in plasmas of specific temperatures. Also, emission line strengths and ratios, as well as line broadening and radial velocities, when measured over the stars' pulsation cycles, offer important atmospheric diagnostics. FUV emission line strengths and ratios (along with changes in radial velocity [RV] during pulsations) can also distinguish between supergiant (Cepheid) atmosphere emissions or unresolved main sequence companions (if present).

Emission line fluxes were measured by integrating the total flux beneath the emission line itself, subtracting off any neighboring continuum radiation (if present), with the *Specview* software tool (http://www.stsci.edu/institute/software_hardware/specview), published by the Space Telescope Science Institute. Emission line centers (for RV studies) were measured by fitting Gaussian profiles to each emission line, using the IRAF *splot* routine.

Although many important emission lines are present in the COS spectra, a few of importance were selected for the current study. Future plans for the UV study are to utilize the full set of observed emission lines to construct a differential emission measure for each spectrum. This details the full distribution of plasma temperatures within the atmosphere, and these distributions can then be compared to those of other cool supergiants and even dwarfs. As discussed later, specific line ratios can also allow the stellar atmospheric density to be determined, but the Cepheid spectra do not display any of the most often used (hence, best understood) line pairings. This has forced us to search for less-commonly used density diagnostics, with the unavoidable consequence that these diagnostics will not be as well-studied. So great care must be taken to find usable (and, above all, reliable) density diagnostics, given the lines present in the COS FUV spectra of the Cepheids.

For this study, the following lines were selected because they represent a wide range of formation temperatures, and are relatively free from contamination via blending with nearby lines, offering a "pure" measurement of the emission line in question. The lines selected are:

- **O I 1358 Å** – selected in favor of the well-known O I triplet at ~1300 Å because the triplet are fluorescent lines excited by H Lyman-β radiation (Koncewicz & Jordan 2007) and also suffer from blending with (primarily) nearby S I lines. Therefore, the flux of the O triplet is directly related to the Lyman flux and not necessarily indicative of plasmas at the O I peak formation temperature of $1 – 2 \times 10^4$ K (Doyle et al. 1997).
- **Si IV 1394/1403 Å** – a well-known doublet with a peak formation temperature of $\sim 5 – 8 \times 10^4$ K (Linsky et al. 1995). This doublet represents an important link between the cooler, O I emitting plasmas and the hotter plasmas responsible for the N V and O V emission lines.
- **N V 1239/1243 Å** – another well-known doublet, but with a higher peak formation temperature of $\sim 1.5 – 2.5 \times 10^5$ K (Linsky et al. 1995). This doublet is very important because the lines are prominent in the spectra and occur at shorter wavelengths where photospheric continuum flux is essentially negligible. The N V doublet is the best measure of higher-temperature plasma variability in the UV range.
- **O V 1218 Å** – although not selected for a variability study due to its location on the red wing of the Lyman-α geocoronal-contaminated emission line, the line is clearly observed in the spectra (Fig. 43) and has a peak formation temperature of $\sim 2 – 4 \times 10^5$ K (Linsky et al. 1995). This places it among the hottest lines observable in the UV spectra of cool stars like the Cepheids, and its presence provides an important link to higher-temperature X-ray emissions.

Progress in the satellite-based high-energy studies has been much slower than for the optical studies. As a result, only δ Cep and β Dor currently have substantial phase coverage with COS, although important phase-gaps still exist in the β Dor dataset, and approved observations are still to be carried out to completely fill in the Cepheid's phase-space. As an illustration of the results of the UV program thus far, the current COS light curves of δ Cep and β Dor are given in Figs. 45 & 46. In looking at these plots, a few features are immediately striking:

**The specific phases at which line emissions begin to increase.** The phase coverage of δ Cep has been more illustrative in terms of the rise in emission flux. However, preliminary conclusions can be extrapolated for β Dor from the currently available data, taking into account the findings of previous UV studies. In comparing the emission line fluxes to the included RV plots for δ Cep, we see good correlation between the phase where UV line emissions begin to peak and a “pre-piston phase” as given by the radial velocities. During this pre-piston phase, the photosphere has almost reached its minimum value, its recession is decelerating, and it is about to begin expanding again. Fokin et al. (1996) have shown that a shock should be propagating through the atmosphere of δ Cep at this phase (~0.8 – 0.85$\phi$). The emission line flux behavior, combined with the results of Fokin et al. (1996), give strong evidence in favor of a shock heating mechanism. As will be discussed later, RVs show that this phase range is also when atmospheric compression begins. This effect would further excite atmospheric plasmas.

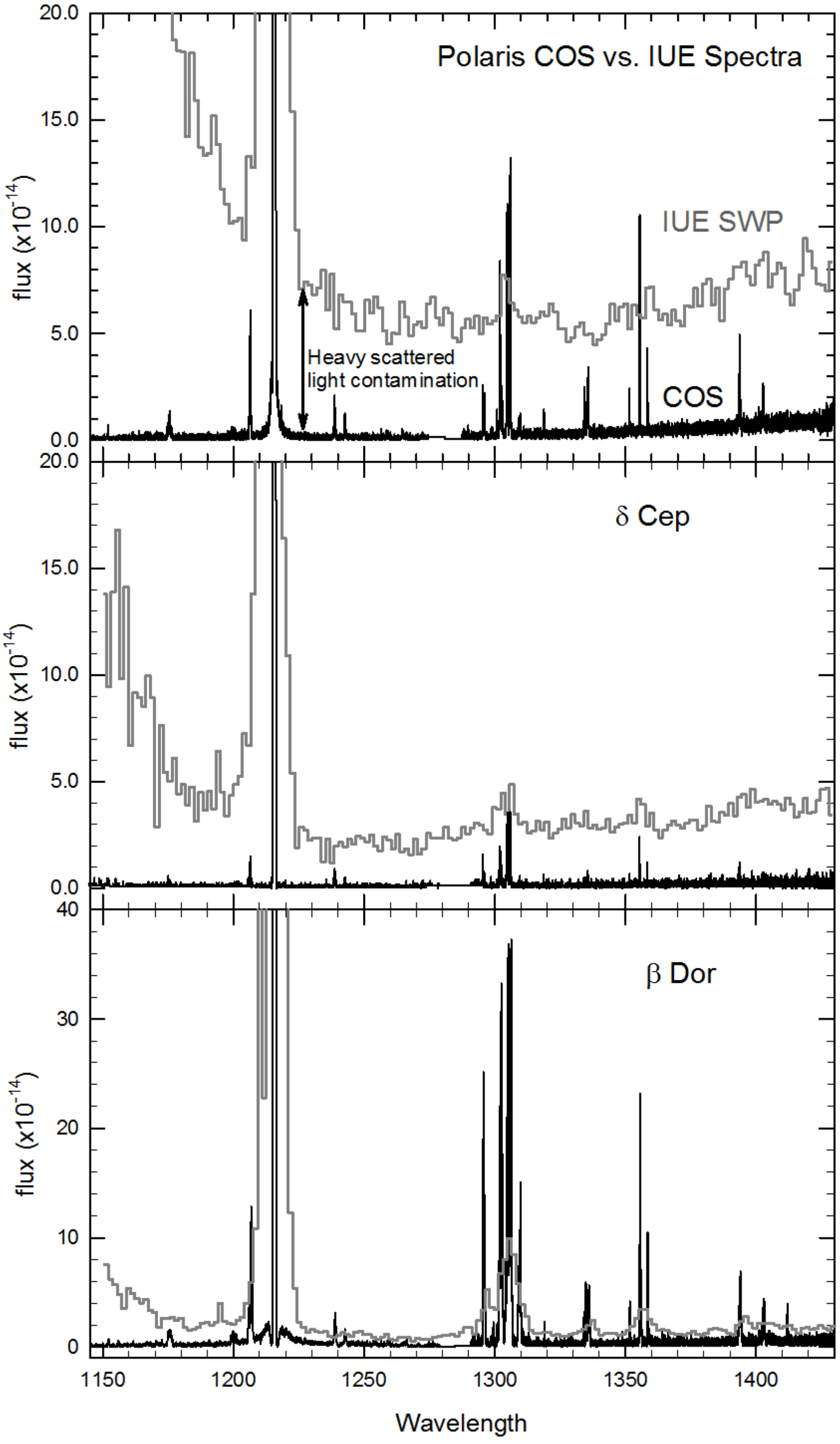

Figure 43 – Comparisons of IUE and HST/COS data for the three Cepheids observed with COS thus far. The *dramatic* improvement in data quality, particularly in the reduction of scattered light contamination, is easy to see.

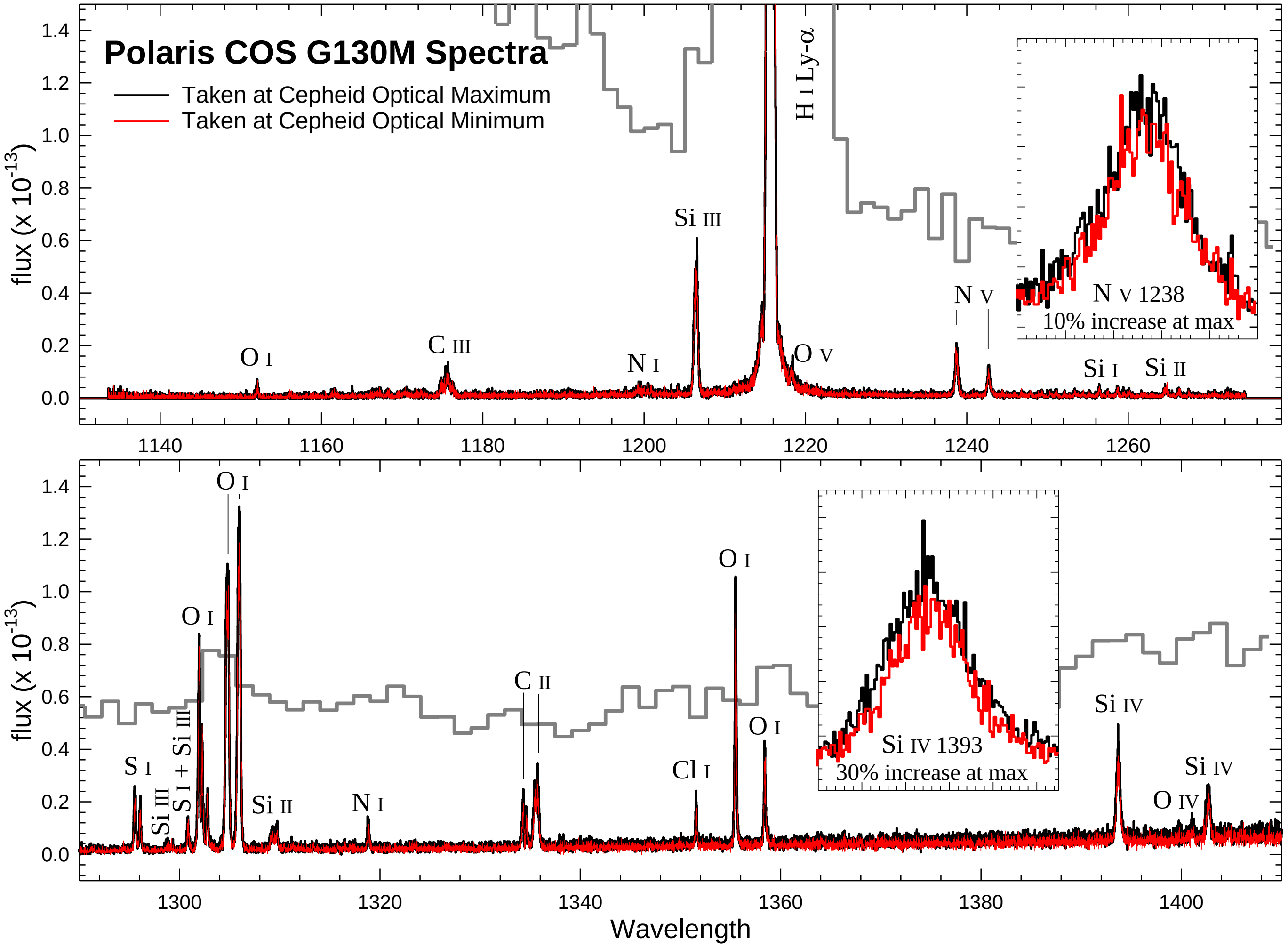

Figure 44 – A detailed view of the Polaris HST/COS FUV (G130M grating) spectra is shown. Again, the IUE data is shown as the dark gray spectrum. All visible emission lines are identified. The low level of emission line variability between the spectra is highlighted by inserts.

For β Dor, as mentioned, the conclusions are a bit more ambiguous, arising from the current (at the time of writing) lack of phase coverage to observe the rise in emission line flux. As seen in Fig. 46, the approach to maximum is partially observed for N V, but the cooler emission lines are only observed during phases of decreasing flux or at phases of quiescent emission. The phase coverage begins just after the piston phase has begun for β Dor, but it is still possible that the rise in emission line flux begins shortly after the deceleration of the photosphere, as it does in δ Cep. Hutchinson et al. (1975) report that a shock would be propagating through the atmosphere around this phase, based on five-filter UV photometry (in the ~1330 – 4250 Å range) taken by the *Wisconsin Experiment Package* (*WEP*) onboard *Orbiting Astronomical Observatory-2* (*OAO-2*).

**The abruptness of the increase** in emissions from δ Cep is indicative of a sudden heating or excitation mechanism. For example, over a phase-space of just 0.08$\phi$, the O I flux increases by ~7× (numbered points 4 – 7 in Fig. 45) and the Si IV flux increases by ~10×. Such an abrupt, strong plasma excitation also points to a shock-related mechanism. The decrease in flux is also rapid (though not as rapid as the increase), as one would expect for a sudden heating event such as a shock. It's important to note that a possibly related behavior occurs in Mira variables (pulsating, asymptotic giant branch [AGB] stars), which have shown "cool" UV emission lines excited by their own pulsation-induced shocks (Wood & Karovska 2000 and references therein). The only UV lines observed in the Miras have cooler formation temperatures than the O I line studied in Cepheids, and no "warmer" lines are observed. This could be understood as a consequence of the Miras producing weaker shocks, since their pulsation periods are much longer than the Cepheids' and their velocity amplitudes are also smaller. It also appears that the line emissions disappear at certain phases, implying that there are no quiescent atmospheric emissions, only those excited by the shocks.

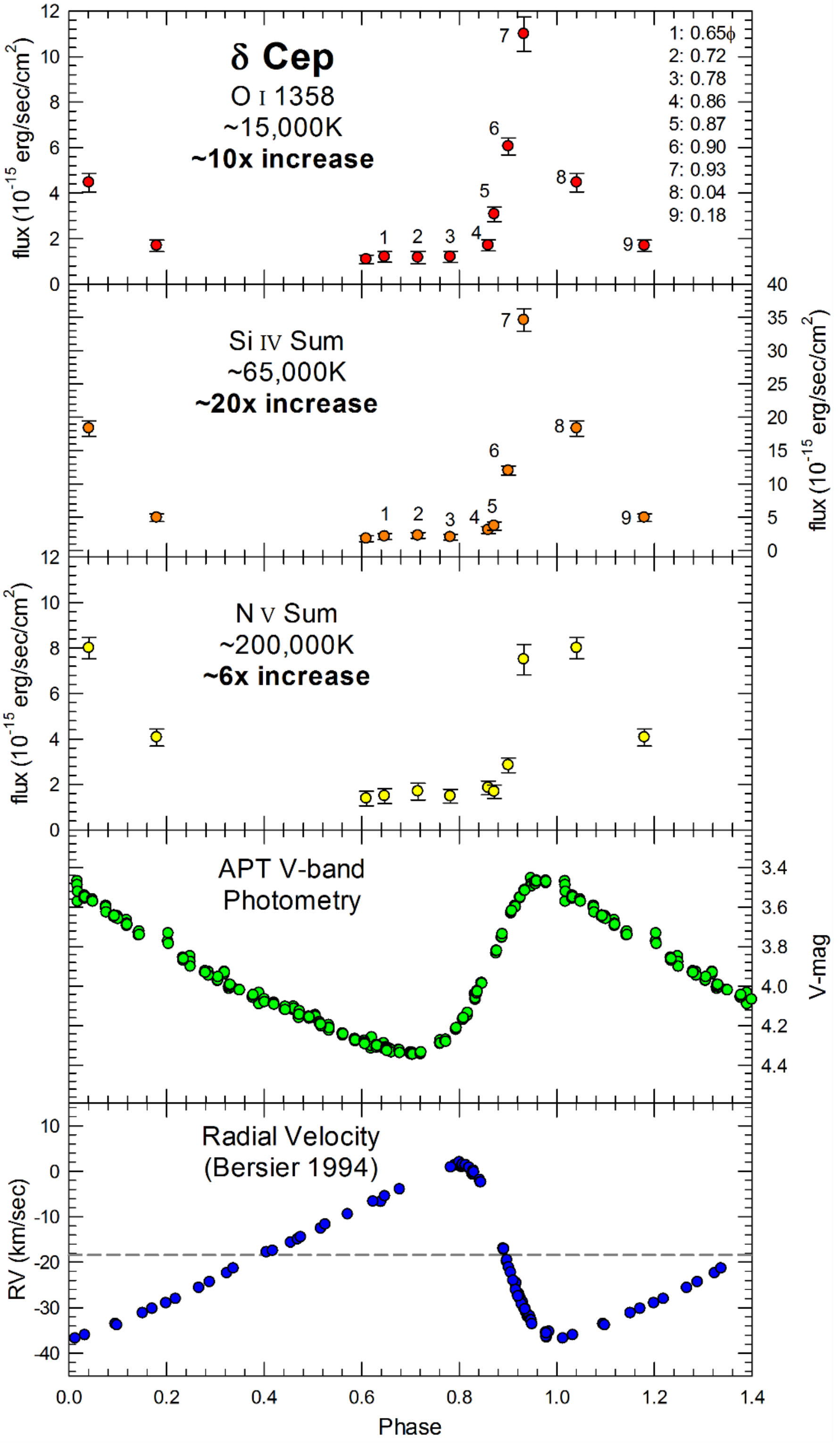

Figure 45 – Integrated fluxes of 3 important emission lines observed in δ Cep with COS - top three panels. The fourth panel (green points) is the V-band light curve from this program. The bottom panel plots the photospheric radial velocities. Spectra referred to later in the thesis are numbered, and the associated phases are given in the top plot.

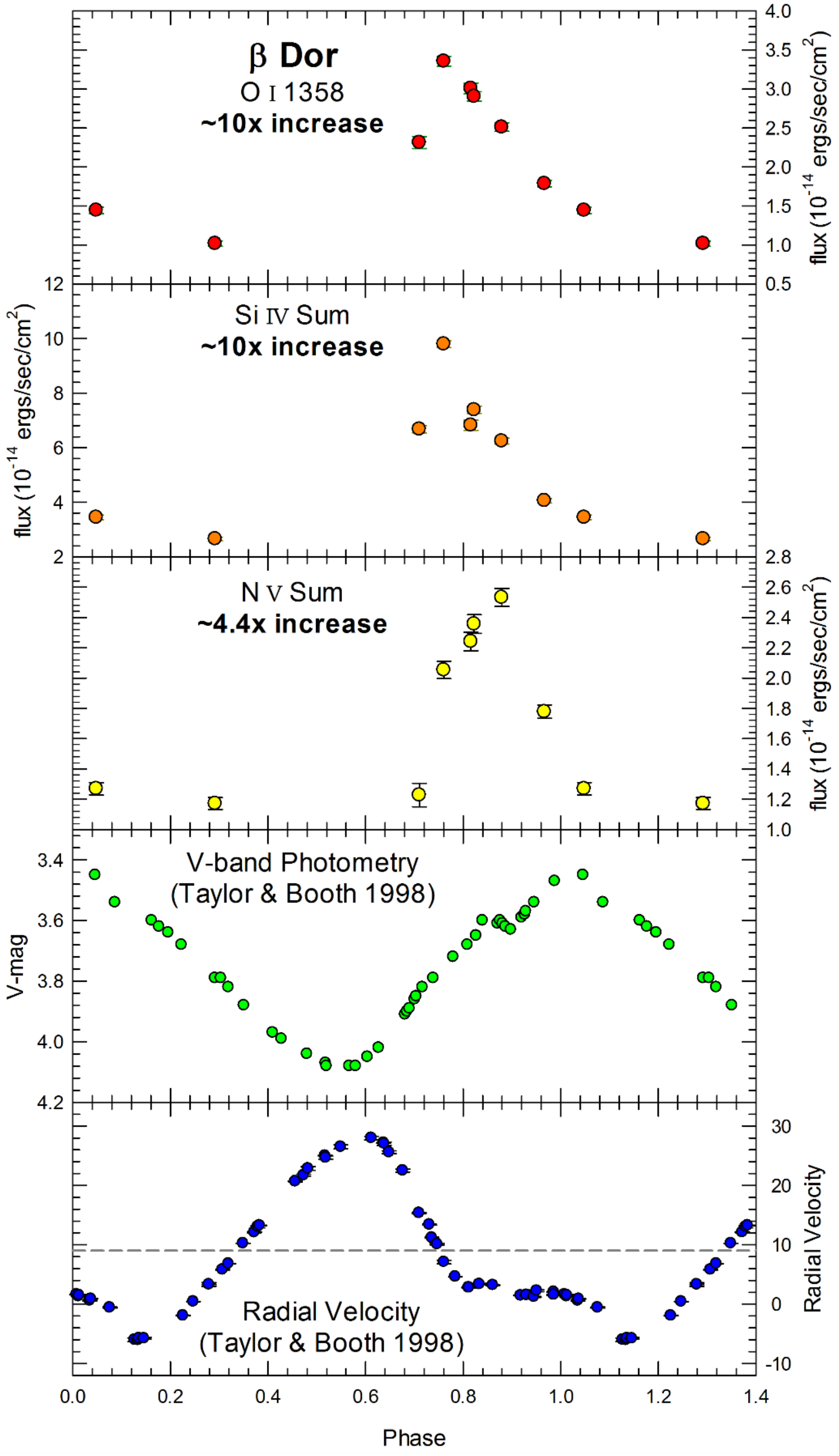

Figure 46 – The same arrangement as in Fig. 45, but for β Dor.

For β Dor, again the conclusions are mitigated by insufficient phase coverage, but the decrease in emission line flux appears much more gradual than in δ Cep. Also, the range of flux variability is much smaller in β Dor. This partially arises because β Dor is more active during quiescent phases than δ Cep. Since β Dor is a longer period Cepheid, and reaches lower temperatures, it is safe to assume that β Dor would possess a deeper, larger convective zone which could be responsible for the stronger quiescent emissions observed, on top of which additional compression/shock heating is present at the appropriate phases.

**The phase-difference between the peak flux** of the most energetic (highest peak formation temperature) emission feature observed – N V ~1240 Å – and the peak fluxes of the two cooler emission features. This aspect of the program requires robust phase-coverage, so neither Cepheid can have strict, quantitative conclusions drawn. However, in both Cepheids the lower temperature plasma emission lines appear to peak earlier than N V, the hottest emission line. The phase difference is a good bit larger in β Dor than it is in δ Cep. However, Böhm-Vitense & Love (1994) theorized that emissions from the hottest plasmas should peak first in the case of shock-heating, followed by line emissions from cooling plasmas in the post-shock regions. This is clearly an interesting behavior, and the newly approved HST-COS Cycle 20 observations will allow us to more strictly define the phase differences.

In a final evaluation of the overall intensity of atmospheric emissions from the Cepheids observed thus far, we plot spectra (scaled to account for differences in distance) representative of the range of emissions found for each star in Fig. 47. For Polaris, whose range of line variability appears small, only the spectrum with stronger emissions is plotted to prevent further crowding of the plot. As we have mentioned before, the true phase of maximum atmospheric emissions from Polaris has most likely been missed by our limited observations. The fluxes that we have measured for Polaris place it within the range observed in δ Cep, but further observations are needed to truly compare the two Cepheids' maximum emissions. To a certain degree, though, the range of emissions from δ Cep and β Dor are understood and can be compared. For N V emissions, β Dor is ~5× as strong as δ Cep at their maximum phases, and ~13× as strong at their respective minima. For Si IV emissions, β Dor is ~3× as strong as δ Cep at their maximum phases, and ~7× as strong at their respective minima. Finally, for O I emissions, β Dor is ~3× as strong as δ Cep at their maximum phases, and ~16× as strong at their respective minima. Since we presently have no reliable estimates of the sizes of the Cepheid atmospheres, we can use their photospheric radii to serve as an approximation of the ratio of their emitting areas. On average, β Dor has ~2.3× the emitting area of δ Cep. As one can see, this is perhaps enough to account for the difference in *maximum* emissions. However, a difference of emitting area is not enough to account for the relative strength of *average* atmospheric emissions in β Dor, which can also indicate different balances of heating mechanisms at play in the Cepheids.

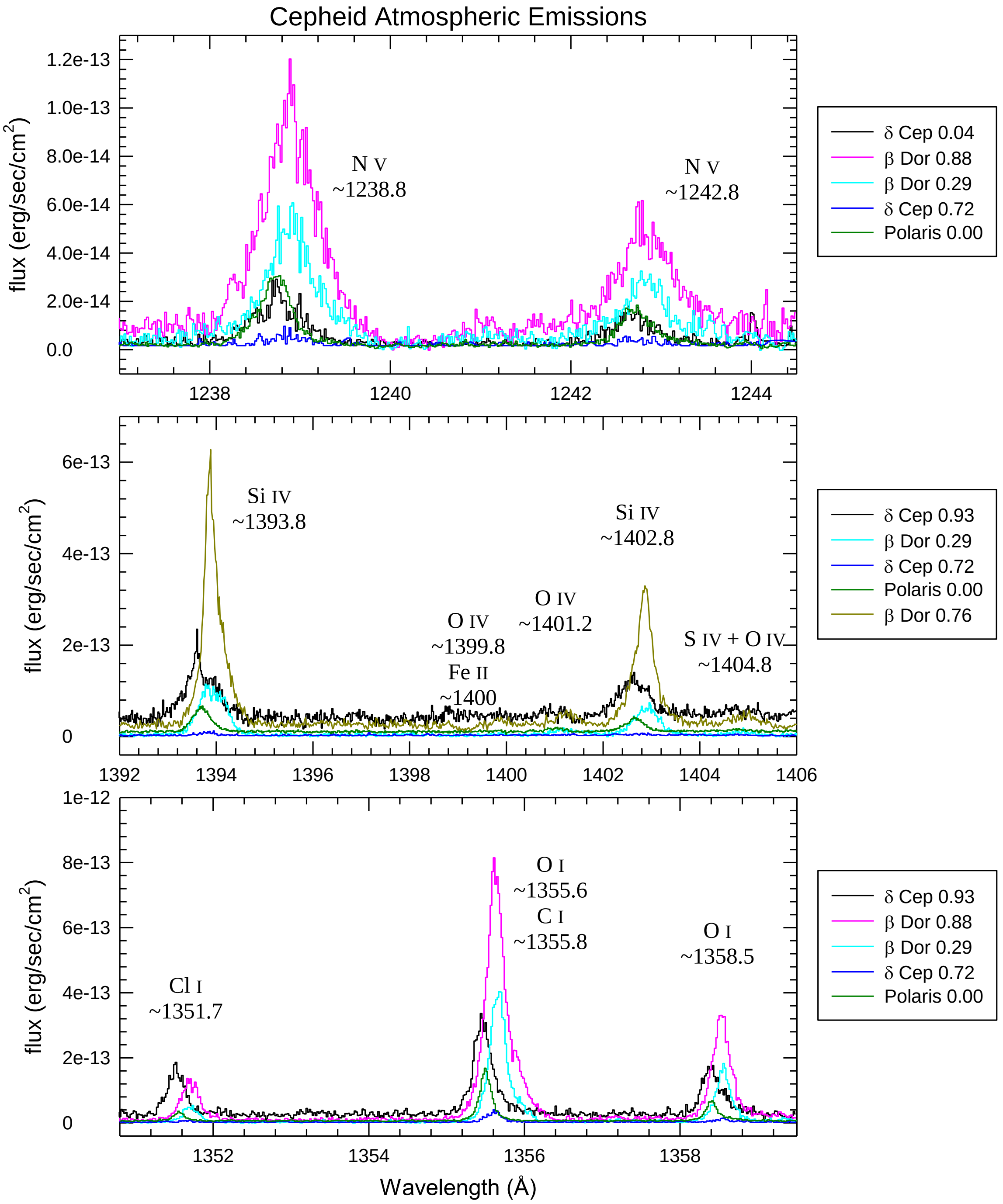

Figure 47 – A comparison of the COS spectra of all three Cepheids, giving an idea of their relative emission strengths and widths. As seen here, β Dor is easily the most active Cepheid. The phase of each Cepheid's spectrum is given in the legends.

In addition to what the emission line fluxes can tell us about the heating mechanism(s) at work in Cepheid atmospheres, the line profiles and radial velocities observed in the COS spectra can provide valuable, complementary information. Fig. 48 gives the profiles of the O I 1358 and Si IV 1394 emission lines observed in several spectra of δ Cep. The most potentially informative characteristic is the heavy asymmetry present in spectra 6, 7 and (for O I) 8, where the lines show a strong, additional blue-shifted emission component. This can be understood as the effect of an expanding shock emerging from the Cepheid photosphere. On the "near" side of the Cepheid atmosphere, the shock is approaching, producing the blue-shifted emission. In spectrum 8, the O I line still shows heavy asymmetry, but the Si IV line shows an extremely broad and even emission profile, indicative of a large velocity distribution but no additional blue-shifted feature. At this phase (0.04$\phi$) we could be observing Si IV emission from a very turbulent post-shock region, where the high turbulence would be responsible for the velocity distribution, meaning that the shock has "passed by" the Si IV emitting region. The difference in O I and Si IV line profiles at this phase indicates that they are likely originating from different regions (heights) within the Cepheid atmosphere. The line profiles offer up further evidence in favor of shock-heating and compression being responsible for the enhanced emissions.

The radial velocities of the UV emission lines can give additional information on the workings of the δ Cep atmosphere, although we note there are COS wavelength calibration issues affecting their absolute accuracy (Aloisi et al. 2010). As such, the velocities can possess a larger than normal uncertainty, but the agreement in overall velocity trends between the three lines plotted gives confidence in the measures. In Fig. 48, the radial velocities (from top to bottom panels) of the O I, Si IV, N V emission lines and photosphere are plotted. As indicated in the figure, the emission line radial velocities have had the phase-specific photospheric radial velocity removed. For spectra where the Si IV and/or O I line showed asymmetry, two Gaussian profiles were fit to the line. The RV of the broad atmospheric emission is plotted, as opposed to the blue-shifted emission component discussed in the previous paragraph. The agreement between the Si IV and O I velocity behaviors and that of the N V line, which maintained a symmetric single-Gaussian profile throughout the observed phases, gives confidence in the double-Gaussian approach. As seen in the plot, the phases of increased flux (0.85 – 0.04$\phi$) correspond to phases where the stellar atmosphere is receding (compressing) relative to the photosphere. Specifics of key, selected phases are:

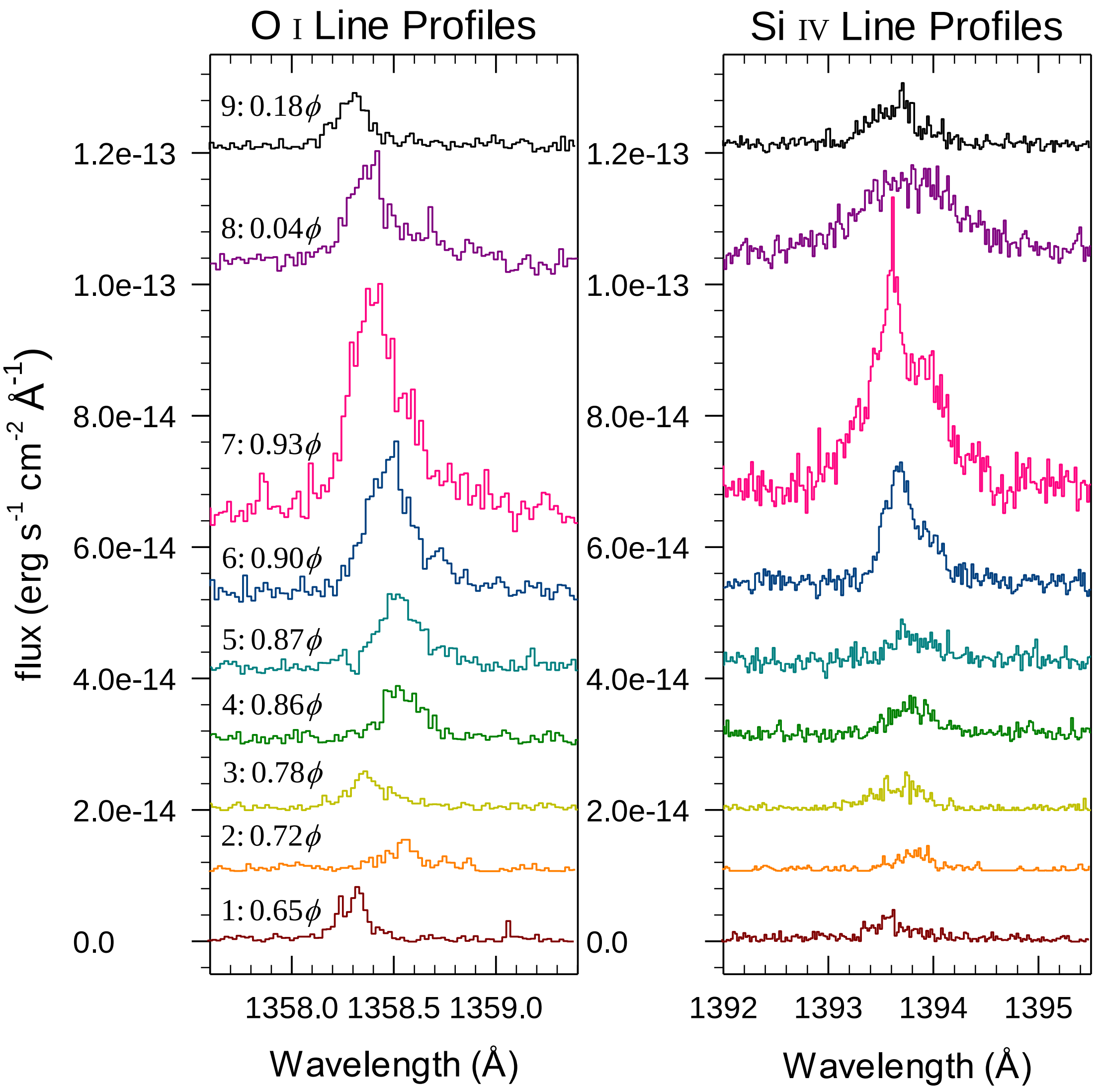

Figure 48 – The line profiles of O I (left) and Si IV (right) for δ Cep are shown. The different emission strengths can be seen, along with the asymmetries present in several phases, caused by the emergence of an additional blueward emission feature during phases where a shock is propagating through the atmosphere.

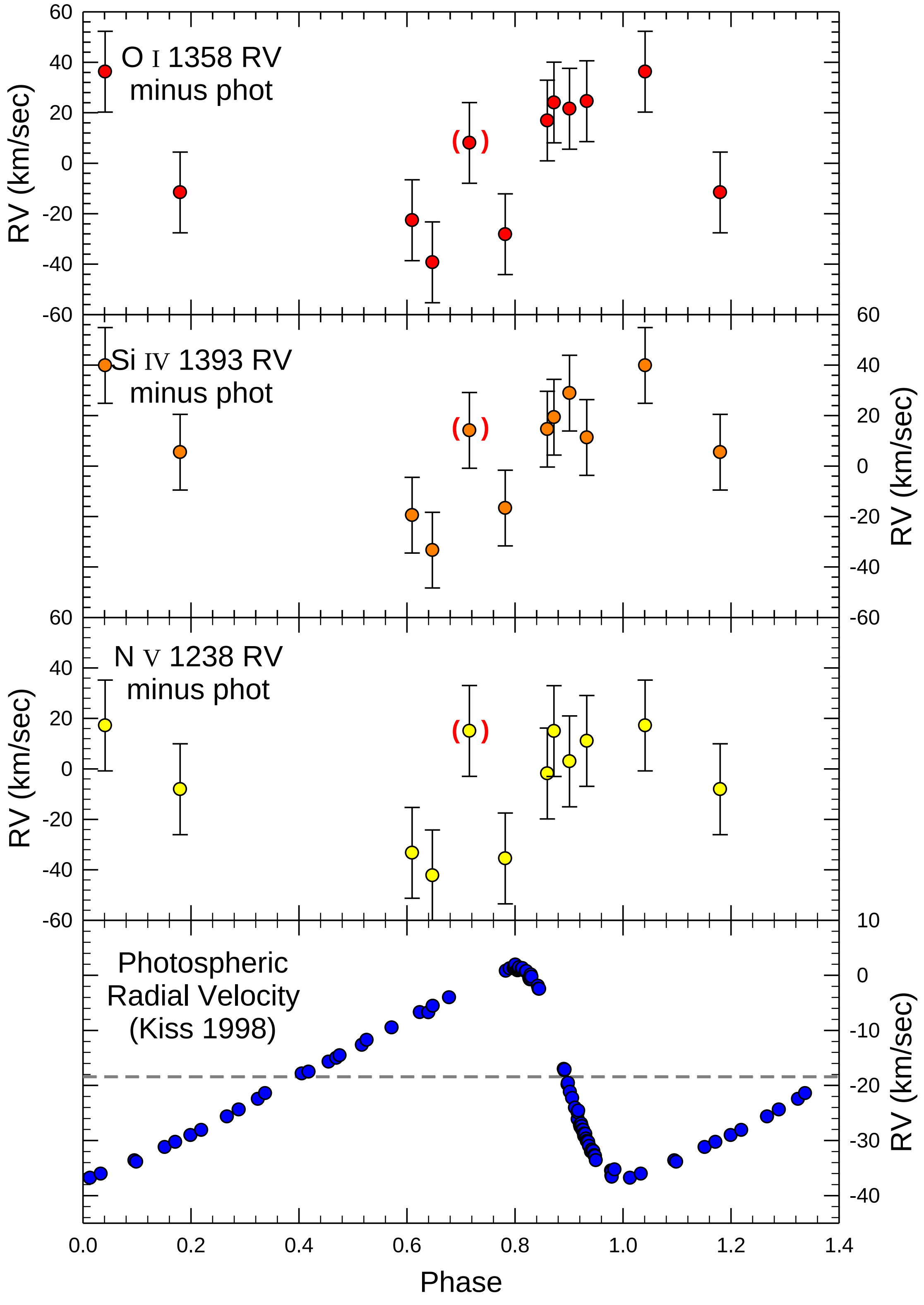

Figure 49 – Radial velocities determined for the COS-observed emission lines of δ Cep are shown vs. phase. The emission line velocities have the photospheric velocities (bottom panel) removed.

The bracketed RV is from a spectrum with a possible wavelength discrepancy, but the lack of continuum flux prevents us from confirming via photospheric or ISM absorption lines.

- $\phi \approx 0.86 - 0.87$: At these phases, the recession of the photosphere begins decelerating as it approaches minimum radius. While the photospheric recession is decelerating, the atmosphere recedes at a faster pace, as indicated by the emission line velocities. Thus, this phase-range is when the atmospheric compression begins. This is also the phase-range where Fokin et al. (1996) predict that a shock should emerge from the photosphere.
- $\phi \approx 0.90$: The photosphere has reached minimum radius (the photospheric velocity is at its mean value here), before expanding again – the start of the piston phase. While the stellar surface is no longer receding, the atmosphere is collapsing down onto it with appreciable velocity. The line profiles broaden at this phase, the result of atmospheric compression. Also, a blueward emission component becomes visible, indicating that an outward propagating shock is now passing by the line formation region of the atmosphere. This phase is most likely rather complex, in terms of atmospheric heating/excitation.
- $\phi \approx 0.93$: The photosphere has now begun expanding again, but the atmosphere continues receding relative to it, leading to even stronger atmospheric compression and line broadening. Line asymmetry is very pronounced at this phase. Line emissions from both the compressed atmospheric plasmas, and the shocked plasmas, are at their strongest levels at this phase.
- $\phi \approx 0.04$: Just after maximum visual brightness, at this phase the photosphere is undergoing its most rapid expansion and while the atmosphere continues to recede. Emission line fluxes are decreasing at this phase, yet line profiles are broader than in any other phase. The O I line profile still displays a strong blueward emission component. However, the Si IV line profile is very even and symmetric. The markedly different emission line profiles (with rather different peak formation temperatures) could indicate that they originate at different levels (heights) within the atmosphere. At this phase, shocked plasmas still exist in the O I line formation region, but we appear to be observing Si IV emissions from a turbulent post-shock region.
- At all phases, the N V lines display symmetric line profiles lacking any additional emission components, though they undergo the same radial velocity changes as the cooler lines. The phase lag, where peak N V emission appears to occur later than peak Si IV emission which, in turn, is slightly later in phase than O I, could indicate the time necessary for the shock to reach the different atmospheric levels, but could also indicate the time required to achieve higher levels of ionization. Finally, cooler O I emission could also occur at multiple heights within the atmosphere, leading to longer observation of shocked material, whereas emissions from the hotter lines are more localized.

Because of the incomplete understanding of the balance of heating mechanisms at work in Cepheid atmospheres, it was decided to place them within the context of somewhat similar stars. The most natural comparison to be made is with the so-called "Hybrid-atmosphere supergiants" (simply *Hybrids* hereafter): stars that display cool winds *and* heated atmospheres. The Hybrids are similar to the Cepheids in spectral type and luminosity, but show no definitive evidence of significant variability or pulsations. To some extent, Hybrids can be seen as an example of how Cepheids might behave if they didn't undergo the radial pulsations which define them as a class of variable star. Fortunately, HST medium- to high-resolution UV spectra have been obtained for a number of Hybrids. We compared these spectra to better illustrate the atmospheres present in each class of supergiant. The Hybrids, for which UV spectra were obtained and plotted, are given in Table 16. Data sources marked as STIS are pipeline-processed spectra obtained from the MAST archive, where STIS STARCAT come from a database of further refined spectra maintained by Thomas Ayres (http://casa.colorado.edu/~ayres/StarCAT/).

**Table 17 – Spectral Types and Data Sources for Hybrids and Cepheids**

| Star | Spectral Type | Data Source |
|---|---|---|
| **Representative Hybrids** | | |
| α TrA | K4 II | STIS |
| α Aqr | G2 Ib | STIS |
| β Aqr | G0 Ib | STIS STARCAT |
| β Cam | G0 Ib | STIS STARCAT |
| β Dra | G2 II | STIS |
| **Cepheids** | | |
| β Dor | F4–G4 Ia–II | COS |
| δ Cep | F5–G1 Ib | COS |
| Polaris | F7–F8 Ib–II | COS |

The results of the comparisons including β Dor are shown in Figs. 50 and 51, and those including δ Cep are shown in Figs. 52 and 53. The "very active supergiants" β Cam and β Dra (Reimers et al. 1996) are easily the strongest emitters when it comes to the hottest plasmas (N V and O IV emissions). No matter what line is being studied, δ Cep and Polaris are much less active than all other supergiants plotted, but β Dor does attain nearly Hybrid-levels of activity from hotter plasmas. For S IV emissions, β Dor ranges from less active than all supergiants at its minimum to more active than the Hybrids but less active than the very active supergiants at maximum. For the cool O I emissions, however, β Dor reaches the highest activity level of the supergiants plotted, due in part to its noticeably broadened profile. This can be taken as further indication of its likely turbulent atmosphere's strong velocity gradient.

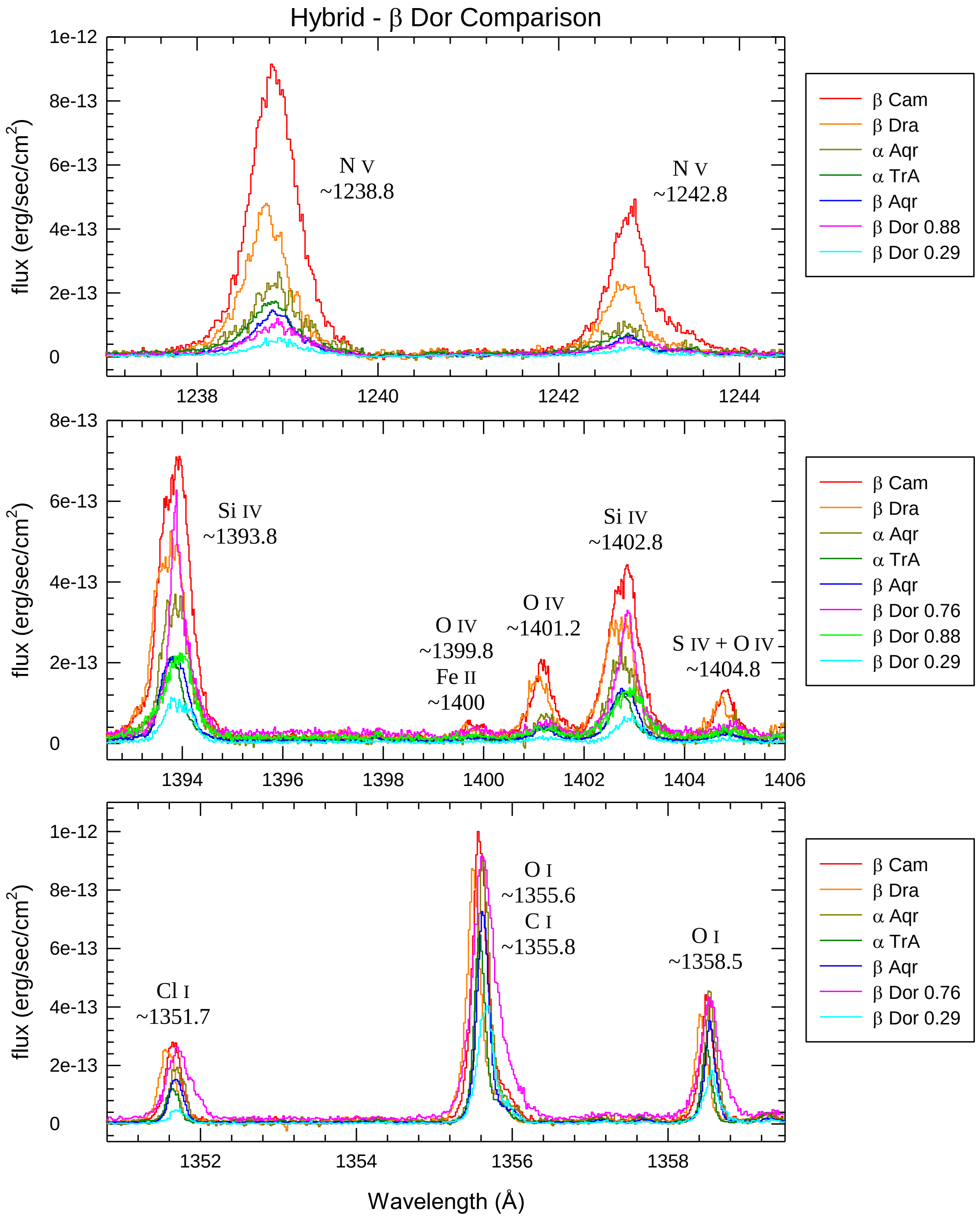

Figure 50 – Comparison of UV spectral regions of: the very active supergiants (β Cam and β Dra), the Hybrids (α and β Aqr and α TrA), and the Cepheid β Dor at its varying emission levels. The phase of each β Dor spectrum is given in the legend.

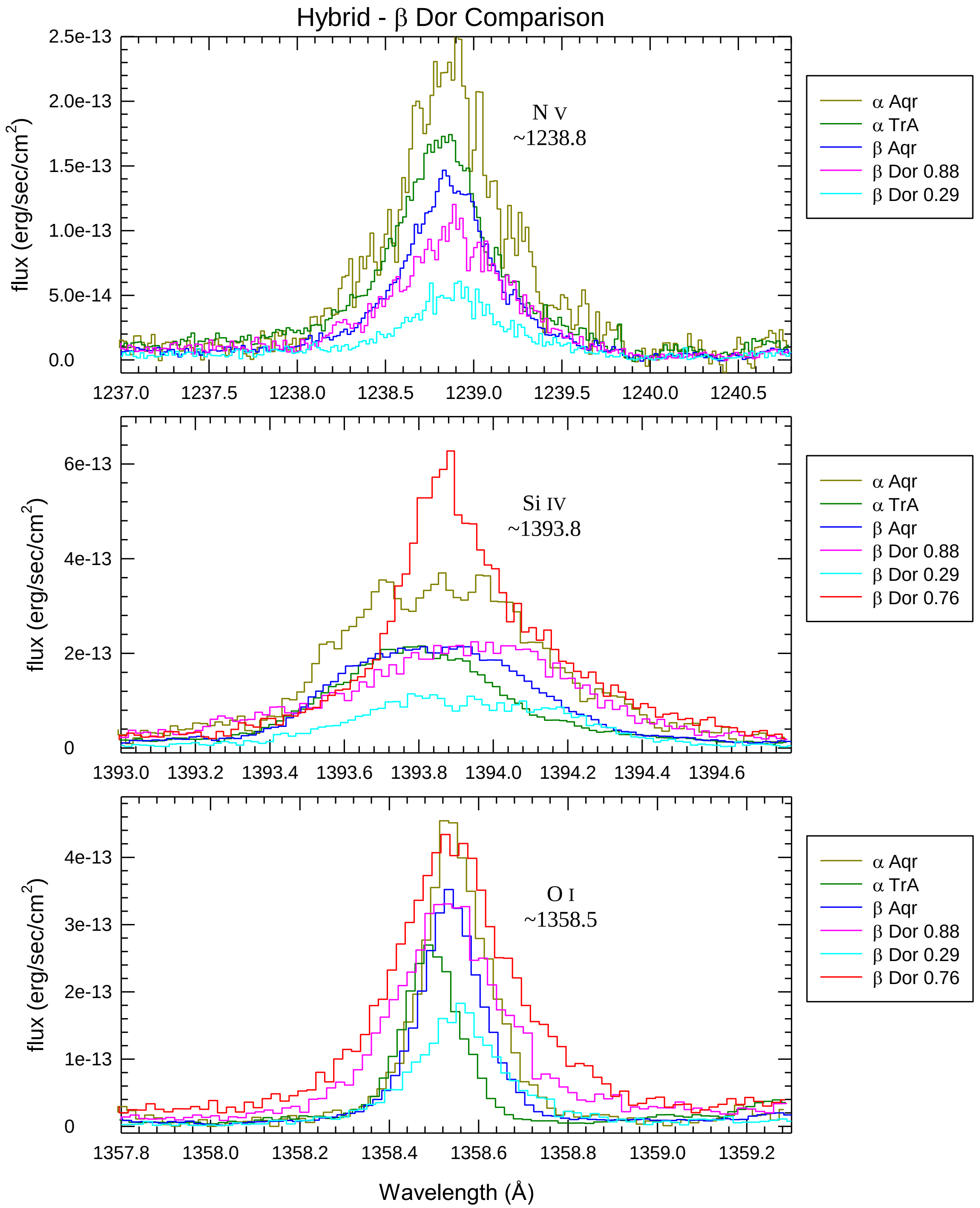

Figure 51 – A closer view of individual emission lines in the supergiants, with the very active supergiants removed for ease of viewing. At or near maximum, one can easily see the broadness of β Dor's O I emissions, and the asymmetry of the O I and Si IV lines when compared to the Hybrids. The phase of each β Dor spectrum is given in the legend.

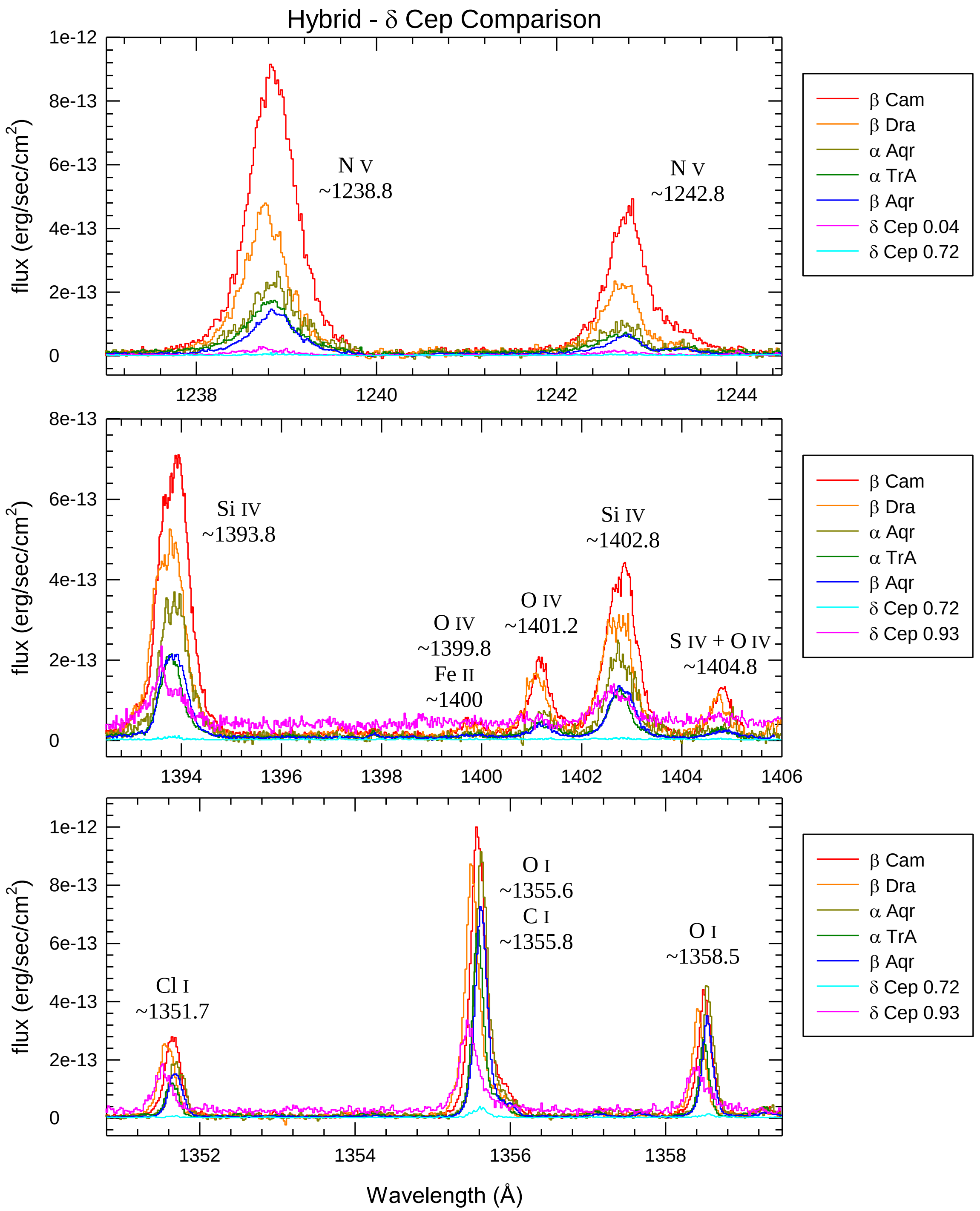

Figure 52 – The same convention as Fig. 50, but with δ Cep instead of β Dor.

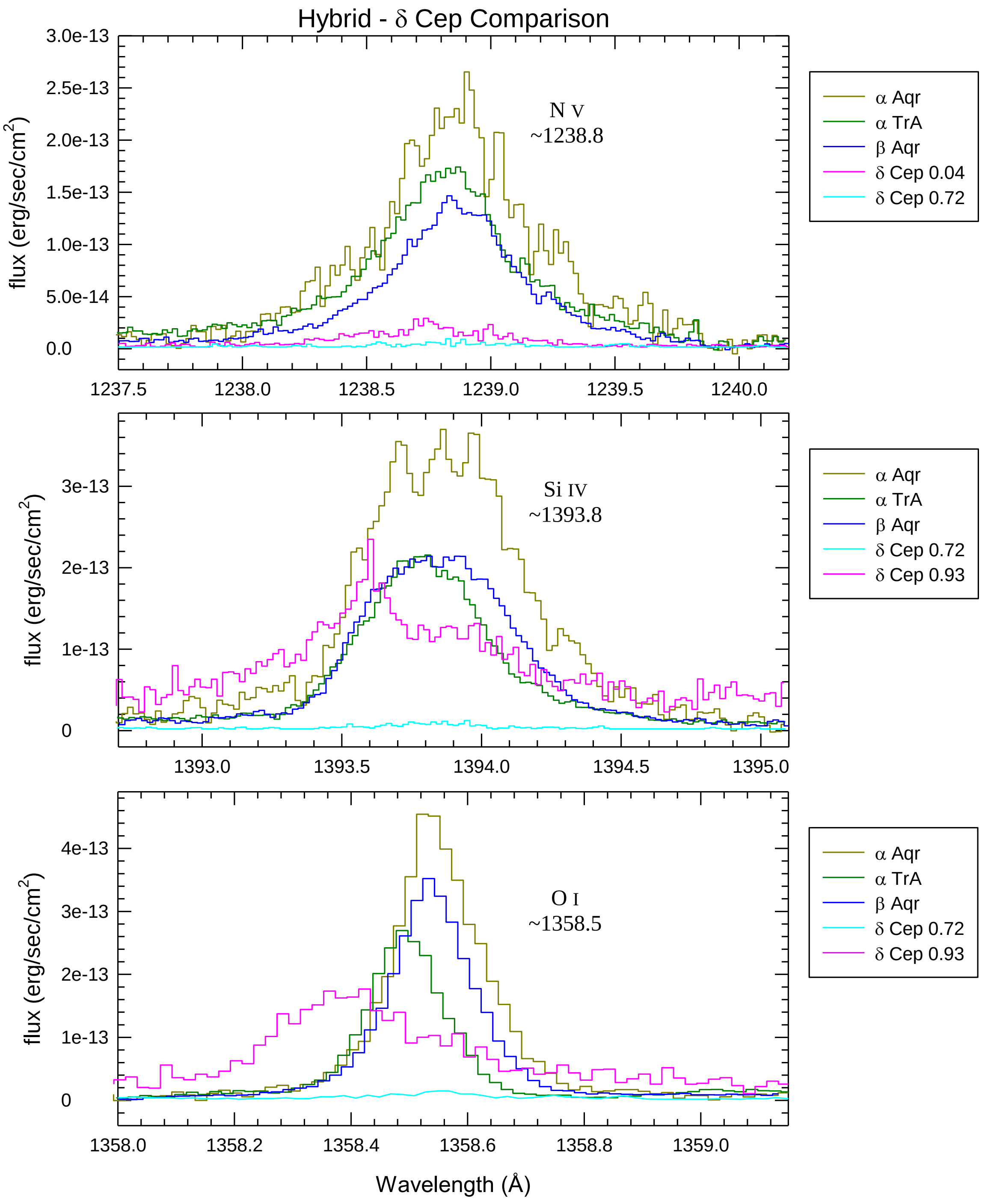

Figure 53 – The same convention as Fig. 51, but with δ Cep instead of β Dor.

When comparing δ Cep to the other supergiants, however, equal comparisons can never be made. It is obvious that, at the plasma temperatures probed by the UV emissions, the atmosphere of δ Cep (and the comparably active Polaris as seen in Fig. 47) is far less active than either the Hybrids or the very active supergiants, no matter which phase of the Cepheid is being observed. At maximum emission, δ Cep does have similarly broad emission lines compared to the Hybrids, but still with much less overall intensity, again in stark contrast to the behavior of β Dor. The Hybrids raise further questions about the specific heating mechanisms at work in the Cepheids.

Most recently, in light of the pulsation-induced variability and likely compression of the atmosphere, attempts have been made to find suitable electron density-sensitive emission line ratios to gain further physical insights into the Cepheid and other supergiant atmospheres. A handful of well-studied ratios exist in the literature, making use of such emission lines as, e.g. C III (1909 Å), Si III (1892 Å) and O IV (~1400 Å). Very unfortunately, either the spectra available for the program stars do not cover the wavelengths at which these emission lines occur (C III 1909 Å or Si III 1892 Å), or the lines are not strong enough to allow an unambiguous measurement (O IV ~1400 Å). Thus, none of the usual density measures could successfully be applied, and a search for less well-known density ratios was carried out. Making use of the latest CHIANTI atomic database (http://www.chiantidatabase.org/) available at the time of writing (v7.1.3), investigations were carried out for a ratio of Si III line fluxes:

$$R_{Si\ III} = \frac{f(1298.894 + 1298.948)}{f(1206.502)}$$

Measurements of all Hybrid spectra show very low densities, as do *almost* all Cepheid spectra. Spectrum 6 of δ Cep (Fig. 45), where the flux is steeply rising, shows a much higher density of $N_e \approx 3.2 \times 10^{10}$ cm$^{-3}$. For reference, at similar plasma temperatures to those probed by the Si III ratio above, quiet regions of the Sun have measured densities of ~5 × 10$^{10}$ cm$^{-3}$ and active regions have densities of ~1 × 10$^{10}$ cm$^{-3}$ (Dupree et al. 1976), and other studies (e.g. Keenan et al. 1989) have found solar densities to match that measured for δ Cep.

It must be noted, though, that issues have been raised with using the Si III 1206.502 Å line in density diagnostics. Dufton et al. (1983) calculated that, for the Sun, the 1206 line would be far too optically thick to give an accurate density. Although we believe that, in the supergiants, the line would essentially be optically thin and suitable in that regard, there is also the temperature sensitivity to take into account when using a ground state transition like 1206 with other subordinate features from higher levels. As such, we view the result as confirmation of increased atmospheric density during the phase of rising flux, but are still investigating the usefulness of the Si III ratio in returning an accurate, numerical density measure.

### 3.2 X-ray Studies with XMM-Newton and the Chandra X-ray Observatory

UV line emissions with formation temperatures of $10^6$ K (MK) and hotter are rare, typically weak, and only appear in the most active of stars. The best example of this would likely be the coronal Fe XXI 1354 Å emission line (see Linsky et al. 1995 for a discussion of the line). This line has a peak formation temperature of ~$1 \times 10^7$ K, but is a relatively weak line and is also blended with a neighboring C I line. None of the COS Cepheid spectra that we obtained unambiguously displayed the Fe XXI emission line. Thus, to detect and study Cepheids at MK temperatures, X-ray observations are needed. Though X-ray activity in Cepheids was considered possible (from, e.g., pulsation-induced shocks) in the mid-1990s (Sasselov & Lester 1994), the failure of previous efforts to detect X-rays with pointed Einstein and ROSAT observations reinforced the theory that Cepheids are not (at least significant) X-ray sources. Even with *log* $L_X \approx 29$ erg/s, the problem with detecting X-rays from Cepheids is that they (except for Polaris at ~133 pc – van Leeuwen 2007) are relatively far away at d > 250 pc. Thus, based on the null detections, Cepheids could only have relatively weak X-ray fluxes ($f_X \leq 10^{-14}$ ergs/s/cm$^{2)}$. Though Polaris was detected on the 3σ level in a *Röntgen Satellite* (ROSAT)/High Resolution Imager (HRI) archival image (the detection was not discovered until several years after the observation was carried out – see Evans et al. 2007), definitive detection of X-rays from Polaris and other "nearby" Cepheids had to wait for the arrival of powerful X-ray observatories such as XMM-Newton (XMM) and Chandra.

Accordingly, we successfully obtained Chandra (PI: Evans) and XMM-Newton (PI: Guinan) observations of multiple Cepheids – Polaris, δ Cep, β Dor, SU Cas and ℓ Car have been observed to date (Table 17). The Chandra data reduction for Polaris is discussed in Evans et al. (2010). The XMM observations were fully re-processed from raw data with XMM-SAS (Scientific Analysis System) and filtered for any background flaring events. The data were then modeled using the *Sherpa* modeling and fitting package (distributed as part of the *Chandra Interactive Analysis of Observations* (*CIAO*) suite). *MEKAL* models (http://cxc.harvard.edu/sherpa/ahelp/xsmekal.html) were used for the final one- and two-temperature (1T and 2T) fitting and flux calculations. The models calculate synthetic X-ray energy distributions for a given plasma temperature, utilizing a database of emission line parameters for various elements, known to prominently occur at X-ray wavelengths. The nearest 3 Cepheids (and so far the only Cepheids detected) – Polaris, δ Cep & β Dor – display X-ray luminosities of *log* $L_X \approx 28.6 – 29.2$ ergs/s (Table 17 gives the relevant information for the detections). Neither the short period Cepheid SU Cas (P = 1.95-d; d = 395 ± 30 pc) nor the long period, luminous Cepheid ℓ Car (P = 35.5-d; d = 498 ± 55 pc: Benedict et al. 2007) were detected. Upper X-ray luminosity limits of *log* $L_X < 29.6$ and 29.5 ergs/s were estimated for ℓ Car and SU Cas, respectively, based on exposure times, background count rates and stellar distances. Therefore, it is still possible that SU Cas and ℓ Car are X-ray sources with similar levels of activity to the Cepheids detected thus far, but are too distant to be detected above the background of the XMM exposures. However, the failure to detect two of our targets underscores a long-standing ambiguity present in the X-ray studies of Cepheids.

Since the initial detections of the three Cepheids at X-ray wavelengths first occurred, the argument has been made that unresolved companions were perhaps responsible for the activity being mistakenly assigned to Cepheids (see Evans et al. 2010). This is a definite possibility that needs to be taken into account, since Cepheids are young stars (~50 – 200 Myr) and any main-sequence G-K-M companions (if present) would be coronal X-ray sources with X-ray luminosities similar to that of the Cepheids (see Guinan & Engle 2009 and references therein). To within the accuracy of the instruments (4-arcsec per pixel for the XMM images), the X-ray detections are centered on the locations of the Cepheids themselves to within a single pixel, as shown in Figs. 55 – 57. Given the spatial resolution of the XMM detectors, this alone does not rule out nearby companions, but it is nevertheless the best positional confirmation that can be achieved with the instrument. Further confirmation comes from our HST/COS results, which show plasmas up to $10^5$ K that are variable with the Cepheids' pulsation periods – indicating that plasmas approaching soft X-ray emitting temperatures exist in Cepheid atmospheres. Also, our X-ray observations of δ Cep show variability possibly correlated with the stars' pulsation periods, as with the FUV emission lines (see Fig. 54). While we feel confident that the Cepheids are indeed producing the detected X-ray activity, further observations will allow us to investigate the phase dependence of the X-ray emissions on pulsation period. In addition to providing more conclusive proof that Cepheids themselves are X-ray sources (and not hypothetical young coronal companions, as previously mentioned), the additional X-ray observations will also shed more light on the origin and variability of hot (MK) Cepheid plasmas, and what its exact relationship is (if any) to the warm UV-emitting plasmas. There are different mechanisms that could produce X-ray variability in Cepheids, and again it can benefit the study to put their high-energy properties into a stellar context.

**Table 18 – Observation Log for Cepheid X-ray Data Used in Here**

| Cepheid | Observation | Start Time UT Julian Date Phase | Stop Time UT Julian Date Phase | Grouping | 1 kT (keV) | log $N_H$ | flux (0.3-2.5 keV) | 2 kT (keV) | Normalizations | flux (0.3-2.5 keV) | 2 kT Lx |
|---|---|---|---|---|---|---|---|---|---|---|---|
| δ Cep | 603740901 | 1/19/2006 18:04 | 1/20/2006 12:37 | 25-channel | 0.609 | 20.5 | 4.261E-15 | 0.151 + 0.655 | 1.329E-6 + 1.343E-6 | 5.342E-15 | 4.737E+28 |
| | XMM-pn | 2455217.253 | 2455218.026 | | | | | | | | |
| | | 0.054 | 0.12 | | | | | | | | |
| δ Cep | 0603741001_I | 1/21/2006 18:05 | 1/22/2006 14:17 | 25-channel | 1.357 | 20.5 | 1.338E-14 | 0.404 + 1.610 | 2.232E-6 + 7.033E-6 | 1.508E-14 | 1.337E+29 |
| | XMM-pn | 2455219.254 | 2455220.095 | | | | | | | | |
| | | 0.43 | 0.50 | | | | | | | | |
| | 0603741001_II | 0.51 | 0.58 | 25-channel | 0.586 | 20.5 | 6.761E-15 | 0.489 + 2.104 | 1.699E-6 + 3.320E-6 | 8.420E-15 | 7.466E+28 |
| δ Cep | 552410401 | 6/4/2004 14:26 | 6/4/2004 21:53 | 25-channel | 0.654 | 20.5 | 5.719E-15 | 0.619 + 4.000 | 1.591E-6 + 2.569E-6 | 6.886E-15 | 6.106E+28 |
| | XMM-pn | 2454623.101 | 2454623.412 | | | | | | | | |
| | | 0.33 | 0.39 | | | | | | | | |
| δ Cep | 0723540301_I | 6/28/2013 6:34 | 6/29/2013 13:49 | 25-channel | 0.963 | 20.5 | 4.460E-15 | 0.321 + 1.328 | 1.435E-6 + 1.473E-6 | 5.634E-15 | 4.996E+28 |
| | XMM-pn | 2456471.774 | 2456473.076 | | | | | | | | |
| | | 0.84 | 0.96 | | | | | | | | |
| | 0723540301_II | 0.96 | 0.08 | 25-channel | 0.486 | 20.5 | 4.374E-15 | 0.408 + 1.378 | 1.392E-6 + 8.998E-7 | 4.921E-15 | 4.363E+28 |
| δ Cep | 0723540401_I | 7/2/2013 6:17 | 7/3/2013 7:51 | 25-channel | 0.613 | 20.5 | 3.825E-15 | 0.613 + 0.741 | 1.388E-6 + 2.138E-14 | 3.825E-15 | 3.392E+28 |
| | XMM-pn | 2456475.762 | 2456476.827 | | | | | | | | |
| | | 0.58 | 0.68 | | | | | | | | |
| | 0723540401_II | 0.68 | 0.78 | 25-channel | 0.468 | 20.5 | 5.526E-15 | 0.016 + 0.482 | 3.202 + 1.999E-6 | 6.112E-15 | 5.419E+28 |
| β Dor | 603740801 | 2/1/2006 9:09 | 2/2/2006 3:54 | 25-channel | 0.297 | 20.4 | 9.545E-15 | 0.266 + 2.129 | 3.121E-6 + 5.609E-6 | 1.306E-14 | 1.576E+29 |
| | XMM-pn | 2455230.25 | 2455230.66 | | | | | | | | |
| | | 0.41 | 0.45 | | | | | | | | |
| | | | | | | | | | | | |
| β Dor | 603741101 | 3/23/2006 4:26 | 3/23/2006 23:40 | 25-channel | 0.325 | 20.4 | 7.954E-15 | 0.298 + 1.302 | 2.865E-6 + 1.946E-6 | 9.544E-15 | 1.152E+29 |
| | XMM-pn | 2455280.08 | 2455280.49 | | | | | | | | |
| | | 0.47 | 0.51 | | | | | | | | |
| β Dor | 552410101 | 6/22/2004 6:35 | 6/22/2004 17:49 | 25-channel | 0.528 | 20.4 | 5.159E-15 | 0.480 + 4.000 | 1.391E-6 + 2.784E-6 | 6.480E-15 | 7.821E+28 |
| | XMM-pn | 2454640.774 | 2454641.242 | | | | | | | | |
| | | 0.52 | 0.56 | | | | | | | | |
| Polaris | 503140101 | 2/23/2008 21:51 | 2/24/2008 2:25 | 25-channel | 0.390 | 20.0 | 2.800E-14 | 0.085 + 0.591 | 3.584E-5 + 9.276E-6 | 3.749E-14 | 7.919E+28 |
| | XMM-pn | 2454520.410 | 2454520.601 | | | | | | | | |
| | | 0.21 | 0.26 | | | | | | | | |
| Polaris | 503140401 | 4/29/2008 8:04 | 4/29/2008 13:00 | 25-channel | 0.477 | 20.0 | 2.975E-14 | 0.107 + 0.543 | 9.282E-6 + 9.812E-6 | 3.313E-14 | 6.998E+28 |
| | XMM-pn | 2454585.836 | 2454586.042 | | | | | | | | |
| | | 0.68 | 0.74 | | | | | | | | |
| Polaris | 654780201 | 5/1/2010 9:12 | 5/1/2010 19:43 | 10-channel | 0.495 | 20.0 | 3.011E-14 | 0.136 + 0.502 | 7.676E-7 + 1.087E-5 | 3.051E-14 | 6.445E+28 |
| | XMM-MOS | 2455317.883 | 2455318.322 | | | | | | | | |
| | | 0.97 | 0.08 | | | | | | | | |
| Polaris | 6431 (Chandra) | 2/9/2006 1:01 | 2/9/2006 3:47 | | | | | 0.57 + 0.12 | | 3.800E-14 | 8.027E+28 |
| | | 2453775.542 | 2453775.658 | | | | | | | | |
| | | 0.71 | 0.73 | | | | | | | | |
| ℓ Car | 603740301 | 4/2/2010 11:38 | 5/2/2010 3:01 | | | | | | | | |
| | | 2455231.985 | 2455232.626 | | | | | | | | |
| | | 0.639 | 0.968 | | | | | | | | |
| SU Cas | 603740501 | 8/2/2010 10:32 | 9/2/2010 2:32 | | | | | | | | |
| | | 2455235.939 | 2455236.606 | | | | | | | | |
| | | 0.932 | 0.951 | | | | | | | | |

*log $N_H$ – number of neutral Hydrogen atoms per cubic cm between us and the target (column density) – calculated using the online tool at http://archive.stsci.edu/euve/ism/ismform.html

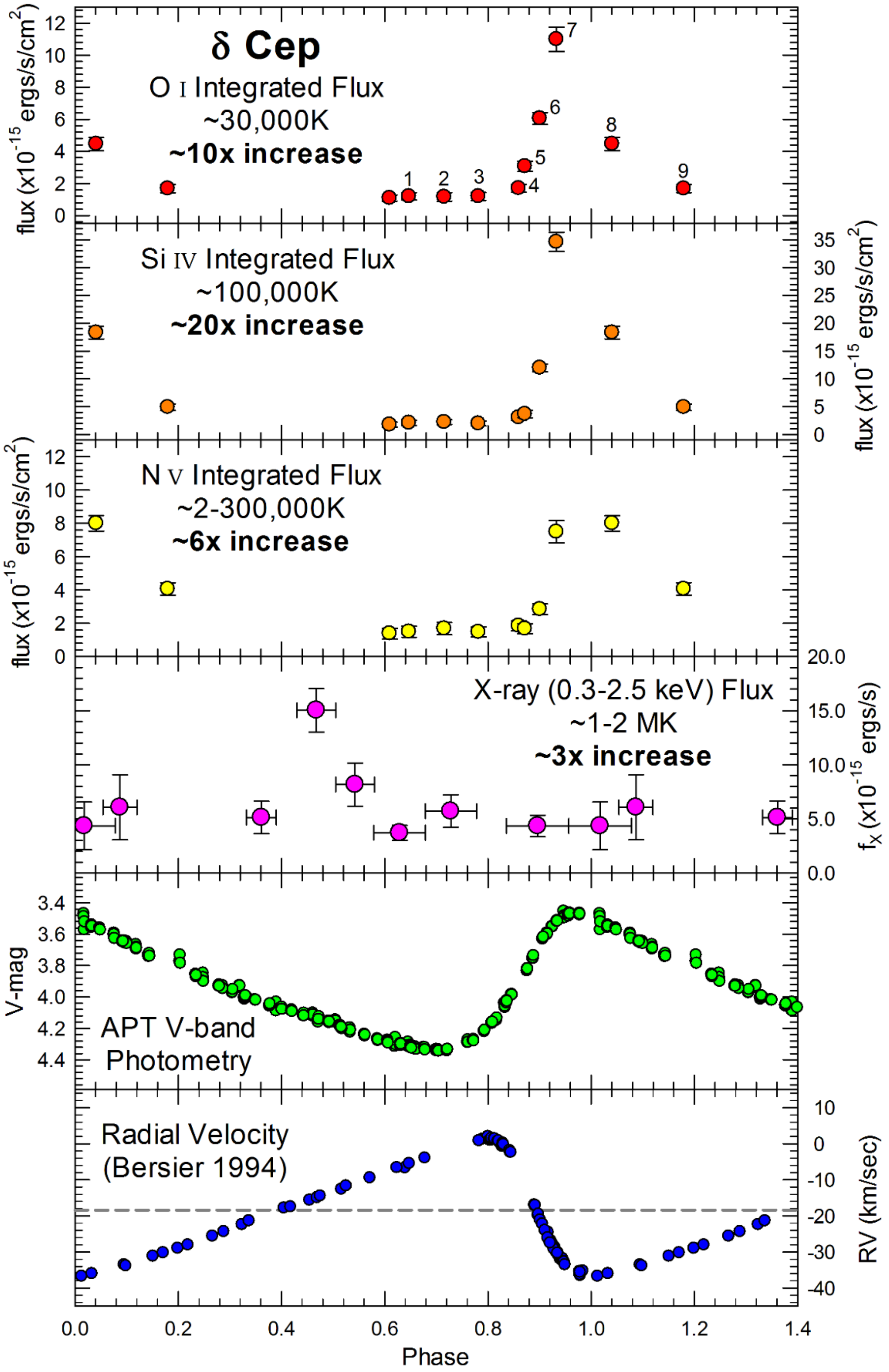

Figure 54 – The top four panels give the UV emission line fluxes and the X-ray fluxes measured vs. pulsational phase for δ Cep. The bottom two panels give the V-band photometry and photospheric radial velocities, as previously seen. The X-ray activity appears to decrease when the UV activity increases which, if confirmed, would give valuable insight into the stellar dynamics.

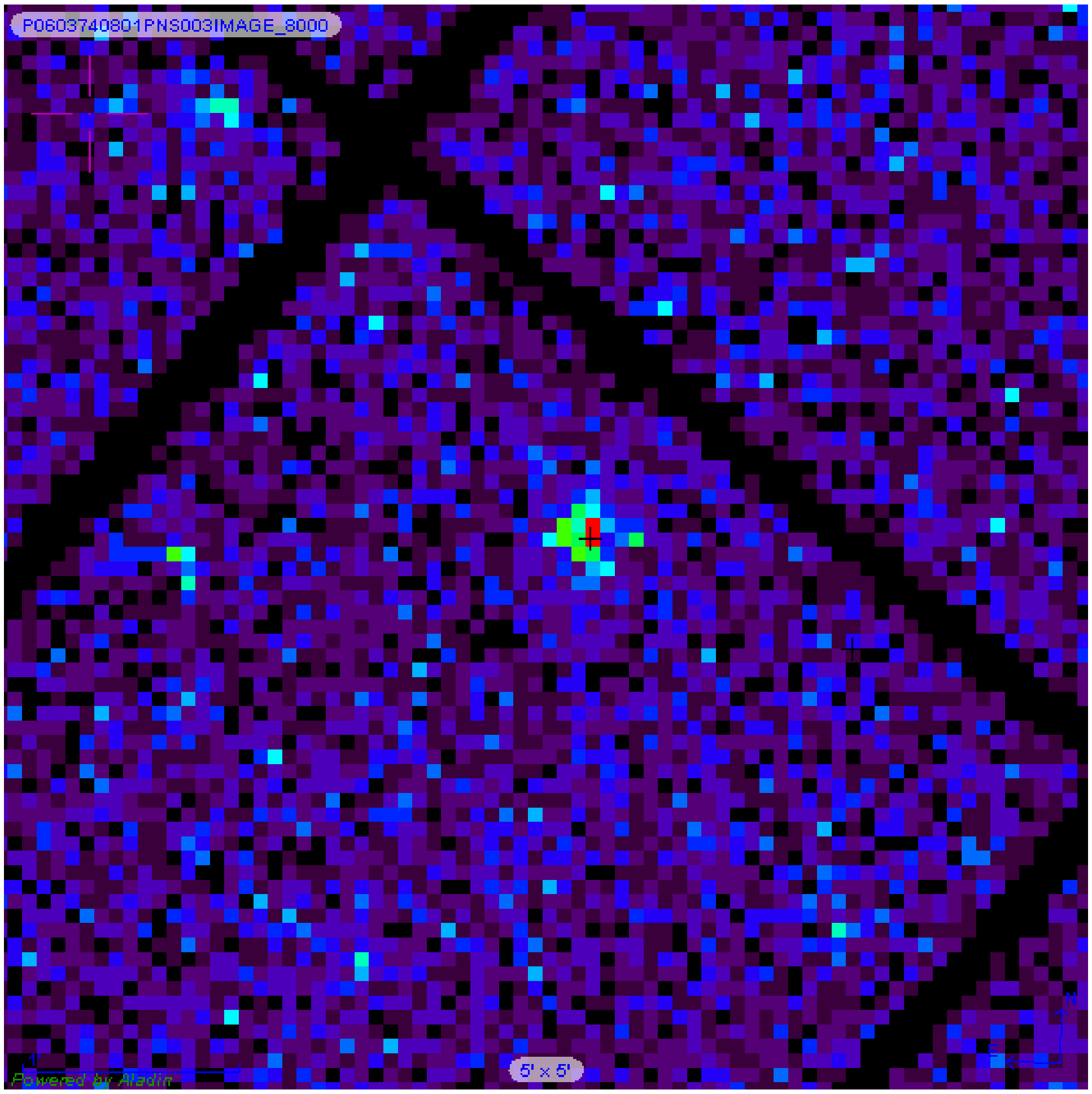

Figure 55 – A 5×5 arcmin section of XMM observation 801 of β Dor is shown. Near the center, the X-ray source can be seen, with the black cross at the center indicating the coordinates of the Cepheid.

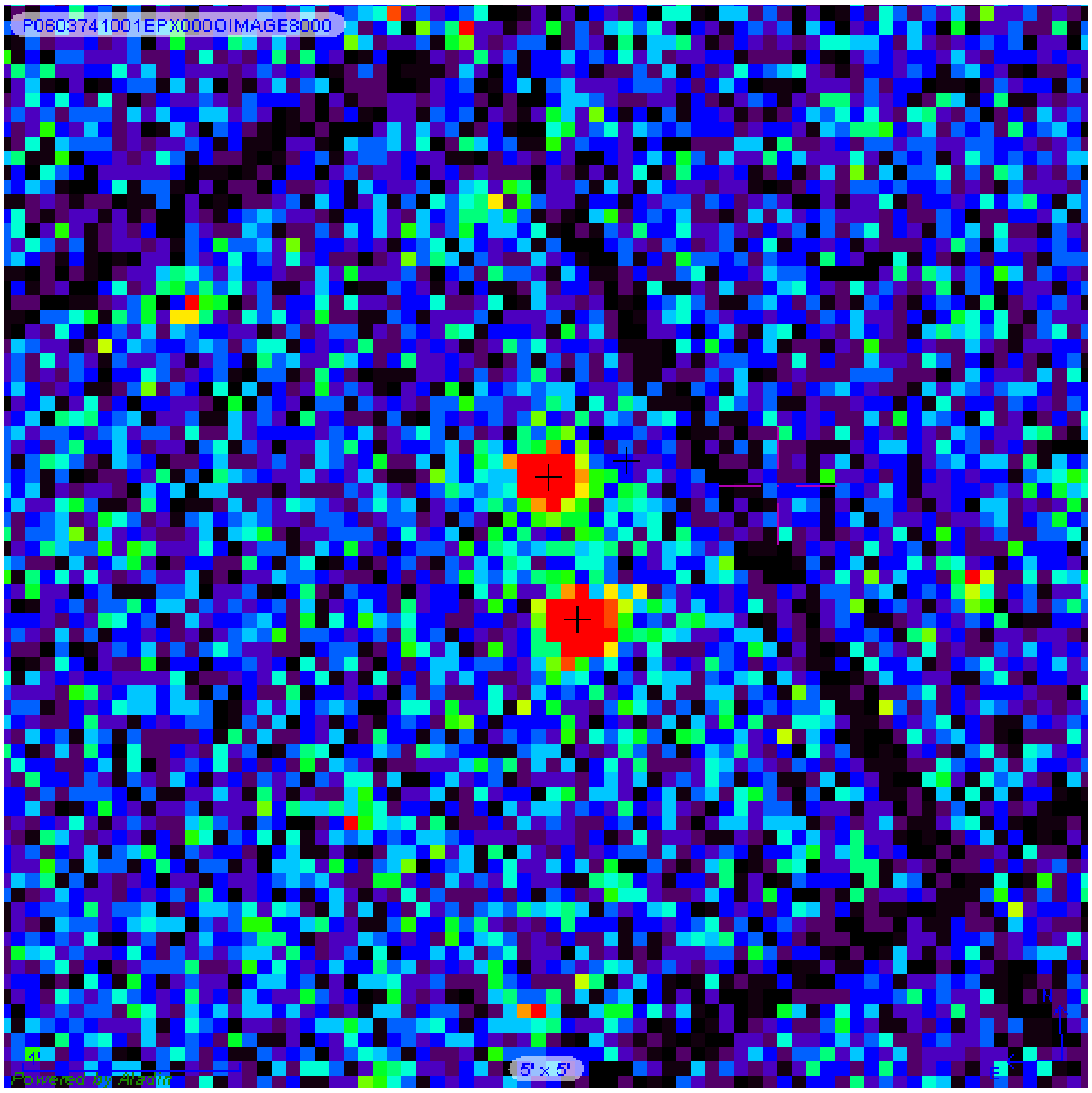

Figure 56 – A 5×5 arcmin section of XMM observation 1001 of δ Cep is shown. Near the center, two X-ray sources are seen, with the black cross at the center of the northern source indicating the coordinates of the Cepheid, and the black cross at the center of the southern source indicating the coordinates of the Cepheid's long-known hotter binary companion (see Section 2.3).

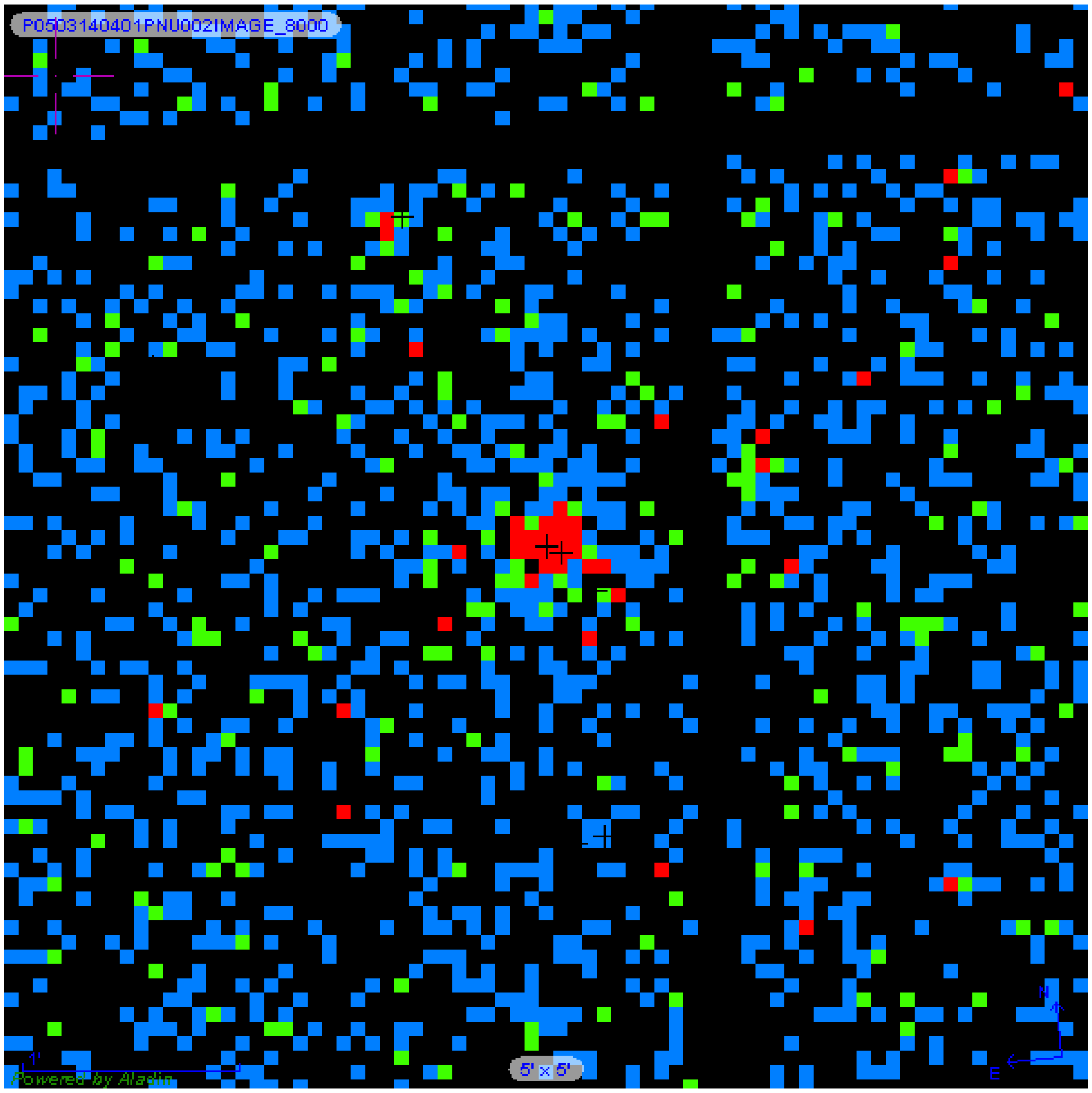

Figure 57 – A 5×5 arcmin section of XMM observation 401 of Polaris is shown. Near the center, the X-ray source can be seen, with the black cross at the center indicating the coordinates of the Cepheid. The black cross below and to the right of the center cross marks the location of a distant background star, originally thought to be a companion to the Cepheid, but recently disproven as such (Evans et al. 2010).

Figs. 58 (Polaris), 59 (β Dor) and 60 (δ Cep) give the energy distributions obtained from our XMM observations, along with the best-fitting 2T MEKAL models for each observation (parameters given in Table 17). Fig. 61 gives only the 2T models of each Cepheid for an easier comparison. As one can see, the background-subtracted count rates that we encounter are somewhat low, due to the distances of the Cepheids. Because of this, uncertainties on the X-ray measures are between ~20 – 40%, so the plasma temperatures and fluxes derived from the data have a bit of "leeway" associated with them. However, the general X-ray properties of the three Cepheids are similar. The energy distributions show all three Cepheids to have peak overall emissions in the 0.6 – 0.8 keV range. By contrast, observations of young, cool main sequence stars show their emissions to consistently peak in the 0.9 – 1.0 keV range. This is a small difference, especially in light of the larger errors associated with the Cepheid observations, but it is noticeable and consistent. The softer X-ray emissions of the Cepheids give evidence that they are the source of the activity, instead of main sequence companions. As with the UV data, X-ray observations of other supergiants can be used to put the Cepheid detections in context.

Fig. 62 shows a plot adapted from Ayres, Brown & Harper (2003), in which various "groups" of cool, X-ray emitting stars are mapped according to their ratios of X-ray and C IV 1550Å luminosities to their bolometric luminosities. The three detected Cepheids have been plotted as the purple, diagonal striped boxes. There does seem to be a division between the two shorter-period Cepheids δ Cep and Polaris – which are also of more similar spectral type – and the longer period, slightly later spectral type Cepheid β Dor. The higher level of UV activity in β Dor is the main segregator between the Cepheids and, of course, observations of additional Cepheids are warranted to determine if this segregation holds true. However, the results are still interesting. Polaris and δ Cep are among the more "UV deficient" stars plotted, and border on Group 5, labeled by Ayres et al. as a region of cool, inactive giants, whereas β Dor lies between Group 5 and the GK supergiants and Hybrids that make up Group 6. The very active supergiants discussed in the UV studies occupy Group 2, and cool, main sequence stars (including the Sun marked by the solar symbol) occupy Group 1. Fig. 63 shows the same general figure (this time adapted from Ayres 2011), but now some of the groups of stars have been replaced with giants and supergiants of different spectral type. Polaris and δ Cep fall very close to the two other F-type supergiants measured, although perhaps a bit X-ray deficient (but confirming this would require measurements of additional stars), whereas β Dor is currently UV deficient when compared to other early-G supergiants. As with O I and Si IV, only the descending branch of the C IV activity curve has been observed for β Dor. However, it would be surprising if the approved observations show the true maximum of C IV activity in β Dor to rival the levels observed in other, non-pulsating (and Hybrid) G-type supergiants. As shown previously in Fig. 51, however, the lower-temperature UV emissions of β Dor can match and even exceed other supergiants. This implies that the pulsations of Cepheids may inhibit an overall atmospheric heating mechanism (perhaps convective strength), and the shocks they generate can excite cooler atmospheric emissions enough to compensate for this deficiency, but the shocks are not strong enough to recover the "missing" higher-temperature activity. Again, in Fig. 63 (as in Fig. 62) the cool, main sequence stars are plotted in the yellow wedge. It can be seen how much more active (when taking into account bolometric luminosity) the cool main sequence stars are, as compared to most cool supergiants and the Cepheids.

As example, from a first glance at the hottest atmospheric emissions, the mean $L_X$ values observed for Cepheids are on the order of 30 – 60× that of the mean solar value (adopting $L_X \approx 1 \times 10^{29}$ ergs/s for Cepheids and average $\langle L_X \rangle_\odot \approx 2 \times 10^{27}$ ergs/s from DeWarf et al. 2010). However, Cepheid surface X-ray fluxes ($F_X$) and $L_X/L_{bol}$ ratios, compared to the Sun, are much smaller (e.g. *log* ($L_X/L_{bol}$) ≈ -7.7 [δ Cep] and -6.3 [Sun]). For further example, in the case of δ Cep (adopting a radius of $R = 44.5\ R_\odot$, $L \approx 2{,}000\ L_\odot$ (Matthews et al. 2012), the average surface X-ray flux is $F_X \approx 8.3 \times 10^2$ ergs/s/cm$^2$. The corresponding mean value for the Sun is $\langle F_X \rangle_\odot \approx 4.1 \times 10^5$ ergs/s/cm$^2$. Thus, the average X-ray surface flux of the Sun is ~500× stronger than that of the Cepheid. There is an even more dramatic difference when the fact that the Sun is a middle-aged main sequence star is taken into account, since the UV and X-ray activity of cool, main sequence stars declines over time. The X-ray surface flux (or $L_X/L_{bol}$ values) of cool main sequence stars at similar ages to those of the Cepheids could be as much as ~100,000× stronger than the Cepheids.

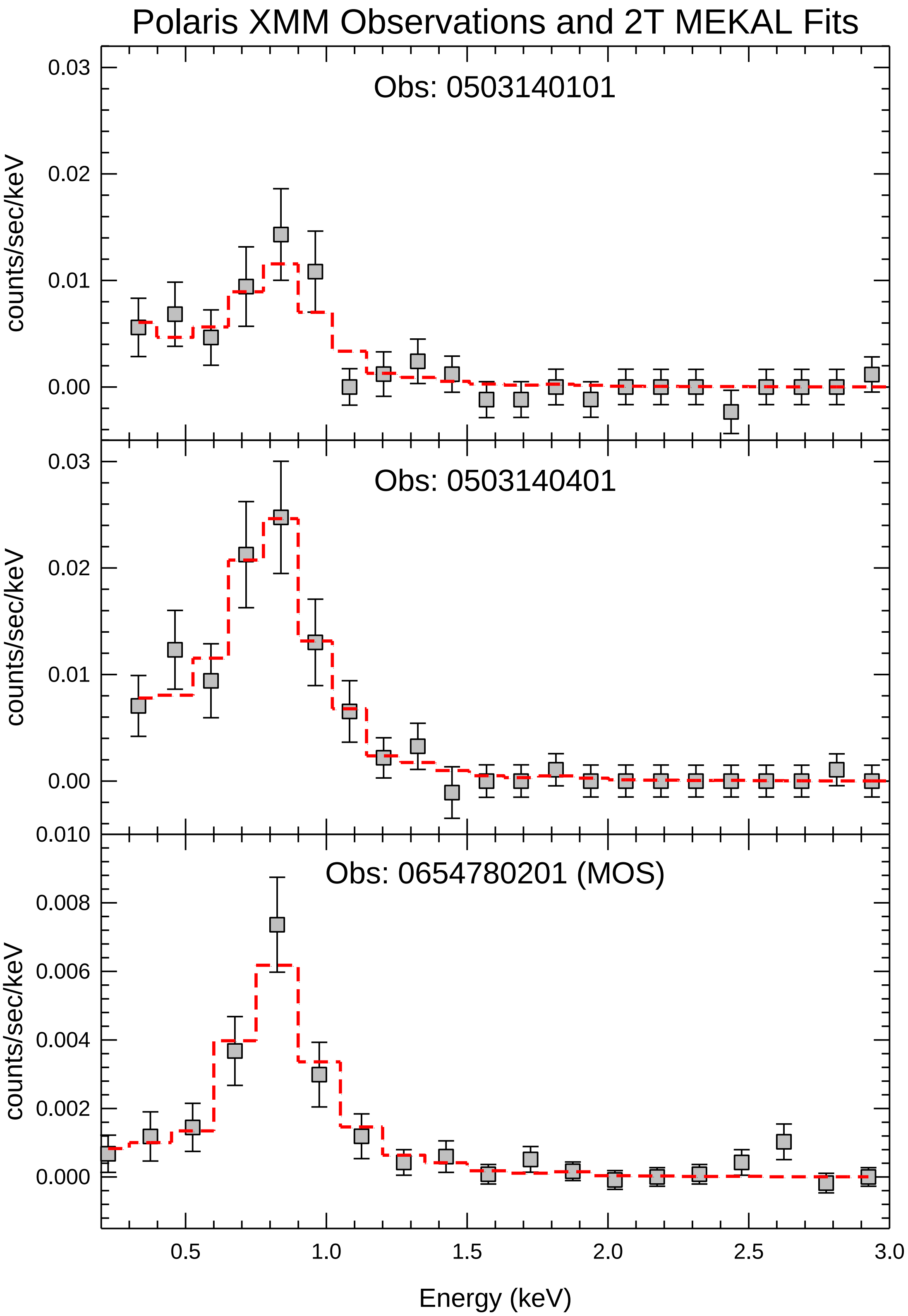

Figure 58 – The three XMM observations of Polaris are shown, along with the best fitting two-temperature MEKAL models.

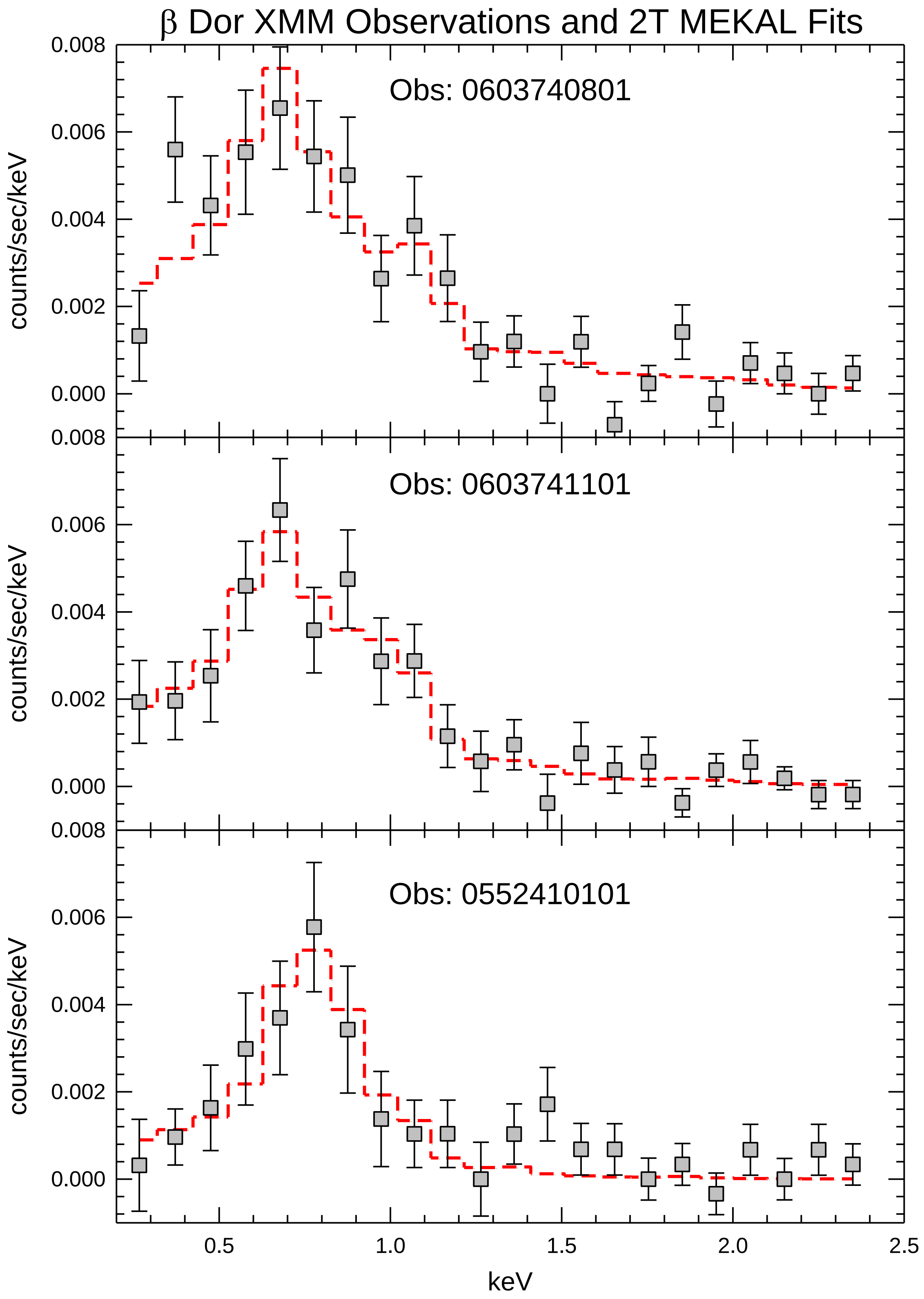

Figure 59 – Same convention as Fig. 58, but for β Dor.

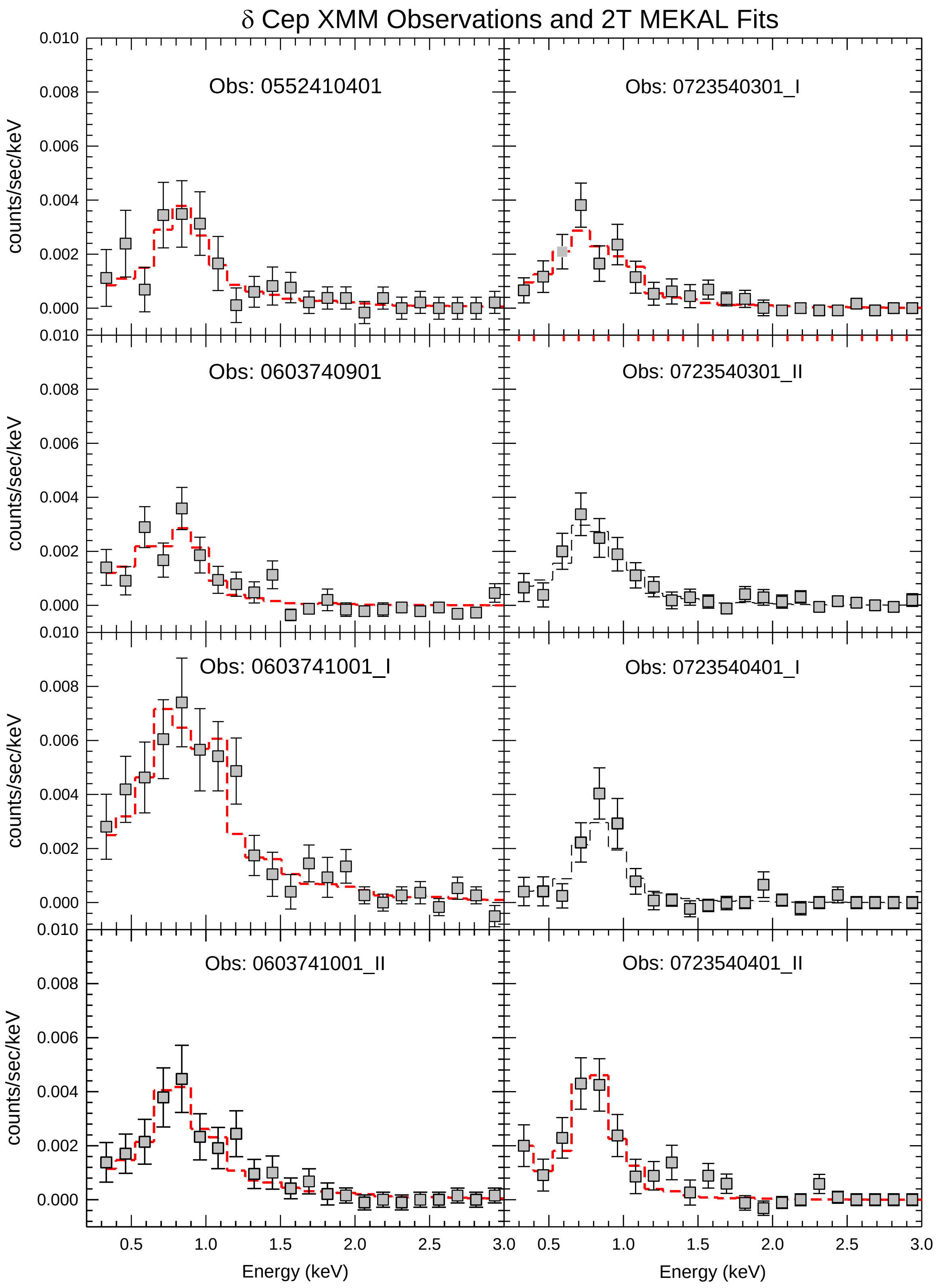

Figure 60 – Same convention as Fig. 58, but for δ Cep.

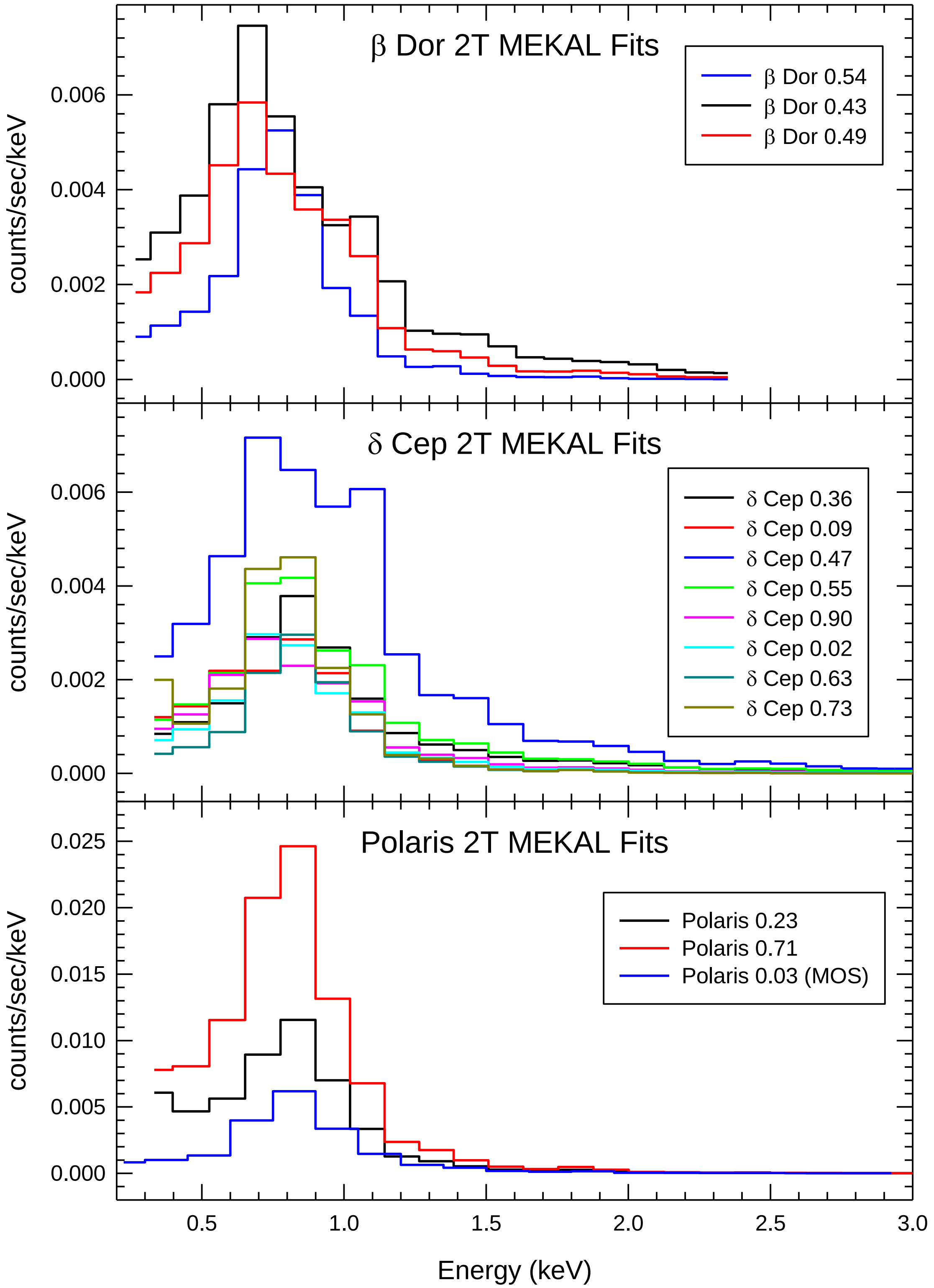

Figure 61 – For ease of comparison, this figure shows only the models from Figs. 58, 59 and 60. The phases of the individual X-ray distributions for each Cepheid are given in the legends.

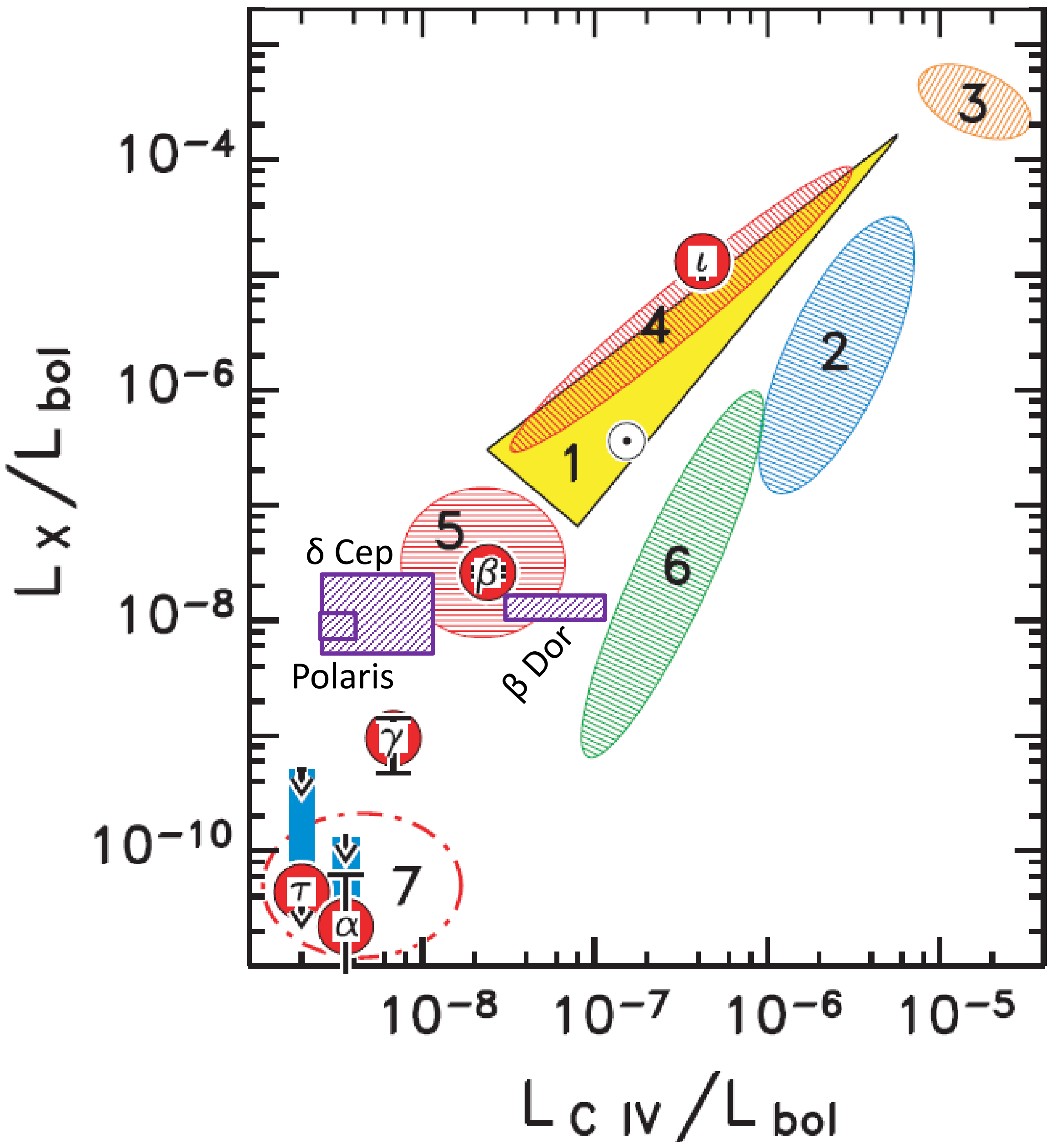

Figure 62 – A cool stars atmospheric comparison plot, adapted from Ayres, Brown & Harper (2003). One can see the separation of β Dor from δ Cep and Polaris, placing it closer to Group 6, which includes the Hybrids. Further details are given in the text. The numbered groups represent (1) GK dwarfs (the circled dot marks the average solar ratio), (2) "X-ray deficient" Hertzsprung gap giants and the "very active supergiants" β Cam and β Dra, (3) hyperactive RS CVn binaries, (4) active clump (G8-K0) giants, (5) inactive but still coronal K0 giants, (6) GK supergiants (the lower activity Hybrids are located here), and (7) noncoronal (≥K1) red giants. Filled red circles mark α Boo ("α" – K1.5 III), τ Tau (" τ " – K5 III), and three comparison stars: ι Cap (" ι " – G8 III), β Gem (" β " K0 III), and γ Dra (" γ " – K5 III). Vertical blue bars connect earlier ROSAT upper limits with newer Chandra measurements.

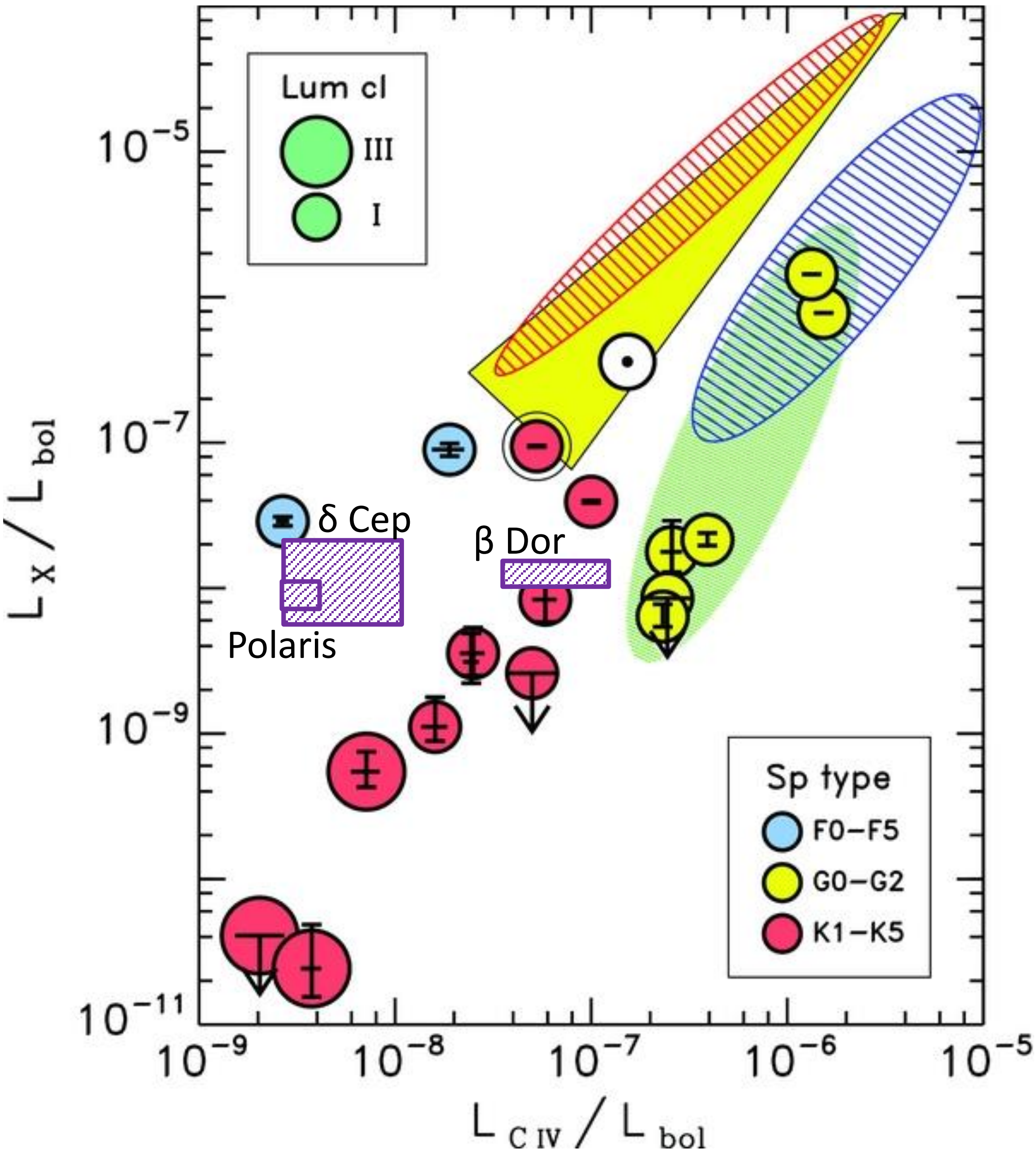

Figure 63 – The same as Fig. 62, except now the separation of supergiant spectral types is shown (adapted from Ayres 2011). Here δ Cep and Polaris are shown to fittingly lie near other F-type supergiants, whereas the somewhat later spectral type β Dor is between the F-type and G-type supergiants, in the range of K-type supergiants.

### 3.3 A Brief Summary of Chapter 3

Multiple, high resolution UV (~1150 – 1750 Å) spectra have been obtained for Polaris, δ Cep and β Dor with *HST*-COS to characterize the effects of the stars' pulsations on their atmospheric emissions from heated plasmas. Only two phases of Polaris' pulsations have been observed thus far, compared to multiple observations of δ Cep and β Dor. The crucial pulsation phases of δ Cep were observed with greater efficiently, though, resulting in a more complete understanding of its atmospheric variability. The emission curve given in Fig. 45 shows an abrupt, strong increase in emission line fluxes beginning at phase = 0.86$\phi$. This is just before the start of the Cepheid's piston phase, where the star begins to expand again. The change from stellar recession to expansion is expected to compress the atmosphere of the Cepheid, and generate a shock, which would propagate through the atmospheric plasmas. The phasing of the increase in flux strongly supports this expectation. In addition, radial velocities show that the atmosphere continues to recede at phases where the photosphere begins expanding, resulting in atmospheric compression. Finally, emission line profiles during the piston phases show an additional, blue-shifted feature originating from excited plasmas at the bow of the propagating shock, and increasing line broadening from phases 0.83 – 0.01$\phi$ show increased emissions from compressed plasmas and also turbulent plasmas in the wake of the shock. The results of our study for δ Cep can also be found in Engle et al. (2014).

For β Dor, the phases of rising emission flux (and possibly maximum flux) have not yet been observed, but important similarities and differences are already apparent when compared to δ Cep. Fig. 46 shows the current emission curve for β Dor. An overall similar behavior to δ Cep is seen, with enhanced atmospheric emissions around the piston phase, but there are specifics that distinguish the two Cepheids. In δ Cep, it appears that N V, the hottest emission line plotted, peaks in activity at a slightly later phase than the cooler lines. For β Dor, however, the phase difference is dramatic, with the peaks separated by ~0.1$\phi$ or more. Preliminary conclusions about the relative activity levels of the Cepheids can also be drawn. Polaris' emission line fluxes are approximate to the average fluxes of δ Cep. This can be expected since the two Cepheids are very similar in terms of both average surface temperature and size, but the more interesting question is the effect of pulsations on the stellar atmospheres, since Polaris' pulsations are so weak in comparison to those of δ Cep. However, with just two COS spectra of Polaris, investigating the effects of pulsation must await further data. The δ Cep and β Dor datasets are more complete, and when comparing maximum O I, Si IV and N V fluxes, β Dor has ~6×, ~3× and ~5× stronger emissions, respectively. With our current understanding of the quiescent levels of O I, Si IV and N V fluxes for β Dor, it is ~16×, ~7× and ~13× more active than δ Cep. Clearly, β Dor possesses a much higher level of persistent atmospheric emissions, yet a smaller amplitude of variability when compared to δ Cep. This would seem to indicate weaker shock activity in β Dor.

Polaris, δ Cep and β Dor have also been successfully detected and measured at X-ray wavelengths. Table 18 gives the observation specifics, the determined plasma temperatures and X-ray activity levels. Figs. 58 – 61 give the reduced data, and show that the three Cepheids display similar X-ray energy distributions, with peak emissions in the 0.6 – 0.9 keV range. As with the UV data, δ Cep has had more complete pulsation phase coverage at X-ray wavelengths, and shows a fairly consistent activity level with the exception of phase ≈ 0.45$\phi$ (Fig. 54). At this phase, there is a ~3× jump in the X-ray flux and the peak X-ray emission broadens out to include higher energies (up to ~1.2 keV). Analyses of this data (background counts, nearby sources…) confirm that the increased flux comes from the location of δ Cep itself, and is not the result of a data artifact or error of some sort. Additional X-ray data have been proposed to confirm the increased activity at this phase, and also cover the missing phases directly beforehand.

# CHAPTER 4 – SUMMARY & CONCLUSIONS

Apart from the scientific conclusions of this program, there is an overall ethos that this study will hopefully convey, and it is that likely no class of object in astronomy has "given up all of its secrets" just yet. For a long time, Cepheids were prized for the stability and predictability of their light curves. Objects like Polaris and V473 Lyr, that displayed either cyclic or long-term amplitude variability, were seen as anomalous. However, this study shows that perhaps these objects are more common than previously thought.

Similarly, in the high-energy (X-ray – UV) regime, the important discoveries of previous IUE studies were never followed up with modern UV instruments, nor were the non-detections with previous-generation X-ray instruments. In fact, even though Polaris *was* detected in a pointed 1997 ROSAT-HRI exposure, the detection went unpublished (and, it would seem, unknown) for roughly a decade, until an archival search by the author "re-discovered" it. Now, recent studies at these wavelengths have the potential to redefine our understanding of supergiant atmospheres under the influence of radial pulsations.

From the two studies contained within this thesis, the conclusions are as follows.

## 4.1 Summary of Findings from the Optical Studies

The several years of regular photometric monitoring carried out as part of this study, when combined with literature data, clearly indicate variations in the light amplitudes of Cepheids on various timescales. This is, of course, in addition to the well-known and documented period changes of Cepheids previously documented. To summarize:

**δ Cep –** The prototype of Cepheids shows a well-documented decreasing period, but also an increasing light amplitude over the past ~100 years. The amplitude change shows an overall linearly increasing trend. Shorter timescale periodicity is hinted at, but the sparseness of the data prevents a conclusive find.

**η Aql –** The first-discovered Cepheid shows a well-documented, increasing period, but its amplitude shows no coherent trend. In all likelihood, to within the errors and calibrations of the data, the amplitude of this Cepheid appears constant over the past ~100 years.

**EU Tau –** This short-period (~2.1-day) Cepheid has recently begun to show period variation, either in the form of a smooth, consistent decrease or in the form of a rapid shift to a shorter pulsation period. The recent lack of regular photometric observations prevents a firm conclusion at this time. The overall trend in the light amplitude measures of EU Tau hints at a possible amplitude decrease over time, however, all measures lie within acceptable errors of each other and the amplitude can easily have remained constant.

**Polaris –** The North Star shows a (mostly) smooth, decreasing pulsation period, except for an apparent abrupt shift in period around 1963 – 1964. This roughly coincides with the beginning of Polaris' decline in light amplitude (visual or *V*-band) from ~0.12- to 0.025-mag. After having reached a minimum amplitude of ~0.025-mag around the year 2000, recent observations show the amplitude to again be increasing. For example, preliminary reductions of the most recent V-band photometry acquired but not included in this thesis, yields a light amplitude of ~0.067-mag, thus continuing the recent trend of increasing light amplitude. Further data will be needed to discern whether we are witnessing the beginning of an amplitude cycle for Polaris, or if it is simply returning to earlier, larger amplitude levels.

**SU Cas –** The shortest-period (~1.95-day) Cepheid studied, the new O-C points indicate that the period is increasing but, as with EU Tau, additional data is needed to confirm a continued period increase, instead of a possible sudden lengthening of the period. The light amplitudes of SU Cas show a possible increase over time, but if the older, less accurate observations are excluded, then the trend becomes rather small, approaching the level of scatter allowed by observational error.

**SV Vul –** The longest-period (~45-day) Cepheid in our study, SV Vul displays a substantial, consistent decrease in period over time. Additionally, SV Vul displays an apparent ~30-year cyclic variability in both light amplitude and pulsation period: an interesting behavior that we will of course continue to monitor.

**SZ Cas –** This Cepheid has for years displayed a well-documented increasing period. On top of this trend, there appeared to be evidence of a lengthy cyclic behavior, as well. However, the more recent data (including those of this study) indicate that the period may now be holding steady. The amplitude behavior of this Cepheid is intriguing, having decreased by over 10% in the span of a couple decades. New data show that the light amplitude may again be increasing.

**SZ Tau –** This Cepheid displays a very complex O-C diagram, likely characterized by constant long-term period decrease, but with substantial additional variability on shorter timescales, with a possible ~59-year period. Measures of SZ Tau's amplitude show that it has steadily increased by ~20% over the past century.

**VY Cyg –** The new O-C points given in this study confirm the recent evidence that the period of this Cepheid is increasing. The quality of the older data makes it hard to determine if this has been a long-term, subtle increase, or if it has only recently begun. VY Cyg is unfortunately deficient in modern light amplitude measures; however, the spread of the existing measures is larger than the observational errors, suggesting that light amplitude variations may be in progress.

**ζ Gem –** This well-studied Cepheid shows a steadily decreasing pulsation period over time with no evidence of additional, secondary period variability. The light amplitudes of this Cepheid display an increase over time, as well as (for the past ~60 years) a possible ~40-year cyclic variability.

**β Dor –** This Cepheid shows a slow, steadily increasing pulsation period over time with no additional coherent period variability. A light amplitude variability study has not yet been conducted for this Cepheid.

### 4.2 Implications of the Optical Study

As part of this study, ten Cepheids have been (and continue to be) observed. Of these, eight of the Cepheids display evidence of light amplitude variability, by way of either long-term increasing/decreasing trends or possible decades-long light amplitude cycles. All ten Cepheids display coherent changes in pulsation period, and three Cepheids show signs of additional cyclic period variability. Of course, the Cepheids studied here represent but a small fraction of the known galactic Cepheids, so this number cannot be taken as indicative of Cepheids as a whole. Changes in the pulsation period of a Cepheid can be expected, as studies of larger samples of Cepheids (e.g. Turner et al. 2006) have shown such changes to be commonplace. However, our findings do indicate that Cepheid light amplitudes may not be as fixed as previously thought.

Light amplitude changes in Cepheids, although still regarded as a rare phenomenon, is not a completely alien concept. As mentioned, the Blazhko effect in RR Lyr stars – cyclic modulations of their pulsational periods and amplitudes – has been known for over a century, and has been observed in an increasing fraction of RR Lyr stars over time. In fact, when all ground- and space-based surveys are taken into account (especially recent studies with the *Kepler* satellite), as much as 50% of all fundamental mode RR Lyr stars display the Blazhko effect (Kolenberg 2012). This is obviously a significant fraction of RR Lyr stars, for whom the κ-mechanism is primarily driving their radial pulsations, as it does with the Cepheids. The pulsations of both classes of variable stars are driven by the same primary mechanism, but many of their stellar properties (Mass, $T_{\rm eff}$, Luminosity, etc.) are very different. However, because the Blazhko effect is a pulsation-related phenomenon, it is not so unreasonable to speculate that Cepheids are also capable of displaying the effect.

At present, no studies have been undertaken to monitor a substantial fraction of galactic Cepheids to search for Blazhko-type variabilities, although the Optical Gravitational Lensing Experiment (*OGLE*) has formed an extensive Cepheid inventory for the Magellanic Clouds. A study of the most recent *completed* phase – OGLE-III – found ~4% of fundamental mode

Cepheids, and ~28% of first overtone Cepheids, to undergo Blazhko-type variability (Soszynski et al. 2008). These percentages are lower than those for RR Lyr stars; however, the longer periods of Cepheids make it more difficult to adequately cover individual pulsation cycles. With multiple pulsations usually being merged together to form a single well-covered light curve, searching for Blazhko-type variability will be much more difficult. Additionally, Cepheids have not benefitted as much as RR Lyr stars have from the modern era of ultra-precise, continuous space-based photometric programs (e.g. there is only one confirmed Cepheid in the entire *Kepler* satellite field of view – V1154 Cyg) which are extremely well-suited to revealing period and amplitude modulations. A near-complete survey of Blazhko-type variability in Cepheids will have to wait for either a future space-based photometric mission or a multi-site, high-cadence ground-based survey program – neither of which are likely in the near future. The recently launched (at the time writing) BRITE-Constellation (http://www.brite-constellation.at/) nano-satellites are designed to carry out very-high-precision photometry, and Cepheid targets have already been approved for observations. However, as the satellites are designed to observe very bright objects, at present only ~6 Cepheids are approved for BRITE-Constellation photometry.

Even with the "high standards" set for Cepheids to reveal Blazhko-type variability, the percentages of "Blazhko Cepheids" found in OGLE-III data is both revealing and encouraging. For RR Lyr stars, the occurrence rate for the effect is ~1.5× higher in the OGLE Galactic Bulge fields than in the LMC. This originally suggested a metallicity dependence, which was found to not be true for Galactic RR Lyr (Smolec 2005), but it is still correct that Blazhko RR Lyr stars occur more frequently in the galaxy than in the LMC. Testing whether this applies to Blazhko Cepheids, as mentioned, will be difficult, to say the least. However, the results of our optical study support the conclusion that Blazhko-type variability in galactic Cepheids may be a more common occurrence than previously thought. The importance of such a finding, given the use of Cepheids as standard candles, easily warrants further investigation.

### 4.3 The High-Energy (UV and X-ray) Study

The UV and X-ray studies of Cepheids have definitely had their ups and downs. On one hand, the quality of the HST-COS spectra and the scientific potential contained within, especially when compared to the previously available IUE data (see Figs. 43 and 44), can be described without exaggeration as amazing. The X-ray data and evidence of variability, when combined with the UV observations, reveal Cepheid atmospheres of such complexity that it appears new ground is being broken, and perhaps Cepheids cannot even be treated as a group, but instead need to be studied and understood on a case-by-case basis, or at least placed into subgroups based on pulsation period and spectral type range (as Figs. 62 and 63 might imply). On the other hand, the competitive nature of applying for satellite observing time, and the number/length of observations required, have combined to slow the progress of the high-energy study. The current empirical findings of the high-energy study are as follows:

**UV fluxes –** Both β Dor and δ Cep display moderate-to-strong variability in the intensities of their atmospheric UV emission lines. Since the observations have been gathered over the course of months, and even years, there can be no doubt that the phasing dictates a pulsation-driven variability (as opposed to a transient event such as a flare, which could possibly be argued for if the data were taken consecutively over a short span of time). Although β Dor has stronger *overall* UV line emissions, δ Cep displays a much larger range of variability. This conclusion is currently mitigated by the poorer phase coverage of β Dor, but future approved HST-COS observations of β Dor have been phase-constrained to further our understanding of this Cepheid's range of UV activity.

**UV phasing –** δ Cep shows a much more abrupt rise in emission flux (although, phases where the flux of the cooler emission lines of β Dor are expected to rise have not yet been observed), which would be in agreement with the photospheric RV curves of the Cepheids. Specifically, δ Cep goes from minimum radius to maximum outward velocity in just ~0.1$\phi$ (~0.54-days), while β Dor takes ~0.4$\phi$ (~3.9-days) to do the same (see Figs. 45 and 46). Based on the current data, β Dor also takes longer to return to quiescent emission levels. The phase at which emission flux in δ Cep begins to increase (~0.85$\phi$) is shortly after the recession of the Cepheid photosphere has begun to decelerate. This theoretically coincides with the formation of a shock in δ Cep and its propagation through the stellar atmosphere. With the present phase coverage, conclusions cannot be drawn for the specific phase at which line emissions begin to strengthen in β Dor. However, phase-lags can be seen in both Cepheids, where the hottest emission feature – the N V 1240Å doublet – peaks later in phase than either of the cooler lines. This behavior could arise from N V forming in a different level of the Cepheid atmosphere, where the lag represents the time needed for the shock to reach the N V emitting region, or it could represent the additional time needed for the shocked plasmas to excite a highly ionized species such as N V. The lag is more pronounced in β Dor, a possible consequence of β Dor having a more extended atmosphere than δ Cep, or perhaps a consequence of specific atmospheric parameters such as density/pressure, relative abundances or shock strengths. Detailed atmospheric modeling is needed to better understand this issue.

**UV line profiles and velocities –** These data have only been shown for δ Cep, since both the rise and fall of its emissions have been observed. The profiles and velocities are very informative, revealing a correlation between atmospheric compression (as indicated by relative atmospheric and photospheric velocities) and a broadening of the line profiles during enhanced emissions, indicating increased density/pressure. The appearance of an additional, blue-shifted emission component in the phases of rising emission flux is clear indication of an outward-propagating shock moving through the atmosphere. Also, increased line width during declining emissions could be a consequence of the high turbulent velocity gradient present in a post-shock atmosphere.

Results from the X-ray study are not as detailed; a consequence of the low signal strength of the Cepheids at these wavelengths, preventing the higher spectral resolution observations and thorough phase coverage from which the UV study has greatly benefitted. Nonetheless, new ground has been broken by the detections of Cepheids at X-ray wavelengths. Five Cepheids were observed, with three detections (Polaris, δ Cep and β Dor) and two non-detections (SU Cas and $\ell$ Car), which can be understood given the greater distances of the undetected targets and the detected Cepheid X-ray luminosities of $\langle \log L_X \rangle \approx 10^{29}$ erg/sec. Although the low count rates prevent highly detailed energy distributions from being built (Figs. 58 – 60), the three detected Cepheids show generally similar X-ray luminosities ($\langle \log L_X \rangle \approx 10^{29}$ erg/sec) and plasma temperatures ($kT \approx 0.3 - 0.7$ keV). When viewing Fig. 61, it could be argued that β Dor has slightly softer X-ray emission than either δ Cep or Polaris, but with the current inventory and quality of data, this conclusion remains somewhat ambiguous. X-ray data from all three Cepheids show variability. For Polaris, the level of variability is low, as would be expected for this low light- and RV-amplitude Cepheid, but within the observational errors, so no formal conclusion on phased variability can yet be made. β Dor has insufficient phase-coverage to conclusively determine variability. δ Cep, on the other hand, shows a level of X-ray variability beyond the errors, and has much better pulsational phase coverage. Further data is needed, and has been proposed for, but at present δ Cep shows evidence of phased X-ray variability, although the active/quiet phases are completely at odds with the UV activity.

**4.4 Implications of the High-Energy Study**

The first question raised is exactly how Cepheids are generating the high-energy activity and variability. A very plausible (and promising) mechanism for the variable, pulsation-phase dependent UV – FUV and observed (possibly phase-dependent) X-ray emissions is discussed by Sasselov & Lester (1994). They concluded that Cepheids should have outer atmospheres heated by acoustic (or magnetic) wave dissipation, in addition to the transient heating by pulsation dynamics. From this theory (and observations of the He I 10,830 Å line), they predict average X-ray surface fluxes (0.05 – 1 keV) in the range of $F_X \approx 600 - 11{,}000$ ergs/cm$^2$/sec. Taking into account the large stellar radii of Cepheids, this suggests average $L_X$ values of ~$10^{29}$ erg/s for Cepheids with P ≤ 10-days, in good agreement with what we have observed in Polaris, β Dor and δ Cep. This study was carried out in 1994, before X-rays were discovered from Cepheids (although, as mentioned, pointed X-ray observations of the brighter (more nearby) Cepheids were carried out by Einstein and ROSAT). Our limited X-ray results seem to confirm these theoretical expectations. Sasselov & Lester (1994) also estimated that the X-ray luminosities of Cepheids could vary by a factor of ~2.5× over the pulsation period, due to propagating shocks within the stellar atmosphere. Indeed, as shown in Fig 54, the X-ray observations to date for δ Cep vary by at least a factor of ~2. Additional phase coverage is needed to fully explore the X-ray maximum (during $\phi \approx 0.5$) and determine the actual range of X-ray emissions. Sasselov & Lester further concluded that Cepheids should not possess solar-type "coronae", as it would likely be difficult to support a hot outer atmosphere overlying the extensive warm chromosphere that the (supergiant) Cepheids would possess.

The dual-mechanism explanation, in which pulsations modify an underlying, persistent heating source, appears more likely when examining the UV variability of the Cepheids. For δ Cep there is a steady quiescent chromosphere at $\phi \approx 0.6 - 0.8$, which is additionally heated by shock/compression behavior. For β Dor, the phase coverage is not yet extensive enough to either support or deny a dual-mechanism explanation. In terms of the X-ray activity and variability, this explanation could be further supported by the non-detection of RR Lyr in a deep Chandra X-ray image (PI: Guinan). The observation places an upper limit of $log\ L_X \approx 28.0$, which is well below the Cepheid detections. There is the small possibility of RR Lyr being a rather strong X-ray variable, but the non-detection is otherwise fairly conclusive. RR Lyr has a much shorter period than the observed Cepheids *and* much larger RV amplitude, which combined would initiate much stronger shocks. However, RR Lyr has a spectral type of A8 – F7, which is earlier than the Cepheids, giving RR Lyr a much shallower convective zone (or none at all). Since many stellar atmospheres in the cool half of the H-R diagram are heated by acoustic/magnetic activity originating in the convective zone, this would imply that a minimum convective zone depth/strength is required to produce the quiescent UV / X-ray activity, which propagating shocks can then modulate. Also, Bono et al. (1999) showed that, for a Cepheid model of similar properties to δ Cep, convective flux increases during phases of compression ($\phi \approx 0.5 - 0.9$) and sharply decreases after. Stronger convection within the Cepheid could lead to a stronger dynamo and increased magnetic activity (and X-ray activity), explaining the observed X-ray variability in δ Cep. However, one can reasonably assume the varying magnetic field would also affect the UV emissions. This does not seem to be the case, although full phase coverage of the Cepheids at UV / X-ray wavelengths would be needed to definitively rule out such behavior. At present, though, the data implies that different mechanisms govern the UV and X-ray variabilities. Proposed XMM observations of δ Cep will provide full phase coverage and help resolve this issue. If the X-ray variability is not necessarily caused by modulation of the magnetic field, then perhaps the structure of the magnetic field is responsible.

Numerous studies have been carried out to detect magnetic fields around Cepheids, and most have either failed or returned ambiguous results. Recently, however, a survey of cool supergiants has detected a weak magnetic field in the Cepheid η Aql (Grunhut et al. 2010). However, neither δ Cep nor ζ Gem were found to have Zeeman signatures, but the high incidence of supergiant magnetic field detections in the study led the authors to conclude that perhaps all cool supergiants possess magnetic fields, but of varying degrees of complexity that could mitigate their detection. This is in accord with an earlier theory suggested by Ayres, Brown & Harper (2003) to explain the atmospheric activity and magnetic fields of red giants, shown in Fig. 64. Ayres suggested that the extended chromospheres of red giants would result in "buried coronae" where the magnetic structures responsible for X-ray activity exist entirely within the extended chromospheres. The low surface gravity Cepheids could also possess extended warm chromospheres, as pointed out by Sasselov & Lester, and their internal structures and convective zone placements could also lead to magnetic structures nearer the stellar surface. The variable ionization and compression of the Cepheid chromospheres would cause differing levels of X-ray absorption, and thus varying levels of X-ray emission. However, this scenario remains speculative; an extrapolation of that proposed for red giants, which happens to loosely fit the observations so far for Cepheids. True confirmation of buried coronae in Cepheids will have to await not only additional data, to get a full picture of X-ray activity and variability, and its relation (if any) to the UV emissions, but also a more detailed model of Cepheid atmospheres and internal structure. This may have to be done on a Cepheid-by-Cepheid basis since Cepheids can have a range of masses, temperatures and luminosities.

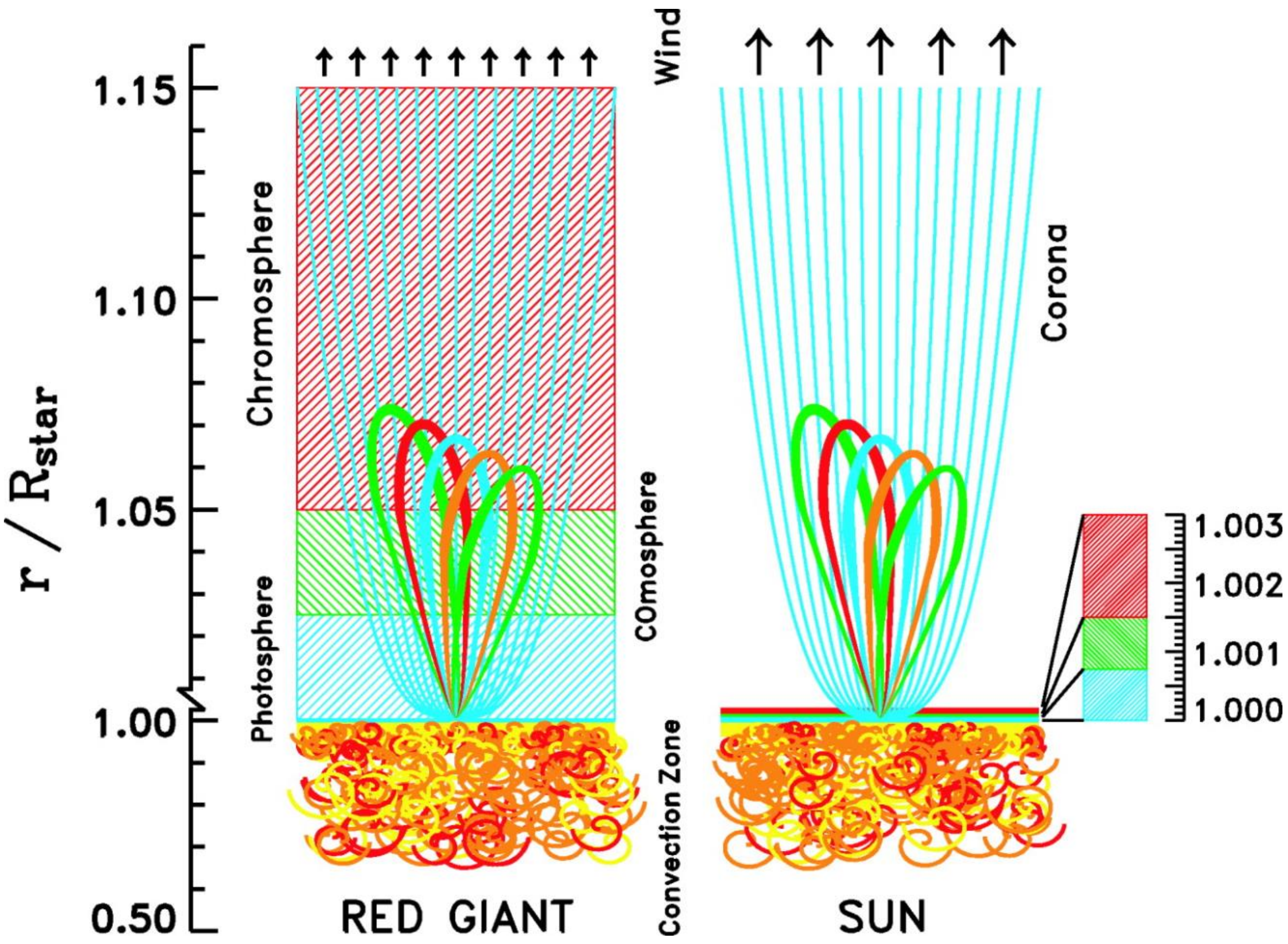

Figure 64 – A cartoon illustration from Ayres, Brown & Harper (2003) depicting the "buried coronae" of red giants. In a cool, main sequence star like the Sun, magnetic structures exist well above the relatively cooler regions of the stellar (solar) atmosphere, and X-ray activity is easily seen. The larger red giants have magnetic structures which scale in height to the Sun, but their low gravities result in much larger, X-ray absorbing chromospheres.

**4.5 Future Work**

The work that will be carried out in the future is, first and foremost, the continued expansion of the Cepheid database at optical, UV and X-ray wavelengths. As seen in Fig. 65, the Cepheids observed here occupy a large range of the instability strip, but they do not uniformly cover that range, and are simply too few to represent Cepheids as a stellar class. All Cepheids for which optical data have been presented here continue to be observed with the aim of obtaining well-covered light curves of the Cepheids as regularly as observational constraints permit. This will allow us to further refine Cepheid period behaviors – e.g. long-term changes and possible cyclic changes – and characterize amplitude variability, and the possibility of Blazhko effects in Cepheids. Although the optical study has shown amplitude variability in several Cepheids, the full characterization of a Blazhko-type effect in Cepheids other than V473 Lyr is the final (hopeful) goal. Unfortunately, many more years of data will likely be required to meet this goal, but it is a challenge that we are up to.

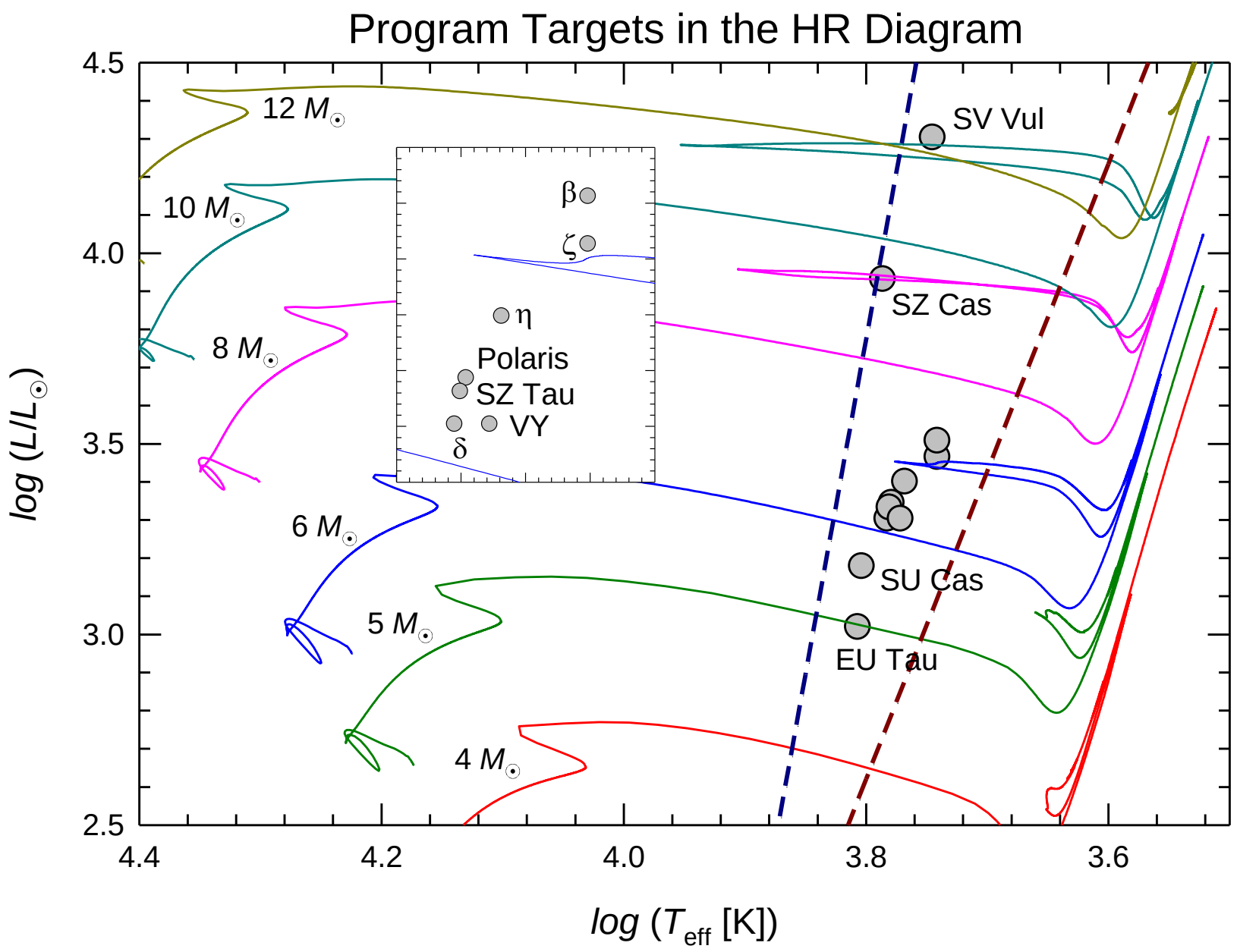

Figure 65 - The Cepheids included in this study are plotted on an HR diagram. SAtlas evolutionary tracks are also plotted (with the initial masses indicated) along with the blue and red edges of the Cepheid instability strip. The sub-plot is included to allow the close group of Cepheids to be spread out for easier identification.

At UV wavelengths, we have recently been approved for additional HST-COS observations of β Dor, to better fill in the activity curve, and for new observations of ℓ Car to precisely understand the large levels of UV activity IUE data have shown for it. HST is currently the only instrument capable of high-resolution UV – FUV spectroscopy, and no future servicing missions will be performed. The only future mission compatible with this study would be the World Space Observatory – Ultraviolet (WSO-UV), currently planned for launch in 2016. However, the possibility of a US Guest Observer program for WSO-UV is uncertain. At present, the pressure is on to observe Cepheids at a variety of periods and pulsation amplitudes so we can begin to understand how UV activity manifests itself and differs across as wide a range of Cepheids as possible. This will give us much greater knowledge of how cool supergiant stellar atmospheres are affected by radial pulsations. Additionally, we plan to observe a sample of non-variable supergiants within the Cepheid instability strip, to provide a more complete understanding of cool supergiant atmospheres as a whole.

The future of the X-ray study is more uncertain. As stated previously, accurate X-ray measures of the Cepheids require long exposure times, and observing time on X-ray satellites such as Chandra and XMM is highly competitive. Also, both are somewhat aged satellites, with Chandra currently in the 13$^{th}$ year of its mission and XMM in the 12$^{th}$. We have recently been approved for a new Chandra observation of β Dor to search for X-ray variability, and have proposed for an additional XMM observation of δ Cep to confirm its X-ray variability. However, the XMM non-detections of two Cepheids pose a possibly serious problem in that the background levels of the XMM detectors may be high enough to prevent the detections of any Cepheids much more distant than β Dor. If true, then Chandra (with its very low background levels) will be the only satellite capable of extending the X-ray inventory to new, more distant Cepheids. A future proposal is planned to carry out exploratory observations of not only new and varied Cepheids but also (as with the UV study) selected non-variable supergiants within the Cepheid instability strip. The only future program-compatible, planned X-ray mission is ASTRO-H, with a proposed launch date of 2015 and operations planned to begin in 2016, but (again) the status of a US Guest Investigator program is unknown. Until ASTRO-H successfully launches, and such a program is approved, it appears time is running short for Cepheid X-ray studies.

Outside of acquiring more data, the current high-energy dataset is of such complexity that it of course warrants further study. The detailed models required to completely exploit the scientific potential of the data is beyond the scope of this thesis. However, recent collaborations have been made to facilitate this endeavor. To better understand the structure of dynamic stellar atmospheres, program collaborator Hilding Neilson (and his own pre-existing collaborators) are developing a new hydrodynamic stellar atmosphere code based on the Sasselov & Raga (1992) hydrodynamic pulsation code and the SAtlas stellar atmospheres code (Lester & Neilson 2008), also taking into account the newest atmospheric emission line data of CHIANTI. This new code will compute the atmospheric structure and emergent radiation of the stars including the effects of pulsation, full LTE opacities and atmospheric extension. This code will model these three aspects together, along with the shock physics necessary to understand Cepheid photospheres. Combinations of two of the three effects have been presented in previous articles, but not all three (Nardetto et al. 2007; Marengo et al. 2002). The code will allow a direct comparison of synthetic and observed spectra as a function of pulsation phase to constrain the role of shocks (as can the comparison of pulsating Cepheid models and spectra to those of non-pulsating supergiants), while differences between synthetic and observed spectra can provide a measure of the role of magnetic heating and shed light on the driving mechanism of Cepheid and supergiant winds. The formation of emission lines in the post-shock regions can be assessed using traditional emission measure analyses (using the CHIANTI IDL procedures), which has more recently been evaluated and improved by considering nonstandard processes including effects associated with the breakdown of statistical equilibria (Judge et al. 1995, and subsequent work). We will also measure the thermal ($\rho\nu T$) and hydrodynamic ($\rho\nu^2$) energy content of the shocks determined by the post-shock temperatures and outflow velocities, respectively. Finally, discrepancies between the shock models and the observations can tell us the role that magnetic heating plays while outflow velocities will hopefully shed light on the important topic of mass loss in Cepheids.

There is still much to be done before we can claim a total understanding of the secret lives of Cepheids. The first step, taken here, is to recognize that these secret lives exist, and can only be understood through dedicated, persistent study and extensive, rewarding collaborations.

# APPENDIX A: CEPHEID PHOTOMETRIC DATA OBTAINED AS PART OF THIS PROGRAM

## Photometry of δ Cep

| HJD | Phase | V-mag | V-err | HJD | Phase | U-B | U-B err | HJD | Phase | B-V | B-V err |
|---|---|---|---|---|---|---|---|---|---|---|---|
| 2454437.5956 | -0.2357 | 4.3136 | 0.0026 | 2454437.5926 | -0.2362 | 0.6518 | 0.0060 | 2454437.5941 | -0.2360 | 0.8588 | 0.0062 |
| 2454439.5928 | 0.1365 | 3.6690 | 0.0056 | 2454439.5941 | 0.1367 | 0.3887 | 0.0216 | 2454439.5928 | 0.1365 | 0.5839 | 0.0104 |
| 2454440.5858 | 0.3215 | 3.9369 | 0.0015 | 2454440.5856 | 0.3215 | 0.5173 | 0.0081 | 2454440.5857 | 0.3215 | 0.7411 | 0.0058 |
| 2454447.5794 | -0.3752 | 4.2614 | 0.0029 | 2454447.5806 | -0.3750 | 0.6512 | 0.0043 | 2454447.5793 | -0.3752 | 0.8752 | 0.0035 |
| 2454449.5797 | -0.0024 | 3.4627 | 0.0206 | 2454449.5795 | -0.0025 | 0.3408 | 0.0429 | 2454449.5797 | -0.0024 | 0.4522 | 0.0270 |
| 2454450.5797 | 0.1839 | 3.7435 | 0.0051 | 2454450.5795 | 0.1839 | 0.4299 | 0.0150 | 2454450.5797 | 0.1839 | 0.6358 | 0.0064 |
| 2454452.6001 | -0.4396 | 4.1927 | 0.0008 | 2454452.6000 | -0.4396 | 0.6393 | 0.0033 | 2454452.6001 | -0.4396 | 0.8575 | 0.0028 |
| 2454454.5902 | -0.0687 | 3.6357 | 0.0080 | 2454454.5914 | -0.0685 | 0.3344 | 0.0223 | 2454454.5901 | -0.0687 | 0.5130 | 0.0192 |
| 2454458.6295 | -0.3160 | 4.3210 | 0.0038 | 2454458.6293 | -0.3160 | 0.6698 | 0.0066 | 2454458.6294 | -0.3160 | 0.8739 | 0.0075 |
| 2454459.6007 | -0.1350 | 4.0087 | 0.0061 | 2454459.6005 | -0.1350 | 0.4212 | 0.0188 | 2454459.6007 | -0.1350 | 0.6829 | 0.0087 |
| 2454460.6012 | 0.0514 | 3.5319 | 0.0044 | 2454460.6010 | 0.0514 | 0.3427 | 0.0241 | 2454460.6011 | 0.0514 | 0.4928 | 0.0098 |
| 2454465.5856 | -0.0197 | 3.4726 | 0.0097 | 2454465.5854 | -0.0197 | 0.3114 | 0.0393 | 2454465.5855 | -0.0197 | 0.4432 | 0.0161 |
| 2454466.5860 | 0.1667 | 3.7179 | 0.0030 | 2454466.5858 | 0.1667 | 0.4135 | 0.0243 | 2454466.5860 | 0.1667 | 0.6152 | 0.0090 |
| 2454623.9115 | 0.4845 | 4.0934 | 0.0069 | 2454623.9125 | 0.4847 | 0.5818 | 0.0075 | 2454623.9114 | 0.4845 | 0.8290 | 0.0102 |
| 2454624.9073 | -0.3299 | 4.2869 | 0.0101 | 2454624.9096 | -0.3295 | 0.6965 | 0.0045 | 2454624.9072 | -0.3299 | 0.8948 | 0.0110 |
| 2454625.9096 | -0.1431 | 4.0377 | 0.0132 | 2454625.9119 | -0.1427 | 0.4678 | 0.0181 | 2454625.9095 | -0.1431 | 0.6995 | 0.0203 |
| 2454626.9007 | 0.0416 | 3.5028 | 0.0736 | 2454626.9005 | 0.0415 | 0.3168 | 0.1136 | 2454626.9006 | 0.0416 | 0.4855 | 0.1044 |
| 2454628.8954 | 0.4133 | 4.0524 | 0.0255 | 2454628.8952 | 0.4133 | 0.5968 | 0.0100 | 2454628.8953 | 0.4133 | 0.7961 | 0.0272 |
| 2454630.8915 | -0.2147 | 4.2666 | 0.0064 | 2454630.8914 | -0.2147 | 0.6220 | 0.0067 | 2454630.8915 | -0.2147 | 0.8435 | 0.0091 |
| 2454631.8927 | -0.0282 | 3.4735 | 0.0338 | 2454631.8925 | -0.0282 | 0.3129 | 0.0547 | 2454631.8926 | -0.0282 | 0.4578 | 0.0472 |
| 2454634.8826 | -0.4710 | 4.1342 | 0.0059 | 2454634.8837 | -0.4708 | 0.6428 | 0.0085 | 2454634.8825 | -0.4710 | 0.8499 | 0.0103 |
| 2454636.8754 | -0.0996 | 3.8125 | 0.0075 | 2454636.8752 | -0.0997 | 0.3607 | 0.0195 | 2454636.8753 | -0.0996 | 0.5974 | 0.0164 |
| 2454637.9466 | 0.1000 | 3.5949 | 0.0221 | 2454637.9464 | 0.1000 | 0.3873 | 0.0198 | 2454637.9465 | 0.1000 | 0.5456 | 0.0281 |
| 2454638.8721 | 0.2725 | 3.8548 | 0.0283 | 2454638.8719 | 0.2724 | 0.4832 | 0.0229 | 2454638.8720 | 0.2724 | 0.6989 | 0.0339 |
| 2454640.8664 | -0.3559 | 4.2744 | 0.0198 | 2454640.8712 | -0.3550 | 0.6568 | 0.0127 | 2454640.8688 | -0.3554 | 0.8617 | 0.0228 |
| 2455098.6395 | -0.0493 | 3.5439 | 0.0031 | 2455098.6394 | -0.0493 | 0.3202 | 0.0300 | 2455098.6395 | -0.0493 | 0.4703 | 0.0115 |
| 2455098.8061 | -0.0182 | 3.4698 | 0.0098 | 2455098.8059 | -0.0183 | 0.3162 | 0.0282 | 2455098.8060 | -0.0182 | 0.4468 | 0.0147 |
| 2455099.6796 | 0.1445 | 3.6712 | 0.0174 | 2455099.6794 | 0.1445 | 0.3933 | 0.0283 | 2455099.6795 | 0.1445 | 0.5873 | 0.0251 |
| 2455099.8098 | 0.1688 | 3.7231 | 0.0083 | 2455099.8096 | 0.1688 | 0.3957 | 0.0177 | 2455099.8097 | 0.1688 | 0.6079 | 0.0111 |
| 2455100.6797 | 0.3309 | 3.9503 | 0.0098 | 2455100.6795 | 0.3309 | 0.4961 | 0.0188 | 2455100.6796 | 0.3309 | 0.7481 | 0.0099 |
| 2455100.8049 | 0.3542 | 3.9894 | 0.0057 | 2455100.8047 | 0.3542 | 0.5408 | 0.0110 | 2455100.8049 | 0.3543 | 0.7592 | 0.0064 |
| 2455101.8009 | -0.4601 | 4.1662 | 0.0051 | 2455101.8007 | -0.4602 | 0.6152 | 0.0068 | 2455101.8008 | -0.4602 | 0.8436 | 0.0056 |
| 2455102.6697 | -0.2982 | 4.3076 | 0.0037 | 2455102.6695 | -0.2983 | 0.6399 | 0.0111 | 2455102.6696 | -0.2983 | 0.8748 | 0.0098 |
| 2455102.7986 | -0.2742 | 4.3196 | 0.0036 | 2455102.7984 | -0.2743 | 0.6272 | 0.0052 | 2455102.7986 | -0.2742 | 0.8799 | 0.0044 |
| 2455106.7864 | 0.4689 | 4.0959 | 0.0071 | 2455106.7862 | 0.4689 | 0.5972 | 0.0116 | 2455106.7863 | 0.4689 | 0.8225 | 0.0090 |
| 2455131.7559 | 0.1220 | 3.6400 | 0.0047 | 2455131.7557 | 0.1220 | 0.3726 | 0.0261 | 2455131.7558 | 0.1220 | 0.5682 | 0.0131 |
| 2455133.7727 | 0.4978 | 4.1171 | 0.0079 | 2455133.7725 | 0.4978 | 0.6057 | 0.0094 | 2455133.7726 | 0.4978 | 0.8312 | 0.0087 |
| 2455134.7810 | -0.3143 | 4.3066 | 0.0058 | 2455134.7808 | -0.3143 | 0.6630 | 0.0104 | 2455134.7810 | -0.3143 | 0.8879 | 0.0103 |
| 2455135.7811 | -0.1279 | 3.9710 | 0.0016 | 2455135.7809 | -0.1279 | 0.4214 | 0.0131 | 2455135.7810 | -0.1279 | 0.6597 | 0.0042 |
| 2455136.7802 | 0.0583 | 3.5396 | 0.0107 | 2455136.7800 | 0.0583 | 0.3263 | 0.0237 | 2455136.7801 | 0.0583 | 0.5041 | 0.0139 |
| 2455143.7567 | 0.3584 | 3.9866 | 0.0033 | 2455143.7565 | 0.3583 | 0.5265 | 0.0077 | 2455143.7566 | 0.3584 | 0.7666 | 0.0068 |
| 2455144.7438 | -0.4577 | 4.1803 | 0.0026 | 2455144.7449 | -0.4575 | 0.6064 | 0.0024 | 2455144.7438 | -0.4577 | 0.8490 | 0.0032 |
| 2455145.7524 | -0.2697 | 4.3230 | 0.0037 | 2455145.7535 | -0.2695 | 0.6619 | 0.0052 | 2455145.7523 | -0.2697 | 0.8749 | 0.0049 |
| 2455146.7367 | -0.0863 | 3.7385 | 0.0118 | 2455146.7378 | -0.0861 | 0.3430 | 0.0261 | 2455146.7366 | -0.0863 | 0.5586 | 0.0177 |
| 2455151.7214 | -0.1574 | 4.1278 | 0.0068 | 2455151.7212 | -0.1574 | 0.4719 | 0.0093 | 2455151.7214 | -0.1574 | 0.7403 | 0.0095 |
| 2455468.8625 | -0.0577 | 3.5880 | 0.0058 | 2455468.8623 | -0.0578 | 0.2636 | 0.0332 | 2455468.8624 | -0.0577 | 0.4836 | 0.0086 |
| 2455469.8524 | 0.1267 | 3.6395 | 0.0091 | 2455469.8522 | 0.1267 | 0.3709 | 0.0277 | 2455469.8523 | 0.1267 | 0.5750 | 0.0101 |
| 2455470.8496 | 0.3126 | 3.9216 | 0.0090 | 2455470.8506 | 0.3128 | 0.4550 | 0.0090 | 2455470.8495 | 0.3126 | 0.7337 | 0.0107 |
| 2455471.8477 | 0.4986 | 4.1325 | 0.0074 | 2455471.8476 | 0.4986 | 0.5523 | 0.0085 | 2455471.8477 | 0.4986 | 0.8203 | 0.0075 |
| 2455476.8328 | 0.4275 | 4.0543 | 0.0066 | 2455476.8326 | 0.4275 | 0.5565 | 0.0078 | 2455476.8328 | 0.4276 | 0.7966 | 0.0079 |
| 2455477.8317 | -0.3863 | 4.2534 | 0.0037 | 2455477.8315 | -0.3863 | 0.6369 | 0.0067 | 2455477.8317 | -0.3863 | 0.8640 | 0.0050 |
| 2455478.8278 | -0.2007 | 4.2579 | 0.0035 | 2455478.8288 | -0.2005 | 0.5306 | 0.0030 | 2455478.8277 | -0.2007 | 0.8274 | 0.0043 |
| 2455479.8259 | -0.0147 | 3.4598 | 0.0058 | 2455479.8258 | -0.0147 | 0.2990 | 0.0374 | 2455479.8259 | -0.0147 | 0.4438 | 0.0124 |
| 2455480.8224 | 0.1710 | 3.7180 | 0.0116 | 2455480.8222 | 0.1710 | 0.3776 | 0.0215 | 2455480.8223 | 0.1710 | 0.6115 | 0.0154 |
| 2455481.8206 | 0.3570 | 3.9785 | 0.0029 | 2455481.8204 | 0.3570 | 0.5003 | 0.0195 | 2455481.8205 | 0.3570 | 0.7636 | 0.0049 |
| 2455482.8167 | -0.4573 | 4.1735 | 0.0006 | 2455482.8165 | -0.4574 | 0.6105 | 0.0096 | 2455482.8166 | -0.4574 | 0.8388 | 0.0063 |
| 2455488.8022 | -0.3419 | 4.2813 | 0.0050 | 2455488.8033 | -0.3417 | 0.6661 | 0.0053 | 2455488.8034 | -0.3417 | 0.8781 | 0.0051 |
| 2455492.8071 | 0.4044 | 4.0341 | 0.0058 | 2455492.8069 | 0.4044 | 0.5646 | 0.0185 | 2455492.8070 | 0.4044 | 0.7843 | 0.0097 |
| 2455493.7961 | -0.4113 | 4.2234 | 0.0036 | 2455493.7959 | -0.4113 | 0.6438 | 0.0093 | 2455493.7960 | -0.4113 | 0.8642 | 0.0054 |
| 2455495.7963 | -0.0386 | 3.5080 | 0.0040 | 2455495.7962 | -0.0386 | 0.3188 | 0.0224 | 2455495.7963 | -0.0386 | 0.4529 | 0.0059 |
| 2455498.7913 | -0.4805 | 4.1397 | 0.0038 | 2455498.7911 | -0.4805 | 0.5898 | 0.0089 | 2455498.7913 | -0.4804 | 0.8403 | 0.0063 |
| 2455499.7861 | -0.2951 | 4.3109 | 0.0060 | 2455499.7872 | -0.2949 | 0.6147 | 0.0060 | 2455499.7860 | -0.2951 | 0.8810 | 0.0075 |
| 2455501.7758 | 0.0757 | 3.5584 | 0.0083 | 2455502.7734 | 0.2616 | 0.4184 | 0.0107 | 2455501.7758 | 0.0757 | 0.5080 | 0.0131 |
| 2455502.7736 | 0.2617 | 3.8492 | 0.0084 | 2455503.7702 | 0.4474 | 0.5650 | 0.0081 | 2455502.7735 | 0.2616 | 0.6962 | 0.0085 |
| 2455503.7704 | 0.4474 | 4.0703 | 0.0041 | 2455504.7687 | -0.3666 | 0.6450 | 0.0064 | 2455503.7703 | 0.4474 | 0.8109 | 0.0073 |
| 2455504.7676 | -0.3668 | 4.2629 | 0.0068 | 2455506.7628 | 0.0051 | 0.3137 | 0.0261 | 2455504.7676 | -0.3668 | 0.8764 | 0.0087 |
| 2455506.7656 | 0.0056 | 3.4644 | 0.0101 | 2455508.7574 | 0.3767 | 0.4885 | 0.0094 | 2455506.7642 | 0.0053 | 0.4493 | 0.0161 |
| 2455508.7576 | 0.3768 | 4.0027 | 0.0015 | 2455510.7521 | -0.2515 | 0.6259 | 0.0038 | 2455508.7575 | 0.3768 | 0.7787 | 0.0063 |
| 2455510.7523 | -0.2515 | 4.3192 | 0.0031 | 2455511.7497 | -0.0656 | 0.3313 | 0.0237 | 2455510.7522 | -0.2515 | 0.8726 | 0.0042 |
| 2455511.7498 | -0.0656 | 3.6148 | 0.0078 | 2455512.7461 | 0.1200 | 0.3340 | 0.0193 | 2455511.7498 | -0.0656 | 0.5041 | 0.0247 |
| 2455512.7463 | 0.1201 | 3.6349 | 0.0040 | 2455513.7464 | 0.3065 | 0.4628 | 0.0073 | 2455512.7462 | 0.1201 | 0.5566 | 0.0100 |
| 2455513.7465 | 0.3065 | 3.9130 | 0.0042 | 2455515.7437 | -0.3213 | 0.6444 | 0.0078 | 2455513.7465 | 0.3065 | 0.7305 | 0.0070 |
| 2455515.7439 | -0.3213 | 4.2962 | 0.0067 | 2455516.7379 | -0.1361 | 0.3991 | 0.0162 | 2455515.7438 | -0.3213 | 0.8734 | 0.0084 |
| 2455516.7381 | -0.1360 | 4.0154 | 0.0058 | 2455532.6060 | -0.1790 | 0.5079 | 0.0084 | 2455516.7381 | -0.1360 | 0.6895 | 0.0087 |
| 2455532.6062 | -0.1790 | 4.1985 | 0.0037 | 2455532.6851 | -0.1643 | 0.4582 | 0.0140 | 2455532.6061 | -0.1790 | 0.7854 | 0.0039 |
| 2455532.6853 | -0.1643 | 4.1480 | 0.0034 | 2455533.5938 | 0.0050 | 0.3076 | 0.0636 | 2455532.6852 | -0.1643 | 0.7549 | 0.0081 |
| 2455533.5940 | 0.0051 | 3.4631 | 0.0044 | 2455862.6716 | 0.3291 | 0.5273 | 0.0073 | 2455533.5939 | 0.0051 | 0.4426 | 0.0056 |
| 2455862.6718 | 0.3292 | 3.9454 | 0.0048 | 2455865.6121 | -0.1229 | 0.4116 | 0.0161 | 2455862.6717 | 0.3291 | 0.7361 | 0.0058 |
| 2455865.6113 | -0.1231 | 3.9479 | 0.0047 | 2455866.5960 | 0.0604 | 0.3642 | 0.0151 | 2455865.6112 | -0.1231 | 0.6473 | 0.0085 |
| 2455866.5962 | 0.0605 | 3.5297 | 0.0046 | 2455867.5953 | 0.2467 | 0.4690 | 0.0202 | 2455866.5962 | 0.0605 | 0.4819 | 0.0056 |
| 2455867.5955 | 0.2467 | 3.8237 | 0.0048 | 2455868.5949 | 0.4329 | 0.5862 | 0.0089 | 2455867.5954 | 0.2467 | 0.6749 | 0.0078 |
| 2455868.5950 | 0.4330 | 4.0575 | 0.0014 | 2455881.5649 | -0.1501 | 0.4587 | 0.0050 | 2455868.5950 | 0.4330 | 0.8013 | 0.0030 |
| 2455881.5651 | -0.1500 | 4.0841 | 0.0023 | 2455881.5798 | -0.1473 | 0.4551 | 0.0141 | 2455881.5651 | -0.1500 | 0.7125 | 0.0035 |
| 2455881.5800 | -0.1473 | 4.0662 | 0.0026 | 2455881.5946 | -0.1445 | 0.4413 | 0.0064 | 2455881.5799 | -0.1473 | 0.7114 | 0.0043 |
| 2455881.5948 | -0.1445 | 4.0545 | 0.0033 | 2455881.6095 | -0.1418 | 0.4362 | 0.0135 | 2455881.5947 | -0.1445 | 0.7006 | 0.0052 |
| 2455881.6097 | -0.1417 | 4.0402 | 0.0042 | 2455881.6244 | -0.1390 | 0.4287 | 0.0077 | 2455881.6096 | -0.1417 | 0.6987 | 0.0069 |
| 2455881.6246 | -0.1390 | 4.0255 | 0.0036 | 2455881.6393 | -0.1362 | 0.4173 | 0.0093 | 2455881.6245 | -0.1390 | 0.6893 | 0.0044 |
| 2455881.6395 | -0.1362 | 4.0084 | 0.0047 | 2455881.6539 | -0.1335 | 0.4129 | 0.0020 | 2455881.6394 | -0.1362 | 0.6861 | 0.0073 |
| 2455881.6541 | -0.1335 | 3.9959 | 0.0056 | 2455881.6685 | -0.1308 | 0.4136 | 0.0092 | 2455881.6541 | -0.1335 | 0.6806 | 0.0058 |
| 2455881.6687 | -0.1307 | 3.9808 | 0.0021 | 2455881.6832 | -0.1280 | 0.4107 | 0.0057 | 2455881.6687 | -0.1307 | 0.6726 | 0.0056 |
| 2455881.6833 | -0.1280 | 3.9622 | 0.0076 | 2455881.6978 | -0.1253 | 0.4064 | 0.0085 | 2455881.6833 | -0.1280 | 0.6657 | 0.0087 |
| 2455881.6980 | -0.1253 | 3.9496 | 0.0025 | 2455881.7125 | -0.1226 | 0.3926 | 0.0164 | 2455881.6980 | -0.1253 | 0.6482 | 0.0061 |
| 2455881.7127 | -0.1225 | 3.9351 | 0.0048 | 2455881.7272 | -0.1198 | 0.3862 | 0.0083 | 2455881.7127 | -0.1225 | 0.6463 | 0.0061 |
| 2455881.7274 | -0.1198 | 3.9171 | 0.0030 | 2455882.7385 | 0.0686 | 0.3482 | 0.0101 | 2455881.7273 | -0.1198 | 0.6407 | 0.0066 |
| 2455882.7387 | 0.0687 | 3.5467 | 0.0059 | 2455888.7206 | 0.1834 | 0.4150 | 0.0201 | 2455882.7386 | 0.0686 | 0.5016 | 0.0065 |
| 2455888.7209 | 0.1834 | 3.7327 | 0.0025 | 2455892.6967 | -0.0757 | 0.3329 | 0.0126 | 2455888.7208 | 0.1834 | 0.6285 | 0.0045 |
| 2455892.6969 | -0.0756 | 3.6782 | 0.0051 | 2455893.7072 | 0.1127 | 0.3696 | 0.0104 | 2455892.6968 | -0.0756 | 0.5418 | 0.0128 |
| 2455893.7074 | 0.1127 | 3.6233 | 0.0062 | 2455900.6873 | 0.4134 | 0.5663 | 0.0066 | 2455893.7073 | 0.1127 | 0.5509 | 0.0072 |
| 2455900.6875 | 0.4134 | 4.0396 | 0.0048 | 2455911.6608 | 0.4583 | 0.5911 | 0.0063 | 2455900.6874 | 0.4134 | 0.7927 | 0.0068 |
| 2455911.6610 | 0.4584 | 4.0806 | 0.0023 | 2455912.6201 | -0.3629 | 0.6577 | 0.0056 | 2455911.6609 | 0.4584 | 0.8116 | 0.0054 |
| 2455912.6203 | -0.3629 | 4.2621 | 0.0061 | 2455919.5647 | -0.0688 | 0.3442 | 0.0165 | 2455912.6202 | -0.3629 | 0.8832 | 0.0077 |
| 2455919.5649 | -0.0687 | 3.6346 | 0.0062 | 2455919.5793 | -0.0660 | 0.3317 | 0.0183 | 2455919.5648 | -0.0687 | 0.5022 | 0.0125 |
| 2455919.5816 | -0.0656 | 3.6206 | 0.0063 | 2455919.5944 | -0.0632 | 0.3344 | 0.0118 | 2455919.5805 | -0.0658 | 0.5072 | 0.0098 |
| 2455919.5946 | -0.0632 | 3.6030 | 0.0111 | 2455919.6066 | -0.0610 | 0.3326 | 0.0126 | 2455919.5946 | -0.0632 | 0.4904 | 0.0152 |
| 2455919.6068 | -0.0609 | 3.5938 | 0.0103 | 2455919.6213 | -0.0582 | 0.3053 | 0.0177 | 2455919.6067 | -0.0609 | 0.4928 | 0.0126 |
| 2455919.6215 | -0.0582 | 3.5778 | 0.0056 | 2455922.6358 | -0.4965 | 0.5951 | 0.0057 | 2455919.6214 | -0.0582 | 0.4911 | 0.0056 |
| 2455922.6360 | -0.4964 | 4.1270 | 0.0054 | 2455926.6210 | 0.2462 | 0.4516 | 0.0113 | 2455922.6359 | -0.4964 | 0.8262 | 0.0065 |
| 2455926.6211 | 0.2462 | 3.8209 | 0.0048 | 2455927.6222 | 0.4328 | 0.5737 | 0.0143 | 2455926.6211 | 0.2462 | 0.6892 | 0.0083 |
| 2455927.6223 | 0.4328 | 4.0375 | 0.0060 | 2455931.6089 | 0.1757 | 0.3997 | 0.0097 | 2455927.6223 | 0.4328 | 0.8152 | 0.0090 |
| 2455931.6091 | 0.1757 | 3.7166 | 0.0021 | | | | | 2455931.6090 | 0.1757 | 0.6286 | 0.0046 |

**Photometry of δ Cep, cont…**

| HJD | Phase | V-R | V-R err | HJD | Phase | R-I | R-I err |
|---|---|---|---|---|---|---|---|
| 2454623.9116 | 0.4846 | 0.8475 | 0.0133 | 2454623.9118 | 0.4846 | 0.4449 | 0.0182 |
| 2454624.9074 | -0.3299 | 0.8906 | 0.0133 | 2454624.9076 | -0.3298 | 0.4568 | 0.0213 |
| 2454625.9097 | -0.1431 | 0.8115 | 0.0306 | 2454625.9099 | -0.1431 | 0.3999 | 0.0298 |
| 2454626.9008 | 0.0416 | 0.6316 | 0.1148 | 2454626.9010 | 0.0416 | 0.3114 | 0.1183 |
| 2454628.8955 | 0.4133 | 0.8242 | 0.0368 | 2454628.8957 | 0.4134 | 0.4223 | 0.0388 |
| 2454630.8917 | -0.2147 | 0.8686 | 0.0130 | 2454630.8919 | -0.2147 | 0.4471 | 0.0157 |
| 2454631.8928 | -0.0281 | 0.6187 | 0.0376 | 2454631.8930 | -0.0281 | 0.3017 | 0.0322 |
| 2454634.8827 | -0.4710 | 0.8436 | 0.0104 | 2454634.8829 | -0.4709 | 0.4573 | 0.0132 |
| 2454636.8755 | -0.0996 | 0.7132 | 0.0159 | 2454636.8757 | -0.0996 | 0.3449 | 0.0187 |
| 2454637.9467 | 0.1000 | 0.6852 | 0.0240 | 2454637.9469 | 0.1001 | 0.3425 | 0.0154 |
| 2454638.8722 | 0.2725 | 0.7897 | 0.0339 | 2454638.8724 | 0.2725 | 0.4029 | 0.0381 |
| 2454640.8665 | -0.3559 | 0.8592 | 0.0332 | 2454640.8667 | -0.3558 | 0.4639 | 0.0394 |
| 2455098.6397 | -0.0492 | 0.6392 | 0.0147 | 2455098.6399 | -0.0492 | 0.3083 | 0.0179 |
| 2455098.8062 | -0.0182 | 0.6168 | 0.0112 | 2455098.8064 | -0.0182 | 0.2922 | 0.0067 |
| 2455099.6797 | 0.1446 | 0.7060 | 0.0267 | 2455099.6799 | 0.1446 | 0.3626 | 0.0325 |
| 2455099.8099 | 0.1688 | 0.7150 | 0.0091 | 2455099.8101 | 0.1689 | 0.3659 | 0.0079 |
| 2455100.6798 | 0.3309 | 0.7976 | 0.0147 | 2455100.6800 | 0.3310 | 0.4048 | 0.0113 |
| 2455100.8050 | 0.3543 | 0.7972 | 0.0098 | 2455100.8052 | 0.3543 | 0.4187 | 0.0084 |
| 2455101.8010 | -0.4601 | 0.8341 | 0.0063 | 2455101.8012 | -0.4601 | 0.4423 | 0.0075 |
| 2455102.6711 | -0.2980 | 0.8716 | 0.0120 | 2455102.6713 | -0.2979 | 0.4458 | 0.0148 |
| 2455102.7988 | -0.2742 | 0.8655 | 0.0043 | 2455102.7989 | -0.2742 | 0.4518 | 0.0037 |
| 2455106.7865 | 0.4689 | 0.8297 | 0.0105 | 2455106.7867 | 0.4690 | 0.4382 | 0.0095 |
| 2455131.7560 | 0.1220 | 0.6839 | 0.0093 | 2455131.7562 | 0.1221 | 0.3430 | 0.0087 |
| 2455133.7728 | 0.4979 | 0.8295 | 0.0112 | 2455133.7730 | 0.4979 | 0.4465 | 0.0096 |
| 2455134.7811 | -0.3142 | 0.8777 | 0.0105 | 2455134.7813 | -0.3142 | 0.4600 | 0.0121 |
| 2455135.7812 | -0.1279 | 0.7493 | 0.0051 | 2455135.7814 | -0.1278 | 0.3802 | 0.0063 |
| 2455136.7803 | 0.0583 | 0.6346 | 0.0162 | 2455136.7805 | 0.0584 | 0.3161 | 0.0164 |
| 2455143.7568 | 0.3584 | 0.7916 | 0.0094 | 2455143.7570 | 0.3584 | 0.4198 | 0.0095 |
| 2455144.7440 | -0.4576 | 0.8426 | 0.0040 | 2455144.7442 | -0.4576 | 0.4436 | 0.0044 |
| 2455145.7525 | -0.2697 | 0.8528 | 0.0057 | 2455145.7527 | -0.2697 | 0.4529 | 0.0047 |
| 2455146.7368 | -0.0863 | 0.6751 | 0.0133 | 2455146.7370 | -0.0862 | 0.3376 | 0.0091 |
| 2455151.7215 | -0.1574 | 0.7938 | 0.0089 | 2455151.7217 | -0.1573 | 0.4094 | 0.0075 |
| 2455468.8626 | -0.0577 | 0.6324 | 0.0083 | 2455468.8628 | -0.0577 | 0.3048 | 0.0098 |
| 2455469.8525 | 0.1268 | 0.6848 | 0.0109 | 2455469.8527 | 0.1268 | 0.3427 | 0.0078 |
| 2455470.8497 | 0.3126 | 0.7778 | 0.0119 | 2455470.8499 | 0.3126 | 0.4063 | 0.0100 |
| 2455471.8479 | 0.4986 | 0.8250 | 0.0085 | 2455471.8481 | 0.4987 | 0.4364 | 0.0092 |
| 2455476.8329 | 0.4276 | 0.8102 | 0.0108 | 2455476.8331 | 0.4276 | 0.4323 | 0.0127 |
| 2455477.8318 | -0.3863 | 0.8584 | 0.0070 | 2455477.8320 | -0.3862 | 0.4547 | 0.0083 |
| 2455478.8279 | -0.2007 | 0.8286 | 0.0038 | 2455478.8281 | -0.2006 | 0.4406 | 0.0048 |
| 2455479.8261 | -0.0146 | 0.6120 | 0.0090 | 2455479.8263 | -0.0146 | 0.2867 | 0.0101 |
| 2455480.8225 | 0.1710 | 0.7142 | 0.0151 | 2455480.8227 | 0.1711 | 0.3588 | 0.0147 |
| 2455481.8207 | 0.3571 | 0.7908 | 0.0090 | 2455481.8209 | 0.3571 | 0.4168 | 0.0099 |
| 2455482.8168 | -0.4573 | 0.8476 | 0.0022 | 2455482.8170 | -0.4573 | 0.4427 | 0.0055 |
| 2455488.8023 | -0.3419 | 0.8648 | 0.0057 | 2455488.8025 | -0.3419 | 0.4536 | 0.0050 |
| 2455492.8072 | 0.4044 | 0.8116 | 0.0069 | 2455492.8074 | 0.4044 | 0.4269 | 0.0051 |
| 2455493.7962 | -0.4113 | 0.8532 | 0.0073 | 2455493.7964 | -0.4113 | 0.4506 | 0.0078 |
| 2455495.7965 | -0.0385 | 0.6257 | 0.0089 | 2455495.7967 | -0.0385 | 0.2863 | 0.0159 |
| 2455498.7914 | -0.4804 | 0.8334 | 0.0042 | 2455498.7916 | -0.4804 | 0.4434 | 0.0021 |
| 2455499.7862 | -0.2950 | 0.8576 | 0.0082 | 2455499.7864 | -0.2950 | 0.4483 | 0.0096 |
| 2455501.7760 | 0.0758 | 0.6490 | 0.0141 | 2455501.7761 | 0.0758 | 0.3168 | 0.0154 |
| 2455502.7737 | 0.2617 | 0.7618 | 0.0091 | 2455502.7739 | 0.2617 | 0.3878 | 0.0057 |
| 2455503.7705 | 0.4474 | 0.8195 | 0.0057 | 2455503.7707 | 0.4475 | 0.4345 | 0.0047 |
| 2455504.7678 | -0.3667 | 0.8593 | 0.0076 | 2455504.7679 | -0.3667 | 0.4492 | 0.0054 |
| 2455506.7644 | 0.0054 | 0.6110 | 0.0125 | 2455506.7633 | 0.0051 | 0.2903 | 0.0178 |
| 2455508.7577 | 0.3768 | 0.8131 | 0.0051 | 2455508.7579 | 0.3768 | 0.4182 | 0.0051 |
| 2455510.7524 | -0.2515 | 0.8705 | 0.0049 | 2455510.7526 | -0.2514 | 0.4505 | 0.0052 |
| 2455511.7500 | -0.0656 | 0.6559 | 0.0093 | 2455511.7502 | -0.0655 | 0.3113 | 0.0094 |
| 2455512.7464 | 0.1201 | 0.6888 | 0.0080 | 2455512.7466 | 0.1201 | 0.3422 | 0.0100 |
| 2455513.7467 | 0.3065 | 0.7853 | 0.0058 | 2455513.7469 | 0.3065 | 0.4056 | 0.0047 |
| 2455515.7440 | -0.3213 | 0.8583 | 0.0084 | 2455515.7442 | -0.3213 | 0.4594 | 0.0066 |
| 2455516.7383 | -0.1360 | 0.7643 | 0.0068 | 2455516.7384 | -0.1360 | 0.3856 | 0.0085 |
| 2455532.6063 | -0.1790 | 0.8205 | 0.0040 | 2455532.6065 | -0.1789 | 0.4245 | 0.0019 |
| 2455532.6854 | -0.1642 | 0.8072 | 0.0039 | 2455532.6856 | -0.1642 | 0.4113 | 0.0024 |
| 2455533.5941 | 0.0051 | 0.6192 | 0.0055 | 2455533.5943 | 0.0051 | 0.2939 | 0.0052 |
| 2455862.6718 | 0.3292 | 0.7970 | 0.0059 | 2455862.6720 | 0.3292 | 0.4151 | 0.0039 |
| 2455865.6113 | -0.1231 | 0.7487 | 0.0054 | 2455865.6114 | -0.1230 | 0.3780 | 0.0073 |
| 2455866.5963 | 0.0605 | 0.6474 | 0.0064 | 2455866.5964 | 0.0605 | 0.3171 | 0.0048 |
| 2455867.5955 | 0.2467 | 0.7607 | 0.0056 | 2455867.5957 | 0.2467 | 0.3860 | 0.0058 |
| 2455868.5951 | 0.4330 | 0.8311 | 0.0035 | 2455868.5952 | 0.4330 | 0.4313 | 0.0053 |
| 2455881.5652 | -0.1500 | 0.7904 | 0.0045 | 2455881.5653 | -0.1500 | 0.4069 | 0.0046 |
| 2455881.5800 | -0.1473 | 0.7815 | 0.0030 | 2455881.5802 | -0.1472 | 0.4031 | 0.0031 |
| 2455881.5949 | -0.1445 | 0.7834 | 0.0067 | 2455881.5950 | -0.1445 | 0.3968 | 0.0064 |
| 2455881.6098 | -0.1417 | 0.7768 | 0.0053 | 2455881.6099 | -0.1417 | 0.3948 | 0.0038 |
| 2455881.6246 | -0.1390 | 0.7713 | 0.0041 | 2455881.6248 | -0.1389 | 0.3922 | 0.0049 |
| 2455881.6396 | -0.1362 | 0.7630 | 0.0063 | 2455881.6397 | -0.1361 | 0.3914 | 0.0066 |
| 2455881.6542 | -0.1334 | 0.7657 | 0.0067 | 2455881.6543 | -0.1334 | 0.3871 | 0.0052 |
| 2455881.6688 | -0.1307 | 0.7597 | 0.0032 | 2455881.6689 | -0.1307 | 0.3864 | 0.0047 |
| 2455881.6834 | -0.1280 | 0.7515 | 0.0079 | 2455881.6835 | -0.1280 | 0.3776 | 0.0049 |
| 2455881.6981 | -0.1253 | 0.7551 | 0.0035 | 2455881.6982 | -0.1252 | 0.3733 | 0.0042 |
| 2455881.7128 | -0.1225 | 0.7531 | 0.0084 | 2455881.7129 | -0.1225 | 0.3776 | 0.0074 |
| 2455881.7274 | -0.1198 | 0.7378 | 0.0052 | 2455881.7276 | -0.1198 | 0.3727 | 0.0050 |
| 2455882.7387 | 0.0687 | 0.6529 | 0.0120 | 2455882.7389 | 0.0687 | 0.3096 | 0.0122 |
| 2455888.7209 | 0.1835 | 0.7202 | 0.0049 | 2455888.7210 | 0.1835 | 0.3609 | 0.0049 |
| 2455892.6969 | -0.0756 | 0.6567 | 0.0080 | 2455892.6971 | -0.0756 | 0.3329 | 0.0099 |
| 2455893.7074 | 0.1127 | 0.6821 | 0.0075 | 2455893.7075 | 0.1127 | 0.3441 | 0.0047 |
| 2455900.6875 | 0.4134 | 0.8189 | 0.0075 | 2455900.6877 | 0.4135 | 0.4316 | 0.0063 |
| 2455911.6610 | 0.4584 | 0.8345 | 0.0057 | 2455911.6611 | 0.4584 | 0.4278 | 0.0082 |
| 2455912.6203 | -0.3629 | 0.8583 | 0.0074 | 2455912.6204 | -0.3628 | 0.4601 | 0.0050 |
| 2455919.5649 | -0.0687 | 0.6601 | 0.0076 | 2455919.5650 | -0.0687 | 0.3246 | 0.0104 |
| 2455919.5806 | -0.0658 | 0.6615 | 0.0101 | 2455919.5796 | -0.0660 | 0.3160 | 0.0112 |
| 2455919.5947 | -0.0632 | 0.6539 | 0.0117 | 2455919.5948 | -0.0632 | 0.3128 | 0.0101 |
| 2455919.6069 | -0.0609 | 0.6510 | 0.0114 | 2455919.6070 | -0.0609 | 0.3162 | 0.0140 |
| 2455919.6215 | -0.0582 | 0.6572 | 0.0092 | 2455919.6217 | -0.0581 | 0.2988 | 0.0087 |
| 2455922.6360 | -0.4964 | 0.8478 | 0.0065 | 2455922.6361 | -0.4964 | 0.4338 | 0.0039 |
| 2455926.6212 | 0.2462 | 0.7505 | 0.0086 | 2455926.6214 | 0.2463 | 0.3917 | 0.0088 |
| 2455927.6224 | 0.4328 | 0.8066 | 0.0098 | 2455927.6225 | 0.4328 | 0.4227 | 0.0108 |
| 2455931.6091 | 0.1757 | 0.7170 | 0.0067 | 2455931.6092 | 0.1757 | 0.3588 | 0.0071 |

**Photometry of η Aql**

| HJD | Phase | V-mag | V-err | HJD | Phase | U-B | U-B err | HJD | Phase | B-V | B-V err |
|---|---|---|---|---|---|---|---|---|---|---|---|
| 2454623.8857 | 0.2292 | 3.7495 | 0.0174 | 2454623.8860 | 0.2292 | 0.4789 | 0.0199 | 2454623.8858 | 0.2292 | 0.8047 | 0.0235 |
| 2454625.8694 | -0.4944 | 4.0584 | 0.0210 | 2454625.8697 | -0.4944 | 0.6830 | 0.0176 | 2454625.8695 | -0.4944 | 0.9514 | 0.0246 |
| 2454626.8750 | -0.3543 | 4.2333 | 0.0040 | 2454626.8752 | -0.3543 | 0.7556 | 0.0047 | 2454626.8750 | -0.3543 | 1.0183 | 0.0046 |
| 2454627.8737 | -0.2151 | 4.1733 | 0.0126 | 2454627.8739 | -0.2151 | 0.5786 | 0.0156 | 2454627.8737 | -0.2151 | 0.9398 | 0.0161 |
| 2454628.8784 | -0.0751 | 3.6784 | 0.0111 | 2454628.8786 | -0.0751 | 0.3893 | 0.0326 | 2454628.8784 | -0.0751 | 0.6897 | 0.0233 |
| 2454630.8659 | 0.2018 | 3.7484 | 0.0115 | 2454630.8661 | 0.2018 | 0.4931 | 0.0238 | 2454630.8660 | 0.2018 | 0.7816 | 0.0161 |
| 2454631.8599 | 0.3403 | 3.7627 | 0.0160 | 2454631.8602 | 0.3403 | 0.5411 | 0.0137 | 2454631.8600 | 0.3403 | 0.8170 | 0.0173 |
| 2454634.9098 | -0.2348 | 4.2332 | 0.0101 | 2454634.9100 | -0.2347 | 0.6800 | 0.0097 | 2454634.9099 | -0.2348 | 0.9591 | 0.0109 |
| 2454636.8961 | 0.0420 | 3.5217 | 0.0227 | 2454636.8963 | 0.0420 | 0.3654 | 0.0306 | 2454636.8961 | 0.0420 | 0.6261 | 0.0344 |
| 2454638.9073 | 0.3222 | 3.7592 | 0.0139 | 2454638.9076 | 0.3223 | 0.5174 | 0.0165 | 2454638.9074 | 0.3222 | 0.8106 | 0.0205 |
| 2454643.9336 | 0.0225 | 3.4871 | 0.0178 | 2454643.9338 | 0.0226 | 0.3725 | 0.0260 | 2454643.9337 | 0.0226 | 0.6116 | 0.0202 |
| 2454648.8659 | -0.2902 | 4.2592 | 0.0054 | 2454648.8661 | -0.2902 | 0.7303 | 0.0105 | 2454648.8660 | -0.2902 | 1.0154 | 0.0061 |
| 2454733.6930 | -0.4710 | 4.0718 | 0.0106 | 2454733.6933 | -0.4709 | 0.7301 | 0.0063 | 2454733.6931 | -0.4709 | 0.9719 | 0.0117 |
| 2454734.6956 | -0.3313 | 4.2543 | 0.0023 | 2454734.6959 | -0.3312 | 0.7674 | 0.0111 | 2454734.6957 | -0.3313 | 1.0093 | 0.0026 |
| 2454737.6951 | 0.0867 | 3.5780 | 0.0030 | 2454737.6953 | 0.0867 | 0.4031 | 0.0320 | 2454737.6951 | 0.0867 | 0.6589 | 0.0042 |
| 2454738.7153 | 0.2288 | 3.7501 | 0.0065 | 2454738.7156 | 0.2289 | 0.4983 | 0.0124 | 2454738.7154 | 0.2288 | 0.8016 | 0.0081 |
| 2454741.6762 | -0.3586 | 4.2165 | 0.0028 | 2454741.6765 | -0.3586 | 0.7782 | 0.0107 | 2454741.6763 | -0.3586 | 1.0125 | 0.0058 |
| 2455098.7215 | 0.3897 | 3.8754 | 0.0055 | 2455098.7218 | 0.3898 | 0.5918 | 0.0207 | 2455098.7216 | 0.3898 | 0.8737 | 0.0091 |
| 2455099.6100 | -0.4865 | 4.0787 | 0.0043 | 2455099.6102 | -0.4864 | 0.6830 | 0.0150 | 2455099.6100 | -0.4865 | 0.9518 | 0.0045 |
| 2455099.6477 | -0.4812 | 4.0746 | 0.0016 | 2455099.6490 | -0.4810 | 0.6877 | 0.0043 | 2455099.6478 | -0.4812 | 0.9655 | 0.0038 |
| 2455099.7313 | -0.4696 | 4.0812 | 0.0075 | 2455099.7316 | -0.4695 | 0.7206 | 0.0086 | 2455099.7314 | -0.4696 | 0.9690 | 0.0082 |
| 2455100.6088 | -0.3473 | 4.2484 | 0.0026 | 2455100.6090 | -0.3473 | 0.7551 | 0.0038 | 2455100.6088 | -0.3473 | 1.0116 | 0.0030 |
| 2455100.7239 | -0.3313 | 4.2476 | 0.0044 | 2455100.7242 | -0.3312 | 0.7770 | 0.0093 | 2455100.7240 | -0.3312 | 1.0163 | 0.0073 |
| 2455101.6848 | -0.1974 | 4.1294 | 0.0062 | 2455101.6851 | -0.1973 | 0.5789 | 0.0150 | 2455101.6849 | -0.1974 | 0.9053 | 0.0112 |
| 2455102.6090 | -0.0686 | 3.6543 | 0.0048 | 2455102.6072 | -0.0689 | 0.3688 | 0.0296 | 2455102.6081 | -0.0687 | 0.6642 | 0.0082 |
| 2455102.7213 | -0.0530 | 3.5866 | 0.0120 | 2455102.7215 | -0.0529 | 0.3544 | 0.0262 | 2455102.7213 | -0.0530 | 0.6262 | 0.0191 |
| 2455133.6527 | 0.2568 | 3.7685 | 0.0029 | 2455133.6529 | 0.2569 | 0.5197 | 0.0085 | 2455133.6527 | 0.2568 | 0.8070 | 0.0044 |
| 2455134.6521 | 0.3961 | 3.9132 | 0.0019 | 2455134.6524 | 0.3961 | 0.6100 | 0.0155 | 2455134.6522 | 0.3961 | 0.8760 | 0.0071 |
| 2455136.6452 | -0.3262 | 4.2535 | 0.0055 | 2455136.6454 | -0.3262 | 0.7575 | 0.0089 | 2455136.6452 | -0.3262 | 1.0197 | 0.0092 |
| 2455137.6444 | -0.1870 | 4.1141 | 0.0029 | 2455137.6447 | -0.1870 | 0.5487 | 0.0053 | 2455137.6445 | -0.1870 | 0.8840 | 0.0045 |
| 2455138.6440 | -0.0477 | 3.5579 | 0.0057 | 2455138.6443 | -0.0477 | 0.3490 | 0.0302 | 2455138.6441 | -0.0477 | 0.6176 | 0.0074 |
| 2455139.6431 | 0.0915 | 3.5839 | 0.0040 | 2455139.6434 | 0.0915 | 0.4044 | 0.0122 | 2455139.6432 | 0.0915 | 0.6634 | 0.0051 |
| 2455141.6423 | 0.3700 | 3.8287 | 0.0086 | 2455141.6426 | 0.3701 | 0.5140 | 0.0543 | 2455141.6424 | 0.3701 | 0.8404 | 0.0117 |
| 2455143.6450 | -0.3509 | 4.2309 | 0.0024 | 2455143.6453 | -0.3509 | 0.8021 | 0.0055 | 2455143.6451 | -0.3509 | 1.0100 | 0.0051 |
| 2455144.6407 | -0.2122 | 4.1795 | 0.0069 | 2455144.6410 | -0.2121 | 0.6452 | 0.0088 | 2455144.6408 | -0.2122 | 0.9338 | 0.0098 |
| 2455146.6396 | 0.0663 | 3.5446 | 0.0096 | 2455146.6398 | 0.0664 | 0.3857 | 0.0281 | 2455146.6396 | 0.0663 | 0.6509 | 0.0136 |
| 2455508.5890 | 0.4980 | 4.0595 | 0.0012 | 2455508.5893 | 0.4981 | 0.6794 | 0.0076 | 2455508.5891 | 0.4980 | 0.9506 | 0.0042 |
| 2455513.5866 | 0.1943 | 3.7499 | 0.0044 | 2455513.5869 | 0.1944 | 0.4950 | 0.0130 | 2455513.5867 | 0.1944 | 0.7727 | 0.0067 |
| 2455516.5857 | -0.3878 | 4.2044 | 0.0008 | 2455516.5860 | -0.3877 | 0.7682 | 0.0077 | 2455516.5858 | -0.3878 | 1.0020 | 0.0052 |
| 2455826.8136 | -0.1626 | 4.0397 | 0.0100 | 2455826.8138 | -0.1626 | 0.5278 | 0.0141 | 2455826.8136 | -0.1626 | 0.7967 | 0.0141 |
| 2455827.8105 | -0.0237 | 3.5147 | 0.0106 | 2455827.8108 | -0.0237 | 0.3887 | 0.0778 | 2455827.8106 | -0.0237 | 0.5870 | 0.0238 |
| 2455828.7796 | 0.1113 | 3.6196 | 0.0205 | 2455828.7798 | 0.1113 | 0.4207 | 0.0269 | 2455828.7796 | 0.1113 | 0.7011 | 0.0280 |
| 2455840.7593 | -0.2195 | 4.2057 | 0.0035 | 2455840.7596 | -0.2195 | 0.6511 | 0.0049 | 2455840.7594 | -0.2195 | 0.9207 | 0.0045 |
| 2455842.7536 | 0.0583 | 3.5371 | 0.0090 | 2455842.7539 | 0.0584 | 0.3983 | 0.0221 | 2455842.7537 | 0.0583 | 0.6427 | 0.0157 |
| 2455843.7499 | 0.1972 | 3.7463 | 0.0072 | 2455843.7502 | 0.1972 | 0.5191 | 0.0056 | 2455843.7500 | 0.1972 | 0.7718 | 0.0080 |
| 2455844.7578 | 0.3376 | 3.7876 | 0.0042 | 2455844.7580 | 0.3376 | 0.5523 | 0.0057 | 2455844.7578 | 0.3376 | 0.8048 | 0.0057 |
| 2455845.7454 | 0.4752 | 4.0403 | 0.0066 | 2455845.7456 | 0.4752 | 0.6979 | 0.0071 | 2455845.7455 | 0.4752 | 0.9298 | 0.0079 |
| 2455846.7423 | -0.3859 | 4.2054 | 0.0032 | 2455846.7426 | -0.3859 | 0.7649 | 0.0071 | 2455846.7424 | -0.3859 | 0.9951 | 0.0054 |
| 2455847.7385 | -0.2471 | 4.2581 | 0.0027 | 2455847.7399 | -0.2469 | 0.7115 | 0.0058 | 2455847.7386 | -0.2471 | 0.9558 | 0.0058 |
| 2455848.7359 | -0.1081 | 3.8300 | 0.0048 | 2455848.7362 | -0.1081 | 0.4342 | 0.0173 | 2455848.7360 | -0.1081 | 0.7312 | 0.0066 |
| 2455849.7336 | 0.0309 | 3.5156 | 0.0066 | 2455849.7339 | 0.0309 | 0.3809 | 0.0230 | 2455849.7337 | 0.0309 | 0.6146 | 0.0149 |
| 2455854.7336 | -0.2725 | 4.2693 | 0.0014 | 2455854.7338 | -0.2724 | 0.7449 | 0.0029 | 2455854.7337 | -0.2724 | 0.9766 | 0.0025 |
| 2455855.5865 | -0.1536 | 3.9784 | 0.0042 | 2455855.5868 | -0.1536 | 0.4577 | 0.0097 | 2455855.5866 | -0.1536 | 0.8074 | 0.0047 |
| 2455856.5858 | -0.0144 | 3.4988 | 0.0042 | 2455856.5861 | -0.0143 | 0.3482 | 0.0129 | 2455856.5859 | -0.0144 | 0.5964 | 0.0134 |
| 2455857.5849 | 0.1248 | 3.6486 | 0.0066 | 2455857.5852 | 0.1249 | 0.4194 | 0.0190 | 2455857.5850 | 0.1248 | 0.7141 | 0.0120 |
| 2455858.5841 | 0.2641 | 3.7700 | 0.0029 | 2455858.5844 | 0.2641 | 0.5074 | 0.0083 | 2455858.5842 | 0.2641 | 0.8048 | 0.0034 |
| 2455861.5818 | -0.3183 | 4.2733 | 0.0025 | 2455861.5821 | -0.3182 | 0.7622 | 0.0044 | 2455861.5819 | -0.3183 | 1.0127 | 0.0026 |
| 2455862.5706 | -0.1805 | 4.0873 | 0.0017 | 2455862.5708 | -0.1805 | 0.5395 | 0.0039 | 2455862.5706 | -0.1805 | 0.8611 | 0.0027 |
| 2455862.5852 | -0.1785 | 4.0754 | 0.0028 | 2455862.5855 | -0.1784 | 0.5325 | 0.0060 | 2455862.5853 | -0.1784 | 0.8635 | 0.0047 |
| 2455862.5999 | -0.1764 | 4.0690 | 0.0039 | 2455862.6001 | -0.1764 | 0.5263 | 0.0073 | 2455862.6000 | -0.1764 | 0.8542 | 0.0067 |
| 2455862.6146 | -0.1744 | 4.0575 | 0.0019 | 2455862.6149 | -0.1743 | 0.5219 | 0.0083 | 2455862.6147 | -0.1744 | 0.8516 | 0.0063 |
| 2455862.6293 | -0.1723 | 4.0524 | 0.0024 | 2455862.6296 | -0.1723 | 0.5239 | 0.0042 | 2455862.6294 | -0.1723 | 0.8462 | 0.0035 |
| 2455864.6530 | 0.1097 | 3.6188 | 0.0032 | 2455864.6532 | 0.1097 | 0.4196 | 0.0121 | 2455864.6531 | 0.1097 | 0.6914 | 0.0110 |
| 2455865.5790 | 0.2387 | 3.7718 | 0.0045 | 2455865.5792 | 0.2387 | 0.4964 | 0.0089 | 2455865.5790 | 0.2387 | 0.7943 | 0.0064 |

**Photometry of η Aql, cont…**

| HJD | Phase | V-R | V-R err | HJD | Phase | R-I | R-I err |
|---|---|---|---|---|---|---|---|
| 2454623.8859 | 0.2292 | 0.6292 | 0.0248 | 2454623.8861 | 0.2293 | 0.4401 | 0.0215 |
| 2454625.8696 | -0.4944 | 0.7119 | 0.0270 | 2454625.8698 | -0.4943 | 0.5025 | 0.0336 |
| 2454626.8751 | -0.3543 | 0.7316 | 0.0042 | 2454626.8753 | -0.3542 | 0.5265 | 0.0056 |
| 2454627.8738 | -0.2151 | 0.7192 | 0.0168 | 2454627.8741 | -0.2151 | 0.5564 | 0.0166 |
| 2454628.8785 | -0.0751 | 0.5682 | 0.0219 | 2454628.8788 | -0.0751 | 0.4037 | 0.0364 |
| 2454630.8660 | 0.2018 | 0.6188 | 0.0229 | 2454630.8663 | 0.2018 | 0.4391 | 0.0339 |
| 2454631.8601 | 0.3403 | 0.6401 | 0.0183 | 2454631.8603 | 0.3403 | 0.4392 | 0.0117 |
| 2454634.9099 | -0.2348 | 0.7137 | 0.0132 | 2454634.9102 | -0.2347 | 0.4863 | 0.0158 |
| 2454636.8962 | 0.0420 | 0.5257 | 0.0304 | 2454636.8965 | 0.0420 | 0.3477 | 0.0265 |
| 2454638.9074 | 0.3222 | 0.6445 | 0.0199 | 2454638.9077 | 0.3223 | 0.4384 | 0.0285 |
| 2454643.9337 | 0.0226 | 0.5055 | 0.0185 | 2454643.9340 | 0.0226 | 0.3579 | 0.0344 |
| 2454648.8660 | -0.2902 | 0.7215 | 0.0079 | 2454648.8663 | -0.2902 | 0.5058 | 0.0130 |
| 2454733.6931 | -0.4709 | 0.7077 | 0.0121 | 2454733.6934 | -0.4709 | 0.4559 | 0.0130 |
| 2454734.6957 | -0.3313 | 0.7466 | 0.0076 | 2454734.6960 | -0.3312 | 0.4812 | 0.0133 |
| 2454737.6952 | 0.0867 | 0.5615 | 0.0066 | 2454737.6955 | 0.0867 | 0.3599 | 0.0123 |
| 2454738.7155 | 0.2288 | 0.6229 | 0.0115 | 2454738.7157 | 0.2289 | 0.4136 | 0.0178 |
| 2454741.6763 | -0.3586 | 0.7413 | 0.0036 | 2454741.6766 | -0.3586 | 0.4806 | 0.0040 |
| 2455098.7217 | 0.3898 | 0.6634 | 0.0109 | 2455098.7219 | 0.3898 | 0.4291 | 0.0155 |
| 2455099.6101 | -0.4865 | 0.7194 | 0.0068 | 2455099.6103 | -0.4864 | 0.4703 | 0.0151 |
| 2455099.6478 | -0.4812 | 0.7162 | 0.0065 | 2455099.6481 | -0.4812 | 0.4621 | 0.0141 |
| 2455099.7315 | -0.4695 | 0.7167 | 0.0101 | 2455099.7317 | -0.4695 | 0.4489 | 0.0202 |
| 2455100.6089 | -0.3473 | 0.7499 | 0.0049 | 2455100.6091 | -0.3473 | 0.4942 | 0.0099 |
| 2455100.7240 | -0.3312 | 0.7307 | 0.0078 | 2455100.7243 | -0.3312 | 0.4805 | 0.0104 |
| 2455101.6850 | -0.1973 | 0.6824 | 0.0084 | 2455101.6852 | -0.1973 | 0.4509 | 0.0109 |
| 2455102.6081 | -0.0687 | 0.5477 | 0.0092 | 2455102.6073 | -0.0688 | 0.3523 | 0.0108 |
| 2455102.7214 | -0.0529 | 0.5226 | 0.0194 | 2455102.7227 | -0.0528 | 0.3667 | 0.0153 |
| 2455133.6528 | 0.2568 | 0.6285 | 0.0061 | 2455133.6531 | 0.2569 | 0.4041 | 0.0131 |
| 2455134.6522 | 0.3961 | 0.6736 | 0.0115 | 2455134.6525 | 0.3961 | 0.4354 | 0.0168 |
| 2455136.6453 | -0.3262 | 0.7412 | 0.0130 | 2455136.6456 | -0.3262 | 0.4794 | 0.0135 |
| 2455137.6445 | -0.1870 | 0.6805 | 0.0055 | 2455138.6444 | -0.0477 | 0.3371 | 0.0192 |
| 2455138.6441 | -0.0477 | 0.5453 | 0.0109 | 2455139.6435 | 0.0915 | 0.3665 | 0.0117 |
| 2455139.6433 | 0.0915 | 0.5562 | 0.0082 | 2455141.6427 | 0.3701 | 0.4192 | 0.0270 |
| 2455141.6425 | 0.3701 | 0.6650 | 0.0143 | 2455143.6454 | -0.3509 | 0.4900 | 0.0101 |
| 2455143.6452 | -0.3509 | 0.7215 | 0.0072 | 2455144.6422 | -0.2120 | 0.4760 | 0.0117 |
| 2455144.6409 | -0.2122 | 0.6951 | 0.0080 | 2455146.6399 | 0.0664 | 0.3543 | 0.0154 |
| 2455146.6397 | 0.0663 | 0.5324 | 0.0122 | 2455508.5894 | 0.4981 | 0.4756 | 0.0136 |
| 2455508.5892 | 0.4980 | 0.7209 | 0.0025 | 2455513.5870 | 0.1944 | 0.4159 | 0.0133 |
| 2455513.5867 | 0.1944 | 0.6329 | 0.0086 | 2455516.5861 | -0.3877 | 0.4943 | 0.0121 |
| 2455516.5858 | -0.3878 | 0.7316 | 0.0070 | 2455826.8139 | -0.1626 | 0.3982 | 0.0141 |
| 2455826.8137 | -0.1626 | 0.7005 | 0.0141 | 2455827.8109 | -0.0237 | 0.3232 | 0.0292 |
| 2455827.8107 | -0.0237 | 0.4908 | 0.0223 | 2455828.7800 | 0.1113 | 0.3589 | 0.0281 |
| 2455828.7797 | 0.1113 | 0.5871 | 0.0293 | 2455840.7597 | -0.2195 | 0.4502 | 0.0123 |
| 2455840.7595 | -0.2195 | 0.7190 | 0.0048 | 2455842.7540 | 0.0584 | 0.3513 | 0.0243 |
| 2455842.7538 | 0.0584 | 0.5082 | 0.0183 | 2455843.7503 | 0.1972 | 0.4025 | 0.0244 |
| 2455843.7500 | 0.1972 | 0.6121 | 0.0187 | 2455844.7581 | 0.3376 | 0.4235 | 0.0058 |
| 2455844.7579 | 0.3376 | 0.6526 | 0.0062 | 2455845.7458 | 0.4752 | 0.4590 | 0.0114 |
| 2455845.7455 | 0.4752 | 0.6863 | 0.0113 | 2455846.7427 | -0.3859 | 0.4704 | 0.0191 |
| 2455846.7424 | -0.3859 | 0.7350 | 0.0101 | 2455847.7400 | -0.2469 | 0.4744 | 0.0053 |
| 2455847.7387 | -0.2471 | 0.7187 | 0.0058 | 2455848.7363 | -0.1081 | 0.3736 | 0.0132 |
| 2455848.7361 | -0.1081 | 0.6041 | 0.0116 | 2455849.7340 | 0.0309 | 0.3408 | 0.0231 |
| 2455849.7338 | 0.0309 | 0.5205 | 0.0162 | 2455854.7340 | -0.2724 | 0.4711 | 0.0069 |
| 2455854.7337 | -0.2724 | 0.7399 | 0.0066 | 2455855.5869 | -0.1536 | 0.4380 | 0.0079 |
| 2455855.5866 | -0.1536 | 0.6528 | 0.0044 | 2455856.5862 | -0.0143 | 0.3537 | 0.0181 |
| 2455856.5860 | -0.0144 | 0.5103 | 0.0050 | 2455857.5853 | 0.1249 | 0.3847 | 0.0207 |
| 2455857.5851 | 0.1249 | 0.5762 | 0.0113 | 2455858.5845 | 0.2641 | 0.4142 | 0.0196 |
| 2455858.5843 | 0.2641 | 0.6305 | 0.0095 | 2455861.5822 | -0.3182 | 0.4739 | 0.0068 |
| 2455861.5819 | -0.3183 | 0.7397 | 0.0036 | 2455862.5710 | -0.1804 | 0.4157 | 0.0077 |
| 2455862.5707 | -0.1805 | 0.6825 | 0.0051 | 2455862.5856 | -0.1784 | 0.4165 | 0.0077 |
| 2455862.5854 | -0.1784 | 0.6705 | 0.0050 | 2455862.6003 | -0.1764 | 0.4135 | 0.0094 |
| 2455862.6000 | -0.1764 | 0.6733 | 0.0057 | 2455862.6150 | -0.1743 | 0.4205 | 0.0125 |
| 2455862.6147 | -0.1744 | 0.6658 | 0.0058 | 2455862.6297 | -0.1723 | 0.4163 | 0.0087 |
| 2455862.6294 | -0.1723 | 0.6615 | 0.0072 | 2455864.6534 | 0.1097 | 0.3630 | 0.0173 |
| 2455864.6531 | 0.1097 | 0.5712 | 0.0080 | 2455865.5793 | 0.2387 | 0.4350 | 0.0191 |
| 2455865.5791 | 0.2387 | 0.6354 | 0.0064 | | | | |

**Photometry of EU Tau**

| HJD | Phase | V-mag | V-err | HJD | Phase | U-B | U-B err | HJD | Phase | B-V | B-V err |
|---|---|---|---|---|---|---|---|---|---|---|---|
| 2454439.8995 | 0.2252 | 8.0471 | 0.0029 | 2454439.8993 | 0.2251 | 0.4397 | 0.0042 | 2454439.8994 | 0.2252 | 0.6788 | 0.0035 |
| 2454440.8767 | -0.3099 | 8.1773 | 0.0018 | 2454440.8765 | -0.3100 | 0.4429 | 0.0049 | 2454440.8767 | -0.3099 | 0.7159 | 0.0021 |
| 2454447.8994 | 0.0305 | 7.9251 | 0.0035 | 2454447.8993 | 0.0305 | 0.4074 | 0.0162 | 2454447.8994 | 0.0305 | 0.6121 | 0.0055 |
| 2454448.8949 | -0.4959 | 8.2249 | 0.0007 | 2454448.8947 | -0.4960 | 0.4688 | 0.0046 | 2454448.8949 | -0.4959 | 0.7532 | 0.0020 |
| 2454449.9033 | -0.0163 | 7.9278 | 0.0024 | 2454449.9031 | -0.0164 | 0.4035 | 0.0115 | 2454449.9032 | -0.0163 | 0.6164 | 0.0057 |
| 2454450.8903 | 0.4532 | 8.2017 | 0.0014 | 2454450.8902 | 0.4532 | 0.4760 | 0.0046 | 2454450.8903 | 0.4532 | 0.7361 | 0.0035 |
| 2454451.9579 | -0.0390 | 7.9093 | 0.0029 | 2454451.9577 | -0.0391 | 0.3848 | 0.0099 | 2454451.9578 | -0.0390 | 0.6124 | 0.0067 |
| 2454452.8615 | 0.3909 | 8.1638 | 0.0003 | 2454452.8614 | 0.3908 | 0.4623 | 0.0050 | 2454452.8615 | 0.3909 | 0.7345 | 0.0033 |
| 2454453.8663 | -0.1312 | 7.9863 | 0.0017 | 2454453.8662 | -0.1312 | 0.4084 | 0.0038 | 2454453.8663 | -0.1312 | 0.6367 | 0.0030 |
| 2454454.8773 | 0.3497 | 8.1365 | 0.0020 | 2454454.8771 | 0.3496 | 0.4593 | 0.0091 | 2454454.8772 | 0.3497 | 0.7227 | 0.0042 |
| 2454458.9014 | 0.2639 | 8.0721 | 0.0053 | 2454458.9012 | 0.2638 | 0.4365 | 0.0114 | 2454458.9013 | 0.2638 | 0.7087 | 0.0074 |
| 2454459.8868 | -0.2674 | 8.1326 | 0.0047 | 2454459.8866 | -0.2675 | 0.4389 | 0.0125 | 2454459.8867 | -0.2675 | 0.7010 | 0.0075 |
| 2454460.8781 | 0.2041 | 8.0252 | 0.0018 | 2454460.8779 | 0.2040 | 0.4237 | 0.0073 | 2454460.8780 | 0.2041 | 0.6768 | 0.0033 |
| 2454461.9041 | -0.3078 | 8.1833 | 0.0041 | 2454461.9039 | -0.3079 | 0.4431 | 0.0070 | 2454461.9040 | -0.3079 | 0.7218 | 0.0069 |
| 2454464.9022 | 0.1183 | 7.9413 | 0.0074 | 2454464.9020 | 0.1182 | 0.4180 | 0.0208 | 2454464.9021 | 0.1182 | 0.6484 | 0.0133 |
| 2454465.8949 | -0.4095 | 8.2130 | 0.0032 | 2454465.8947 | -0.4096 | 0.4683 | 0.0062 | 2454465.8949 | -0.4095 | 0.7586 | 0.0033 |
| 2454475.8352 | 0.3188 | 8.1091 | 0.0050 | 2454475.8350 | 0.3187 | 0.4449 | 0.0092 | 2454475.8351 | 0.3187 | 0.7119 | 0.0053 |
| 2454476.8260 | -0.2100 | 8.0741 | 0.0038 | 2454476.8258 | -0.2100 | 0.4187 | 0.0061 | 2454476.8259 | -0.2100 | 0.6810 | 0.0053 |
| 2454480.8116 | -0.3141 | 8.1924 | 0.0033 | 2454480.8115 | -0.3142 | 0.4517 | 0.0069 | 2454480.8116 | -0.3141 | 0.7183 | 0.0062 |
| 2454481.8149 | 0.1631 | 7.9995 | 0.0050 | 2454481.8147 | 0.1630 | 0.4332 | 0.0204 | 2454481.8148 | 0.1631 | 0.6629 | 0.0100 |
| 2454483.8133 | 0.1137 | 7.9537 | 0.0131 | 2454483.8131 | 0.1136 | 0.4145 | 0.0169 | 2454483.8132 | 0.1136 | 0.6372 | 0.0168 |
| 2454484.8017 | -0.4162 | 8.2347 | 0.0028 | 2454484.8015 | -0.4162 | 0.4639 | 0.0062 | 2454484.8016 | -0.4162 | 0.7383 | 0.0039 |
| 2454485.8403 | 0.0779 | 7.9416 | 0.0049 | 2454485.8401 | 0.0778 | 0.4146 | 0.0048 | 2454485.8403 | 0.0779 | 0.6338 | 0.0056 |
| 2454486.7951 | -0.4680 | 8.2285 | 0.0089 | 2454486.7949 | -0.4680 | 0.4719 | 0.0071 | 2454486.7950 | -0.4680 | 0.7515 | 0.0103 |
| 2454494.7853 | 0.3327 | 8.1165 | 0.0095 | 2454494.7851 | 0.3326 | 0.4552 | 0.0089 | 2454494.7853 | 0.3327 | 0.7258 | 0.0103 |
| 2454495.8039 | -0.1827 | 8.0293 | 0.0077 | 2454495.8038 | -0.1828 | 0.4132 | 0.0087 | 2454495.8039 | -0.1827 | 0.6606 | 0.0103 |
| 2454496.8152 | 0.2983 | 8.0838 | 0.0048 | 2454496.8151 | 0.2983 | 0.4469 | 0.0089 | 2454496.8152 | 0.2983 | 0.7081 | 0.0064 |
| 2454503.7894 | -0.3843 | 8.2003 | 0.0043 | 2454503.7892 | -0.3844 | 0.4603 | 0.0081 | 2454503.7893 | -0.3843 | 0.7558 | 0.0060 |
| 2454504.7525 | 0.0738 | 7.9219 | 0.0164 | 2454504.7523 | 0.0737 | 0.4157 | 0.0199 | 2454504.7524 | 0.0738 | 0.6262 | 0.0223 |
| 2454505.7879 | -0.4337 | 8.2149 | 0.0111 | 2454505.7877 | -0.4337 | 0.4736 | 0.0100 | 2454505.7878 | -0.4337 | 0.7603 | 0.0137 |
| 2454506.7576 | 0.0276 | 7.9064 | 0.0082 | 2454506.7575 | 0.0276 | 0.4135 | 0.0090 | 2454506.7576 | 0.0276 | 0.6216 | 0.0111 |
| 2454508.7604 | -0.0197 | 7.9072 | 0.0019 | 2454508.7602 | -0.0198 | 0.3982 | 0.0180 | 2454508.7603 | -0.0198 | 0.6065 | 0.0102 |
| 2454509.7758 | 0.4633 | 8.1900 | 0.0048 | 2454509.7756 | 0.4632 | 0.4490 | 0.0085 | 2454509.7757 | 0.4632 | 0.7569 | 0.0080 |
| 2454838.7356 | -0.0605 | 7.9349 | 0.0015 | 2454838.7354 | -0.0606 | 0.3964 | 0.0081 | 2454838.7355 | -0.0606 | 0.6184 | 0.0032 |
| 2454838.8983 | 0.0169 | 7.8909 | 0.0100 | 2454838.8981 | 0.0168 | 0.4206 | 0.0141 | 2454838.8983 | 0.0169 | 0.6465 | 0.0141 |
| 2454839.7321 | 0.4135 | 8.1848 | 0.0028 | 2454839.7319 | 0.4134 | 0.4575 | 0.0073 | 2454839.7321 | 0.4135 | 0.7362 | 0.0067 |
| 2454839.8892 | 0.4882 | 8.2355 | 0.0021 | 2454839.8880 | 0.4877 | 0.4476 | 0.0142 | 2454839.8892 | 0.4882 | 0.7502 | 0.0061 |
| 2454840.7334 | -0.1102 | 7.9546 | 0.0042 | 2454840.7332 | -0.1103 | 0.4100 | 0.0085 | 2454840.7333 | -0.1103 | 0.6195 | 0.0067 |
| 2454841.7330 | 0.3653 | 8.1555 | 0.0089 | 2454841.7328 | 0.3652 | 0.4532 | 0.0167 | 2454841.7330 | 0.3653 | 0.7246 | 0.0106 |
| 2454842.7285 | -0.1612 | 8.0283 | 0.0020 | 2454842.7283 | -0.1613 | 0.4208 | 0.0047 | 2454842.7284 | -0.1613 | 0.6399 | 0.0034 |
| 2454842.8937 | -0.0826 | 7.9450 | 0.0034 | 2454842.8935 | -0.0827 | 0.3917 | 0.0106 | 2454842.8936 | -0.0827 | 0.6122 | 0.0059 |
| 2454844.7177 | -0.2150 | 8.1004 | 0.0066 | 2454844.7175 | -0.2151 | 0.4060 | 0.0062 | 2454844.7177 | -0.2150 | 0.6723 | 0.0071 |
| 2454844.8817 | -0.1370 | 7.9935 | 0.0028 | 2454844.8815 | -0.1371 | 0.3971 | 0.0111 | 2454844.8816 | -0.1370 | 0.6530 | 0.0029 |
| 2454845.8660 | 0.3312 | 8.1191 | 0.0017 | 2454845.8658 | 0.3311 | 0.4505 | 0.0081 | 2454845.8659 | 0.3312 | 0.7163 | 0.0021 |
| 2454846.7192 | -0.2630 | 8.1261 | 0.0047 | 2454846.7190 | -0.2630 | 0.4207 | 0.0074 | 2454846.7191 | -0.2630 | 0.6950 | 0.0051 |
| 2454846.8826 | -0.1852 | 8.0245 | 0.0033 | 2454846.8824 | -0.1853 | 0.4038 | 0.0040 | 2454846.8825 | -0.1853 | 0.6572 | 0.0050 |
| 2454847.7108 | 0.2087 | 8.0293 | 0.0019 | 2454847.7106 | 0.2086 | 0.4287 | 0.0034 | 2454847.7107 | 0.2087 | 0.6689 | 0.0038 |
| 2454847.8734 | 0.2861 | 8.0840 | 0.0020 | 2454847.8742 | 0.2864 | 0.4365 | 0.0053 | 2454847.8733 | 0.2860 | 0.7040 | 0.0029 |
| 2454849.7037 | 0.1567 | 7.9978 | 0.0019 | 2454849.7035 | 0.1566 | 0.4252 | 0.0025 | 2454849.7036 | 0.1566 | 0.6558 | 0.0021 |
| 2454849.8797 | 0.2404 | 8.0506 | 0.0100 | 2454849.8796 | 0.2404 | 0.4345 | 0.0141 | 2454849.8797 | 0.2404 | 0.7031 | 0.0141 |
| 2454856.7018 | 0.4855 | 8.2144 | 0.0010 | 2454856.7016 | 0.4854 | 0.4738 | 0.0047 | 2454856.7017 | 0.4854 | 0.7495 | 0.0027 |
| 2454856.8472 | -0.4454 | 8.1989 | 0.0084 | 2454856.8470 | -0.4455 | 0.4779 | 0.0080 | 2454856.8471 | -0.4454 | 0.7617 | 0.0106 |
| 2454859.8448 | -0.0195 | 7.9180 | 0.0100 | 2454859.8405 | -0.0215 | 0.3962 | 0.0078 | 2454859.8427 | -0.0205 | 0.6080 | 0.0113 |
| 2454860.8363 | 0.4521 | 8.1730 | 0.0013 | 2454860.8341 | 0.4511 | 0.4608 | 0.0047 | 2454860.8352 | 0.4516 | 0.7539 | 0.0035 |
| 2454861.8356 | -0.0725 | 7.9299 | 0.0023 | 2454861.8354 | -0.0726 | 0.3919 | 0.0071 | 2454861.8356 | -0.0725 | 0.6189 | 0.0023 |
| 2454862.8307 | 0.4008 | 8.1718 | 0.0009 | 2454862.8285 | 0.3998 | 0.4613 | 0.0051 | 2454862.8296 | 0.4003 | 0.7387 | 0.0009 |
| 2454863.8321 | -0.1229 | 7.9739 | 0.0005 | 2454863.8308 | -0.1235 | 0.3819 | 0.0039 | 2454863.8320 | -0.1229 | 0.6444 | 0.0005 |
| 2454864.8193 | 0.3467 | 8.1118 | 0.0037 | 2454864.8170 | 0.3456 | 0.4469 | 0.0045 | 2454864.8182 | 0.3462 | 0.7364 | 0.0057 |
| 2454865.8245 | -0.1751 | 8.0258 | 0.0032 | 2454865.8244 | -0.1752 | 0.3985 | 0.0157 | 2454865.8245 | -0.1751 | 0.6515 | 0.0113 |
| 2454876.7912 | 0.0414 | 7.9174 | 0.0045 | 2454876.7921 | 0.0418 | 0.4137 | 0.0089 | 2454876.7912 | 0.0414 | 0.6212 | 0.0051 |
| 2455614.7979 | 0.0888 | 7.9333 | 0.0016 | 2455618.7943 | -0.0102 | 0.3827 | 0.0053 | 2455614.7978 | 0.0888 | 0.6354 | 0.0095 |
| 2455615.7929 | -0.4379 | 8.2173 | 0.0033 | 2455622.7536 | -0.1269 | 0.3913 | 0.0131 | 2455615.7929 | -0.4379 | 0.7637 | 0.0038 |
| 2455618.7945 | -0.0101 | 7.9099 | 0.0062 | 2455624.7560 | -0.1744 | 0.4048 | 0.0127 | 2455618.7944 | -0.0102 | 0.6081 | 0.0082 |
| 2455622.7538 | -0.1268 | 7.9846 | 0.0005 | 2455625.7375 | 0.2925 | 0.4372 | 0.0119 | 2455622.7537 | -0.1268 | 0.6370 | 0.0028 |
| 2455624.7562 | -0.1743 | 8.0129 | 0.0053 | 2455629.7458 | 0.1991 | 0.4297 | 0.0072 | 2455624.7562 | -0.1743 | 0.6566 | 0.0059 |
| 2455625.7377 | 0.2926 | 8.0840 | 0.0015 | 2455630.7451 | -0.3256 | 0.4304 | 0.0091 | 2455625.7377 | 0.2926 | 0.7137 | 0.0017 |
| 2455629.7460 | 0.1992 | 8.0050 | 0.0041 | 2455631.7340 | 0.1448 | 0.4158 | 0.0090 | 2455629.7460 | 0.1992 | 0.6858 | 0.0076 |
| 2455630.7453 | -0.3255 | 8.1811 | 0.0112 | 2455634.7478 | -0.4216 | 0.4520 | 0.0030 | 2455630.7453 | -0.3255 | 0.7562 | 0.0117 |
| 2455631.7342 | 0.1449 | 7.9795 | 0.0067 | 2455637.7228 | -0.0065 | 0.4081 | 0.0126 | 2455631.7341 | 0.1449 | 0.6583 | 0.0074 |
| 2455634.7480 | -0.4215 | 8.2188 | 0.0003 | 2455638.7248 | 0.4701 | 0.4734 | 0.0092 | 2455634.7479 | -0.4216 | 0.7639 | 0.0009 |
| 2455637.7230 | -0.0064 | 7.8886 | 0.0164 | 2455639.7268 | -0.0532 | 0.3660 | 0.0101 | 2455637.7229 | -0.0064 | 0.6280 | 0.0174 |
| 2455638.7240 | 0.4698 | 8.1885 | 0.0081 | 2455643.7082 | -0.1594 | 0.3957 | 0.0162 | 2455638.7239 | 0.4697 | 0.7522 | 0.0088 |
| 2455639.7260 | -0.0536 | 7.9234 | 0.0012 | 2455644.7087 | 0.3165 | 0.4389 | 0.0132 | 2455639.7259 | -0.0537 | 0.6324 | 0.0054 |
| 2455643.7084 | -0.1593 | 7.9997 | 0.0043 | 2455646.7128 | 0.2698 | 0.4143 | 0.0300 | 2455643.7083 | -0.1594 | 0.6553 | 0.0072 |
| 2455644.7089 | 0.3166 | 8.1140 | 0.0033 | 2455647.7101 | -0.2558 | 0.4162 | 0.0124 | 2455644.7088 | 0.3165 | 0.7231 | 0.0051 |
| 2455646.7130 | 0.2699 | 8.0732 | 0.0001 | 2455648.7064 | 0.2181 | 0.4294 | 0.0102 | 2455646.7129 | 0.2698 | 0.7076 | 0.0057 |
| 2455647.7103 | -0.2557 | 8.1199 | 0.0124 | 2455649.7033 | -0.3077 | 0.4369 | 0.0072 | 2455647.7102 | -0.2558 | 0.7155 | 0.0140 |
| 2455648.7065 | 0.2181 | 8.0324 | 0.0030 | 2455652.6974 | 0.1165 | 0.4293 | 0.0161 | 2455648.7065 | 0.2181 | 0.6798 | 0.0034 |
| 2455649.7035 | -0.3076 | 8.1682 | 0.0042 | 2455663.6668 | 0.3343 | 0.4471 | 0.0060 | 2455649.7034 | -0.3077 | 0.7331 | 0.0055 |
| 2455652.6975 | 0.1165 | 7.9559 | 0.0128 | 2455664.6616 | -0.1925 | 0.4007 | 0.0053 | 2455652.6975 | 0.1165 | 0.6433 | 0.0174 |
| 2455663.6670 | 0.3344 | 8.1164 | 0.0127 | 2455672.6445 | -0.3953 | 0.4573 | 0.0141 | 2455663.6670 | 0.3344 | 0.7205 | 0.0136 |
| 2455664.6619 | -0.1924 | 8.0567 | 0.0017 | 2455944.8622 | 0.0904 | 0.4203 | 0.0080 | 2455664.6619 | -0.1924 | 0.6714 | 0.0017 |
| 2455672.6448 | -0.3951 | 8.2150 | 0.0100 | | | | | 2455672.6447 | -0.3952 | 0.7553 | 0.0141 |
| 2455944.8625 | 0.0906 | 7.9251 | 0.0082 | | | | | 2455944.8625 | 0.0906 | 0.6411 | 0.0106 |

**Photometry of EU Tau, cont…**

| HJD | Phase | V-R | V-R err | HJD | Phase | R-I | R-I err |
|---|---|---|---|---|---|---|---|
| 2454439.8995 | 0.2252 | 0.8739 | 0.0039 | 2454439.8997 | 0.2253 | 0.4438 | 0.0049 |
| 2454440.8768 | -0.3099 | 0.9100 | 0.0025 | 2454440.8769 | -0.3099 | 0.4613 | 0.0021 |
| 2454447.8995 | 0.0306 | 0.8383 | 0.0038 | 2454447.8996 | 0.0306 | 0.4139 | 0.0019 |
| 2454448.8950 | -0.4959 | 0.9196 | 0.0019 | 2454448.8951 | -0.4958 | 0.4815 | 0.0041 |
| 2454449.9034 | -0.0162 | 0.8347 | 0.0030 | 2454449.9035 | -0.0162 | 0.4191 | 0.0028 |
| 2454450.8904 | 0.4533 | 0.9177 | 0.0027 | 2454450.8905 | 0.4533 | 0.4696 | 0.0035 |
| 2454451.9579 | -0.0390 | 0.8308 | 0.0050 | 2454451.9581 | -0.0389 | 0.4145 | 0.0064 |
| 2454452.8616 | 0.3909 | 0.9222 | 0.0047 | 2454452.8617 | 0.3910 | 0.4734 | 0.0050 |
| 2454453.8664 | -0.1311 | 0.8542 | 0.0025 | 2454453.8665 | -0.1311 | 0.4299 | 0.0033 |
| 2454454.8773 | 0.3497 | 0.9082 | 0.0039 | 2454454.8775 | 0.3498 | 0.4665 | 0.0045 |
| 2454458.9014 | 0.2639 | 0.8923 | 0.0091 | 2454458.9016 | 0.2640 | 0.4564 | 0.0097 |
| 2454459.8868 | -0.2674 | 0.9023 | 0.0075 | 2454459.8869 | -0.2674 | 0.4622 | 0.0116 |
| 2454460.8782 | 0.2042 | 0.8786 | 0.0048 | 2454460.8783 | 0.2042 | 0.4473 | 0.0070 |
| 2454461.9042 | -0.3078 | 0.9105 | 0.0043 | 2454461.9043 | -0.3078 | 0.4701 | 0.0024 |
| 2454464.9023 | 0.1183 | 0.8570 | 0.0103 | 2454464.9024 | 0.1184 | 0.4292 | 0.0131 |
| 2454465.8950 | -0.4095 | 0.9314 | 0.0042 | 2454465.8951 | -0.4095 | 0.4823 | 0.0049 |
| 2454475.8352 | 0.3188 | 0.8978 | 0.0070 | 2454475.8353 | 0.3188 | 0.4629 | 0.0064 |
| 2454476.8260 | -0.2100 | 0.8828 | 0.0079 | 2454476.8262 | -0.2099 | 0.4460 | 0.0091 |
| 2454480.8117 | -0.3141 | 0.9154 | 0.0040 | 2454480.8118 | -0.3140 | 0.4679 | 0.0073 |
| 2454481.8149 | 0.1631 | 0.8705 | 0.0066 | 2454481.8151 | 0.1632 | 0.4418 | 0.0058 |
| 2454483.8133 | 0.1137 | 0.8610 | 0.0137 | 2454483.8134 | 0.1137 | 0.4333 | 0.0092 |
| 2454484.8018 | -0.4161 | 0.9177 | 0.0061 | 2454484.8019 | -0.4161 | 0.4756 | 0.0103 |
| 2454485.8404 | 0.0779 | 0.8485 | 0.0088 | 2454485.8405 | 0.0780 | 0.4334 | 0.0105 |
| 2454486.7951 | -0.4680 | 0.9318 | 0.0094 | 2454486.7952 | -0.4679 | 0.4785 | 0.0040 |
| 2454494.7854 | 0.3328 | 0.9041 | 0.0136 | 2454494.7855 | 0.3328 | 0.4621 | 0.0124 |
| 2454495.8040 | -0.1827 | 0.8755 | 0.0102 | 2454495.8041 | -0.1826 | 0.4361 | 0.0112 |
| 2454496.8153 | 0.2984 | 0.8982 | 0.0066 | 2454496.8154 | 0.2984 | 0.4616 | 0.0074 |
| 2454503.7894 | -0.3843 | 0.9179 | 0.0099 | 2454503.7896 | -0.3842 | 0.4770 | 0.0120 |
| 2454504.7525 | 0.0738 | 0.8467 | 0.0247 | 2454504.7527 | 0.0739 | 0.4214 | 0.0217 |
| 2454505.7880 | -0.4336 | 0.9281 | 0.0163 | 2454505.7881 | -0.4336 | 0.4826 | 0.0161 |
| 2454506.7577 | 0.0277 | 0.8432 | 0.0097 | 2454506.7578 | 0.0277 | 0.4201 | 0.0103 |
| 2454508.7604 | -0.0197 | 0.8424 | 0.0031 | 2454508.7606 | -0.0196 | 0.4225 | 0.0128 |
| 2454509.7758 | 0.4633 | 0.9258 | 0.0058 | 2454509.7760 | 0.4634 | 0.4751 | 0.0070 |
| 2454838.7356 | -0.0605 | 0.8404 | 0.0015 | 2454838.7357 | -0.0605 | 0.4186 | 0.0062 |
| 2454838.8984 | 0.0169 | 0.8457 | 0.0141 | 2454838.8985 | 0.0170 | 0.4280 | 0.0141 |
| 2454839.7322 | 0.4135 | 0.9196 | 0.0055 | 2454839.7323 | 0.4136 | 0.4761 | 0.0087 |
| 2454839.8883 | 0.4878 | 0.9328 | 0.0043 | 2454839.8884 | 0.4878 | 0.4676 | 0.0040 |
| 2454840.7334 | -0.1102 | 0.8447 | 0.0085 | 2454840.7336 | -0.1101 | 0.4172 | 0.0105 |
| 2454841.7331 | 0.3653 | 0.9093 | 0.0104 | 2454841.7332 | 0.3654 | 0.4698 | 0.0069 |
| 2454842.7285 | -0.1612 | 0.8551 | 0.0051 | 2454842.7287 | -0.1611 | 0.4292 | 0.0055 |
| 2454842.8937 | -0.0826 | 0.8423 | 0.0078 | 2454842.8939 | -0.0825 | 0.4157 | 0.0092 |
| 2454844.7178 | -0.2150 | 0.8792 | 0.0099 | 2454844.7179 | -0.2149 | 0.4452 | 0.0100 |
| 2454844.8828 | -0.1365 | 0.8649 | 0.0091 | 2454844.8829 | -0.1364 | 0.4298 | 0.0122 |
| 2454845.8660 | 0.3312 | 0.9108 | 0.0034 | 2454845.8662 | 0.3313 | 0.4717 | 0.0060 |
| 2454846.7192 | -0.2630 | 0.8958 | 0.0056 | 2454846.7194 | -0.2629 | 0.4510 | 0.0094 |
| 2454846.8827 | -0.1852 | 0.8609 | 0.0040 | 2454846.8828 | -0.1851 | 0.4378 | 0.0031 |
| 2454847.7108 | 0.2087 | 0.8764 | 0.0044 | 2454847.7109 | 0.2088 | 0.4453 | 0.0049 |
| 2454847.8734 | 0.2861 | 0.9005 | 0.0023 | 2454847.8736 | 0.2862 | 0.4586 | 0.0041 |
| 2454849.7037 | 0.1567 | 0.8682 | 0.0026 | 2454849.7039 | 0.1568 | 0.4430 | 0.0118 |
| 2454849.8788 | 0.2400 | 0.8837 | 0.0100 | 2454849.8768 | 0.2390 | 0.4620 | 0.0042 |
| 2454856.7018 | 0.4855 | 0.9215 | 0.0057 | 2454856.7019 | 0.4855 | 0.4835 | 0.0071 |
| 2454856.8473 | -0.4453 | 0.9327 | 0.0107 | 2454856.8474 | -0.4453 | 0.4824 | 0.0110 |
| 2454859.8428 | -0.0205 | 0.8562 | 0.0104 | 2454859.8409 | -0.0214 | 0.4243 | 0.0037 |
| 2454860.8354 | 0.4517 | 0.9276 | 0.0031 | 2454860.8345 | 0.4513 | 0.4771 | 0.0063 |
| 2454861.8357 | -0.0725 | 0.8485 | 0.0048 | 2454861.8358 | -0.0724 | 0.4190 | 0.0077 |
| 2454862.8308 | 0.4008 | 0.9271 | 0.0029 | 2454862.8309 | 0.4009 | 0.4802 | 0.0046 |
| 2454863.8321 | -0.1229 | 0.8612 | 0.0014 | 2454863.8312 | -0.1233 | 0.4289 | 0.0066 |
| 2454864.8183 | 0.3462 | 0.8963 | 0.0085 | 2454864.8174 | 0.3458 | 0.4708 | 0.0089 |
| 2454865.8246 | -0.1751 | 0.8739 | 0.0054 | 2454865.8247 | -0.1750 | 0.4363 | 0.0059 |
| 2454876.7923 | 0.0419 | 0.8531 | 0.0059 | 2454876.7924 | 0.0420 | 0.4219 | 0.0070 |
| 2455614.7979 | 0.0888 | 0.8430 | 0.0053 | 2455618.7947 | -0.0100 | 0.4213 | 0.0096 |
| 2455615.7940 | -0.4374 | 0.9377 | 0.0039 | 2455622.7540 | -0.1267 | 0.4239 | 0.0081 |
| 2455618.7945 | -0.0101 | 0.8362 | 0.0110 | 2455624.7564 | -0.1742 | 0.4450 | 0.0185 |
| 2455622.7539 | -0.1267 | 0.8508 | 0.0045 | 2455625.7379 | 0.2927 | 0.4608 | 0.0106 |
| 2455624.7563 | -0.1743 | 0.8693 | 0.0108 | 2455629.7462 | 0.1993 | 0.4571 | 0.0050 |
| 2455625.7378 | 0.2926 | 0.8980 | 0.0029 | 2455630.7476 | -0.3244 | 0.4656 | 0.0051 |
| 2455629.7461 | 0.1992 | 0.8660 | 0.0053 | 2455631.7343 | 0.1450 | 0.4323 | 0.0151 |
| 2455630.7464 | -0.3249 | 0.8992 | 0.0123 | 2455634.7482 | -0.4214 | 0.4736 | 0.0052 |
| 2455631.7342 | 0.1449 | 0.8769 | 0.0107 | 2455637.7232 | -0.0063 | 0.4199 | 0.0065 |
| 2455634.7480 | -0.4215 | 0.9536 | 0.0049 | 2455638.7252 | 0.4703 | 0.4872 | 0.0057 |
| 2455637.7230 | -0.0064 | 0.8418 | 0.0166 | 2455639.7262 | -0.0535 | 0.4274 | 0.0114 |
| 2455638.7240 | 0.4698 | 0.9383 | 0.0095 | 2455643.7086 | -0.1592 | 0.4362 | 0.0107 |
| 2455639.7260 | -0.0536 | 0.8549 | 0.0094 | 2455644.7091 | 0.3167 | 0.4703 | 0.0076 |
| 2455643.7084 | -0.1593 | 0.8571 | 0.0048 | 2455646.7131 | 0.2699 | 0.4485 | 0.0096 |
| 2455644.7089 | 0.3166 | 0.9063 | 0.0043 | 2455647.7105 | -0.2556 | 0.4635 | 0.0075 |
| 2455646.7130 | 0.2699 | 0.8987 | 0.0060 | 2455648.7067 | 0.2182 | 0.4511 | 0.0139 |
| 2455647.7104 | -0.2557 | 0.9083 | 0.0130 | 2455649.7037 | -0.3075 | 0.4739 | 0.0162 |
| 2455648.7066 | 0.2182 | 0.8958 | 0.0109 | 2455652.6977 | 0.1166 | 0.4369 | 0.0061 |
| 2455649.7036 | -0.3076 | 0.9067 | 0.0137 | 2455663.6672 | 0.3345 | 0.4511 | 0.0297 |
| 2455652.6976 | 0.1166 | 0.8724 | 0.0129 | 2455664.6621 | -0.1923 | 0.4672 | 0.0067 |
| 2455663.6671 | 0.3344 | 0.9415 | 0.0299 | 2455672.6450 | -0.3951 | 0.4868 | 0.0141 |
| 2455664.6620 | -0.1923 | 0.8745 | 0.0017 | 2455944.8627 | 0.0907 | 0.4325 | 0.0107 |
| 2455672.6448 | -0.3951 | 0.9257 | 0.0141 | | | | |
| 2455944.8626 | 0.0906 | 0.8546 | 0.0119 | | | | |

**Photometry of SU Cas**

| HJD | Phase | V-mag | V-err | HJD | Phase | b-y | b-y err |
|---|---|---|---|---|---|---|---|
| 2453665.8510 | 0.1845 | 5.8525 | 0.0019 | 2453665.8510 | 0.1845 | 0.4764 | 0.0037 |
| 2453666.7856 | -0.3361 | 6.1195 | 0.0015 | 2453666.7855 | -0.3361 | 0.5281 | 0.0031 |
| 2453667.8076 | 0.1882 | 5.8478 | 0.0049 | 2453667.8076 | 0.1882 | 0.4757 | 0.0059 |
| 2453668.8027 | -0.3013 | 6.1061 | 0.0042 | 2453668.8026 | -0.3014 | 0.5248 | 0.0052 |
| 2453669.7148 | 0.1666 | 5.8366 | 0.0047 | 2453669.7148 | 0.1666 | 0.4719 | 0.0062 |
| 2453669.8349 | 0.2282 | 5.8966 | 0.0041 | 2453669.8348 | 0.2281 | 0.4885 | 0.0057 |
| 2453670.7114 | -0.3222 | 6.1176 | 0.0040 | 2453670.7113 | -0.3222 | 0.5404 | 0.0042 |
| 2453670.8371 | -0.2577 | 6.0577 | 0.0029 | 2453670.8370 | -0.2577 | 0.5175 | 0.0045 |
| 2453671.7097 | 0.1900 | 5.8619 | 0.0020 | 2453671.7096 | 0.1899 | 0.4747 | 0.0055 |
| 2453671.9694 | 0.3232 | 5.9882 | 0.0031 | 2453671.9693 | 0.3231 | 0.5163 | 0.0051 |
| 2453672.7050 | -0.2994 | 6.0928 | 0.0023 | 2453672.7049 | -0.2995 | 0.5225 | 0.0044 |
| 2453672.8326 | -0.2340 | 6.0141 | 0.0030 | 2453672.8325 | -0.2340 | 0.4975 | 0.0047 |
| 2453673.7028 | 0.2124 | 5.8762 | 0.0047 | 2453673.7028 | 0.2124 | 0.4793 | 0.0062 |
| 2453673.8241 | 0.2747 | 5.9389 | 0.0014 | 2453673.8240 | 0.2746 | 0.4983 | 0.0021 |
| 2453673.9641 | 0.3465 | 6.0028 | 0.0031 | 2453673.9641 | 0.3465 | 0.5206 | 0.0052 |
| 2453674.7041 | -0.2739 | 6.0697 | 0.0019 | 2453674.7041 | -0.2739 | 0.5220 | 0.0042 |
| 2453674.8296 | -0.2095 | 5.9690 | 0.0059 | 2453674.8295 | -0.2096 | 0.4861 | 0.0079 |
| 2453674.9622 | -0.1415 | 5.8467 | 0.0060 | 2453674.9621 | -0.1416 | 0.4652 | 0.0073 |
| 2453675.8243 | 0.3007 | 5.9636 | 0.0021 | 2453675.8243 | 0.3007 | 0.5032 | 0.0025 |
| 2453677.8180 | 0.3235 | 5.9827 | 0.0017 | 2453677.8180 | 0.3235 | 0.5117 | 0.0032 |
| 2453678.8119 | -0.1666 | 5.9046 | 0.0061 | 2453678.8118 | -0.1667 | 0.4785 | 0.0087 |
| 2453679.6847 | 0.2811 | 5.9470 | 0.0012 | 2453679.6846 | 0.2811 | 0.4980 | 0.0045 |
| 2453679.8058 | 0.3432 | 6.0022 | 0.0008 | 2453679.8058 | 0.3432 | 0.5190 | 0.0019 |
| 2453680.6845 | -0.2060 | 5.9601 | 0.0048 | 2453680.6844 | -0.2060 | 0.4925 | 0.0061 |
| 2453680.8061 | -0.1436 | 5.8550 | 0.0060 | 2453680.8060 | -0.1437 | 0.4649 | 0.0085 |
| 2453684.6734 | -0.1597 | 5.8973 | 0.0034 | 2453684.6734 | -0.1597 | 0.4673 | 0.0055 |
| 2453684.7917 | -0.0990 | 5.8003 | 0.0059 | 2453684.7916 | -0.0991 | 0.4573 | 0.0068 |
| 2453688.6583 | -0.1155 | 5.8080 | 0.0018 | 2453688.6583 | -0.1155 | 0.4497 | 0.0033 |
| 2453688.7819 | -0.0520 | 5.7464 | 0.0048 | 2453688.7819 | -0.0521 | 0.4395 | 0.0073 |
| 2453688.8870 | 0.0019 | 5.7383 | 0.0026 | 2453688.8869 | 0.0018 | 0.4409 | 0.0067 |
| 2453689.6550 | 0.3959 | 6.0376 | 0.0027 | 2453689.6549 | 0.3958 | 0.5235 | 0.0036 |
| 2453690.6530 | -0.0922 | 5.7930 | 0.0041 | 2453690.6529 | -0.0922 | 0.4418 | 0.0105 |
| 2453692.7715 | -0.0054 | 5.7379 | 0.0036 | 2453692.7714 | -0.0055 | 0.4425 | 0.0060 |
| 2453692.9067 | 0.0640 | 5.7529 | 0.0049 | 2453692.9066 | 0.0639 | 0.4532 | 0.0085 |
| 2453693.6520 | 0.4463 | 6.0747 | 0.0030 | 2453693.6520 | 0.4463 | 0.5371 | 0.0038 |
| 2453693.7711 | -0.4926 | 6.1068 | 0.0019 | 2453693.7710 | -0.4927 | 0.5448 | 0.0022 |
| 2453694.6439 | -0.0449 | 5.7402 | 0.0043 | 2453694.6438 | -0.0449 | 0.4522 | 0.0049 |
| 2453694.7688 | 0.0192 | 5.7439 | 0.0035 | 2453694.7687 | 0.0192 | 0.4409 | 0.0039 |
| 2453695.6426 | 0.4675 | 6.0847 | 0.0036 | 2453695.6425 | 0.4674 | 0.5357 | 0.0062 |
| 2453695.7646 | -0.4699 | 6.1123 | 0.0026 | 2453695.7645 | -0.4700 | 0.5397 | 0.0028 |
| 2453695.8943 | -0.4034 | 6.1338 | 0.0022 | 2453695.8942 | -0.4035 | 0.5451 | 0.0033 |
| 2453699.6324 | -0.4858 | 6.1172 | 0.0042 | 2453699.6324 | -0.4858 | 0.5352 | 0.0052 |
| 2453699.7511 | -0.4249 | 6.1401 | 0.0025 | 2453699.7511 | -0.4249 | 0.5393 | 0.0038 |
| 2453699.8871 | -0.3551 | 6.1351 | 0.0046 | 2453699.8871 | -0.3551 | 0.5396 | 0.0069 |
| 2453700.7588 | 0.0921 | 5.7691 | 0.0043 | 2453700.7588 | 0.0921 | 0.4567 | 0.0067 |
| 2453701.6358 | -0.4580 | 6.1256 | 0.0063 | 2453701.6357 | -0.4581 | 0.5303 | 0.0086 |
| 2453702.6603 | 0.0675 | 5.7582 | 0.0074 | 2453702.6603 | 0.0675 | 0.4442 | 0.0077 |
| 2453702.7466 | 0.1118 | 5.7910 | 0.0045 | 2453702.7466 | 0.1118 | 0.4570 | 0.0064 |
| 2453702.8927 | 0.1868 | 5.8570 | 0.0022 | 2453702.8927 | 0.1867 | 0.4793 | 0.0056 |
| 2453703.6492 | -0.4252 | 6.1618 | 0.0029 | 2453703.6491 | -0.4252 | 0.5452 | 0.0067 |
| 2453703.7597 | -0.3685 | 6.1297 | 0.0026 | 2453703.7596 | -0.3685 | 0.5423 | 0.0041 |
| 2453705.6176 | -0.4154 | 6.1392 | 0.0029 | 2453705.6176 | -0.4154 | 0.5356 | 0.0031 |
| 2453705.7344 | -0.3555 | 6.1287 | 0.0017 | 2453705.7343 | -0.3555 | 0.5393 | 0.0021 |
| 2453705.8799 | -0.2808 | 6.0799 | 0.0018 | 2453705.8798 | -0.2809 | 0.5163 | 0.0019 |
| 2453706.6176 | 0.0976 | 5.7762 | 0.0042 | 2453706.6176 | 0.0976 | 0.4533 | 0.0056 |
| 2453706.7320 | 0.1563 | 5.8203 | 0.0005 | 2453706.7319 | 0.1562 | 0.4700 | 0.0009 |
| 2453706.8692 | 0.2267 | 5.8857 | 0.0047 | 2453706.8691 | 0.2266 | 0.4812 | 0.0055 |
| 2454764.7114 | -0.1037 | 5.8109 | 0.0085 | 2454764.7114 | -0.1037 | 0.4489 | 0.0090 |
| 2454766.7058 | -0.0806 | 5.7760 | 0.0035 | 2454766.7057 | -0.0806 | 0.4448 | 0.0067 |
| 2454766.9464 | 0.0429 | 5.7541 | 0.0107 | 2454766.9464 | 0.0429 | 0.4288 | 0.0109 |
| 2454767.7043 | 0.4317 | 6.0665 | 0.0091 | 2454767.7042 | 0.4316 | 0.5338 | 0.0115 |
| 2454767.9270 | -0.4541 | 6.1236 | 0.0024 | 2454767.9269 | -0.4542 | 0.5458 | 0.0036 |
| 2454768.9566 | 0.0741 | 5.7587 | 0.0031 | 2454768.9566 | 0.0741 | 0.4402 | 0.0044 |
| 2454770.9604 | 0.1020 | 5.7689 | 0.0052 | 2454770.9603 | 0.1020 | 0.4528 | 0.0067 |
| 2454771.6923 | 0.4775 | 6.1027 | 0.0092 | 2454771.6922 | 0.4774 | 0.5350 | 0.0106 |
| 2454771.9508 | -0.3899 | 6.1407 | 0.0014 | 2454771.9507 | -0.3900 | 0.5448 | 0.0037 |
| 2454772.6892 | -0.0111 | 5.7439 | 0.0058 | 2454772.6891 | -0.0112 | 0.4441 | 0.0072 |
| 2454775.6898 | -0.4718 | 6.1143 | 0.0067 | 2454775.6897 | -0.4719 | 0.5390 | 0.0083 |
| 2454775.9058 | -0.3610 | 6.1332 | 0.0046 | 2454775.9058 | -0.3610 | 0.5349 | 0.0062 |
| 2454778.6774 | 0.0608 | 5.7552 | 0.0010 | 2454778.6773 | 0.0608 | 0.4429 | 0.0040 |
| 2454783.6670 | -0.3795 | 6.1528 | 0.0011 | 2454783.6670 | -0.3795 | 0.5347 | 0.0044 |
| 2454785.6565 | -0.3589 | 6.1438 | 0.0038 | 2454785.6564 | -0.3590 | 0.5473 | 0.0046 |
| 2454788.6458 | 0.1746 | 5.8520 | 0.0066 | 2454788.6457 | 0.1745 | 0.4631 | 0.0078 |
| 2454789.6458 | -0.3124 | 6.1045 | 0.0024 | 2454789.6457 | -0.3125 | 0.5326 | 0.0045 |
| 2454790.6407 | 0.1980 | 5.8691 | 0.0087 | 2454790.6406 | 0.1979 | 0.4802 | 0.0106 |
| 2454791.6447 | -0.2870 | 6.0899 | 0.0126 | 2454791.6446 | -0.2870 | 0.5253 | 0.0149 |
| 2454794.8744 | 0.3698 | 6.0257 | 0.0012 | 2454794.8743 | 0.3698 | 0.5246 | 0.0014 |
| 2454801.8492 | -0.0521 | 5.7472 | 0.0024 | 2454801.8491 | -0.0522 | 0.4440 | 0.0044 |
| 2454802.6127 | 0.3396 | 6.0041 | 0.0020 | 2454802.6127 | 0.3396 | 0.5194 | 0.0022 |
| 2454802.8480 | 0.4603 | 6.0909 | 0.0028 | 2454802.8480 | 0.4603 | 0.5340 | 0.0033 |
| 2454803.6090 | -0.1493 | 5.8661 | 0.0080 | 2454803.6089 | -0.1494 | 0.4757 | 0.0090 |
| 2454810.6095 | 0.4419 | 6.0772 | 0.0045 | 2454810.6094 | 0.4418 | 0.5476 | 0.0082 |
| 2454811.6064 | -0.0467 | 5.7539 | 0.0025 | 2454811.6063 | -0.0468 | 0.4470 | 0.0031 |
| 2454821.8059 | 0.1856 | 5.8528 | 0.0008 | 2454821.8059 | 0.1856 | 0.4760 | 0.0047 |
| 2454822.7736 | -0.3180 | 6.1108 | 0.0042 | 2454822.7735 | -0.3180 | 0.5391 | 0.0043 |
| 2454829.7872 | 0.2800 | 5.9431 | 0.0043 | 2454829.7871 | 0.2799 | 0.4919 | 0.0052 |
| 2454830.7767 | -0.2124 | 5.9800 | 0.0042 | 2454830.7766 | -0.2125 | 0.4962 | 0.0051 |
| 2455184.6074 | 0.3016 | 5.9739 | 0.0027 | 2455184.6074 | 0.3016 | 0.5124 | 0.0032 |
| 2455184.7715 | 0.3858 | 6.0463 | 0.0025 | 2455184.7715 | 0.3858 | 0.5222 | 0.0029 |
| 2455189.6040 | -0.1351 | 5.8349 | 0.0022 | 2455189.6040 | -0.1352 | 0.4663 | 0.0046 |
| 2455189.7569 | -0.0567 | 5.7524 | 0.0085 | 2455189.7569 | -0.0567 | 0.4342 | 0.0091 |
| 2455191.7584 | -0.0299 | 5.7319 | 0.0008 | 2455191.7584 | -0.0300 | 0.4382 | 0.0024 |
| 2455195.7657 | 0.0258 | 5.7420 | 0.0019 | 2455195.7656 | 0.0257 | 0.4399 | 0.0033 |
| 2455197.5788 | -0.0441 | 5.7368 | 0.0017 | 2455197.5787 | -0.0442 | 0.4388 | 0.0029 |
| 2455197.7377 | 0.0374 | 5.7343 | 0.0054 | 2455197.7377 | 0.0374 | 0.4431 | 0.0061 |
| 2455198.5790 | 0.4690 | 6.0911 | 0.0006 | 2455198.5790 | 0.4690 | 0.5320 | 0.0017 |
| 2455198.7473 | -0.4447 | 6.1295 | 0.0008 | 2455198.7481 | -0.4443 | 0.5399 | 0.0009 |
| 2455199.5797 | -0.0176 | 5.7385 | 0.0052 | 2455199.5797 | -0.0177 | 0.4320 | 0.0052 |
| 2455199.7440 | 0.0666 | 5.7474 | 0.0029 | 2455199.7440 | 0.0666 | 0.4552 | 0.0039 |
| 2455200.5800 | 0.4955 | 6.1086 | 0.0013 | 2455200.5799 | 0.4954 | 0.5427 | 0.0022 |
| 2455200.7444 | -0.4202 | 6.1407 | 0.0014 | 2455200.7443 | -0.4202 | 0.5449 | 0.0026 |
| 2455202.5809 | -0.4780 | 6.1126 | 0.0027 | 2455202.5808 | -0.4781 | 0.5488 | 0.0037 |
| 2455202.7218 | -0.4058 | 6.1368 | 0.0008 | 2455202.7227 | -0.4053 | 0.5480 | 0.0009 |
| 2455204.7204 | -0.3805 | 6.1310 | 0.0015 | 2455204.7204 | -0.3805 | 0.5500 | 0.0023 |
| 2455208.5857 | -0.3976 | 6.1403 | 0.0025 | 2455208.5847 | -0.3981 | 0.5539 | 0.0042 |
| 2455208.7056 | -0.3361 | 6.1320 | 0.0011 | 2455208.7055 | -0.3362 | 0.5402 | 0.0019 |
| 2455209.7018 | 0.1750 | 5.8517 | 0.0012 | 2455209.7018 | 0.1749 | 0.4680 | 0.0031 |
| 2455210.6985 | -0.3137 | 6.1129 | 0.0032 | 2455210.6984 | -0.3138 | 0.5364 | 0.0032 |
| 2455211.6990 | 0.1995 | 5.8713 | 0.0034 | 2455211.6990 | 0.1995 | 0.4734 | 0.0075 |
| 2455212.5938 | -0.3415 | 6.1273 | 0.0014 | 2455212.5937 | -0.3415 | 0.5315 | 0.0015 |
| 2455213.7139 | 0.2332 | 5.8968 | 0.0021 | 2455213.7139 | 0.2331 | 0.5001 | 0.0080 |
| 2455236.6477 | -0.0019 | 5.7316 | 0.0018 | 2455236.6476 | -0.0019 | 0.4336 | 0.0055 |
| 2455240.6386 | 0.0454 | 5.7484 | 0.0050 | 2455240.6386 | 0.0454 | 0.4499 | 0.0065 |
| 2455241.6321 | -0.4449 | 6.1300 | 0.0034 | 2455241.6320 | -0.4450 | 0.5456 | 0.0038 |
| 2455532.6529 | -0.1522 | 5.8759 | 0.0021 | 2455532.6528 | -0.1522 | 0.4671 | 0.0058 |
| 2455535.8084 | 0.4666 | 6.0855 | 0.0041 | 2455535.8084 | 0.4666 | 0.5419 | 0.0041 |
| 2455536.6790 | -0.0868 | 5.7863 | 0.0001 | 2455536.6789 | -0.0869 | 0.4454 | 0.0080 |
| 2455536.8099 | -0.0196 | 5.7402 | 0.0084 | 2455536.8098 | -0.0197 | 0.4330 | 0.0108 |
| 2455538.8048 | 0.0037 | 5.7401 | 0.0035 | 2455538.8047 | 0.0037 | 0.4326 | 0.0052 |
| 2455539.7994 | -0.4860 | 6.1185 | 0.0020 | 2455539.7994 | -0.4860 | 0.5384 | 0.0020 |
| 2455543.6258 | 0.4769 | 6.0907 | 0.0012 | 2455543.6257 | 0.4768 | 0.5430 | 0.0019 |
| 2455543.7913 | -0.4382 | 6.1277 | 0.0034 | 2455543.7913 | -0.4382 | 0.5499 | 0.0041 |
| 2455544.6206 | -0.0128 | 5.7405 | 0.0019 | 2455544.6205 | -0.0128 | 0.4391 | 0.0021 |
| 2455544.7884 | 0.0733 | 5.7516 | 0.0020 | 2455544.7884 | 0.0733 | 0.4521 | 0.0038 |
| 2455545.6177 | 0.4987 | 6.1043 | 0.0026 | 2455545.6177 | 0.4987 | 0.5492 | 0.0035 |
| 2455545.7850 | -0.4154 | 6.1418 | 0.0008 | 2455545.7850 | -0.4155 | 0.5447 | 0.0022 |
| 2455546.7825 | 0.0963 | 5.7800 | 0.0040 | 2455546.7824 | 0.0962 | 0.4516 | 0.0048 |
| 2455580.7065 | 0.4992 | 6.1068 | 0.0012 | 2455580.7065 | 0.4992 | 0.5396 | 0.0030 |
| 2455581.7078 | 0.0128 | 5.7283 | 0.0022 | 2455581.7077 | 0.0128 | 0.4464 | 0.0093 |
| 2455582.7082 | -0.4740 | 6.1231 | 0.0010 | 2455582.7081 | -0.4740 | 0.5400 | 0.0068 |
| 2455583.6969 | 0.0332 | 5.7468 | 0.0018 | 2455583.6969 | 0.0332 | 0.4388 | 0.0032 |
| 2455584.7000 | -0.4522 | 6.1297 | 0.0035 | 2455584.6999 | -0.4522 | 0.5468 | 0.0055 |
| 2455585.6913 | 0.0564 | 5.7563 | 0.0045 | 2455585.6913 | 0.0563 | 0.4451 | 0.0048 |
| 2455588.6864 | -0.4072 | 6.1322 | 0.0062 | 2455588.6864 | -0.4072 | 0.5443 | 0.0064 |
| 2455597.6949 | 0.2142 | 5.8724 | 0.0127 | 2455597.6948 | 0.2141 | 0.4931 | 0.0130 |
| 2455598.6434 | -0.2993 | 6.1087 | 0.0064 | 2455598.6433 | -0.2993 | 0.5304 | 0.0087 |
| 2455603.6376 | 0.2628 | 5.9313 | 0.0014 | 2455603.6376 | 0.2627 | 0.4946 | 0.0032 |
| 2455604.6451 | -0.2204 | 5.9986 | 0.0057 | 2455604.6451 | -0.2204 | 0.5038 | 0.0067 |
| 2455605.6299 | 0.2848 | 5.9524 | 0.0027 | 2455605.6299 | 0.2848 | 0.5017 | 0.0031 |
| 2455607.6265 | 0.3090 | 5.9741 | 0.0060 | 2455607.6265 | 0.3090 | 0.5156 | 0.0103 |
| 2455608.6316 | -0.1753 | 5.9161 | 0.0006 | 2455608.6315 | -0.1754 | 0.4790 | 0.0032 |
| 2455883.9138 | 0.0435 | 5.7536 | 0.0040 | 2455883.9137 | 0.0435 | 0.4437 | 0.0057 |
| 2455888.9019 | -0.3976 | 6.1408 | 0.0021 | 2455888.9018 | -0.3976 | 0.5482 | 0.0034 |
| 2455893.8869 | 0.1597 | 5.8280 | 0.0019 | 2455893.8878 | 0.1602 | 0.4720 | 0.0020 |
| 2455900.8273 | -0.2799 | 6.0813 | 0.0043 | 2455900.8272 | -0.2800 | 0.5223 | 0.0066 |
| 2455911.8080 | 0.3532 | 6.0152 | 0.0018 | 2455911.8079 | 0.3531 | 0.5111 | 0.0027 |
| 2455921.7735 | 0.4654 | 6.0860 | 0.0023 | 2455921.7735 | 0.4654 | 0.5452 | 0.0026 |
| 2455922.7676 | -0.0246 | 5.7440 | 0.0014 | 2455922.7675 | -0.0246 | 0.4393 | 0.0035 |
| 2455925.7620 | -0.4885 | 6.1185 | 0.0020 | 2455925.7619 | -0.4885 | 0.5320 | 0.0032 |
| 2455926.7580 | 0.0225 | 5.7427 | 0.0039 | 2455926.7579 | 0.0224 | 0.4352 | 0.0045 |
| 2455927.7547 | -0.4662 | 6.1308 | 0.0014 | 2455927.7546 | -0.4663 | 0.5416 | 0.0036 |
| 2455944.6660 | 0.2092 | 5.8776 | 0.0042 | 2455944.6659 | 0.2092 | 0.4788 | 0.0072 |

**Photometry of SU Cas, cont…**

| HJD | Phase | c1 | c1 err | HJD | Phase | m1 | m1 err |
|---|---|---|---|---|---|---|---|
| 2453665.8508 | 0.1844 | 0.9410 | 0.0064 | 2453665.8509 | 0.1844 | 0.1576 | 0.0058 |
| 2453666.7853 | -0.3362 | 0.7814 | 0.0055 | 2453666.7854 | -0.3362 | 0.1813 | 0.0049 |
| 2453667.8074 | 0.1881 | 0.9510 | 0.0088 | 2453667.8075 | 0.1881 | 0.1557 | 0.0075 |
| 2453668.8025 | -0.3014 | 0.7840 | 0.0066 | 2453668.8026 | -0.3014 | 0.1750 | 0.0062 |
| 2453669.7152 | 0.1668 | 0.9575 | 0.0147 | 2453669.7147 | 0.1665 | 0.1587 | 0.0098 |
| 2453669.8346 | 0.2280 | 0.9125 | 0.0092 | 2453669.8347 | 0.2281 | 0.1597 | 0.0074 |
| 2453670.7111 | -0.3223 | 0.7726 | 0.0099 | 2453670.7113 | -0.3222 | 0.1640 | 0.0059 |
| 2453670.8369 | -0.2578 | 0.8071 | 0.0069 | 2453670.8370 | -0.2577 | 0.1685 | 0.0066 |
| 2453671.7094 | 0.1898 | 0.9386 | 0.0095 | 2453671.7096 | 0.1899 | 0.1626 | 0.0080 |
| 2453671.9691 | 0.3230 | 0.8452 | 0.0067 | 2453671.9693 | 0.3231 | 0.1584 | 0.0073 |
| 2453672.7048 | -0.2996 | 0.7923 | 0.0071 | 2453672.7049 | -0.2995 | 0.1750 | 0.0071 |
| 2453672.8323 | -0.2341 | 0.8425 | 0.0082 | 2453672.8325 | -0.2340 | 0.1735 | 0.0065 |
| 2453673.7026 | 0.2123 | 0.9231 | 0.0097 | 2453673.7027 | 0.2124 | 0.1651 | 0.0086 |
| 2453673.8238 | 0.2745 | 0.8884 | 0.0097 | 2453673.8240 | 0.2746 | 0.1613 | 0.0062 |
| 2453673.9639 | 0.3464 | 0.8300 | 0.0062 | 2453673.9640 | 0.3464 | 0.1615 | 0.0069 |
| 2453674.7039 | -0.2740 | 0.8051 | 0.0077 | 2453674.7040 | -0.2740 | 0.1714 | 0.0066 |
| 2453674.8294 | -0.2096 | 0.8689 | 0.0105 | 2453674.8295 | -0.2096 | 0.1708 | 0.0098 |
| 2453674.9619 | -0.1417 | 0.9456 | 0.0120 | 2453674.9620 | -0.1416 | 0.1523 | 0.0112 |
| 2453675.8241 | 0.3006 | 0.8677 | 0.0063 | 2453675.8242 | 0.3007 | 0.1651 | 0.0045 |
| 2453677.8178 | 0.3234 | 0.8500 | 0.0101 | 2453677.8179 | 0.3235 | 0.1654 | 0.0056 |
| 2453678.8117 | -0.1667 | 0.9079 | 0.0130 | 2453678.8118 | -0.1667 | 0.1570 | 0.0119 |
| 2453679.6844 | 0.2810 | 0.8749 | 0.0057 | 2453679.6846 | 0.2811 | 0.1686 | 0.0064 |
| 2453679.8056 | 0.3431 | 0.8364 | 0.0082 | 2453679.8057 | 0.3432 | 0.1622 | 0.0052 |
| 2453680.6842 | -0.2061 | 0.8842 | 0.0091 | 2453680.6844 | -0.2060 | 0.1597 | 0.0081 |
| 2453680.8058 | -0.1438 | 0.9411 | 0.0102 | 2453680.8060 | -0.1437 | 0.1565 | 0.0112 |
| 2453684.6732 | -0.1598 | 0.9096 | 0.0096 | 2453684.6733 | -0.1598 | 0.1694 | 0.0084 |
| 2453684.7914 | -0.0992 | 0.9698 | 0.0068 | 2453684.7915 | -0.0991 | 0.1485 | 0.0080 |
| 2453688.6581 | -0.1156 | 0.9620 | 0.0064 | 2453688.6582 | -0.1155 | 0.1606 | 0.0056 |
| 2453688.7817 | -0.0522 | 1.0055 | 0.0111 | 2453688.7818 | -0.0521 | 0.1563 | 0.0099 |
| 2453688.8867 | 0.0017 | 1.0176 | 0.0122 | 2453688.8868 | 0.0018 | 0.1498 | 0.0115 |
| 2453689.6547 | 0.3957 | 0.8161 | 0.0042 | 2453689.6549 | 0.3958 | 0.1703 | 0.0049 |
| 2453690.6527 | -0.0923 | 0.9652 | 0.0159 | 2453690.6529 | -0.0922 | 0.1658 | 0.0167 |
| 2453692.7712 | -0.0056 | 1.0009 | 0.0082 | 2453692.7714 | -0.0055 | 0.1579 | 0.0083 |
| 2453692.9064 | 0.0638 | 1.0160 | 0.0143 | 2453692.9066 | 0.0639 | 0.1411 | 0.0125 |
| 2453693.6518 | 0.4462 | 0.7740 | 0.0083 | 2453693.6519 | 0.4462 | 0.1729 | 0.0071 |
| 2453693.7715 | -0.4924 | 0.7609 | 0.0045 | 2453693.7710 | -0.4927 | 0.1734 | 0.0038 |
| 2453694.6437 | -0.0450 | 1.0099 | 0.0115 | 2453694.6438 | -0.0449 | 0.1385 | 0.0078 |
| 2453694.7685 | 0.0191 | 1.0182 | 0.0112 | 2453694.7686 | 0.0191 | 0.1525 | 0.0069 |
| 2453695.6424 | 0.4674 | 0.7766 | 0.0088 | 2453695.6425 | 0.4674 | 0.1766 | 0.0087 |
| 2453695.7643 | -0.4701 | 0.7731 | 0.0030 | 2453695.7645 | -0.4700 | 0.1763 | 0.0031 |
| 2453695.8947 | -0.4032 | 0.7729 | 0.0044 | 2453695.8942 | -0.4035 | 0.1662 | 0.0044 |
| 2453699.6322 | -0.4859 | 0.7659 | 0.0089 | 2453699.6323 | -0.4858 | 0.1832 | 0.0072 |
| 2453699.7509 | -0.4250 | 0.7570 | 0.0051 | 2453699.7510 | -0.4249 | 0.1815 | 0.0053 |
| 2453699.8869 | -0.3552 | 0.7739 | 0.0121 | 2453699.8870 | -0.3552 | 0.1678 | 0.0110 |
| 2453700.7586 | 0.0920 | 0.9925 | 0.0144 | 2453700.7587 | 0.0920 | 0.1531 | 0.0112 |
| 2453701.6356 | -0.4581 | 0.7567 | 0.0092 | 2453701.6357 | -0.4581 | 0.1894 | 0.0106 |
| 2453702.6601 | 0.0674 | 1.0088 | 0.0062 | 2453702.6602 | 0.0675 | 0.1605 | 0.0081 |
| 2453702.7464 | 0.1117 | 0.9897 | 0.0092 | 2453702.7465 | 0.1117 | 0.1558 | 0.0082 |
| 2453702.8925 | 0.1866 | 0.9314 | 0.0106 | 2453702.8926 | 0.1867 | 0.1546 | 0.0088 |
| 2453703.6766 | -0.4111 | 0.7274 | 0.0079 | 2453703.6490 | -0.4253 | 0.1711 | 0.0092 |
| 2453705.6180 | -0.4152 | 0.7655 | 0.0066 | 2453703.7596 | -0.3685 | 0.1705 | 0.0056 |
| 2453705.7342 | -0.3556 | 0.7676 | 0.0044 | 2453705.6175 | -0.4154 | 0.1810 | 0.0036 |
| 2453705.8796 | -0.2810 | 0.8006 | 0.0067 | 2453705.7343 | -0.3555 | 0.1731 | 0.0028 |
| 2453706.6174 | 0.0975 | 0.9970 | 0.0073 | 2453705.8798 | -0.2809 | 0.1746 | 0.0051 |
| 2453706.7318 | 0.1562 | 0.9674 | 0.0097 | 2453706.6175 | 0.0976 | 0.1547 | 0.0073 |
| 2453706.8690 | 0.2266 | 0.9180 | 0.0079 | 2453706.7319 | 0.1562 | 0.1544 | 0.0026 |
| 2454764.7112 | -0.1038 | 0.9631 | 0.0163 | 2453706.8691 | 0.2266 | 0.1621 | 0.0076 |
| 2454766.7056 | -0.0807 | 0.9718 | 0.0079 | 2454764.7113 | -0.1037 | 0.1656 | 0.0101 |
| 2454766.9462 | 0.0428 | 0.9982 | 0.0135 | 2454766.7057 | -0.0806 | 0.1672 | 0.0090 |
| 2454767.7041 | 0.4316 | 0.7893 | 0.0195 | 2454766.9463 | 0.0428 | 0.1745 | 0.0124 |
| 2454767.9267 | -0.4543 | 0.7418 | 0.0076 | 2454767.7042 | 0.4316 | 0.1726 | 0.0151 |
| 2454768.9564 | 0.0740 | 1.0028 | 0.0079 | 2454767.9268 | -0.4542 | 0.1764 | 0.0050 |
| 2454770.9601 | 0.1019 | 1.0052 | 0.0061 | 2454768.9565 | 0.0740 | 0.1627 | 0.0063 |
| 2454771.6920 | 0.4773 | 0.7743 | 0.0129 | 2454770.9603 | 0.1020 | 0.1514 | 0.0085 |
| 2454771.9505 | -0.3901 | 0.7536 | 0.0051 | 2454771.6922 | 0.4774 | 0.1802 | 0.0141 |
| 2454772.6889 | -0.0113 | 1.0093 | 0.0075 | 2454771.9507 | -0.3900 | 0.1741 | 0.0054 |
| 2454775.6895 | -0.4720 | 0.7688 | 0.0085 | 2454772.6891 | -0.0112 | 0.1527 | 0.0085 |
| 2454775.9056 | -0.3611 | 0.7571 | 0.0095 | 2454775.6896 | -0.4719 | 0.1779 | 0.0108 |
| 2454778.6771 | 0.0607 | 1.0029 | 0.0084 | 2454775.9057 | -0.3611 | 0.1824 | 0.0095 |
| 2454783.6668 | -0.3796 | 0.7849 | 0.0078 | 2454778.6772 | 0.0607 | 0.1634 | 0.0066 |
| 2454785.6562 | -0.3591 | 0.7557 | 0.0074 | 2454783.6669 | -0.3796 | 0.1690 | 0.0069 |
| 2454788.6455 | 0.1744 | 0.9404 | 0.0089 | 2454785.6563 | -0.3590 | 0.1674 | 0.0071 |
| 2454789.6455 | -0.3126 | 0.7842 | 0.0080 | 2454788.6456 | 0.1745 | 0.1748 | 0.0097 |
| 2454790.6404 | 0.1978 | 0.9433 | 0.0221 | 2454789.6456 | -0.3125 | 0.1683 | 0.0066 |
| 2454791.6444 | -0.2872 | 0.7666 | 0.0225 | 2454790.6406 | 0.1979 | 0.1552 | 0.0158 |
| 2454794.8741 | 0.3697 | 0.8248 | 0.0082 | 2454791.6446 | -0.2870 | 0.1810 | 0.0217 |
| 2454801.8489 | -0.0523 | 0.9888 | 0.0083 | 2454794.8742 | 0.3697 | 0.1629 | 0.0038 |
| 2454802.6125 | 0.3395 | 0.8346 | 0.0110 | 2454801.8490 | -0.0522 | 0.1528 | 0.0060 |
| 2454802.8491 | 0.4608 | 0.7776 | 0.0042 | 2454802.6126 | 0.3395 | 0.1669 | 0.0040 |
| 2454803.6087 | -0.1495 | 0.9301 | 0.0089 | 2454802.8483 | 0.4604 | 0.1795 | 0.0038 |
| 2454810.6093 | 0.4418 | 0.7990 | 0.0127 | 2454803.6088 | -0.1495 | 0.1478 | 0.0103 |
| 2454811.6062 | -0.0468 | 0.9917 | 0.0139 | 2454810.6094 | 0.4418 | 0.1497 | 0.0112 |
| 2454821.8057 | 0.1855 | 0.9509 | 0.0210 | 2454811.6063 | -0.0468 | 0.1505 | 0.0101 |
| 2454822.7733 | -0.3181 | 0.7670 | 0.0093 | 2454821.8058 | 0.1856 | 0.1574 | 0.0119 |
| 2454829.7870 | 0.2799 | 0.8661 | 0.0096 | 2454822.7734 | -0.3181 | 0.1718 | 0.0046 |
| 2454830.7764 | -0.2126 | 0.8605 | 0.0094 | 2454829.7871 | 0.2799 | 0.1762 | 0.0072 |
| 2455184.6072 | 0.3015 | 0.8727 | 0.0086 | 2454830.7766 | -0.2125 | 0.1604 | 0.0079 |
| 2455184.7713 | 0.3857 | 0.8158 | 0.0095 | 2455184.6073 | 0.3016 | 0.1513 | 0.0068 |
| 2455189.6038 | -0.1353 | 0.9626 | 0.0053 | 2455184.7714 | 0.3857 | 0.1653 | 0.0068 |
| 2455189.7567 | -0.0568 | 0.9956 | 0.0150 | 2455189.6039 | -0.1352 | 0.1462 | 0.0064 |
| 2455191.7582 | -0.0301 | 1.0080 | 0.0078 | 2455189.7568 | -0.0568 | 0.1668 | 0.0115 |
| 2455195.7654 | 0.0256 | 1.0218 | 0.0098 | 2455191.7583 | -0.0300 | 0.1539 | 0.0048 |
| 2455197.5785 | -0.0443 | 1.0120 | 0.0064 | 2455195.7655 | 0.0257 | 0.1539 | 0.0051 |
| 2455197.7375 | 0.0373 | 1.0057 | 0.0109 | 2455197.5786 | -0.0442 | 0.1559 | 0.0042 |
| 2455198.5788 | 0.4689 | 0.7821 | 0.0076 | 2455197.7376 | 0.0374 | 0.1557 | 0.0089 |
| 2455198.7479 | -0.4444 | 0.7651 | 0.0049 | 2455198.5789 | 0.4689 | 0.1806 | 0.0031 |
| 2455199.5795 | -0.0178 | 1.0195 | 0.0089 | 2455198.7481 | -0.4443 | 0.1776 | 0.0024 |
| 2455199.7438 | 0.0665 | 1.0195 | 0.0056 | 2455199.5796 | -0.0177 | 0.1594 | 0.0076 |
| 2455200.5798 | 0.4954 | 0.7738 | 0.0040 | 2455199.7439 | 0.0666 | 0.1397 | 0.0048 |
| 2455200.7441 | -0.4203 | 0.7444 | 0.0049 | 2455200.5799 | 0.4954 | 0.1707 | 0.0031 |
| 2455202.5806 | -0.4782 | 0.7514 | 0.0068 | 2455200.7442 | -0.4203 | 0.1783 | 0.0038 |
| 2455202.7220 | -0.4057 | 0.7497 | 0.0074 | 2455202.5807 | -0.4782 | 0.1707 | 0.0062 |
| 2455204.7202 | -0.3806 | 0.7482 | 0.0049 | 2455202.7226 | -0.4054 | 0.1725 | 0.0015 |
| 2455208.5841 | -0.3984 | 0.7545 | 0.0042 | 2455204.7203 | -0.3805 | 0.1663 | 0.0031 |
| 2455208.7053 | -0.3363 | 0.7561 | 0.0071 | 2455208.5842 | -0.3984 | 0.1639 | 0.0057 |
| 2455209.7016 | 0.1748 | 0.9401 | 0.0133 | 2455208.7054 | -0.3362 | 0.1746 | 0.0046 |
| 2455210.6982 | -0.3139 | 0.7572 | 0.0053 | 2455209.7017 | 0.1749 | 0.1667 | 0.0064 |
| 2455211.6988 | 0.1994 | 0.9157 | 0.0170 | 2455210.6984 | -0.3138 | 0.1728 | 0.0039 |
| 2455212.5935 | -0.3416 | 0.7606 | 0.0106 | 2455211.6989 | 0.1995 | 0.1673 | 0.0124 |
| 2455213.7137 | 0.2330 | 0.9190 | 0.0203 | 2455212.5936 | -0.3416 | 0.1811 | 0.0059 |
| 2455236.6474 | -0.0020 | 1.0227 | 0.0150 | 2455213.7138 | 0.2331 | 0.1453 | 0.0160 |
| 2455240.6384 | 0.0453 | 1.0100 | 0.0107 | 2455236.6475 | -0.0020 | 0.1570 | 0.0081 |
| 2455241.6318 | -0.4451 | 0.7663 | 0.0075 | 2455240.6385 | 0.0454 | 0.1436 | 0.0082 |
| 2455532.6526 | -0.1523 | 0.9327 | 0.0111 | 2455241.6319 | -0.4450 | 0.1670 | 0.0050 |
| 2455535.8082 | 0.4665 | 0.7763 | 0.0032 | 2455532.6527 | -0.1523 | 0.1575 | 0.0095 |
| 2455536.6787 | -0.0870 | 0.9760 | 0.0099 | 2455535.8083 | 0.4665 | 0.1637 | 0.0044 |
| 2455536.8097 | -0.0198 | 1.0150 | 0.0160 | 2455536.6789 | -0.0869 | 0.1577 | 0.0117 |
| 2455538.8045 | 0.0036 | 1.0060 | 0.0128 | 2455536.8098 | -0.0197 | 0.1604 | 0.0141 |
| 2455539.7992 | -0.4862 | 0.7583 | 0.0043 | 2455538.8047 | 0.0037 | 0.1656 | 0.0086 |
| 2455543.6255 | 0.4767 | 0.7674 | 0.0063 | 2455539.7993 | -0.4861 | 0.1806 | 0.0030 |
| 2455543.7911 | -0.4383 | 0.7645 | 0.0030 | 2455543.6256 | 0.4768 | 0.1733 | 0.0030 |
| 2455544.6203 | -0.0129 | 1.0075 | 0.0083 | 2455543.7912 | -0.4383 | 0.1671 | 0.0048 |
| 2455544.7882 | 0.0732 | 1.0109 | 0.0117 | 2455544.6204 | -0.0129 | 0.1576 | 0.0053 |
| 2455545.6175 | 0.4986 | 0.7515 | 0.0053 | 2455544.7883 | 0.0732 | 0.1479 | 0.0064 |
| 2455545.7848 | -0.4156 | 0.7295 | 0.0059 | 2455545.6176 | 0.4987 | 0.1720 | 0.0053 |
| 2455546.7822 | 0.0961 | 0.9861 | 0.0066 | 2455545.7849 | -0.4155 | 0.1815 | 0.0037 |
| 2455580.7063 | 0.4991 | 0.7720 | 0.0038 | 2455546.7823 | 0.0962 | 0.1634 | 0.0055 |
| 2455581.7080 | 0.0129 | 1.0247 | 0.0124 | 2455580.7064 | 0.4991 | 0.1744 | 0.0042 |
| 2455582.7079 | -0.4741 | 0.7637 | 0.0088 | 2455581.7076 | 0.0127 | 0.1357 | 0.0132 |
| 2455583.6967 | 0.0331 | 1.0079 | 0.0164 | 2455582.7081 | -0.4740 | 0.1778 | 0.0096 |
| 2455584.6997 | -0.4523 | 0.7550 | 0.0068 | 2455583.6968 | 0.0332 | 0.1616 | 0.0085 |
| 2455585.6911 | 0.0562 | 1.0070 | 0.0164 | 2455584.6999 | -0.4522 | 0.1706 | 0.0074 |
| 2455588.6862 | -0.4073 | 0.7643 | 0.0060 | 2455585.6912 | 0.0563 | 0.1530 | 0.0116 |
| 2455597.6947 | 0.2141 | 0.9339 | 0.0107 | 2455588.6863 | -0.4072 | 0.1725 | 0.0071 |
| 2455598.6431 | -0.2994 | 0.7850 | 0.0087 | 2455597.6948 | 0.2141 | 0.1439 | 0.0150 |
| 2455603.6374 | 0.2626 | 0.8888 | 0.0073 | 2455598.6433 | -0.2993 | 0.1687 | 0.0107 |
| 2455604.6449 | -0.2205 | 0.8422 | 0.0071 | 2455603.6375 | 0.2627 | 0.1651 | 0.0051 |
| 2455605.6297 | 0.2847 | 0.8818 | 0.0073 | 2455604.6450 | -0.2205 | 0.1638 | 0.0078 |
| 2455607.6263 | 0.3089 | 0.8667 | 0.0131 | 2455605.6298 | 0.2847 | 0.1648 | 0.0041 |
| 2455608.6314 | -0.1755 | 0.9042 | 0.0084 | 2455607.6264 | 0.3090 | 0.1503 | 0.0147 |
| 2455883.9136 | 0.0434 | 1.0158 | 0.0091 | 2455608.6315 | -0.1754 | 0.1538 | 0.0046 |
| 2455888.9016 | -0.3977 | 0.7581 | 0.0092 | 2455883.9137 | 0.0435 | 0.1519 | 0.0088 |
| 2455893.8880 | 0.1603 | 0.9531 | 0.0085 | 2455888.9018 | -0.3976 | 0.1693 | 0.0058 |
| 2455900.8270 | -0.2801 | 0.8003 | 0.0061 | 2455893.8881 | 0.1603 | 0.1577 | 0.0031 |
| 2455911.8077 | 0.3530 | 0.8293 | 0.0098 | 2455900.8271 | -0.2800 | 0.1703 | 0.0083 |
| 2455921.7733 | 0.4653 | 0.7846 | 0.0068 | 2455911.8078 | 0.3531 | 0.1788 | 0.0054 |
| 2455922.7673 | -0.0248 | 1.0192 | 0.0168 | 2455921.7734 | 0.4654 | 0.1598 | 0.0033 |
| 2455925.7617 | -0.4886 | 0.7597 | 0.0107 | 2455922.7674 | -0.0247 | 0.1505 | 0.0079 |
| 2455926.7577 | 0.0223 | 1.0125 | 0.0085 | 2455925.7618 | -0.4886 | 0.1893 | 0.0047 |
| 2455927.7544 | -0.4664 | 0.7582 | 0.0082 | 2455926.7579 | 0.0224 | 0.1626 | 0.0062 |
| 2455944.6657 | 0.2091 | 0.9222 | 0.0086 | 2455927.7546 | -0.4663 | 0.1775 | 0.0060 |
| | | | | 2455944.6659 | 0.2092 | 0.1658 | 0.0096 |

## Photometry of SV Vul

| HJD | Phase | V-mag | V-err | HJD | Phase | b-y | b-y err |
|---|---|---|---|---|---|---|---|
| 2453302.6632 | 0.1866 | 6.9934 | 0.0044 | 2453302.6632 | 0.1866 | 0.9925 | 0.0138 |
| 2453303.6658 | 0.2089 | 6.9912 | 0.0100 | 2453303.6652 | 0.2089 | 1.0316 | 0.0112 |
| 2453309.6579 | 0.3421 | 7.1843 | 0.0039 | 2453309.6578 | 0.3421 | 1.1531 | 0.0046 |
| 2453310.6572 | 0.3643 | 7.2132 | 0.0026 | 2453310.6571 | 0.3643 | 1.1607 | 0.0033 |
| 2453311.6572 | 0.3865 | 7.2451 | 0.0013 | 2453311.6572 | 0.3865 | 1.1854 | 0.0016 |
| 2453312.6421 | 0.4084 | 7.2717 | 0.0044 | 2453312.6421 | 0.4084 | 1.1907 | 0.0044 |
| 2453313.6312 | 0.4304 | 7.3027 | 0.0028 | 2453313.6311 | 0.4304 | 1.2001 | 0.0037 |
| 2453318.6246 | -0.4586 | 7.4440 | 0.0013 | 2453318.6246 | -0.4586 | 1.2383 | 0.0021 |
| 2453319.6429 | -0.4360 | 7.4748 | 0.0009 | 2453319.6428 | -0.4360 | 1.2275 | 0.0012 |
| 2453320.5875 | -0.4150 | 7.5030 | 0.0020 | 2453320.5874 | -0.4150 | 1.2326 | 0.0044 |
| 2453336.6094 | -0.0589 | 6.8461 | 0.0028 | 2453336.6093 | -0.0589 | 0.8590 | 0.0063 |
| 2453340.5663 | 0.0291 | 6.7230 | 0.0050 | 2453340.5663 | 0.0291 | 0.8585 | 0.0059 |
| 2453341.5662 | 0.0513 | 6.7477 | 0.0017 | 2453341.5662 | 0.0513 | 0.8826 | 0.0025 |
| 2453348.5663 | 0.2069 | 6.9681 | 0.0024 | 2453348.5662 | 0.2069 | 1.0558 | 0.0054 |
| 2453349.5665 | 0.2291 | 7.0026 | 0.0052 | 2453349.5665 | 0.2291 | 1.0652 | 0.0057 |
| 2453350.5667 | 0.2513 | 7.0366 | 0.0017 | 2453350.5666 | 0.2513 | 1.0941 | 0.0038 |
| 2453351.5668 | 0.2736 | 7.0787 | 0.0042 | 2453351.5667 | 0.2735 | 1.0938 | 0.0068 |
| 2453354.5741 | 0.3404 | 7.1765 | 0.0018 | 2453354.5740 | 0.3404 | 1.1387 | 0.0022 |
| 2453355.5744 | 0.3626 | 7.1966 | 0.0100 | 2453355.5744 | 0.3626 | 1.1649 | 0.0141 |
| 2453357.5681 | 0.4069 | 7.2471 | 0.0100 | 2453357.5681 | 0.4069 | 1.2166 | 0.0141 |
| 2453462.9773 | -0.2503 | 7.7311 | 0.0010 | 2453462.9772 | -0.2503 | 1.2344 | 0.0015 |
| 2453464.9716 | -0.2060 | 7.7431 | 0.0027 | 2453464.9716 | -0.2060 | 1.2132 | 0.0035 |
| 2453466.9764 | -0.1614 | 7.6350 | 0.0006 | 2453466.9764 | -0.1614 | 1.1825 | 0.0023 |
| 2453476.9552 | 0.0604 | 6.7665 | 0.0008 | 2453476.9551 | 0.0604 | 0.8825 | 0.0039 |
| 2453480.9372 | 0.1489 | 6.9038 | 0.0065 | 2453480.9372 | 0.1489 | 0.9812 | 0.0077 |
| 2453481.9453 | 0.1713 | 6.9353 | 0.0027 | 2453481.9452 | 0.1713 | 1.0017 | 0.0067 |
| 2453483.9402 | 0.2156 | 6.9976 | 0.0071 | 2453483.9401 | 0.2156 | 1.0550 | 0.0076 |
| 2453487.9229 | 0.3042 | 7.1333 | 0.0023 | 2453487.9229 | 0.3042 | 1.1212 | 0.0044 |
| 2453488.9163 | 0.3262 | 7.1641 | 0.0005 | 2453488.9162 | 0.3262 | 1.1364 | 0.0034 |
| 2453490.9110 | 0.3706 | 7.2254 | 0.0052 | 2453490.9109 | 0.3706 | 1.1735 | 0.0053 |
| 2453494.8978 | 0.4592 | 7.3432 | 0.0036 | 2453494.8977 | 0.4592 | 1.1980 | 0.0060 |
| 2453496.9678 | -0.4948 | 7.4016 | 0.0015 | 2453496.9678 | -0.4948 | 1.2169 | 0.0027 |
| 2453500.8863 | -0.4077 | 7.5140 | 0.0032 | 2453500.8863 | -0.4077 | 1.2295 | 0.0036 |
| 2453501.8999 | -0.3852 | 7.5393 | 0.0006 | 2453501.8998 | -0.3852 | 1.2340 | 0.0007 |
| 2453502.9547 | -0.3618 | 7.5746 | 0.0012 | 2453502.9546 | -0.3618 | 1.2441 | 0.0014 |
| 2453503.8878 | -0.3410 | 7.6132 | 0.0024 | 2453503.8878 | -0.3410 | 1.2362 | 0.0040 |
| 2453504.9458 | -0.3175 | 7.6496 | 0.0022 | 2453504.9458 | -0.3175 | 1.2445 | 0.0023 |
| 2453506.9488 | -0.2730 | 7.7124 | 0.0012 | 2453506.9488 | -0.2730 | 1.2423 | 0.0019 |
| 2453507.8752 | -0.2524 | 7.7372 | 0.0031 | 2453507.8752 | -0.2524 | 1.2364 | 0.0033 |
| 2453509.8785 | -0.2079 | 7.7467 | 0.0010 | 2453509.8776 | -0.2079 | 1.2203 | 0.0018 |
| 2453510.8642 | -0.1860 | 7.7258 | 0.0018 | 2453510.8642 | -0.1860 | 1.1932 | 0.0020 |
| 2453511.8510 | -0.1640 | 7.6810 | 0.0018 | 2453511.8509 | -0.1640 | 1.1630 | 0.0022 |
| 2453512.8515 | -0.1418 | 7.5859 | 0.0032 | 2453512.8515 | -0.1418 | 1.1238 | 0.0035 |
| 2453513.8455 | -0.1197 | 7.4293 | 0.0017 | 2453513.8454 | -0.1197 | 1.0546 | 0.0066 |
| 2453520.8344 | 0.0356 | 6.7168 | 0.0125 | 2453520.8343 | 0.0356 | 0.8578 | 0.0188 |
| 2453524.8253 | 0.1243 | 6.8527 | 0.0041 | 2453524.8253 | 0.1243 | 0.9528 | 0.0087 |
| 2453525.8204 | 0.1465 | 6.8943 | 0.0040 | 2453525.8203 | 0.1465 | 0.9804 | 0.0080 |
| 2453526.8126 | 0.1685 | 6.9267 | 0.0006 | 2453526.8125 | 0.1685 | 0.9988 | 0.0018 |
| 2453527.8153 | 0.1908 | 6.9546 | 0.0004 | 2453527.8152 | 0.1908 | 1.0259 | 0.0028 |
| 2453528.8092 | 0.2129 | 6.9865 | 0.0037 | 2453528.8091 | 0.2129 | 1.0453 | 0.0039 |
| 2453529.8140 | 0.2352 | 7.0190 | 0.0024 | 2453529.8140 | 0.2352 | 1.0658 | 0.0039 |
| 2453530.8089 | 0.2573 | 7.0503 | 0.0044 | 2453530.8088 | 0.2573 | 1.0876 | 0.0046 |
| 2453531.8032 | 0.2794 | 7.0889 | 0.0018 | 2453531.8031 | 0.2794 | 1.1018 | 0.0023 |
| 2453532.8025 | 0.3016 | 7.1177 | 0.0008 | 2453532.8025 | 0.3016 | 1.1199 | 0.0025 |
| 2453533.7979 | 0.3238 | 7.1431 | 0.0025 | 2453533.7978 | 0.3238 | 1.1409 | 0.0027 |
| 2453534.7927 | 0.3459 | 7.1735 | 0.0038 | 2453534.7926 | 0.3459 | 1.1607 | 0.0046 |
| 2453535.7990 | 0.3682 | 7.2110 | 0.0031 | 2453535.7989 | 0.3682 | 1.1684 | 0.0045 |
| 2453536.7925 | 0.3903 | 7.2482 | 0.0025 | 2453536.7925 | 0.3903 | 1.1844 | 0.0041 |
| 2453537.7868 | 0.4124 | 7.2666 | 0.0020 | 2453537.7868 | 0.4124 | 1.1990 | 0.0056 |
| 2453538.7815 | 0.4345 | 7.3081 | 0.0026 | 2453538.7815 | 0.4345 | 1.2003 | 0.0030 |
| 2453539.7809 | 0.4567 | 7.3398 | 0.0029 | 2453539.7809 | 0.4567 | 1.2130 | 0.0039 |
| 2453543.7881 | -0.4542 | 7.4395 | 0.0007 | 2453543.7880 | -0.4542 | 1.2483 | 0.0034 |
| 2453547.7617 | -0.3659 | 7.5786 | 0.0027 | 2453547.7616 | -0.3659 | 1.2433 | 0.0033 |
| 2453548.7608 | -0.3437 | 7.6137 | 0.0047 | 2453548.7608 | -0.3437 | 1.2486 | 0.0059 |
| 2453549.7485 | -0.3217 | 7.6590 | 0.0008 | 2453549.7484 | -0.3217 | 1.2433 | 0.0038 |
| 2453550.7515 | -0.2994 | 7.6795 | 0.0026 | 2453550.7515 | -0.2994 | 1.2514 | 0.0047 |
| 2453551.7555 | -0.2771 | 7.7155 | 0.0028 | 2453551.7554 | -0.2771 | 1.2386 | 0.0031 |
| 2453552.7501 | -0.2550 | 7.7364 | 0.0011 | 2453552.7500 | -0.2550 | 1.2474 | 0.0015 |
| 2453554.7388 | -0.2108 | 7.7196 | 0.0062 | 2453554.7387 | -0.2108 | 1.2291 | 0.0082 |
| 2453555.7409 | -0.1885 | 7.7406 | 0.0039 | 2453555.7409 | -0.1885 | 1.2007 | 0.0060 |
| 2453556.7318 | -0.1665 | 7.6970 | 0.0017 | 2453556.7317 | -0.1665 | 1.1755 | 0.0025 |
| 2453557.7370 | -0.1442 | 7.6023 | 0.0003 | 2453557.7369 | -0.1442 | 1.1379 | 0.0009 |
| 2453558.7268 | -0.1222 | 7.4514 | 0.0017 | 2453558.7267 | -0.1222 | 1.0830 | 0.0019 |
| 2453560.7539 | -0.0771 | 6.9863 | 0.0056 | 2453560.7539 | -0.0771 | 0.9163 | 0.0065 |
| 2453561.7239 | -0.0556 | 6.7990 | 0.0017 | 2453561.7238 | -0.0556 | 0.8589 | 0.0025 |
| 2453563.7169 | -0.0113 | 6.6999 | 0.0068 | 2453563.7168 | -0.0113 | 0.8417 | 0.0112 |
| 2453565.7177 | 0.0332 | 6.7204 | 0.0069 | 2453565.7177 | 0.0332 | 0.8634 | 0.0097 |
| 2453567.7025 | 0.0773 | 6.7799 | 0.0163 | 2453567.7024 | 0.0773 | 0.9075 | 0.0205 |
| 2453627.6464 | 0.4096 | 7.2619 | 0.0020 | 2453627.6463 | 0.4096 | 1.2077 | 0.0026 |
| 2453628.7497 | 0.4341 | 7.3004 | 0.0017 | 2453628.7496 | 0.4341 | 1.2129 | 0.0029 |
| 2453629.6303 | 0.4537 | 7.3208 | 0.0029 | 2453629.6303 | 0.4537 | 1.2241 | 0.0038 |
| 2453631.6282 | 0.4981 | 7.3762 | 0.0010 | 2453631.6281 | 0.4981 | 1.2334 | 0.0023 |
| 2453632.7195 | -0.4776 | 7.4145 | 0.0040 | 2453632.7195 | -0.4776 | 1.2360 | 0.0041 |
| 2453634.7084 | -0.4334 | 7.4810 | 0.0066 | 2453634.7083 | -0.4334 | 1.2446 | 0.0074 |
| 2453636.7089 | -0.3890 | 7.5382 | 0.0032 | 2453636.7089 | -0.3890 | 1.2413 | 0.0033 |
| 2453637.6230 | -0.3687 | 7.5588 | 0.0064 | 2453637.6229 | -0.3687 | 1.2590 | 0.0066 |
| 2453638.7148 | -0.3444 | 7.6104 | 0.0019 | 2453638.7148 | -0.3444 | 1.2495 | 0.0022 |
| 2453639.7616 | -0.3211 | 7.6448 | 0.0076 | 2453639.7615 | -0.3211 | 1.2616 | 0.0085 |
| 2453641.6193 | -0.2798 | 7.6981 | 0.0043 | 2453641.6193 | -0.2798 | 1.2550 | 0.0045 |
| 2453642.7127 | -0.2555 | 7.7192 | 0.0023 | 2453642.7126 | -0.2555 | 1.2647 | 0.0034 |
| 2453643.6357 | -0.2350 | 7.7363 | 0.0033 | 2453643.6356 | -0.2350 | 1.2537 | 0.0035 |
| 2453644.7072 | -0.2112 | 7.7402 | 0.0026 | 2453644.7072 | -0.2112 | 1.2439 | 0.0032 |
| 2453648.6665 | -0.1232 | 7.4535 | 0.0050 | 2453648.6665 | -0.1232 | 1.0706 | 0.0053 |
| 2453652.6993 | -0.0336 | 6.7282 | 0.0055 | 2453652.6992 | -0.0336 | 0.8326 | 0.0057 |
| 2453653.6256 | -0.0130 | 6.7023 | 0.0096 | 2453653.6256 | -0.0130 | 0.8275 | 0.0112 |
| 2453655.6238 | 0.0314 | 6.7201 | 0.0062 | 2453655.6238 | 0.0314 | 0.8666 | 0.0077 |
| 2453657.6127 | 0.0756 | 6.7929 | 0.0021 | 2453657.6126 | 0.0756 | 0.9016 | 0.0031 |
| 2453667.6150 | 0.2979 | 7.1261 | 0.0028 | 2453667.6149 | 0.2979 | 1.1153 | 0.0054 |
| 2453668.6583 | 0.3211 | 7.1528 | 0.0019 | 2453668.6582 | 0.3211 | 1.1463 | 0.0025 |
| 2453669.6445 | 0.3430 | 7.1879 | 0.0023 | 2453669.6444 | 0.3430 | 1.1558 | 0.0031 |
| 2453671.6471 | 0.3876 | 7.2494 | 0.0030 | 2453671.6471 | 0.3876 | 1.1835 | 0.0038 |
| 2453672.5955 | 0.4086 | 7.2842 | 0.0019 | 2453672.5954 | 0.4086 | 1.1880 | 0.0030 |
| 2453673.5947 | 0.4308 | 7.3113 | 0.0033 | 2453673.5947 | 0.4308 | 1.2042 | 0.0033 |
| 2453674.5943 | 0.4531 | 7.3385 | 0.0013 | 2453674.5943 | 0.4531 | 1.2133 | 0.0026 |
| 2453679.5913 | -0.4359 | 7.4825 | 0.0004 | 2453679.5912 | -0.4359 | 1.2295 | 0.0029 |
| 2453680.5908 | -0.4137 | 7.5130 | 0.0027 | 2453680.5908 | -0.4137 | 1.2342 | 0.0031 |
| 2453688.5870 | -0.2359 | 7.7380 | 0.0022 | 2453688.5870 | -0.2359 | 1.2432 | 0.0033 |
| 2453690.5866 | -0.1915 | 7.7338 | 0.0008 | 2453690.5866 | -0.1915 | 1.2117 | 0.0014 |
| 2453691.6006 | -0.1690 | 7.6826 | 0.0054 | 2453691.6005 | -0.1690 | 1.1876 | 0.0060 |
| 2453694.5691 | -0.1030 | 7.2445 | 0.0013 | 2453694.5690 | -0.1030 | 1.0013 | 0.0035 |
| 2453695.5686 | -0.0808 | 7.0063 | 0.0044 | 2453695.5686 | -0.0808 | 0.9107 | 0.0071 |
| 2454257.8701 | 0.4168 | 7.2550 | 0.0125 | 2454257.8700 | 0.4168 | 1.1777 | 0.0133 |
| 2454258.8574 | 0.4387 | 7.2907 | 0.0048 | 2454258.8573 | 0.4387 | 1.1756 | 0.0062 |
| 2454259.8671 | 0.4611 | 7.3125 | 0.0031 | 2454259.8670 | 0.4611 | 1.1865 | 0.0059 |
| 2454260.8501 | 0.4830 | 7.3427 | 0.0022 | 2454260.8500 | 0.4830 | 1.1982 | 0.0035 |
| 2454261.8621 | -0.4945 | 7.3750 | 0.0068 | 2454261.8621 | -0.4945 | 1.2060 | 0.0069 |
| 2454264.8406 | -0.4283 | 7.4560 | 0.0037 | 2454264.8405 | -0.4283 | 1.2222 | 0.0043 |
| 2454265.8428 | -0.4060 | 7.4816 | 0.0030 | 2454265.8427 | -0.4060 | 1.2257 | 0.0044 |
| 2454266.8502 | -0.3837 | 7.5178 | 0.0046 | 2454266.8501 | -0.3837 | 1.2209 | 0.0054 |
| 2454268.8487 | -0.3392 | 7.5921 | 0.0066 | 2454268.8498 | -0.3392 | 1.2197 | 0.0067 |
| 2454269.8522 | -0.3169 | 7.6347 | 0.0079 | 2454269.8522 | -0.3169 | 1.2310 | 0.0091 |
| 2454270.8360 | -0.2951 | 7.6616 | 0.0025 | 2454270.8359 | -0.2951 | 1.2432 | 0.0032 |
| 2454272.8346 | -0.2507 | 7.7064 | 0.0075 | 2454272.8345 | -0.2507 | 1.2363 | 0.0082 |
| 2454273.8510 | -0.2281 | 7.7276 | 0.0007 | 2454273.8510 | -0.2281 | 1.2266 | 0.0031 |
| 2454275.8470 | -0.1837 | 7.7264 | 0.0077 | 2454275.8469 | -0.1837 | 1.2100 | 0.0089 |
| 2454276.8256 | -0.1619 | 7.6957 | 0.0055 | 2454276.8256 | -0.1619 | 1.1850 | 0.0067 |
| 2454277.8396 | -0.1394 | 7.6208 | 0.0115 | 2454277.8396 | -0.1394 | 1.1507 | 0.0131 |
| 2454278.8156 | -0.1177 | 7.5031 | 0.0050 | 2454278.8155 | -0.1177 | 1.0943 | 0.0077 |
| 2454279.8272 | -0.0952 | 7.3089 | 0.0049 | 2454279.8272 | -0.0952 | 1.0183 | 0.0056 |
| 2454280.8169 | -0.0732 | 7.0855 | 0.0157 | 2454280.8169 | -0.0732 | 0.9343 | 0.0175 |
| 2454281.8222 | -0.0509 | 6.8751 | 0.0162 | 2454281.8222 | -0.0509 | 0.8591 | 0.0257 |
| 2454282.8111 | -0.0289 | 6.7653 | 0.0038 | 2454282.8110 | -0.0289 | 0.8318 | 0.0040 |
| 2454284.8274 | 0.0159 | 6.7045 | 0.0082 | 2454284.8273 | 0.0159 | 0.8286 | 0.0119 |
| 2454285.8187 | 0.0379 | 6.7247 | 0.0083 | 2454285.8186 | 0.0379 | 0.8516 | 0.0132 |
| 2454370.7160 | -0.0752 | 7.1357 | 0.0025 | 2454370.7159 | -0.0752 | 0.9541 | 0.0041 |
| 2454373.6457 | -0.0101 | 6.7150 | 0.0005 | 2454373.6457 | -0.0101 | 0.8218 | 0.0041 |
| 2454380.6932 | 0.1466 | 6.8899 | 0.0158 | 2454380.6931 | 0.1466 | 0.9644 | 0.0198 |
| 2454381.6383 | 0.1676 | 6.8632 | 0.0070 | 2454381.6383 | 0.1676 | 0.9759 | 0.0320 |
| 2454382.6818 | 0.1908 | 6.9469 | 0.0219 | 2454382.6817 | 0.1908 | 0.9877 | 0.0459 |
| 2454388.6703 | 0.3239 | 7.1199 | 0.0027 | 2454388.6703 | 0.3239 | 1.1263 | 0.0035 |
| 2454390.6573 | 0.3680 | 7.1828 | 0.0043 | 2454390.6573 | 0.3680 | 1.1492 | 0.0046 |
| 2454391.6192 | 0.3894 | 7.2103 | 0.0009 | 2454391.6191 | 0.3894 | 1.1747 | 0.0029 |
| 2454392.6555 | 0.4125 | 7.2429 | 0.0022 | 2454392.6554 | 0.4124 | 1.1788 | 0.0041 |
| 2454393.6177 | 0.4338 | 7.2740 | 0.0011 | 2454393.6176 | 0.4338 | 1.1880 | 0.0016 |
| 2454394.6452 | 0.4567 | 7.3044 | 0.0019 | 2454394.6452 | 0.4567 | 1.1989 | 0.0021 |
| 2454737.6343 | 0.0798 | 6.7605 | 0.0057 | 2454737.6343 | 0.0798 | 0.8905 | 0.0098 |
| 2454747.6247 | 0.3019 | 7.0937 | 0.0025 | 2454747.6247 | 0.3019 | 1.1105 | 0.0036 |
| 2454749.6423 | 0.3467 | 7.1517 | 0.0045 | 2454749.6423 | 0.3467 | 1.1506 | 0.0047 |
| 2454753.6465 | 0.4357 | 7.2807 | 0.0024 | 2454753.6464 | 0.4357 | 1.1999 | 0.0030 |
| 2454755.6287 | 0.4798 | 7.3479 | 0.0017 | 2454755.6286 | 0.4798 | 1.2099 | 0.0023 |
| 2454757.6114 | -0.4762 | 7.3996 | 0.0011 | 2454757.6113 | -0.4762 | 1.2273 | 0.0020 |
| 2454759.6099 | -0.4317 | 7.4600 | 0.0036 | 2454759.6098 | -0.4317 | 1.2322 | 0.0043 |
| 2454761.6228 | -0.3870 | 7.5045 | 0.0024 | 2454761.6227 | -0.3870 | 1.2556 | 0.0027 |
| 2454763.6215 | -0.3426 | 7.5708 | 0.0003 | 2454763.6214 | -0.3426 | 1.2631 | 0.0007 |
| 2454765.6290 | -0.2980 | 7.6529 | 0.0008 | 2454765.6289 | -0.2980 | 1.2456 | 0.0012 |
| 2454767.6280 | -0.2535 | 7.7087 | 0.0024 | 2454767.6279 | -0.2535 | 1.2475 | 0.0033 |
| 2454771.6388 | -0.1644 | 7.7200 | 0.0014 | 2454771.6387 | -0.1644 | 1.1983 | 0.0030 |
| 2454775.6235 | -0.0758 | 7.1640 | 0.0083 | 2454775.6234 | -0.0758 | 0.9559 | 0.0122 |
| 2454785.5945 | 0.1458 | 6.8738 | 0.0039 | 2454785.5945 | 0.1458 | 0.9667 | 0.0057 |
| 2454985.8817 | -0.4027 | 7.5200 | 0.0049 | 2454985.8816 | -0.4027 | 1.2222 | 0.0074 |
| 2454989.8728 | -0.3140 | 7.6321 | 0.0036 | 2454989.8719 | -0.3140 | 1.2272 | 0.0062 |
| 2454997.8479 | -0.1367 | 7.6041 | 0.0034 | 2454997.8478 | -0.1367 | 1.1406 | 0.0038 |
| 2454998.8419 | -0.1146 | 7.4731 | 0.0034 | 2454998.8418 | -0.1146 | 1.0885 | 0.0047 |
| 2454999.9182 | -0.0907 | 7.2777 | 0.0070 | 2454999.9181 | -0.0907 | 0.9993 | 0.0112 |
| 2455004.9125 | 0.0203 | 6.7376 | 0.0168 | 2455004.9125 | 0.0203 | 0.8320 | 0.0277 |
| 2455006.8737 | 0.0639 | 6.7883 | 0.0057 | 2455006.8737 | 0.0639 | 0.8691 | 0.0157 |
| 2455098.7335 | 0.1055 | 6.8588 | 0.0102 | 2455098.7334 | 0.1055 | 0.9119 | 0.0259 |
| 2455099.7450 | 0.1280 | 6.8789 | 0.0144 | 2455099.7458 | 0.1280 | 0.9442 | 0.0151 |
| 2455100.7376 | 0.1501 | 6.9014 | 0.0063 | 2455100.7375 | 0.1501 | 0.9703 | 0.0111 |
| 2455101.7538 | 0.1726 | 6.9383 | 0.0041 | 2455101.7529 | 0.1726 | 0.9946 | 0.0069 |
| 2455102.7367 | 0.1945 | 6.9770 | 0.0006 | 2455102.7358 | 0.1945 | 1.0116 | 0.0116 |
| 2455106.7264 | 0.2832 | 7.0876 | 0.0111 | 2455106.7263 | 0.2832 | 1.0896 | 0.0124 |
| 2455109.6904 | 0.3490 | 7.1759 | 0.0125 | 2455109.6904 | 0.3490 | 1.1551 | 0.0263 |
| 2455133.5951 | -0.1197 | 7.5187 | 0.0020 | 2455133.5951 | -0.1197 | 1.1047 | 0.0031 |
| 2455134.5945 | -0.0975 | 7.3416 | 0.0070 | 2455134.5944 | -0.0975 | 1.0364 | 0.0106 |
| 2455136.5932 | -0.0530 | 6.9060 | 0.0055 | 2455136.5932 | -0.0530 | 0.8717 | 0.0074 |
| 2455137.5925 | -0.0308 | 6.7807 | 0.0044 | 2455137.5924 | -0.0308 | 0.8425 | 0.0048 |
| 2455138.5921 | -0.0086 | 6.7359 | 0.0036 | 2455138.5920 | -0.0086 | 0.8303 | 0.0058 |
| 2455139.5913 | 0.0136 | 6.7254 | 0.0046 | 2455139.5913 | 0.0136 | 0.8353 | 0.0076 |
| 2455143.5931 | 0.1025 | 6.8414 | 0.0085 | 2455143.5930 | 0.1025 | 0.9053 | 0.0117 |
| 2455144.5889 | 0.1247 | 6.8619 | 0.0064 | 2455144.5888 | 0.1247 | 0.9367 | 0.0095 |
| 2455146.5878 | 0.1691 | 6.9180 | 0.0053 | 2455146.5878 | 0.1691 | 0.9880 | 0.0072 |
| 2455151.5863 | 0.2802 | 7.0672 | 0.0032 | 2455151.5871 | 0.2802 | 1.0945 | 0.0033 |
| 2455295.9908 | 0.4897 | 7.3662 | 0.0011 | 2455295.9907 | 0.4897 | 1.2148 | 0.0014 |
| 2455296.9876 | -0.4882 | 7.3994 | 0.0025 | 2455296.9875 | -0.4882 | 1.2158 | 0.0028 |
| 2455297.9867 | -0.4660 | 7.4334 | 0.0054 | 2455297.9857 | -0.4660 | 1.2175 | 0.0063 |
| 2455298.9823 | -0.4438 | 7.4599 | 0.0036 | 2455298.9822 | -0.4438 | 1.2260 | 0.0042 |
| 2455299.9800 | -0.4216 | 7.4820 | 0.0042 | 2455299.9808 | -0.4216 | 1.2275 | 0.0049 |
| 2455300.9778 | -0.3995 | 7.5054 | 0.0056 | 2455300.9777 | -0.3995 | 1.2386 | 0.0060 |
| 2455301.9763 | -0.3773 | 7.5379 | 0.0034 | 2455301.9754 | -0.3773 | 1.2352 | 0.0047 |
| 2455314.9776 | -0.0883 | 7.2735 | 0.0072 | 2455314.9775 | -0.0883 | 1.0023 | 0.0183 |
| 2455315.9661 | -0.0663 | 7.0509 | 0.0078 | 2455315.9660 | -0.0663 | 0.9176 | 0.0136 |
| 2455317.9495 | -0.0223 | 6.7400 | 0.0070 | 2455317.9494 | -0.0223 | 0.8315 | 0.0091 |
| 2455319.9868 | 0.0230 | 6.7074 | 0.0053 | 2455319.9868 | 0.0230 | 0.8426 | 0.0087 |
| 2455320.9495 | 0.0444 | 6.7288 | 0.0072 | 2455320.9494 | 0.0444 | 0.8577 | 0.0108 |
| 2455321.9213 | 0.0660 | 6.7627 | 0.0068 | 2455321.9213 | 0.0660 | 0.8799 | 0.0094 |
| 2455323.9637 | 0.1114 | 6.8290 | 0.0043 | 2455323.9636 | 0.1114 | 0.9329 | 0.0051 |
| 2455326.9085 | 0.1769 | 6.9518 | 0.0069 | 2455326.9084 | 0.1769 | 0.9903 | 0.0107 |
| 2455328.9404 | 0.2220 | 6.9933 | 0.0048 | 2455328.9412 | 0.2220 | 1.0390 | 0.0055 |
| 2455335.9252 | 0.3773 | 7.2165 | 0.0025 | 2455335.9251 | 0.3773 | 1.1639 | 0.0029 |
| 2455337.9468 | 0.4222 | 7.2777 | 0.0034 | 2455337.9468 | 0.4222 | 1.1916 | 0.0041 |
| 2455463.7800 | 0.2189 | 7.0197 | 0.0047 | 2455463.7791 | 0.2189 | 1.0396 | 0.0072 |
| 2455464.7555 | 0.2406 | 7.0505 | 0.0111 | 2455464.7563 | 0.2406 | 1.0697 | 0.0131 |
| 2455465.7727 | 0.2632 | 7.0813 | 0.0081 | 2455465.7727 | 0.2632 | 1.0833 | 0.0102 |
| 2455495.6278 | -0.0732 | 7.0557 | 0.0057 | 2455495.6278 | -0.0732 | 0.9152 | 0.0071 |
| 2455498.6708 | -0.0056 | 6.7114 | 0.0162 | 2455498.6708 | -0.0056 | 0.8345 | 0.0210 |
| 2455499.5942 | 0.0149 | 6.7032 | 0.0027 | 2455499.5942 | 0.0149 | 0.8464 | 0.0078 |
| 2455501.5933 | 0.0593 | 6.7574 | 0.0041 | 2455501.5933 | 0.0593 | 0.8807 | 0.0057 |
| 2455502.5926 | 0.0816 | 6.7912 | 0.0020 | 2455502.5925 | 0.0816 | 0.9022 | 0.0052 |
| 2455503.5922 | 0.1038 | 6.8284 | 0.0049 | 2455503.5922 | 0.1038 | 0.9223 | 0.0057 |
| 2455504.5915 | 0.1260 | 6.8635 | 0.0060 | 2455504.5914 | 0.1260 | 0.9553 | 0.0064 |
| 2455508.6042 | 0.2152 | 6.9982 | 0.0074 | 2455508.6042 | 0.2152 | 1.0504 | 0.0093 |
| 2455509.6056 | 0.2374 | 7.0416 | 0.0025 | 2455509.6056 | 0.2374 | 1.0639 | 0.0035 |
| 2455510.6050 | 0.2596 | 7.0663 | 0.0017 | 2455510.6041 | 0.2596 | 1.0813 | 0.0029 |
| 2455511.6027 | 0.2818 | 7.0993 | 0.0043 | 2455511.6027 | 0.2818 | 1.1049 | 0.0045 |
| 2455513.6019 | 0.3262 | 7.1526 | 0.0021 | 2455513.6019 | 0.3262 | 1.1379 | 0.0025 |
| 2455516.6011 | 0.3929 | 7.2605 | 0.0049 | 2455516.6010 | 0.3929 | 1.1733 | 0.0060 |
| 2455517.6006 | 0.4151 | 7.2786 | 0.0050 | 2455517.6006 | 0.4151 | 1.1889 | 0.0079 |
| 2455518.6005 | 0.4373 | 7.2990 | 0.0030 | 2455518.6013 | 0.4374 | 1.2007 | 0.0031 |
| 2455519.6001 | 0.4596 | 7.3328 | 0.0042 | 2455519.6000 | 0.4596 | 1.2093 | 0.0068 |
| 2455523.5762 | -0.4521 | 7.4399 | 0.0029 | 2455523.5761 | -0.4521 | 1.2315 | 0.0033 |
| 2455530.5747 | -0.2965 | 7.6619 | 0.0018 | 2455530.5747 | -0.2965 | 1.2400 | 0.0022 |
| 2455532.5747 | -0.2521 | 7.7109 | 0.0013 | 2455532.5746 | -0.2521 | 1.2429 | 0.0021 |
| 2455702.8369 | -0.4679 | 7.4495 | 0.0060 | 2455702.8368 | -0.4679 | 1.2227 | 0.0068 |
| 2455704.8441 | -0.4233 | 7.5057 | 0.0031 | 2455704.8440 | -0.4233 | 1.2445 | 0.0053 |
| 2455706.8238 | -0.3793 | 7.5578 | 0.0007 | 2455706.8237 | -0.3793 | 1.2544 | 0.0021 |
| 2455710.8259 | -0.2903 | 7.7228 | 0.0139 | 2455710.8257 | -0.2903 | 1.2493 | 0.0144 |
| 2455712.8106 | -0.2462 | 7.7208 | 0.0030 | 2455712.8105 | -0.2462 | 1.2564 | 0.0046 |
| 2455716.7980 | -0.1576 | 7.6094 | 0.0063 | 2455716.7979 | -0.1576 | 1.1609 | 0.0069 |
| 2455722.7974 | -0.0242 | 6.7362 | 0.0050 | 2455722.7973 | -0.0242 | 0.8404 | 0.0093 |
| 2455730.7643 | 0.1528 | 6.9670 | 0.0087 | 2455730.7642 | 0.1528 | 1.0222 | 0.0166 |
| 2455732.7539 | 0.1970 | 7.0250 | 0.0073 | 2455732.7538 | 0.1970 | 1.0480 | 0.0113 |
| 2455734.7651 | 0.2417 | 7.0985 | 0.0040 | 2455734.7650 | 0.2417 | 1.0896 | 0.0056 |
| 2455736.7470 | 0.2858 | 7.1448 | 0.0166 | 2455736.7468 | 0.2858 | 1.1193 | 0.0174 |
| 2455738.9257 | 0.3342 | 7.1801 | 0.0045 | 2455738.9256 | 0.3342 | 1.1715 | 0.0050 |
| 2455740.7563 | 0.3749 | 7.2763 | 0.0085 | 2455740.7561 | 0.3749 | 1.1765 | 0.0110 |
| 2455824.6474 | 0.2394 | 7.0542 | 0.0036 | 2455824.6473 | 0.2394 | 1.0885 | 0.0038 |
| 2455830.8020 | 0.3762 | 7.2298 | 0.0100 | 2455830.8019 | 0.3762 | 1.1891 | 0.0141 |
| 2455835.7882 | 0.4871 | 7.3889 | 0.0100 | 2455835.7881 | 0.4870 | 1.2055 | 0.0141 |
| 2455839.7766 | -0.4243 | 7.4973 | 0.0100 | 2455839.7765 | -0.4243 | 1.2433 | 0.0141 |
| 2455840.7742 | -0.4021 | 7.5422 | 0.0100 | 2455840.7741 | -0.4021 | 1.2399 | 0.0141 |
| 2455842.7686 | -0.3578 | 7.6073 | 0.0100 | 2455842.7685 | -0.3578 | 1.2511 | 0.0141 |
| 2455843.7649 | -0.3357 | 7.6276 | 0.0100 | 2455843.7648 | -0.3357 | 1.2459 | 0.0141 |
| 2455845.7604 | -0.2913 | 7.6749 | 0.0100 | 2455845.7603 | -0.2913 | 1.2721 | 0.0141 |
| 2455846.7574 | -0.2692 | 7.7045 | 0.0100 | 2455846.7572 | -0.2692 | 1.2529 | 0.0141 |
| 2455847.7537 | -0.2470 | 7.7154 | 0.0100 | 2455847.7536 | -0.2470 | 1.2381 | 0.0141 |
| 2455848.7510 | -0.2248 | 7.7144 | 0.0100 | 2455848.7509 | -0.2248 | 1.2449 | 0.0141 |
| 2455849.7488 | -0.2027 | 7.7043 | 0.0100 | 2455849.7487 | -0.2027 | 1.2324 | 0.0141 |
| 2455851.7430 | -0.1583 | 7.5850 | 0.0100 | 2455851.7428 | -0.1583 | 1.1435 | 0.0141 |
| 2455854.5877 | -0.0951 | 7.0493 | 0.0026 | 2455854.5876 | -0.0951 | 0.9236 | 0.0040 |
| 2455855.6018 | -0.0726 | 6.8380 | 0.0028 | 2455855.6016 | -0.0726 | 0.8566 | 0.0056 |
| 2455856.6010 | -0.0504 | 6.7450 | 0.0030 | 2455856.6009 | -0.0504 | 0.8250 | 0.0039 |
| 2455857.6295 | -0.0275 | 6.7166 | 0.0049 | 2455857.6294 | -0.0275 | 0.8356 | 0.0125 |
| 2455858.5994 | -0.0060 | 6.7137 | 0.0054 | 2455858.5993 | -0.0060 | 0.8410 | 0.0069 |
| 2455861.5971 | 0.0607 | 6.8006 | 0.0077 | 2455861.5970 | 0.0607 | 0.8985 | 0.0079 |
| 2455862.6551 | 0.0842 | 6.8340 | 0.0072 | 2455862.6550 | 0.0842 | 0.9291 | 0.0091 |
| 2455864.6685 | 0.1289 | 6.8998 | 0.0024 | 2455864.6684 | 0.1289 | 0.9833 | 0.0038 |
| 2455865.5944 | 0.1495 | 6.9345 | 0.0029 | 2455865.5943 | 0.1495 | 1.0070 | 0.0054 |

**Photometry of SV Vul, cont…**

| HJD | Phase | c1 | c1 err | HJD | Phase | m1 | m1 err |
|---|---|---|---|---|---|---|---|
| 2453302.6651 | 0.1867 | 0.6529 | 0.0217 | 2453302.6638 | 0.1866 | 0.1908 | 0.0215 |
| 2453303.6644 | 0.2089 | 0.5637 | 0.0368 | 2453303.6648 | 0.2089 | 0.2051 | 0.0124 |
| 2453309.6576 | 0.3421 | 0.4044 | 0.0035 | 2453309.6578 | 0.3421 | 0.2669 | 0.0054 |
| 2453310.6570 | 0.3643 | 0.3922 | 0.0091 | 2453310.6571 | 0.3643 | 0.2875 | 0.0064 |
| 2453311.6570 | 0.3865 | 0.3620 | 0.0123 | 2453311.6571 | 0.3865 | 0.2782 | 0.0064 |
| 2453312.6419 | 0.4084 | 0.3442 | 0.0061 | 2453312.6420 | 0.4084 | 0.2997 | 0.0049 |
| 2453313.6309 | 0.4304 | 0.3653 | 0.0106 | 2453313.6311 | 0.4304 | 0.2988 | 0.0057 |
| 2453318.6244 | -0.4586 | 0.3248 | 0.0054 | 2453318.6245 | -0.4586 | 0.3134 | 0.0028 |
| 2453319.6427 | -0.4360 | 0.3203 | 0.0030 | 2453319.6428 | -0.4360 | 0.3357 | 0.0025 |
| 2453320.5873 | -0.4150 | 0.2897 | 0.0067 | 2453320.5874 | -0.4150 | 0.3379 | 0.0061 |
| 2453336.6091 | -0.0589 | 0.8501 | 0.0240 | 2453336.6092 | -0.0589 | 0.1116 | 0.0175 |
| 2453340.5661 | 0.0291 | 0.9467 | 0.0114 | 2453340.5662 | 0.0291 | 0.0834 | 0.0083 |
| 2453341.5660 | 0.0513 | 0.8567 | 0.0212 | 2453341.5661 | 0.0513 | 0.1052 | 0.0104 |
| 2453348.5660 | 0.2069 | 0.5394 | 0.0133 | 2453348.5661 | 0.2069 | 0.2029 | 0.0101 |
| 2453349.5663 | 0.2291 | 0.4881 | 0.0050 | 2453349.5664 | 0.2291 | 0.2323 | 0.0065 |
| 2453350.5664 | 0.2513 | 0.4848 | 0.0164 | 2453350.5666 | 0.2513 | 0.2301 | 0.0106 |
| 2453351.5666 | 0.2735 | 0.4733 | 0.0189 | 2453351.5667 | 0.2735 | 0.2476 | 0.0098 |
| 2453354.5739 | 0.3404 | 0.4002 | 0.0064 | 2453354.5740 | 0.3404 | 0.2885 | 0.0042 |
| 2453355.5742 | 0.3626 | 0.3518 | 0.0200 | 2453355.5743 | 0.3626 | 0.2869 | 0.0200 |
| 2453357.5679 | 0.4069 | 0.3915 | 0.0200 | 2453357.5680 | 0.4069 | 0.2645 | 0.0200 |
| 2453462.9770 | -0.2503 | 0.3009 | 0.0031 | 2453462.9772 | -0.2503 | 0.3456 | 0.0022 |
| 2453464.9714 | -0.2060 | 0.3091 | 0.0073 | 2453464.9715 | -0.2060 | 0.3366 | 0.0042 |
| 2453466.9758 | -0.1614 | 0.3921 | 0.0095 | 2453466.9763 | -0.1614 | 0.2763 | 0.0042 |
| 2453476.9549 | 0.0604 | 0.8597 | 0.0141 | 2453476.9551 | 0.0604 | 0.1036 | 0.0059 |
| 2453480.9370 | 0.1489 | 0.6823 | 0.0122 | 2453480.9371 | 0.1489 | 0.1579 | 0.0096 |
| 2453481.9450 | 0.1713 | 0.6325 | 0.0101 | 2453481.9452 | 0.1713 | 0.1819 | 0.0103 |
| 2453483.9399 | 0.2156 | 0.5556 | 0.0112 | 2453483.9400 | 0.2156 | 0.2053 | 0.0106 |
| 2453487.9227 | 0.3042 | 0.4352 | 0.0060 | 2453487.9228 | 0.3042 | 0.2546 | 0.0062 |
| 2453488.9161 | 0.3262 | 0.4223 | 0.0042 | 2453488.9162 | 0.3262 | 0.2674 | 0.0050 |
| 2453490.9108 | 0.3706 | 0.3967 | 0.0124 | 2453490.9109 | 0.3706 | 0.2735 | 0.0067 |
| 2453494.8975 | 0.4592 | 0.3051 | 0.0085 | 2453494.8976 | 0.4592 | 0.3287 | 0.0085 |
| 2453496.9676 | -0.4948 | 0.3137 | 0.0034 | 2453496.9677 | -0.4948 | 0.3287 | 0.0037 |
| 2453500.8861 | -0.4077 | 0.3160 | 0.0045 | 2453500.8862 | -0.4077 | 0.3340 | 0.0049 |
| 2453501.8997 | -0.3852 | 0.2833 | 0.0047 | 2453501.8998 | -0.3852 | 0.3568 | 0.0019 |
| 2453502.9544 | -0.3618 | 0.3086 | 0.0017 | 2453502.9545 | -0.3618 | 0.3351 | 0.0019 |
| 2453503.8876 | -0.3410 | 0.2843 | 0.0093 | 2453503.8877 | -0.3410 | 0.3583 | 0.0061 |
| 2453504.9456 | -0.3175 | 0.3013 | 0.0076 | 2453504.9457 | -0.3175 | 0.3473 | 0.0054 |
| 2453506.9494 | -0.2730 | 0.2567 | 0.0029 | 2453506.9491 | -0.2730 | 0.3530 | 0.0029 |
| 2453507.8754 | -0.2524 | 0.3035 | 0.0026 | 2453507.8751 | -0.2524 | 0.3393 | 0.0036 |
| 2453509.8766 | -0.2079 | 0.3075 | 0.0037 | 2453509.8771 | -0.2079 | 0.3178 | 0.0027 |
| 2453510.8644 | -0.1860 | 0.3094 | 0.0056 | 2453510.8641 | -0.1860 | 0.3318 | 0.0035 |
| 2453511.8507 | -0.1640 | 0.3353 | 0.0064 | 2453511.8508 | -0.1640 | 0.3122 | 0.0049 |
| 2453512.8513 | -0.1418 | 0.4029 | 0.0069 | 2453512.8514 | -0.1418 | 0.2646 | 0.0046 |
| 2453513.8452 | -0.1197 | 0.4869 | 0.0140 | 2453513.8454 | -0.1197 | 0.2320 | 0.0097 |
| 2453520.8342 | 0.0356 | 0.9315 | 0.0317 | 2453520.8343 | 0.0356 | 0.0850 | 0.0285 |
| 2453524.8251 | 0.1243 | 0.7212 | 0.0151 | 2453524.8252 | 0.1243 | 0.1436 | 0.0142 |
| 2453525.8201 | 0.1464 | 0.7090 | 0.0151 | 2453525.8202 | 0.1465 | 0.1427 | 0.0118 |
| 2453526.8123 | 0.1685 | 0.6546 | 0.0118 | 2453526.8125 | 0.1685 | 0.1633 | 0.0076 |
| 2453527.8150 | 0.1908 | 0.6207 | 0.0216 | 2453527.8151 | 0.1908 | 0.1798 | 0.0062 |
| 2453528.8089 | 0.2129 | 0.5639 | 0.0203 | 2453528.8090 | 0.2129 | 0.2032 | 0.0070 |
| 2453529.8138 | 0.2352 | 0.5231 | 0.0161 | 2453529.8139 | 0.2352 | 0.2168 | 0.0063 |
| 2453530.8086 | 0.2573 | 0.5106 | 0.0080 | 2453530.8087 | 0.2573 | 0.2280 | 0.0061 |
| 2453531.8030 | 0.2794 | 0.4901 | 0.0130 | 2453531.8031 | 0.2794 | 0.2376 | 0.0063 |
| 2453532.8023 | 0.3016 | 0.4409 | 0.0075 | 2453532.8024 | 0.3016 | 0.2530 | 0.0054 |
| 2453533.7976 | 0.3238 | 0.4046 | 0.0091 | 2453533.7977 | 0.3238 | 0.2701 | 0.0050 |
| 2453535.7988 | 0.3682 | 0.3935 | 0.0044 | 2453535.7989 | 0.3682 | 0.2837 | 0.0058 |
| 2453536.7923 | 0.3903 | 0.3728 | 0.0054 | 2453536.7924 | 0.3903 | 0.2831 | 0.0060 |
| 2453537.7866 | 0.4124 | 0.3214 | 0.0105 | 2453537.7867 | 0.4124 | 0.3184 | 0.0079 |
| 2453538.7817 | 0.4345 | 0.3648 | 0.0036 | 2453538.7814 | 0.4345 | 0.3058 | 0.0038 |
| 2453539.7807 | 0.4567 | 0.3382 | 0.0037 | 2453539.7808 | 0.4567 | 0.3078 | 0.0049 |
| 2453543.7878 | -0.4542 | 0.2143 | 0.0065 | 2453543.7880 | -0.4542 | 0.3142 | 0.0060 |
| 2453547.7614 | -0.3659 | 0.3082 | 0.0075 | 2453547.7616 | -0.3659 | 0.3443 | 0.0042 |
| 2453548.7606 | -0.3437 | 0.2690 | 0.0052 | 2453548.7607 | -0.3437 | 0.3532 | 0.0071 |
| 2453549.7482 | -0.3217 | 0.2958 | 0.0058 | 2453549.7484 | -0.3217 | 0.3454 | 0.0054 |
| 2453550.7513 | -0.2994 | 0.3588 | 0.0041 | 2453550.7514 | -0.2994 | 0.3341 | 0.0061 |
| 2453551.7552 | -0.2771 | 0.3109 | 0.0045 | 2453551.7553 | -0.2771 | 0.3596 | 0.0036 |
| 2453552.7499 | -0.2550 | 0.2540 | 0.0058 | 2453552.7500 | -0.2550 | 0.3556 | 0.0026 |
| 2453554.7385 | -0.2108 | 0.2872 | 0.0088 | 2453554.7387 | -0.2108 | 0.3553 | 0.0103 |
| 2453555.7407 | -0.1885 | 0.3331 | 0.0059 | 2453555.7408 | -0.1885 | 0.3274 | 0.0076 |
| 2453556.7324 | -0.1665 | 0.3666 | 0.0081 | 2453556.7321 | -0.1665 | 0.2981 | 0.0040 |
| 2453557.7367 | -0.1442 | 0.3761 | 0.0081 | 2453557.7369 | -0.1442 | 0.2713 | 0.0032 |
| 2453558.7265 | -0.1222 | 0.4859 | 0.0114 | 2453558.7267 | -0.1222 | 0.2182 | 0.0054 |
| 2453560.7537 | -0.0771 | 0.7752 | 0.0113 | 2453560.7538 | -0.0771 | 0.1141 | 0.0093 |
| 2453561.7241 | -0.0556 | 0.9075 | 0.0222 | 2453561.7238 | -0.0556 | 0.0864 | 0.0130 |
| 2453563.7166 | -0.0113 | 0.9649 | 0.0280 | 2453563.7167 | -0.0113 | 0.0601 | 0.0165 |
| 2453565.7175 | 0.0332 | 0.9158 | 0.0206 | 2453565.7176 | 0.0332 | 0.0848 | 0.0145 |
| 2453567.7022 | 0.0773 | 0.8068 | 0.0263 | 2453567.7023 | 0.0773 | 0.1203 | 0.0241 |
| 2453627.6461 | 0.4096 | 0.3341 | 0.0073 | 2453627.6462 | 0.4096 | 0.2865 | 0.0057 |
| 2453628.7494 | 0.4341 | 0.3195 | 0.0092 | 2453628.7495 | 0.4341 | 0.2986 | 0.0044 |
| 2453629.6301 | 0.4537 | 0.3308 | 0.0067 | 2453629.6302 | 0.4537 | 0.3025 | 0.0059 |
| 2453631.6279 | 0.4981 | 0.2765 | 0.0075 | 2453631.6281 | 0.4981 | 0.3206 | 0.0056 |
| 2453632.7193 | -0.4776 | 0.2540 | 0.0046 | 2453632.7194 | -0.4776 | 0.3372 | 0.0051 |
| 2453634.7081 | -0.4334 | 0.3045 | 0.0055 | 2453634.7083 | -0.4334 | 0.3258 | 0.0082 |
| 2453636.7087 | -0.3890 | 0.2782 | 0.0031 | 2453636.7088 | -0.3890 | 0.3568 | 0.0036 |
| 2453637.6227 | -0.3687 | 0.2320 | 0.0086 | 2453637.6229 | -0.3687 | 0.3458 | 0.0083 |
| 2453638.7146 | -0.3444 | 0.2482 | 0.0054 | 2453638.7147 | -0.3444 | 0.3587 | 0.0031 |
| 2453639.7613 | -0.3211 | 0.2332 | 0.0042 | 2453639.7615 | -0.3211 | 0.3462 | 0.0093 |
| 2453641.6191 | -0.2798 | 0.2550 | 0.0053 | 2453641.6192 | -0.2798 | 0.3512 | 0.0051 |
| 2453642.7125 | -0.2555 | 0.2323 | 0.0031 | 2453642.7126 | -0.2555 | 0.3475 | 0.0043 |
| 2453643.6354 | -0.2350 | 0.2524 | 0.0043 | 2453643.6356 | -0.2350 | 0.3347 | 0.0039 |
| 2453644.7070 | -0.2112 | 0.2439 | 0.0062 | 2453644.7071 | -0.2112 | 0.3360 | 0.0048 |
| 2453648.6663 | -0.1232 | 0.4884 | 0.0080 | 2453648.6664 | -0.1232 | 0.2246 | 0.0060 |
| 2453652.6991 | -0.0336 | 0.9467 | 0.0261 | 2453652.6992 | -0.0336 | 0.0798 | 0.0110 |
| 2453653.6254 | -0.0130 | 0.9542 | 0.0286 | 2453653.6255 | -0.0130 | 0.0863 | 0.0199 |
| 2453655.6236 | 0.0314 | 0.9128 | 0.0181 | 2453655.6237 | 0.0314 | 0.0826 | 0.0148 |
| 2453657.6124 | 0.0756 | 0.8440 | 0.0141 | 2453657.6125 | 0.0756 | 0.1091 | 0.0048 |
| 2453667.6148 | 0.2979 | 0.4499 | 0.0092 | 2453667.6149 | 0.2979 | 0.2612 | 0.0075 |
| 2453668.6581 | 0.3211 | 0.4228 | 0.0119 | 2453668.6582 | 0.3211 | 0.2572 | 0.0078 |
| 2453669.6442 | 0.3430 | 0.4043 | 0.0090 | 2453669.6444 | 0.3430 | 0.2750 | 0.0045 |
| 2453671.6469 | 0.3875 | 0.3928 | 0.0067 | 2453671.6470 | 0.3875 | 0.2858 | 0.0062 |
| 2453672.5952 | 0.4086 | 0.3605 | 0.0039 | 2453672.5953 | 0.4086 | 0.3035 | 0.0041 |
| 2453673.5945 | 0.4308 | 0.3497 | 0.0054 | 2453673.5946 | 0.4308 | 0.3027 | 0.0041 |
| 2453674.5941 | 0.4531 | 0.3237 | 0.0075 | 2453674.5942 | 0.4531 | 0.3160 | 0.0046 |
| 2453679.5910 | -0.4359 | 0.2913 | 0.0051 | 2453679.5912 | -0.4359 | 0.3468 | 0.0050 |
| 2453680.5906 | -0.4137 | 0.2794 | 0.0039 | 2453680.5907 | -0.4137 | 0.3452 | 0.0037 |
| 2453688.5868 | -0.2360 | 0.2882 | 0.0042 | 2453688.5869 | -0.2360 | 0.3333 | 0.0042 |
| 2453690.5876 | -0.1915 | 0.3032 | 0.0064 | 2453690.5871 | -0.1915 | 0.3215 | 0.0026 |
| 2453691.6003 | -0.1690 | 0.3140 | 0.0070 | 2453691.6004 | -0.1690 | 0.3029 | 0.0070 |
| 2453694.5699 | -0.1030 | 0.5867 | 0.0121 | 2453694.5690 | -0.1030 | 0.1769 | 0.0063 |
| 2453695.5684 | -0.0808 | 0.7691 | 0.0166 | 2453695.5685 | -0.0808 | 0.1326 | 0.0113 |
| 2454257.8698 | 0.4167 | 0.3801 | 0.0129 | 2454257.8699 | 0.4168 | 0.3025 | 0.0159 |
| 2454258.8572 | 0.4387 | 0.3455 | 0.0092 | 2454258.8573 | 0.4387 | 0.3183 | 0.0082 |
| 2454259.8668 | 0.4611 | 0.3748 | 0.0124 | 2454259.8670 | 0.4611 | 0.3209 | 0.0078 |
| 2454260.8498 | 0.4830 | 0.3653 | 0.0106 | 2454260.8500 | 0.4830 | 0.3288 | 0.0084 |
| 2454261.8619 | -0.4945 | 0.3581 | 0.0032 | 2454261.8620 | -0.4945 | 0.3318 | 0.0071 |
| 2454264.8404 | -0.4283 | 0.3215 | 0.0053 | 2454264.8405 | -0.4283 | 0.3334 | 0.0059 |
| 2454265.8425 | -0.4061 | 0.3152 | 0.0111 | 2454265.8426 | -0.4061 | 0.3447 | 0.0089 |
| 2454266.8500 | -0.3837 | 0.3071 | 0.0079 | 2454266.8501 | -0.3837 | 0.3504 | 0.0071 |
| 2454268.8502 | -0.3392 | 0.2944 | 0.0025 | 2454268.8503 | -0.3392 | 0.3671 | 0.0069 |
| 2454269.8520 | -0.3169 | 0.2977 | 0.0106 | 2454269.8521 | -0.3169 | 0.3623 | 0.0106 |
| 2454270.8363 | -0.2951 | 0.3041 | 0.0044 | 2454270.8358 | -0.2951 | 0.3497 | 0.0047 |
| 2454272.8343 | -0.2507 | 0.3001 | 0.0038 | 2454272.8344 | -0.2507 | 0.3485 | 0.0089 |
| 2454273.8508 | -0.2281 | 0.3142 | 0.0076 | 2454273.8509 | -0.2281 | 0.3594 | 0.0059 |
| 2454275.8467 | -0.1837 | 0.3374 | 0.0048 | 2454275.8468 | -0.1837 | 0.3195 | 0.0100 |
| 2454276.8260 | -0.1619 | 0.3483 | 0.0050 | 2454276.8255 | -0.1619 | 0.3021 | 0.0081 |
| 2454277.8394 | -0.1394 | 0.4252 | 0.0129 | 2454277.8395 | -0.1394 | 0.2685 | 0.0166 |
| 2454278.8153 | -0.1177 | 0.4545 | 0.0116 | 2454278.8154 | -0.1177 | 0.2419 | 0.0102 |
| 2454279.8270 | -0.0952 | 0.5954 | 0.0137 | 2454279.8271 | -0.0952 | 0.1837 | 0.0065 |
| 2454280.8167 | -0.0732 | 0.7082 | 0.0156 | 2454280.8168 | -0.0732 | 0.1439 | 0.0209 |
| 2454281.8220 | -0.0509 | 0.8568 | 0.0328 | 2454281.8221 | -0.0509 | 0.1092 | 0.0367 |
| 2454282.8108 | -0.0289 | 0.9316 | 0.0234 | 2454282.8109 | -0.0289 | 0.0924 | 0.0090 |
| 2454284.8271 | 0.0159 | 0.9333 | 0.0520 | 2454284.8272 | 0.0159 | 0.1098 | 0.0365 |
| 2454285.8185 | 0.0379 | 0.9413 | 0.0322 | 2454285.8186 | 0.0379 | 0.0894 | 0.0220 |
| 2454370.7158 | -0.0752 | 0.6839 | 0.0122 | 2454370.7159 | -0.0752 | 0.1579 | 0.0052 |
| 2454373.6455 | -0.0101 | 0.9664 | 0.0201 | 2454373.6456 | -0.0101 | 0.0924 | 0.0132 |
| 2454380.6936 | 0.1466 | 0.8301 | 0.1829 | 2454380.6931 | 0.1466 | 0.1327 | 0.0333 |
| 2454381.6381 | 0.1676 | 0.8026 | 0.2436 | 2454381.6382 | 0.1676 | 0.0668 | 0.0742 |
| 2454388.6701 | 0.3239 | 0.4599 | 0.0060 | 2454382.6816 | 0.1908 | 0.1977 | 0.2139 |
| 2454390.6571 | 0.3680 | 0.4095 | 0.0111 | 2454388.6702 | 0.3239 | 0.2523 | 0.0043 |
| 2454391.6190 | 0.3894 | 0.4038 | 0.0066 | 2454390.6572 | 0.3680 | 0.2819 | 0.0050 |
| 2454392.6552 | 0.4124 | 0.3741 | 0.0073 | 2454391.6191 | 0.3894 | 0.2732 | 0.0050 |
| 2454393.6174 | 0.4338 | 0.3456 | 0.0037 | 2454392.6554 | 0.4124 | 0.2956 | 0.0067 |
| 2454394.6450 | 0.4567 | 0.3444 | 0.0092 | 2454393.6175 | 0.4338 | 0.3121 | 0.0031 |
| 2454737.6341 | 0.0798 | 0.8774 | 0.0249 | 2454394.6451 | 0.4567 | 0.3124 | 0.0038 |
| 2454747.6245 | 0.3019 | 0.4779 | 0.0113 | 2454737.6342 | 0.0798 | 0.0976 | 0.0176 |
| 2454749.6421 | 0.3467 | 0.3797 | 0.0095 | 2454747.6246 | 0.3019 | 0.2384 | 0.0065 |
| 2454753.6467 | 0.4357 | 0.3897 | 0.0085 | 2454749.6422 | 0.3467 | 0.2785 | 0.0056 |
| 2454755.6284 | 0.4798 | 0.3373 | 0.0032 | 2454753.6464 | 0.4357 | 0.2882 | 0.0040 |
| 2454757.6128 | -0.4761 | 0.3482 | 0.0017 | 2454755.6286 | 0.4798 | 0.3241 | 0.0031 |
| 2454759.6097 | -0.4317 | 0.3112 | 0.0069 | 2454757.6117 | -0.4762 | 0.3235 | 0.0026 |
| 2454761.6225 | -0.3870 | 0.3096 | 0.0066 | 2454759.6098 | -0.4317 | 0.3379 | 0.0064 |
| 2454763.6212 | -0.3426 | 0.2761 | 0.0017 | 2454761.6227 | -0.3870 | 0.3268 | 0.0049 |
| 2454765.6309 | -0.2979 | 0.3447 | 0.0020 | 2454763.6213 | -0.3426 | 0.3354 | 0.0014 |
| 2454767.6278 | -0.2535 | 0.3015 | 0.0068 | 2454765.6297 | -0.2979 | 0.3474 | 0.0016 |
| 2454771.6390 | -0.1644 | 0.3356 | 0.0052 | 2454767.6279 | -0.2535 | 0.3464 | 0.0049 |
| 2454775.6237 | -0.0758 | 0.6693 | 0.0122 | 2454771.6387 | -0.1644 | 0.3237 | 0.0051 |
| 2454785.5943 | 0.1458 | 0.7205 | 0.0128 | 2454775.6234 | -0.0758 | 0.1708 | 0.0159 |
| 2454985.8815 | -0.4027 | 0.3048 | 0.0062 | 2454785.5944 | 0.1458 | 0.1432 | 0.0096 |
| 2454989.8708 | -0.3140 | 0.3332 | 0.0081 | 2454985.8816 | -0.4027 | 0.3443 | 0.0095 |
| 2454997.8476 | -0.1367 | 0.3830 | 0.0070 | 2454989.8714 | -0.3140 | 0.3668 | 0.0091 |
| 2454998.8416 | -0.1147 | 0.4862 | 0.0073 | 2454997.8477 | -0.1367 | 0.2746 | 0.0063 |
| 2454999.9180 | -0.0907 | 0.5918 | 0.0181 | 2454998.8417 | -0.1147 | 0.2220 | 0.0069 |
| 2455004.9123 | 0.0203 | 0.9457 | 0.0387 | 2454999.9181 | -0.0907 | 0.1919 | 0.0166 |
| 2455006.8735 | 0.0639 | 0.8858 | 0.0304 | 2455004.9124 | 0.0203 | 0.0906 | 0.0396 |
| 2455098.7333 | 0.1055 | 0.8109 | 0.0420 | 2455006.8736 | 0.0639 | 0.1047 | 0.0257 |
| 2455099.7461 | 0.1280 | 0.7537 | 0.0153 | 2455098.7334 | 0.1055 | 0.1273 | 0.0414 |
| 2455100.7374 | 0.1500 | 0.7049 | 0.0169 | 2455099.7462 | 0.1280 | 0.1324 | 0.0183 |
| 2455101.7527 | 0.1726 | 0.6788 | 0.0097 | 2455100.7375 | 0.1501 | 0.1520 | 0.0160 |
| 2455102.7352 | 0.1945 | 0.6262 | 0.0180 | 2455101.7528 | 0.1726 | 0.1571 | 0.0097 |
| 2455106.7262 | 0.2832 | 0.5239 | 0.0108 | 2455102.7353 | 0.1945 | 0.1846 | 0.0190 |
| 2455109.6902 | 0.3490 | 0.4064 | 0.0321 | 2455106.7263 | 0.2832 | 0.2411 | 0.0138 |
| 2455133.5949 | -0.1197 | 0.4444 | 0.0072 | 2455109.6903 | 0.3490 | 0.2589 | 0.0368 |
| 2455134.5942 | -0.0975 | 0.5568 | 0.0156 | 2455133.5950 | -0.1197 | 0.2568 | 0.0052 |
| 2455136.5930 | -0.0530 | 0.8658 | 0.0187 | 2455134.5944 | -0.0975 | 0.2041 | 0.0154 |
| 2455137.5922 | -0.0308 | 0.9384 | 0.0125 | 2455136.5931 | -0.0530 | 0.1131 | 0.0094 |
| 2455138.5918 | -0.0086 | 0.9403 | 0.0228 | 2455137.5924 | -0.0308 | 0.0872 | 0.0091 |
| 2455139.5915 | 0.0136 | 0.9422 | 0.0087 | 2455138.5919 | -0.0086 | 0.0936 | 0.0150 |
| 2455143.5929 | 0.1025 | 0.8038 | 0.0178 | 2455139.5912 | 0.0136 | 0.0935 | 0.0098 |
| 2455144.5886 | 0.1247 | 0.7623 | 0.0144 | 2455143.5930 | 0.1025 | 0.1313 | 0.0170 |
| 2455146.5876 | 0.1691 | 0.6716 | 0.0103 | 2455144.5888 | 0.1247 | 0.1379 | 0.0122 |
| 2455151.5865 | 0.2802 | 0.5084 | 0.0064 | 2455146.5877 | 0.1691 | 0.1639 | 0.0098 |
| 2455295.9905 | 0.4897 | 0.3552 | 0.0036 | 2455151.5871 | 0.2802 | 0.2255 | 0.0053 |
| 2455296.9873 | -0.4882 | 0.3235 | 0.0117 | 2455295.9907 | 0.4897 | 0.3158 | 0.0029 |
| 2455297.9847 | -0.4660 | 0.3097 | 0.0087 | 2455296.9874 | -0.4882 | 0.3346 | 0.0042 |
| 2455298.9820 | -0.4438 | 0.3412 | 0.0061 | 2455297.9852 | -0.4660 | 0.3395 | 0.0078 |
| 2455299.9802 | -0.4216 | 0.3032 | 0.0065 | 2455298.9821 | -0.4438 | 0.3286 | 0.0057 |
| 2455300.9776 | -0.3995 | 0.3501 | 0.0107 | 2455299.9808 | -0.4216 | 0.3420 | 0.0063 |
| 2455301.9747 | -0.3773 | 0.3298 | 0.0074 | 2455300.9777 | -0.3995 | 0.3307 | 0.0078 |
| 2455314.9773 | -0.0883 | 0.6219 | 0.0552 | 2455301.9749 | -0.3773 | 0.3435 | 0.0064 |
| 2455315.9658 | -0.0664 | 0.7575 | 0.0251 | 2455314.9774 | -0.0883 | 0.1947 | 0.0400 |
| 2455317.9492 | -0.0223 | 0.9505 | 0.0164 | 2455315.9660 | -0.0663 | 0.1390 | 0.0198 |
| 2455320.9492 | 0.0444 | 0.9244 | 0.0218 | 2455317.9493 | -0.0223 | 0.0905 | 0.0115 |
| 2455321.9211 | 0.0660 | 0.8821 | 0.0125 | 2455320.9494 | 0.0444 | 0.0943 | 0.0174 |
| 2455323.9635 | 0.1114 | 0.7912 | 0.0178 | 2455321.9212 | 0.0660 | 0.0982 | 0.0123 |
| 2455326.9082 | 0.1768 | 0.6518 | 0.0153 | 2455323.9636 | 0.1114 | 0.1232 | 0.0117 |
| 2455328.9406 | 0.2220 | 0.5766 | 0.0104 | 2455326.9084 | 0.1769 | 0.1745 | 0.0140 |
| 2455335.9249 | 0.3773 | 0.3963 | 0.0090 | 2455328.9412 | 0.2220 | 0.2105 | 0.0076 |
| 2455337.9470 | 0.4222 | 0.3738 | 0.0033 | 2455335.9251 | 0.3773 | 0.2803 | 0.0064 |
| 2455463.7780 | 0.2189 | 0.5697 | 0.0225 | 2455337.9467 | 0.4222 | 0.2970 | 0.0048 |
| 2455464.7557 | 0.2406 | 0.5359 | 0.0213 | 2455463.7786 | 0.2189 | 0.2043 | 0.0092 |
| 2455465.7725 | 0.2632 | 0.5071 | 0.0200 | 2455464.7562 | 0.2406 | 0.2077 | 0.0148 |
| 2455495.6276 | -0.0732 | 0.7492 | 0.0118 | 2455465.7726 | 0.2632 | 0.2335 | 0.0149 |
| 2455498.6706 | -0.0056 | 0.9431 | 0.0277 | 2455495.6277 | -0.0732 | 0.1371 | 0.0093 |
| 2455499.5940 | 0.0149 | 0.9705 | 0.0132 | 2455498.6707 | -0.0056 | 0.0825 | 0.0258 |
| 2455501.5931 | 0.0593 | 0.8941 | 0.0116 | 2455499.5941 | 0.0149 | 0.0761 | 0.0128 |
| 2455502.5923 | 0.0815 | 0.8562 | 0.0100 | 2455501.5932 | 0.0593 | 0.0977 | 0.0094 |
| 2455503.5920 | 0.1038 | 0.7852 | 0.0129 | 2455502.5925 | 0.0816 | 0.1132 | 0.0086 |
| 2455504.5913 | 0.1260 | 0.7421 | 0.0099 | 2455503.5921 | 0.1038 | 0.1355 | 0.0087 |
| 2455508.6040 | 0.2152 | 0.5622 | 0.0073 | 2455504.5914 | 0.1260 | 0.1372 | 0.0083 |
| 2455509.6041 | 0.2374 | 0.5413 | 0.0099 | 2455508.6041 | 0.2152 | 0.1889 | 0.0113 |
| 2455510.6031 | 0.2596 | 0.5027 | 0.0077 | 2455509.6051 | 0.2374 | 0.2035 | 0.0069 |
| 2455511.6025 | 0.2818 | 0.5001 | 0.0153 | 2455510.6036 | 0.2596 | 0.2260 | 0.0050 |
| 2455513.6017 | 0.3262 | 0.4116 | 0.0110 | 2455511.6026 | 0.2818 | 0.2285 | 0.0057 |
| 2455516.6008 | 0.3929 | 0.3769 | 0.0120 | 2455513.6018 | 0.3262 | 0.2611 | 0.0074 |
| 2455517.6004 | 0.4151 | 0.3816 | 0.0109 | 2455516.6009 | 0.3929 | 0.2838 | 0.0078 |
| 2455518.6007 | 0.4373 | 0.3581 | 0.0111 | 2455517.6005 | 0.4151 | 0.2874 | 0.0106 |
| 2455519.5998 | 0.4596 | 0.3815 | 0.0165 | 2455518.6013 | 0.4374 | 0.3029 | 0.0061 |
| 2455523.5759 | -0.4521 | 0.3272 | 0.0070 | 2455519.5999 | 0.4596 | 0.2927 | 0.0101 |
| 2455530.5745 | -0.2965 | 0.3365 | 0.0037 | 2455523.5761 | -0.4521 | 0.3262 | 0.0058 |
| 2455532.5745 | -0.2521 | 0.3033 | 0.0048 | 2455530.5746 | -0.2965 | 0.3566 | 0.0031 |
| 2455702.8364 | -0.4679 | 0.2877 | 0.0046 | 2455532.5746 | -0.2521 | 0.3483 | 0.0028 |
| 2455704.8436 | -0.4233 | 0.2732 | 0.0055 | 2455702.8367 | -0.4679 | 0.3540 | 0.0078 |
| 2455710.8254 | -0.2903 | 0.2727 | 0.0119 | 2455704.8439 | -0.4233 | 0.3379 | 0.0072 |
| 2455712.8101 | -0.2462 | 0.2753 | 0.0142 | 2455706.8236 | -0.3793 | 0.3364 | 0.0043 |
| 2455716.7975 | -0.1576 | 0.3627 | 0.0102 | 2455710.8256 | -0.2903 | 0.3522 | 0.0166 |
| 2455722.7974 | -0.0242 | 0.9229 | 0.0193 | 2455712.8103 | -0.2462 | 0.3288 | 0.0080 |
| 2455730.7621 | 0.1528 | 0.6016 | 0.0212 | 2455716.7978 | -0.1576 | 0.2788 | 0.0085 |
| 2455732.7534 | 0.1970 | 0.5169 | 0.0204 | 2455722.7972 | -0.0243 | 0.0824 | 0.0170 |
| 2455734.7646 | 0.2417 | 0.4792 | 0.0238 | 2455730.7635 | 0.1528 | 0.1431 | 0.0223 |
| 2455736.7470 | 0.2858 | 0.4046 | 0.0178 | 2455732.7537 | 0.1970 | 0.2098 | 0.0182 |
| 2455738.9235 | 0.3342 | 0.4032 | 0.0102 | 2455734.7649 | 0.2417 | 0.2311 | 0.0086 |
| 2455740.7558 | 0.3749 | 0.3428 | 0.0164 | 2455736.7467 | 0.2858 | 0.2692 | 0.0204 |
| 2455824.6469 | 0.2394 | 0.5320 | 0.0072 | 2455738.9249 | 0.3342 | 0.2597 | 0.0073 |
| 2455830.8015 | 0.3762 | 0.4123 | 0.0200 | 2455740.7560 | 0.3749 | 0.3081 | 0.0136 |
| 2455835.7877 | 0.4870 | 0.3268 | 0.0200 | 2455824.6472 | 0.2394 | 0.2218 | 0.0059 |
| 2455839.7761 | -0.4243 | 0.3343 | 0.0200 | 2455830.8018 | 0.3762 | 0.2795 | 0.0200 |
| 2455840.7737 | -0.4021 | 0.2856 | 0.0200 | 2455835.7880 | 0.4870 | 0.3492 | 0.0200 |
| 2455842.7681 | -0.3578 | 0.3207 | 0.0200 | 2455839.7764 | -0.4243 | 0.3280 | 0.0200 |
| 2455843.7644 | -0.3357 | 0.3415 | 0.0200 | 2455840.7740 | -0.4021 | 0.3451 | 0.0200 |
| 2455845.7599 | -0.2913 | 0.3569 | 0.0200 | 2455842.7684 | -0.3578 | 0.3446 | 0.0200 |
| 2455846.7568 | -0.2692 | 0.2521 | 0.0200 | 2455843.7646 | -0.3357 | 0.3659 | 0.0200 |
| 2455847.7532 | -0.2470 | 0.4683 | 0.0200 | 2455845.7602 | -0.2913 | 0.3205 | 0.0200 |
| 2455848.7505 | -0.2249 | 0.3675 | 0.0200 | 2455846.7571 | -0.2692 | 0.3474 | 0.0200 |
| 2455849.7483 | -0.2027 | 0.4196 | 0.0200 | 2455847.7534 | -0.2470 | 0.3401 | 0.0200 |
| 2455851.7425 | -0.1584 | 0.4005 | 0.0200 | 2455848.7508 | -0.2248 | 0.3197 | 0.0200 |
| 2455854.5872 | -0.0951 | 0.7642 | 0.0157 | 2455849.7485 | -0.2027 | 0.2985 | 0.0200 |
| 2455855.6013 | -0.0726 | 0.9196 | 0.0113 | 2455851.7427 | -0.1583 | 0.2807 | 0.0200 |
| 2455856.6016 | -0.0504 | 0.9559 | 0.0065 | 2455854.5875 | -0.0951 | 0.1333 | 0.0091 |
| 2455857.6290 | -0.0275 | 0.9657 | 0.0242 | 2455855.6015 | -0.0726 | 0.0912 | 0.0080 |
| 2455858.5989 | -0.0060 | 0.9647 | 0.0122 | 2455856.6013 | -0.0504 | 0.0907 | 0.0048 |
| 2455861.5966 | 0.0607 | 0.8515 | 0.0086 | 2455857.6293 | -0.0275 | 0.0730 | 0.0218 |
| 2455862.6546 | 0.0842 | 0.7840 | 0.0092 | 2455858.5992 | -0.0060 | 0.0779 | 0.0105 |
| 2455864.6680 | 0.1289 | 0.6843 | 0.0126 | 2455861.5969 | 0.0607 | 0.1118 | 0.0095 |
| 2455865.5939 | 0.1495 | 0.6700 | 0.0097 | 2455862.6548 | 0.0842 | 0.1272 | 0.0113 |
| | | | | 2455864.6682 | 0.1289 | 0.1523 | 0.0059 |
| | | | | 2455865.5941 | 0.1495 | 0.1536 | 0.0090 |

**Photometry of SZ Cas**

| HJD | Phase | V-mag | V-err | HJD | Phase | U-B | U-B err | HJD | Phase | B-V | B-V err |
|---|---|---|---|---|---|---|---|---|---|---|---|
| 2454075.7771 | 0.1252 | 9.6756 | 0.0060 | 2454075.7780 | 0.1253 | 0.9934 | 0.0178 | 2454075.7781 | 0.1253 | 1.4263 | 0.0141 |
| 2454076.5892 | 0.1848 | 9.7412 | 0.0057 | 2454076.5880 | 0.1847 | 1.1206 | 0.0194 | 2454076.5881 | 0.1847 | 1.4446 | 0.0095 |
| 2454077.7807 | 0.2722 | 9.8028 | 0.0039 | 2454077.7805 | 0.2721 | 1.1752 | 0.0144 | 2454077.7807 | 0.2721 | 1.5039 | 0.0067 |
| 2454080.5875 | 0.4780 | 9.9544 | 0.0022 | 2454080.5873 | 0.4779 | 1.2594 | 0.0162 | 2454080.5874 | 0.4779 | 1.5792 | 0.0064 |
| 2454081.7701 | -0.4353 | 9.9837 | 0.0050 | 2454081.7699 | -0.4354 | 1.2080 | 0.0145 | 2454081.7700 | -0.4353 | 1.5445 | 0.0091 |
| 2454083.7645 | -0.2891 | 9.9433 | 0.0035 | 2454083.7643 | -0.2891 | 1.1389 | 0.0071 | 2454083.7645 | -0.2891 | 1.4903 | 0.0050 |
| 2454085.7763 | -0.1416 | 9.7580 | 0.0061 | 2454085.7761 | -0.1416 | 1.0628 | 0.0289 | 2454085.7762 | -0.1416 | 1.3948 | 0.0101 |
| 2454086.7809 | -0.0679 | 9.6395 | 0.0040 | 2454086.7807 | -0.0679 | 0.9726 | 0.0071 | 2454086.7809 | -0.0679 | 1.3451 | 0.0076 |
| 2454088.7576 | 0.0770 | 9.6411 | 0.0037 | 2454088.7574 | 0.0770 | 1.0281 | 0.0168 | 2454088.7575 | 0.0770 | 1.3839 | 0.0038 |
| 2454090.7653 | 0.2243 | 9.7670 | 0.0034 | 2454090.7651 | 0.2242 | 1.1269 | 0.0183 | 2454090.7652 | 0.2242 | 1.4747 | 0.0092 |
| 2454094.7579 | -0.4830 | 9.9725 | 0.0032 | 2454094.7577 | -0.4830 | 1.2240 | 0.0143 | 2454094.7578 | -0.4830 | 1.5628 | 0.0068 |
| 2454100.7467 | -0.0438 | 9.6152 | 0.0019 | 2454100.7465 | -0.0439 | 1.0161 | 0.0116 | 2454100.7466 | -0.0439 | 1.3488 | 0.0064 |
| 2454101.6352 | 0.0213 | 9.6141 | 0.0021 | 2454101.6350 | 0.0213 | 1.0247 | 0.0170 | 2454101.6351 | 0.0213 | 1.3659 | 0.0124 |
| 2454102.7394 | 0.1023 | 9.6540 | 0.0089 | 2454102.7392 | 0.1022 | 1.0902 | 0.0233 | 2454102.7393 | 0.1022 | 1.3979 | 0.0113 |
| 2454103.7247 | 0.1745 | 9.7090 | 0.0026 | 2454103.7245 | 0.1745 | 1.1124 | 0.0317 | 2454103.7246 | 0.1745 | 1.4440 | 0.0096 |
| 2454104.7220 | 0.2476 | 9.7766 | 0.0025 | 2454104.7229 | 0.2477 | 1.3032 | 0.0169 | 2454104.7219 | 0.2476 | 1.4968 | 0.0068 |
| 2454105.7205 | 0.3209 | 9.8499 | 0.0053 | 2454105.7203 | 0.3208 | 1.2239 | 0.0269 | 2454105.7204 | 0.3208 | 1.5178 | 0.0090 |
| 2454107.7137 | 0.4670 | 9.9587 | 0.0025 | 2454107.7135 | 0.4670 | 1.2831 | 0.0232 | 2454107.7136 | 0.4670 | 1.5597 | 0.0052 |
| 2454108.6997 | -0.4607 | 9.9825 | 0.0027 | 2454108.6996 | -0.4607 | 1.2412 | 0.0140 | 2454108.6997 | -0.4607 | 1.5498 | 0.0049 |
| 2454115.6975 | 0.0524 | 9.6356 | 0.0016 | 2454115.6974 | 0.0524 | 1.0262 | 0.0153 | 2454115.6975 | 0.0524 | 1.3665 | 0.0049 |
| 2454116.6998 | 0.1259 | 9.6876 | 0.0050 | 2454116.6996 | 0.1259 | 1.0830 | 0.0216 | 2454116.6998 | 0.1259 | 1.4055 | 0.0078 |
| 2454117.6902 | 0.1985 | 9.7413 | 0.0020 | 2454117.6879 | 0.1984 | 1.1289 | 0.0186 | 2454117.6891 | 0.1985 | 1.4548 | 0.0068 |
| 2454124.6677 | -0.2898 | 9.9429 | 0.0073 | 2454124.6675 | -0.2899 | 1.1173 | 0.0140 | 2454124.6677 | -0.2898 | 1.4989 | 0.0100 |
| 2454125.6705 | -0.2163 | 9.8793 | 0.0006 | 2454125.6668 | -0.2166 | 1.1066 | 0.0326 | 2454125.6686 | -0.2164 | 1.4475 | 0.0134 |
| 2454127.6588 | -0.0705 | 9.6470 | 0.0042 | 2454127.6586 | -0.0705 | 1.0109 | 0.0281 | 2454127.6588 | -0.0705 | 1.3467 | 0.0081 |
| 2454128.6606 | 0.0030 | 9.6137 | 0.0034 | 2454128.6604 | 0.0029 | 0.9987 | 0.0140 | 2454128.6606 | 0.0029 | 1.3425 | 0.0071 |
| 2454134.6501 | 0.4421 | 9.9424 | 0.0032 | 2454134.6499 | 0.4421 | 1.3050 | 0.0316 | 2454134.6500 | 0.4421 | 1.5568 | 0.0059 |
| 2454135.6882 | -0.4817 | 9.9745 | 0.0052 | 2454135.6880 | -0.4818 | 1.2415 | 0.0255 | 2454135.6881 | -0.4818 | 1.5530 | 0.0081 |
| 2454136.6511 | -0.4111 | 9.9880 | 0.0047 | 2454136.6509 | -0.4112 | 1.2291 | 0.0186 | 2454136.6510 | -0.4112 | 1.5396 | 0.0058 |
| 2454137.7027 | -0.3340 | 9.9690 | 0.0031 | 2454137.7025 | -0.3341 | 1.1478 | 0.0150 | 2454137.7026 | -0.3340 | 1.5052 | 0.0068 |
| 2454138.6521 | -0.2644 | 9.9355 | 0.0039 | 2454138.6519 | -0.2644 | 1.1499 | 0.0124 | 2454138.6520 | -0.2644 | 1.4710 | 0.0106 |
| 2454140.6473 | -0.1181 | 9.7168 | 0.0040 | 2454140.6472 | -0.1181 | 0.9821 | 0.0167 | 2454140.6473 | -0.1181 | 1.3721 | 0.0107 |
| 2454144.6743 | 0.1772 | 9.7181 | 0.0032 | 2454144.6720 | 0.1770 | 1.0893 | 0.0184 | 2454144.6732 | 0.1771 | 1.4405 | 0.0093 |
| 2454146.6737 | 0.3238 | 9.8536 | 0.0051 | 2454146.6735 | 0.3238 | 1.2135 | 0.0276 | 2454146.6737 | 0.3238 | 1.5144 | 0.0079 |
| 2454147.6709 | 0.3969 | 9.9117 | 0.0034 | 2454147.6718 | 0.3970 | 1.2936 | 0.0141 | 2454147.6709 | 0.3969 | 1.5338 | 0.0064 |
| 2454148.6590 | 0.4694 | 9.9627 | 0.0057 | 2454148.6588 | 0.4693 | 1.2511 | 0.0130 | 2454148.6590 | 0.4693 | 1.5350 | 0.0086 |
| 2454152.6565 | -0.2375 | 9.9000 | 0.0033 | 2454152.6563 | -0.2376 | 1.1102 | 0.0149 | 2454152.6564 | -0.2375 | 1.4515 | 0.0080 |
| 2454153.6452 | -0.1650 | 9.7957 | 0.0026 | 2454153.6450 | -0.1651 | 1.0536 | 0.0231 | 2454153.6451 | -0.1650 | 1.4140 | 0.0078 |
| 2454154.6505 | -0.0913 | 9.6750 | 0.0044 | 2454154.6503 | -0.0913 | 1.0362 | 0.0368 | 2454154.6504 | -0.0913 | 1.3575 | 0.0059 |
| 2454160.6236 | 0.3467 | 9.8674 | 0.0041 | 2454160.6234 | 0.3466 | 1.2250 | 0.0128 | 2454160.6235 | 0.3466 | 1.5364 | 0.0051 |
| 2454166.6268 | -0.2131 | 9.8654 | 0.0039 | 2454166.6267 | -0.2132 | 1.1295 | 0.0295 | 2454166.6268 | -0.2132 | 1.4469 | 0.0085 |
| 2454764.9770 | -0.3387 | 9.9772 | 0.0035 | 2454764.9768 | -0.3387 | 1.1832 | 0.0096 | 2454764.9769 | -0.3387 | 1.5084 | 0.0076 |
| 2454765.9667 | -0.2661 | 9.9349 | 0.0017 | 2454765.9665 | -0.2661 | 1.0784 | 0.0119 | 2454765.9666 | -0.2661 | 1.4761 | 0.0038 |
| 2454766.9550 | -0.1936 | 9.8557 | 0.0053 | 2454766.9558 | -0.1936 | 1.0930 | 0.0128 | 2454766.9549 | -0.1936 | 1.4252 | 0.0128 |
| 2454767.9376 | -0.1216 | 9.7304 | 0.0003 | 2454767.9384 | -0.1215 | 1.0317 | 0.0103 | 2454767.9375 | -0.1216 | 1.3777 | 0.0056 |
| 2454768.9672 | -0.0461 | 9.6223 | 0.0023 | 2454768.9649 | -0.0463 | 1.0014 | 0.0179 | 2454768.9661 | -0.0462 | 1.3505 | 0.0054 |
| 2454769.9714 | 0.0276 | 9.6126 | 0.0053 | 2454769.9723 | 0.0276 | 1.0127 | 0.0094 | 2454769.9714 | 0.0275 | 1.3571 | 0.0059 |
| 2454770.9688 | 0.1007 | 9.6537 | 0.0034 | 2454770.9697 | 0.1007 | 1.0940 | 0.0082 | 2454770.9687 | 0.1007 | 1.3957 | 0.0068 |
| 2454771.9592 | 0.1733 | 9.7275 | 0.0022 | 2454771.9591 | 0.1733 | 1.1244 | 0.0163 | 2454771.9592 | 0.1733 | 1.4348 | 0.0140 |
| 2454773.9404 | 0.3186 | 9.8530 | 0.0016 | 2454773.9402 | 0.3186 | 1.1738 | 0.0150 | 2454773.9404 | 0.3186 | 1.5203 | 0.0057 |
| 2454774.9434 | 0.3921 | 9.9107 | 0.0014 | 2454774.9433 | 0.3921 | 1.2650 | 0.0114 | 2454774.9434 | 0.3921 | 1.5369 | 0.0030 |
| 2454775.9241 | 0.4640 | 9.9653 | 0.0009 | 2454775.9218 | 0.4639 | 1.2309 | 0.0099 | 2454775.9230 | 0.4640 | 1.5518 | 0.0031 |
| 2454778.9197 | -0.3163 | 9.9651 | 0.0018 | 2454778.9174 | -0.3165 | 1.1639 | 0.0064 | 2454778.9186 | -0.3164 | 1.5042 | 0.0046 |
| 2454779.9204 | -0.2429 | 9.9089 | 0.0027 | 2454779.9224 | -0.2428 | 1.1406 | 0.0204 | 2454779.9214 | -0.2429 | 1.4654 | 0.0058 |
| 2454782.8901 | -0.0252 | 9.6163 | 0.0060 | 2454782.8910 | -0.0251 | 0.9382 | 0.0450 | 2454782.8912 | -0.0251 | 1.3562 | 0.0218 |
| 2454794.8954 | -0.1449 | 9.7691 | 0.0013 | 2454794.8932 | -0.1450 | 1.0286 | 0.0077 | 2454794.8943 | -0.1450 | 1.3937 | 0.0049 |
| 2454801.8681 | 0.3664 | 9.8949 | 0.0027 | 2454801.8679 | 0.3664 | 1.2399 | 0.0113 | 2454801.8680 | 0.3664 | 1.5364 | 0.0036 |
| 2454802.8669 | 0.4397 | 9.9467 | 0.0022 | 2454802.8667 | 0.4396 | 1.2308 | 0.0223 | 2454802.8668 | 0.4396 | 1.5527 | 0.0043 |
| 2454803.8751 | -0.4864 | 9.9783 | 0.0029 | 2454803.8750 | -0.4864 | 1.2122 | 0.0164 | 2454803.8751 | -0.4864 | 1.5545 | 0.0044 |
| 2454821.8219 | -0.1705 | 9.8193 | 0.0024 | 2454821.8217 | -0.1705 | 1.0485 | 0.0224 | 2454821.8218 | -0.1705 | 1.4191 | 0.0050 |
| 2454829.8030 | 0.4148 | 9.9318 | 0.0024 | 2454829.8029 | 0.4147 | 1.2542 | 0.0125 | 2454829.8030 | 0.4147 | 1.5540 | 0.0038 |
| 2454830.7925 | 0.4873 | 9.9730 | 0.0019 | 2454830.7923 | 0.4873 | 1.2412 | 0.0203 | 2454830.7924 | 0.4873 | 1.5592 | 0.0058 |
| 2454831.7919 | -0.4394 | 9.9885 | 0.0020 | 2454831.7917 | -0.4394 | 1.2542 | 0.0130 | 2454831.7919 | -0.4394 | 1.5435 | 0.0033 |
| 2454832.7855 | -0.3665 | 9.9845 | 0.0023 | 2454832.7853 | -0.3666 | 1.1996 | 0.0211 | 2454832.7854 | -0.3666 | 1.5259 | 0.0053 |
| 2455097.8466 | 0.0693 | 9.6377 | 0.0118 | 2455097.8464 | 0.0692 | 1.0314 | 0.0323 | 2455097.8465 | 0.0692 | 1.3797 | 0.0119 |
| 2455098.8423 | 0.1423 | 9.6880 | 0.0053 | 2455098.8421 | 0.1422 | 1.0658 | 0.0193 | 2455098.8422 | 0.1422 | 1.4270 | 0.0148 |
| 2455099.8383 | 0.2153 | 9.7526 | 0.0082 | 2455099.8381 | 0.2153 | 1.1475 | 0.0132 | 2455099.8383 | 0.2153 | 1.4702 | 0.0105 |
| 2455100.8413 | 0.2888 | 9.8338 | 0.0078 | 2455100.8421 | 0.2889 | 1.2291 | 0.0230 | 2455100.8412 | 0.2888 | 1.5072 | 0.0132 |
| 2455102.0099 | 0.3745 | 9.9039 | 0.0024 | 2455102.0097 | 0.3745 | 1.2427 | 0.0168 | 2455102.0098 | 0.3745 | 1.5517 | 0.0067 |
| 2455102.8700 | 0.4376 | 9.9409 | 0.0038 | 2455102.8699 | 0.4376 | 1.2757 | 0.0112 | 2455102.8700 | 0.4376 | 1.5558 | 0.0060 |
| 2455106.8238 | -0.2725 | 9.9291 | 0.0071 | 2455131.9464 | -0.4304 | 1.2081 | 0.0119 | 2455106.8238 | -0.2725 | 1.4655 | 0.0104 |
| 2455131.9466 | -0.4303 | 9.9910 | 0.0045 | 2455134.9361 | -0.2111 | 1.1307 | 0.0108 | 2455131.9466 | -0.4303 | 1.5427 | 0.0049 |
| 2455134.9363 | -0.2111 | 9.8733 | 0.0035 | 2455135.7449 | -0.1518 | 1.1058 | 0.0252 | 2455134.9363 | -0.2111 | 1.4452 | 0.0085 |
| 2455135.7441 | -0.1519 | 9.7710 | 0.0057 | 2455136.7441 | -0.0786 | 1.1015 | 0.0142 | 2455135.7440 | -0.1519 | 1.4113 | 0.0113 |
| 2455136.7432 | -0.0786 | 9.6443 | 0.0042 | 2455137.7351 | -0.0059 | 0.9842 | 0.0171 | 2455136.7443 | -0.0786 | 1.3567 | 0.0049 |
| 2455137.7353 | -0.0059 | 9.6135 | 0.0036 | 2455151.6985 | 0.0180 | 1.0486 | 0.0173 | 2455137.7352 | -0.0059 | 1.3475 | 0.0145 |
| 2455151.7009 | 0.0182 | 9.6039 | 0.0051 | 2455152.6959 | 0.0911 | 1.0402 | 0.0071 | 2455151.6997 | 0.0181 | 1.3573 | 0.0072 |
| 2455152.6950 | 0.0911 | 9.6406 | 0.0060 | 2455153.6980 | 0.1646 | 1.1001 | 0.0300 | 2455152.6961 | 0.0911 | 1.4044 | 0.0072 |
| 2455153.6971 | 0.1645 | 9.7100 | 0.0029 | 2455154.6951 | 0.2377 | 1.1485 | 0.0100 | 2455153.6970 | 0.1645 | 1.4439 | 0.0030 |
| 2455154.6963 | 0.2378 | 9.7770 | 0.0040 | 2455155.6959 | 0.3111 | 1.2011 | 0.0186 | 2455154.6963 | 0.2378 | 1.4796 | 0.0067 |
| 2455155.6983 | 0.3113 | 9.8393 | 0.0056 | 2455156.6853 | 0.3836 | 1.2384 | 0.0082 | 2455155.6972 | 0.3112 | 1.5282 | 0.0116 |
| 2455156.6855 | 0.3837 | 9.9127 | 0.0038 | 2455581.7189 | -0.4505 | 1.1300 | 0.0117 | 2455156.6854 | 0.3836 | 1.5416 | 0.0061 |
| 2455581.7191 | -0.4504 | 9.9912 | 0.0024 | 2455584.7110 | -0.2311 | 1.1167 | 0.0203 | 2455581.7190 | -0.4504 | 1.5528 | 0.0111 |
| 2455584.7112 | -0.2310 | 9.9034 | 0.0034 | 2455598.6310 | -0.2104 | 1.1232 | 0.0080 | 2455584.7111 | -0.2310 | 1.4518 | 0.0076 |
| 2455598.6312 | -0.2103 | 9.8796 | 0.0033 | 2455599.6179 | -0.1380 | 1.0702 | 0.0090 | 2455598.6311 | -0.2104 | 1.4501 | 0.0061 |
| 2455599.6159 | -0.1381 | 9.7658 | 0.0038 | 2455603.6230 | 0.1557 | 1.1046 | 0.0106 | 2455599.6169 | -0.1381 | 1.4010 | 0.0054 |
| 2455603.6232 | 0.1557 | 9.7270 | 0.0055 | 2455604.6315 | 0.2296 | 1.1884 | 0.0134 | 2455603.6231 | 0.1557 | 1.4399 | 0.0070 |
| 2455604.6306 | 0.2296 | 9.7911 | 0.0045 | 2455605.6153 | 0.3018 | 1.2555 | 0.0100 | 2455604.6306 | 0.2296 | 1.4737 | 0.0059 |
| 2455605.6155 | 0.3018 | 9.8566 | 0.0039 | 2455607.6130 | 0.4483 | 1.3417 | 0.0105 | 2455605.6154 | 0.3018 | 1.5145 | 0.0056 |
| 2455607.6121 | 0.4482 | 9.9614 | 0.0028 | 2455608.6179 | -0.4781 | 1.3577 | 0.0101 | 2455607.6120 | 0.4482 | 1.5676 | 0.0050 |
| 2455608.6170 | -0.4781 | 9.9849 | 0.0027 | 2455900.8705 | -0.0484 | 0.9982 | 0.0161 | 2455608.6170 | -0.4781 | 1.5556 | 0.0072 |
| 2455900.8707 | -0.0484 | 9.6300 | 0.0066 | 2455911.8321 | -0.2447 | 1.1428 | 0.0127 | 2455900.8706 | -0.0484 | 1.3433 | 0.0131 |
| 2455911.8301 | -0.2448 | 9.9067 | 0.0030 | 2455921.7965 | 0.4860 | 1.2880 | 0.0207 | 2455911.8311 | -0.2447 | 1.4707 | 0.0039 |
| 2455921.7978 | 0.4861 | 9.9890 | 0.0033 | 2455922.7892 | -0.4412 | 1.2577 | 0.0412 | 2455921.7977 | 0.4861 | 1.5645 | 0.0041 |
| 2455922.7894 | -0.4412 | 9.9967 | 0.0021 | 2455924.8143 | -0.2927 | 1.1575 | 0.0437 | 2455922.7893 | -0.4412 | 1.5649 | 0.0052 |
| 2455924.8134 | -0.2928 | 9.9506 | 0.0020 | 2455925.7848 | -0.2216 | 1.0915 | 0.0188 | 2455924.8145 | -0.2927 | 1.5068 | 0.0046 |
| 2455925.7839 | -0.2216 | 9.8969 | 0.0032 | 2455926.7808 | -0.1485 | 1.0995 | 0.0108 | 2455925.7849 | -0.2216 | 1.4521 | 0.0059 |
| 2455926.7799 | -0.1486 | 9.7802 | 0.0038 | 2455927.7764 | -0.0755 | 0.9817 | 0.0397 | 2455926.7798 | -0.1486 | 1.4022 | 0.0095 |
| 2455927.7765 | -0.0755 | 9.6663 | 0.0022 | | | | | 2455927.7765 | -0.0755 | 1.3540 | 0.0083 |

**Photometry of SZ Cas, cont…**

| HJD | Phase | V-R | V-R err | HJD | Phase | R-I | R-I err |
|---|---|---|---|---|---|---|---|
| 2454075.7802 | 0.1254 | 1.2295 | 0.0066 | 2454075.7803 | 0.1255 | 0.9102 | 0.0089 |
| 2454076.5893 | 0.1848 | 1.2552 | 0.0069 | 2454076.5884 | 0.1847 | 0.9276 | 0.0104 |
| 2454077.7808 | 0.2721 | 1.2724 | 0.0043 | 2454077.7809 | 0.2722 | 0.9385 | 0.0032 |
| 2454080.5876 | 0.4780 | 1.3124 | 0.0053 | 2454080.5877 | 0.4780 | 0.9619 | 0.0049 |
| 2454081.7702 | -0.4353 | 1.3206 | 0.0053 | 2454081.7703 | -0.4353 | 0.9542 | 0.0064 |
| 2454083.7646 | -0.2891 | 1.2853 | 0.0037 | 2454083.7647 | -0.2891 | 0.9346 | 0.0045 |
| 2454085.7763 | -0.1416 | 1.2301 | 0.0086 | 2454085.7765 | -0.1416 | 0.8997 | 0.0080 |
| 2454086.7810 | -0.0679 | 1.1917 | 0.0042 | 2454086.7811 | -0.0679 | 0.8850 | 0.0060 |
| 2454088.7587 | 0.0771 | 1.2110 | 0.0054 | 2454088.7599 | 0.0772 | 0.9006 | 0.0081 |
| 2454090.7653 | 0.2242 | 1.2666 | 0.0045 | 2454090.7654 | 0.2243 | 0.9293 | 0.0039 |
| 2454094.7579 | -0.4830 | 1.3008 | 0.0033 | 2454094.7580 | -0.4830 | 0.9729 | 0.0031 |
| 2454100.7468 | -0.0439 | 1.2075 | 0.0036 | 2454100.7469 | -0.0438 | 0.8778 | 0.0059 |
| 2454101.6363 | 0.0214 | 1.2077 | 0.0044 | 2454101.6364 | 0.0214 | 0.8905 | 0.0049 |
| 2454102.7394 | 0.1023 | 1.2280 | 0.0103 | 2454102.7396 | 0.1023 | 0.8988 | 0.0072 |
| 2454103.7248 | 0.1745 | 1.2512 | 0.0061 | 2454103.7249 | 0.1745 | 0.9154 | 0.0087 |
| 2454104.7220 | 0.2476 | 1.2705 | 0.0046 | 2454104.7221 | 0.2476 | 0.9368 | 0.0058 |
| 2454105.7225 | 0.3210 | 1.3135 | 0.0064 | 2454105.7226 | 0.3210 | 0.9309 | 0.0063 |
| 2454107.7137 | 0.4670 | 1.3221 | 0.0034 | 2454107.7139 | 0.4670 | 0.9480 | 0.0074 |
| 2454108.6998 | -0.4607 | 1.3256 | 0.0032 | 2454108.6999 | -0.4607 | 0.9487 | 0.0034 |
| 2454115.6976 | 0.0524 | 1.2173 | 0.0018 | 2454115.6977 | 0.0524 | 0.8897 | 0.0026 |
| 2454116.6999 | 0.1259 | 1.2381 | 0.0060 | 2454116.7000 | 0.1259 | 0.9097 | 0.0053 |
| 2454117.6903 | 0.1985 | 1.2545 | 0.0040 | 2454117.6893 | 0.1985 | 0.9237 | 0.0069 |
| 2454124.6696 | -0.2897 | 1.2878 | 0.0078 | 2454124.6697 | -0.2897 | 0.9282 | 0.0083 |
| 2454125.6688 | -0.2164 | 1.2548 | 0.0030 | 2454125.6671 | -0.2166 | 0.9239 | 0.0110 |
| 2454127.6589 | -0.0705 | 1.2032 | 0.0042 | 2454127.6590 | -0.0705 | 0.8860 | 0.0032 |
| 2454128.6607 | 0.0030 | 1.1998 | 0.0062 | 2454128.6608 | 0.0030 | 0.8918 | 0.0063 |
| 2454134.6501 | 0.4421 | 1.3212 | 0.0036 | 2454134.6502 | 0.4421 | 0.9579 | 0.0041 |
| 2454135.6882 | -0.4818 | 1.3170 | 0.0060 | 2454135.6884 | -0.4817 | 0.9583 | 0.0042 |
| 2454136.6511 | -0.4111 | 1.3119 | 0.0051 | 2454136.6513 | -0.4111 | 0.9544 | 0.0042 |
| 2454137.7028 | -0.3340 | 1.3012 | 0.0032 | 2454137.7029 | -0.3340 | 0.9460 | 0.0019 |
| 2454138.6521 | -0.2644 | 1.2857 | 0.0042 | 2454138.6533 | -0.2643 | 0.9413 | 0.0098 |
| 2454140.6474 | -0.1181 | 1.2127 | 0.0050 | 2454140.6475 | -0.1181 | 0.9005 | 0.0059 |
| 2454144.6733 | 0.1771 | 1.2508 | 0.0045 | 2454144.6724 | 0.1770 | 0.9150 | 0.0051 |
| 2454146.6756 | 0.3239 | 1.3024 | 0.0060 | 2454146.6757 | 0.3239 | 0.9387 | 0.0080 |
| 2454147.6710 | 0.3969 | 1.3196 | 0.0035 | 2454147.6711 | 0.3969 | 0.9433 | 0.0036 |
| 2454148.6591 | 0.4693 | 1.3279 | 0.0079 | 2454148.6592 | 0.4694 | 0.9603 | 0.0073 |
| 2454152.6566 | -0.2375 | 1.2797 | 0.0037 | 2454152.6567 | -0.2375 | 0.9237 | 0.0047 |
| 2454153.6453 | -0.1650 | 1.2401 | 0.0037 | 2454153.6454 | -0.1650 | 0.9100 | 0.0029 |
| 2454154.6506 | -0.0913 | 1.2060 | 0.0073 | 2454154.6507 | -0.0913 | 0.8882 | 0.0063 |
| 2454160.6237 | 0.3467 | 1.3028 | 0.0045 | 2454160.6238 | 0.3467 | 0.9404 | 0.0025 |
| 2454166.6269 | -0.2131 | 1.2599 | 0.0052 | 2454166.6270 | -0.2131 | 0.9190 | 0.0050 |
| 2454764.9771 | -0.3387 | 1.2944 | 0.0040 | 2454764.9772 | -0.3387 | 0.9556 | 0.0032 |
| 2454765.9678 | -0.2660 | 1.2913 | 0.0026 | 2454765.9679 | -0.2660 | 0.9305 | 0.0035 |
| 2454766.9561 | -0.1936 | 1.2538 | 0.0058 | 2454766.9562 | -0.1935 | 0.9262 | 0.0074 |
| 2454767.9376 | -0.1216 | 1.2191 | 0.0030 | 2454767.9378 | -0.1216 | 0.9032 | 0.0048 |
| 2454768.9662 | -0.0462 | 1.1867 | 0.0068 | 2454768.9653 | -0.0462 | 0.8879 | 0.0087 |
| 2454769.9715 | 0.0276 | 1.2044 | 0.0061 | 2454769.9716 | 0.0276 | 0.8921 | 0.0038 |
| 2454770.9689 | 0.1007 | 1.2156 | 0.0060 | 2454770.9690 | 0.1007 | 0.9053 | 0.0076 |
| 2454771.9593 | 0.1733 | 1.2484 | 0.0030 | 2454771.9594 | 0.1733 | 0.9272 | 0.0033 |
| 2454773.9405 | 0.3186 | 1.2963 | 0.0058 | 2454773.9406 | 0.3186 | 0.9480 | 0.0061 |
| 2454774.9435 | 0.3921 | 1.2993 | 0.0017 | 2454774.9436 | 0.3921 | 0.9615 | 0.0067 |
| 2454775.9231 | 0.4640 | 1.3179 | 0.0013 | 2454775.9222 | 0.4639 | 0.9710 | 0.0017 |
| 2454778.9198 | -0.3163 | 1.2973 | 0.0020 | 2454778.9188 | -0.3164 | 0.9413 | 0.0028 |
| 2454779.9205 | -0.2429 | 1.2691 | 0.0027 | 2454779.9206 | -0.2429 | 0.9293 | 0.0031 |
| 2454782.8913 | -0.0251 | 1.1755 | 0.0092 | 2454782.8914 | -0.0251 | 0.8967 | 0.0120 |
| 2454794.8955 | -0.1449 | 1.2292 | 0.0030 | 2454794.8946 | -0.1449 | 0.9073 | 0.0052 |
| 2454801.8682 | 0.3664 | 1.3036 | 0.0038 | 2454801.8683 | 0.3664 | 0.9588 | 0.0038 |
| 2454802.8670 | 0.4396 | 1.3177 | 0.0026 | 2454802.8671 | 0.4396 | 0.9618 | 0.0026 |
| 2454803.8752 | -0.4864 | 1.3144 | 0.0034 | 2454803.8753 | -0.4864 | 0.9634 | 0.0029 |
| 2454821.8219 | -0.1705 | 1.2464 | 0.0026 | 2454821.8221 | -0.1705 | 0.9165 | 0.0066 |
| 2454829.8031 | 0.4148 | 1.3196 | 0.0030 | 2454829.8032 | 0.4148 | 0.9497 | 0.0031 |
| 2454830.7925 | 0.4873 | 1.3163 | 0.0032 | 2454830.7926 | 0.4873 | 0.9629 | 0.0050 |
| 2454831.7920 | -0.4394 | 1.3155 | 0.0021 | 2454831.7921 | -0.4394 | 0.9608 | 0.0030 |
| 2454832.7855 | -0.3666 | 1.3076 | 0.0035 | 2454832.7857 | -0.3665 | 0.9550 | 0.0044 |
| 2455097.8467 | 0.0693 | 1.2225 | 0.0170 | 2455097.8468 | 0.0693 | 0.9087 | 0.0145 |
| 2455098.8423 | 0.1423 | 1.2363 | 0.0127 | 2455098.8425 | 0.1423 | 0.9190 | 0.0153 |
| 2455099.8384 | 0.2153 | 1.2707 | 0.0136 | 2455099.8385 | 0.2153 | 0.9272 | 0.0129 |
| 2455100.8413 | 0.2888 | 1.2924 | 0.0095 | 2455100.8414 | 0.2888 | 0.9504 | 0.0080 |
| 2455102.0089 | 0.3744 | 1.3079 | 0.0076 | 2455102.0079 | 0.3744 | 0.9646 | 0.0089 |
| 2455102.8701 | 0.4376 | 1.3216 | 0.0068 | 2455102.8702 | 0.4376 | 0.9689 | 0.0078 |
| 2455106.8239 | -0.2725 | 1.2756 | 0.0103 | 2455106.8240 | -0.2725 | 0.9467 | 0.0103 |
| 2455131.9467 | -0.4303 | 1.3131 | 0.0072 | 2455131.9468 | -0.4303 | 0.9606 | 0.0069 |
| 2455134.9364 | -0.2111 | 1.2624 | 0.0038 | 2455134.9365 | -0.2111 | 0.9257 | 0.0037 |
| 2455135.7441 | -0.1519 | 1.2264 | 0.0090 | 2455135.7442 | -0.1519 | 0.9181 | 0.0073 |
| 2455136.7433 | -0.0786 | 1.1993 | 0.0093 | 2455136.7434 | -0.0786 | 0.8989 | 0.0132 |
| 2455137.7353 | -0.0059 | 1.2086 | 0.0062 | 2455137.7355 | -0.0059 | 0.8952 | 0.0101 |
| 2455151.7009 | 0.0182 | 1.1888 | 0.0123 | 2455151.7000 | 0.0181 | 0.8915 | 0.0148 |
| 2455152.6951 | 0.0911 | 1.2114 | 0.0076 | 2455152.6952 | 0.0911 | 0.9130 | 0.0063 |
| 2455153.6961 | 0.1645 | 1.2548 | 0.0048 | 2455153.6952 | 0.1644 | 0.9203 | 0.0042 |
| 2455154.6953 | 0.2377 | 1.2809 | 0.0098 | 2455154.6944 | 0.2377 | 0.9398 | 0.0098 |
| 2455155.6984 | 0.3113 | 1.2997 | 0.0103 | 2455155.6985 | 0.3113 | 0.9540 | 0.0130 |
| 2455156.6855 | 0.3837 | 1.3095 | 0.0085 | 2455156.6857 | 0.3837 | 0.9571 | 0.0113 |
| 2455581.7191 | -0.4504 | 1.3180 | 0.0035 | 2455581.7193 | -0.4504 | 0.9738 | 0.0042 |
| 2455584.7112 | -0.2310 | 1.2783 | 0.0065 | 2455584.7114 | -0.2310 | 0.9378 | 0.0069 |
| 2455598.6312 | -0.2103 | 1.2553 | 0.0038 | 2455598.6303 | -0.2104 | 0.9422 | 0.0029 |
| 2455599.6171 | -0.1381 | 1.2206 | 0.0070 | 2455599.6172 | -0.1380 | 0.9260 | 0.0063 |
| 2455603.6232 | 0.1557 | 1.2496 | 0.0057 | 2455603.6234 | 0.1557 | 0.9314 | 0.0095 |
| 2455604.6307 | 0.2296 | 1.2672 | 0.0061 | 2455604.6308 | 0.2296 | 0.9452 | 0.0056 |
| 2455605.6156 | 0.3018 | 1.2947 | 0.0044 | 2455605.6157 | 0.3018 | 0.9631 | 0.0032 |
| 2455607.6132 | 0.4483 | 1.3254 | 0.0029 | 2455607.6133 | 0.4483 | 0.9719 | 0.0021 |
| 2455608.6171 | -0.4781 | 1.3150 | 0.0045 | 2455608.6172 | -0.4781 | 0.9755 | 0.0044 |
| 2455900.8707 | -0.0484 | 1.2060 | 0.0066 | 2455900.8709 | -0.0484 | 0.9042 | 0.0022 |
| 2455911.8301 | -0.2448 | 1.2708 | 0.0039 | 2455911.8303 | -0.2448 | 0.9459 | 0.0041 |
| 2455921.7978 | 0.4861 | 1.3245 | 0.0034 | 2455921.7969 | 0.4860 | 0.9839 | 0.0038 |
| 2455922.7895 | -0.4412 | 1.3182 | 0.0025 | 2455922.7896 | -0.4412 | 0.9779 | 0.0045 |
| 2455924.8146 | -0.2927 | 1.2930 | 0.0031 | 2455924.8158 | -0.2926 | 0.9509 | 0.0040 |
| 2455925.7839 | -0.2216 | 1.2693 | 0.0041 | 2455925.7840 | -0.2216 | 0.9455 | 0.0041 |
| 2455926.7799 | -0.1486 | 1.2281 | 0.0044 | 2455926.7801 | -0.1486 | 0.9286 | 0.0033 |
| 2455927.7766 | -0.0755 | 1.2076 | 0.0092 | 2455927.7767 | -0.0755 | 0.9095 | 0.0098 |

**Photometry of SZ Tau**

| HJD | Phase | V-mag | V-err | HJD | Phase | b-y | b-y err |
|---|---|---|---|---|---|---|---|
| 2453415.6326 | 0.4852 | 6.6865 | 0.0029 | 2453415.6325 | 0.4852 | 0.6462 | 0.0038 |
| 2453415.7360 | -0.4820 | 6.7009 | 0.0038 | 2453415.7359 | -0.4820 | 0.6357 | 0.0045 |
| 2453432.6914 | -0.0973 | 6.3786 | 0.0043 | 2453432.6914 | -0.0973 | 0.5278 | 0.0064 |
| 2453433.6865 | 0.2187 | 6.4925 | 0.0044 | 2453433.6865 | 0.2187 | 0.5845 | 0.0053 |
| 2453437.6724 | 0.4845 | 6.7001 | 0.0072 | 2453437.6723 | 0.4845 | 0.6477 | 0.0081 |
| 2453438.6689 | -0.1990 | 6.4760 | 0.0061 | 2453438.6689 | -0.1990 | 0.5629 | 0.0108 |
| 2453439.6657 | 0.1176 | 6.3922 | 0.0052 | 2453439.6656 | 0.1175 | 0.5586 | 0.0082 |
| 2453442.6592 | 0.0682 | 6.3661 | 0.0063 | 2453442.6591 | 0.0682 | 0.5399 | 0.0100 |
| 2453444.6523 | -0.2988 | 6.5952 | 0.0047 | 2453444.6522 | -0.2988 | 0.5882 | 0.0075 |
| 2453445.6511 | 0.0184 | 6.3479 | 0.0064 | 2453445.6510 | 0.0184 | 0.5330 | 0.0081 |
| 2453446.6432 | 0.3335 | 6.6051 | 0.0058 | 2453446.6432 | 0.3335 | 0.6194 | 0.0071 |
| 2453451.6333 | -0.0818 | 6.3675 | 0.0073 | 2453451.6332 | -0.0818 | 0.5377 | 0.0081 |
| 2453452.6335 | 0.2359 | 6.5085 | 0.0038 | 2453452.6335 | 0.2358 | 0.5896 | 0.0059 |
| 2453696.8202 | -0.2160 | 6.4917 | 0.0008 | 2453696.8202 | -0.2160 | 0.5529 | 0.0063 |
| 2453700.8185 | 0.0537 | 6.3541 | 0.0057 | 2453700.8184 | 0.0537 | 0.5338 | 0.0087 |
| 2453708.9033 | -0.3787 | 6.6695 | 0.0040 | 2453708.9032 | -0.3788 | 0.6032 | 0.0059 |
| 2453721.8860 | -0.2557 | 6.5245 | 0.0037 | 2453721.8859 | -0.2557 | 0.5669 | 0.0055 |
| 2453724.8611 | -0.3109 | 6.5967 | 0.0028 | 2453724.8610 | -0.3109 | 0.5934 | 0.0052 |
| 2453734.8014 | -0.1541 | 6.4094 | 0.0085 | 2453734.8014 | -0.1541 | 0.5353 | 0.0103 |
| 2453735.8035 | 0.1642 | 6.4366 | 0.0031 | 2453735.8034 | 0.1641 | 0.5621 | 0.0042 |
| 2453753.8014 | -0.1201 | 6.3931 | 0.0044 | 2453753.8013 | -0.1201 | 0.5311 | 0.0062 |
| 2453756.7866 | -0.1721 | 6.4399 | 0.0042 | 2453756.7866 | -0.1721 | 0.5460 | 0.0046 |
| 2453758.7836 | 0.4621 | 6.6947 | 0.0027 | 2453758.7835 | 0.4621 | 0.6311 | 0.0031 |
| 2453760.7618 | 0.0904 | 6.3696 | 0.0050 | 2453760.7618 | 0.0904 | 0.5411 | 0.0080 |
| 2453765.7714 | -0.3187 | 6.6145 | 0.0236 | 2453765.7714 | -0.3187 | 0.5974 | 0.0317 |
| 2453766.7650 | -0.0032 | 6.3523 | 0.0195 | 2453766.7650 | -0.0032 | 0.5269 | 0.0261 |
| 2453767.7605 | 0.3130 | 6.6030 | 0.0089 | 2453767.7604 | 0.3130 | 0.6127 | 0.0135 |
| 2453768.7543 | -0.3714 | 6.6714 | 0.0088 | 2453768.7543 | -0.3714 | 0.6061 | 0.0099 |
| 2453782.7068 | 0.0596 | 6.3797 | 0.0263 | 2453782.7046 | 0.0589 | 0.5156 | 0.0316 |
| 2453808.6491 | 0.2983 | 6.6020 | 0.0155 | 2453808.6490 | 0.2982 | 0.6027 | 0.0182 |
| 2453812.6387 | -0.4347 | 6.7158 | 0.0133 | 2453812.6386 | -0.4348 | 0.6209 | 0.0182 |
| 2454396.9625 | 0.1332 | 6.3987 | 0.0045 | 2454396.9625 | 0.1332 | 0.5616 | 0.0101 |
| 2454400.0371 | 0.1096 | 6.3923 | 0.0045 | 2454400.0370 | 0.1096 | 0.5479 | 0.0058 |
| 2454400.9371 | 0.3954 | 6.6462 | 0.0049 | 2454400.9370 | 0.3954 | 0.6328 | 0.0059 |
| 2454401.9561 | -0.2810 | 6.5675 | 0.0038 | 2454401.9560 | -0.2810 | 0.5764 | 0.0054 |
| 2454403.9495 | 0.3521 | 6.6217 | 0.0022 | 2454403.9494 | 0.3521 | 0.6216 | 0.0037 |
| 2454405.9981 | 0.0027 | 6.3409 | 0.0091 | 2454405.9980 | 0.0026 | 0.5224 | 0.0100 |
| 2454407.9825 | -0.3671 | 6.6564 | 0.0057 | 2454407.9824 | -0.3672 | 0.6068 | 0.0058 |
| 2454409.9745 | 0.2655 | 6.5359 | 0.0045 | 2454409.9745 | 0.2655 | 0.5960 | 0.0060 |
| 2454412.9409 | 0.2076 | 6.4737 | 0.0024 | 2454412.9408 | 0.2075 | 0.5816 | 0.0052 |
| 2454413.9908 | -0.4590 | 6.6973 | 0.0032 | 2454413.9908 | -0.4590 | 0.6323 | 0.0044 |
| 2454414.8294 | -0.1927 | 6.4564 | 0.0034 | 2454414.8293 | -0.1927 | 0.5499 | 0.0036 |
| 2454415.9433 | 0.1610 | 6.4346 | 0.0046 | 2454415.9433 | 0.1610 | 0.5549 | 0.0093 |
| 2454417.9842 | -0.1908 | 6.4472 | 0.0059 | 2454417.9841 | -0.1908 | 0.5414 | 0.0078 |
| 2454418.9805 | 0.1256 | 6.4017 | 0.0132 | 2454418.9805 | 0.1256 | 0.5437 | 0.0192 |
| 2454422.9781 | 0.3951 | 6.6656 | 0.0125 | 2454422.9781 | 0.3951 | 0.6193 | 0.0139 |
| 2454423.9740 | -0.2886 | 6.5665 | 0.0089 | 2454423.9739 | -0.2886 | 0.5850 | 0.0145 |
| 2454425.8896 | 0.3198 | 6.5932 | 0.0016 | 2454425.8895 | 0.3197 | 0.6044 | 0.0031 |
| 2454456.8777 | 0.1609 | 6.4465 | 0.0061 | 2454456.8777 | 0.1609 | 0.5558 | 0.0079 |
| 2454458.8531 | -0.2118 | 6.4622 | 0.0114 | 2454458.8530 | -0.2118 | 0.5522 | 0.0163 |
| 2454459.8457 | 0.1034 | 6.3785 | 0.0159 | 2454459.8456 | 0.1034 | 0.5426 | 0.0216 |
| 2454460.8369 | 0.4182 | 6.6883 | 0.0106 | 2454460.8368 | 0.4182 | 0.6236 | 0.0119 |
| 2454461.8643 | -0.2555 | 6.5313 | 0.0082 | 2454461.8642 | -0.2555 | 0.5554 | 0.0108 |
| 2454464.8621 | -0.3035 | 6.6140 | 0.0171 | 2454464.8621 | -0.3035 | 0.5810 | 0.0224 |
| 2454465.8611 | 0.0138 | 6.3469 | 0.0196 | 2454465.8610 | 0.0138 | 0.5192 | 0.0278 |
| 2454466.6580 | 0.2669 | 6.5498 | 0.0197 | 2454466.6580 | 0.2669 | 0.5894 | 0.0254 |
| 2454475.7866 | 0.1659 | 6.4314 | 0.0062 | 2454475.7865 | 0.1659 | 0.5640 | 0.0085 |
| 2454476.7773 | 0.4805 | 6.6977 | 0.0052 | 2454476.7772 | 0.4805 | 0.6301 | 0.0089 |
| 2454479.7751 | 0.4326 | 6.6778 | 0.0070 | 2454479.7750 | 0.4325 | 0.6252 | 0.0091 |
| 2454480.7635 | -0.2535 | 6.5188 | 0.0065 | 2454480.7634 | -0.2536 | 0.5655 | 0.0118 |
| 2454481.7662 | 0.0649 | 6.3614 | 0.0137 | 2454481.7661 | 0.0649 | 0.5282 | 0.0181 |
| 2454482.7604 | 0.3806 | 6.6490 | 0.0109 | 2454482.7603 | 0.3806 | 0.6215 | 0.0132 |
| 2454483.7644 | -0.3005 | 6.5837 | 0.0091 | 2454483.7643 | -0.3006 | 0.5810 | 0.0137 |
| 2454486.7468 | -0.3534 | 6.6345 | 0.0091 | 2454484.7534 | 0.0136 | 0.5308 | 0.0111 |
| 2454494.7366 | 0.1840 | 6.4533 | 0.0146 | 2454486.7467 | -0.3534 | 0.6019 | 0.0146 |
| 2454495.7622 | -0.4903 | 6.7171 | 0.0087 | 2454494.7365 | 0.1840 | 0.5783 | 0.0175 |
| 2454496.7734 | -0.1692 | 6.4372 | 0.0117 | 2454495.7622 | -0.4903 | 0.6238 | 0.0124 |
| 2454498.7373 | 0.4545 | 6.7006 | 0.0096 | 2454496.7733 | -0.1692 | 0.5321 | 0.0151 |
| 2454499.7294 | -0.2304 | 6.5084 | 0.0133 | 2454498.7372 | 0.4545 | 0.6258 | 0.0115 |
| 2454503.7197 | 0.0368 | 6.3521 | 0.0182 | 2454499.7293 | -0.2304 | 0.5547 | 0.0191 |
| 2454504.7031 | 0.3491 | 6.6205 | 0.0123 | 2454503.7217 | 0.0375 | 0.5253 | 0.0262 |
| 2454505.7264 | -0.3259 | 6.6180 | 0.0088 | 2454504.7030 | 0.3491 | 0.6183 | 0.0166 |
| 2454508.7052 | -0.3799 | 6.6852 | 0.0129 | 2454505.7264 | -0.3259 | 0.5873 | 0.0098 |
| 2454509.6379 | -0.0837 | 6.3810 | 0.0170 | 2454508.7052 | -0.3799 | 0.5935 | 0.0135 |
| 2454821.7261 | 0.0284 | 6.3536 | 0.0054 | 2454509.6378 | -0.0837 | 0.5287 | 0.0203 |
| 2454821.8608 | 0.0712 | 6.3585 | 0.0055 | 2454821.7260 | 0.0284 | 0.5274 | 0.0079 |
| 2454822.7074 | 0.3400 | 6.6191 | 0.0049 | 2454821.8607 | 0.0711 | 0.5391 | 0.0057 |
| 2454829.7078 | -0.4368 | 6.7114 | 0.0031 | 2454822.7074 | 0.3400 | 0.6152 | 0.0063 |
| 2454829.8500 | -0.3916 | 6.6925 | 0.0051 | 2454829.7078 | -0.4368 | 0.6352 | 0.0034 |
| 2454830.6982 | -0.1223 | 6.3912 | 0.0086 | 2454829.8499 | -0.3917 | 0.6175 | 0.0053 |
| 2454831.7021 | 0.1966 | 6.4692 | 0.0066 | 2454830.6982 | -0.1223 | 0.5287 | 0.0092 |
| 2454832.8252 | -0.4468 | 6.7095 | 0.0021 | 2454831.7021 | 0.1965 | 0.5700 | 0.0103 |
| 2454838.6990 | 0.4186 | 6.6626 | 0.0146 | 2454832.8252 | -0.4468 | 0.6397 | 0.0037 |
| 2454838.8025 | 0.4515 | 6.6907 | 0.0020 | 2454838.6989 | 0.4186 | 0.6410 | 0.0186 |
| 2454839.6749 | -0.2715 | 6.5423 | 0.0036 | 2454838.8025 | 0.4515 | 0.6245 | 0.0049 |
| 2454839.8002 | -0.2317 | 6.4983 | 0.0062 | 2454839.6748 | -0.2715 | 0.5757 | 0.0049 |
| 2454840.6760 | 0.0465 | 6.3471 | 0.0089 | 2454839.7989 | -0.2321 | 0.5619 | 0.0066 |
| 2454842.6770 | -0.3181 | 6.5992 | 0.0080 | 2454840.6759 | 0.0464 | 0.5308 | 0.0090 |
| 2454844.6764 | 0.3169 | 6.6035 | 0.0037 | 2454842.6770 | -0.3181 | 0.5956 | 0.0081 |
| 2454844.7861 | 0.3517 | 6.6249 | 0.0085 | 2454844.6763 | 0.3169 | 0.6131 | 0.0044 |
| 2454846.7920 | -0.0112 | 6.3419 | 0.0080 | 2454844.7861 | 0.3517 | 0.6289 | 0.0091 |
| 2454847.7780 | 0.3019 | 6.5774 | 0.0072 | 2454846.7919 | -0.0113 | 0.5252 | 0.0144 |
| 2454856.7780 | 0.1601 | 6.4342 | 0.0045 | 2454847.7779 | 0.3018 | 0.6113 | 0.0091 |
| 2454865.7522 | 0.0101 | 6.3546 | 0.0053 | 2454856.7780 | 0.1601 | 0.5618 | 0.0073 |
| 2454867.7471 | -0.3564 | 6.6404 | 0.0034 | 2454865.7522 | 0.0101 | 0.5209 | 0.0085 |
| 2454873.7360 | -0.4544 | 6.7113 | 0.0032 | 2454867.7471 | -0.3564 | 0.6038 | 0.0038 |
| 2454875.7292 | 0.1786 | 6.4518 | 0.0043 | 2454873.7360 | -0.4545 | 0.6383 | 0.0041 |
| 2455213.7664 | -0.4686 | 6.7097 | 0.0023 | 2454875.7291 | 0.1785 | 0.5681 | 0.0053 |
| 2455221.6137 | 0.0236 | 6.3462 | 0.0074 | 2455213.7663 | -0.4686 | 0.6304 | 0.0039 |
| 2455221.7454 | 0.0654 | 6.3577 | 0.0069 | 2455221.6136 | 0.0235 | 0.5185 | 0.0074 |
| 2455222.6172 | 0.3423 | 6.5949 | 0.0028 | 2455221.7441 | 0.0650 | 0.5450 | 0.0148 |
| 2455236.5997 | -0.2172 | 6.4874 | 0.0028 | 2455222.6172 | 0.3423 | 0.6250 | 0.0056 |
| 2455236.7093 | -0.1824 | 6.4513 | 0.0086 | 2455236.5996 | -0.2173 | 0.5609 | 0.0050 |
| 2455240.7001 | 0.0850 | 6.3773 | 0.0093 | 2455236.7093 | -0.1824 | 0.5446 | 0.0095 |
| 2455241.6910 | 0.3997 | 6.6651 | 0.0059 | 2455240.7001 | 0.0850 | 0.5382 | 0.0130 |
| 2455243.7210 | 0.0443 | 6.3504 | 0.0013 | 2455241.6909 | 0.3996 | 0.6232 | 0.0069 |
| 2455253.6765 | 0.2060 | 6.4893 | 0.0087 | 2455243.7209 | 0.0443 | 0.5291 | 0.0066 |
| 2455258.6939 | -0.2006 | 6.4597 | 0.0067 | 2455253.6765 | 0.2060 | 0.5773 | 0.0121 |
| 2455259.6738 | 0.1106 | 6.3833 | 0.0065 | 2455258.6939 | -0.2006 | 0.5520 | 0.0107 |
| 2455268.6784 | -0.0298 | 6.3340 | 0.0063 | 2455259.6737 | 0.1105 | 0.5547 | 0.0100 |
| 2455270.6585 | -0.4009 | 6.6783 | 0.0050 | 2455268.6771 | -0.0302 | 0.5285 | 0.0134 |
| 2455271.6707 | -0.0795 | 6.3546 | 0.0115 | 2455270.6584 | -0.4010 | 0.6146 | 0.0074 |
| 2455273.6607 | -0.4475 | 6.7071 | 0.0052 | 2455271.6694 | -0.0799 | 0.5254 | 0.0148 |
| 2455274.6539 | -0.1321 | 6.3898 | 0.0037 | 2455273.6607 | -0.4475 | 0.6288 | 0.0068 |
| 2455275.6540 | 0.1855 | 6.4557 | 0.0094 | 2455274.6526 | -0.1325 | 0.5294 | 0.0089 |
| 2455580.5975 | 0.0286 | 6.3546 | 0.0040 | 2455275.6527 | 0.1851 | 0.5732 | 0.0122 |
| 2455580.7695 | 0.0833 | 6.3726 | 0.0035 | 2455580.5974 | 0.0286 | 0.5277 | 0.0055 |
| 2455581.5900 | 0.3438 | 6.6172 | 0.0051 | 2455580.7682 | 0.0828 | 0.5395 | 0.0097 |
| 2455581.7675 | 0.4002 | 6.6626 | 0.0050 | 2455581.5912 | 0.3442 | 0.6232 | 0.0054 |
| 2455582.5902 | -0.3385 | 6.6230 | 0.0039 | 2455581.7675 | 0.4002 | 0.6197 | 0.0065 |
| 2455582.7598 | -0.2847 | 6.5691 | 0.0034 | 2455582.5901 | -0.3386 | 0.6045 | 0.0058 |
| 2455583.5908 | -0.0208 | 6.3461 | 0.0029 | 2455582.7597 | -0.2847 | 0.5834 | 0.0063 |
| 2455583.7591 | 0.0327 | 6.3546 | 0.0067 | 2455583.5908 | -0.0208 | 0.5316 | 0.0040 |
| 2455584.5914 | 0.2970 | 6.5779 | 0.0031 | 2455583.7590 | 0.0326 | 0.5356 | 0.0075 |
| 2455584.7597 | 0.3504 | 6.6231 | 0.0016 | 2455584.5914 | 0.2970 | 0.5993 | 0.0047 |
| 2455585.7536 | -0.3339 | 6.6276 | 0.0054 | 2455584.7596 | 0.3504 | 0.6167 | 0.0038 |
| 2455588.5931 | -0.4322 | 6.6976 | 0.0043 | 2455585.7535 | -0.3340 | 0.6028 | 0.0060 |
| 2455588.7452 | -0.3838 | 6.6790 | 0.0057 | 2455588.5931 | -0.4322 | 0.6288 | 0.0070 |
| 2455589.7527 | -0.0639 | 6.3592 | 0.0059 | 2455588.7451 | -0.3839 | 0.6150 | 0.0066 |
| 2455591.7465 | -0.4307 | 6.7012 | 0.0019 | 2455589.7526 | -0.0639 | 0.5286 | 0.0086 |
| 2455592.5954 | -0.1611 | 6.4307 | 0.0055 | 2455591.7464 | -0.4307 | 0.6274 | 0.0055 |
| 2455595.7600 | -0.1561 | 6.4257 | 0.0073 | 2455592.5953 | -0.1612 | 0.5422 | 0.0079 |
| 2455596.7518 | 0.1589 | 6.4324 | 0.0052 | 2455595.7600 | -0.1561 | 0.5400 | 0.0086 |
| 2455597.7477 | 0.4751 | 6.6969 | 0.0051 | 2455596.7517 | 0.1588 | 0.5583 | 0.0074 |
| 2455599.6913 | 0.0924 | 6.3791 | 0.0036 | 2455597.7489 | 0.4755 | 0.6339 | 0.0071 |
| 2455600.7450 | 0.4270 | 6.6791 | 0.0035 | 2455599.6913 | 0.0924 | 0.5493 | 0.0055 |
| 2455601.7416 | -0.2565 | 6.5422 | 0.0060 | 2455600.7449 | 0.4270 | 0.6252 | 0.0043 |
| 2455602.7339 | 0.0586 | 6.3604 | 0.0059 | 2455601.7415 | -0.2565 | 0.5757 | 0.0097 |
| 2455603.6898 | 0.3622 | 6.6342 | 0.0026 | 2455602.7338 | 0.0586 | 0.5403 | 0.0102 |
| 2455604.7298 | -0.3075 | 6.6021 | 0.0036 | 2455603.6897 | 0.3622 | 0.6183 | 0.0047 |
| 2455605.7062 | 0.0026 | 6.3299 | 0.0004 | 2455604.7298 | -0.3075 | 0.5871 | 0.0052 |
| 2455607.7047 | -0.3627 | 6.6527 | 0.0054 | 2455605.7062 | 0.0026 | 0.5317 | 0.0051 |
| 2455608.7165 | -0.0414 | 6.3567 | 0.0030 | 2455607.7046 | -0.3628 | 0.6148 | 0.0068 |
| 2455614.7079 | -0.1387 | 6.4068 | 0.0086 | 2455608.7164 | -0.0415 | 0.5273 | 0.0034 |
| 2455615.6930 | 0.1742 | 6.4451 | 0.0048 | 2455614.7079 | -0.1387 | 0.5326 | 0.0115 |
| 2455616.6944 | 0.4922 | 6.7051 | 0.0028 | 2455615.6929 | 0.1741 | 0.5665 | 0.0059 |
| 2455617.6769 | -0.1958 | 6.4735 | 0.0038 | 2455616.6943 | 0.4921 | 0.6358 | 0.0034 |
| 2455621.6878 | 0.0780 | 6.3669 | 0.0080 | 2455617.6756 | -0.1962 | 0.5582 | 0.0081 |
| 2455944.7763 | -0.3165 | 6.6133 | 0.0053 | 2455621.6877 | 0.0779 | 0.5343 | 0.0101 |
| | | | | 2455944.7762 | -0.3165 | 0.5931 | 0.0064 |

**Photometry of SZ Tau, cont…**

| HJD | Phase | c1 | c1 err | HJD | Phase | m1 | m1 err |
|---|---|---|---|---|---|---|---|
| 2453415.6323 | 0.4851 | 0.5927 | 0.0078 | 2453415.6325 | 0.4852 | 0.2166 | 0.0063 |
| 2453415.7357 | -0.4821 | 0.5923 | 0.0080 | 2453415.7359 | -0.4820 | 0.2227 | 0.0065 |
| 2453432.6912 | -0.0974 | 0.8143 | 0.0223 | 2453432.6913 | -0.0973 | 0.1823 | 0.0148 |
| 2453433.6863 | 0.2186 | 0.7066 | 0.0165 | 2453433.6864 | 0.2187 | 0.1975 | 0.0104 |
| 2453437.6722 | 0.4845 | 0.5923 | 0.0219 | 2453437.6723 | 0.4845 | 0.2155 | 0.0165 |
| 2453438.6687 | -0.1991 | 0.7374 | 0.0181 | 2453438.6688 | -0.1990 | 0.1784 | 0.0151 |
| 2453439.6654 | 0.1175 | 0.7909 | 0.0130 | 2453439.6655 | 0.1175 | 0.1765 | 0.0121 |
| 2453442.6589 | 0.0681 | 0.8096 | 0.0225 | 2453442.6591 | 0.0682 | 0.1848 | 0.0175 |
| 2453444.6520 | -0.2989 | 0.6712 | 0.0175 | 2453444.6522 | -0.2988 | 0.2034 | 0.0143 |
| 2453445.6508 | 0.0183 | 0.8274 | 0.0202 | 2453445.6509 | 0.0183 | 0.1808 | 0.0141 |
| 2453446.6430 | 0.3334 | 0.6470 | 0.0143 | 2453446.6431 | 0.3334 | 0.2100 | 0.0106 |
| 2453451.6331 | -0.0819 | 0.8235 | 0.0311 | 2453451.6332 | -0.0818 | 0.1711 | 0.0163 |
| 2453452.6333 | 0.2358 | 0.6969 | 0.0121 | 2453452.6334 | 0.2358 | 0.2017 | 0.0098 |
| 2453696.8200 | -0.2161 | 0.7104 | 0.0088 | 2453696.8201 | -0.2161 | 0.2165 | 0.0091 |
| 2453700.8183 | 0.0537 | 0.8476 | 0.0123 | 2453700.8184 | 0.0537 | 0.1759 | 0.0130 |
| 2453708.9030 | -0.3788 | 0.6413 | 0.0132 | 2453708.9032 | -0.3788 | 0.2329 | 0.0097 |
| 2453721.8857 | -0.2558 | 0.7222 | 0.0093 | 2453721.8858 | -0.2558 | 0.1927 | 0.0083 |
| 2453724.8609 | -0.3110 | 0.6877 | 0.0119 | 2453724.8610 | -0.3109 | 0.1968 | 0.0099 |
| 2453734.8012 | -0.1541 | 0.7891 | 0.0108 | 2453734.8013 | -0.1541 | 0.1904 | 0.0124 |
| 2453735.8032 | 0.1641 | 0.7795 | 0.0093 | 2453735.8033 | 0.1641 | 0.1861 | 0.0064 |
| 2453753.8011 | -0.1202 | 0.8186 | 0.0164 | 2453753.8012 | -0.1202 | 0.1823 | 0.0103 |
| 2453756.7864 | -0.1721 | 0.7718 | 0.0107 | 2453756.7865 | -0.1721 | 0.1833 | 0.0072 |
| 2453758.7833 | 0.4620 | 0.6215 | 0.0068 | 2453758.7834 | 0.4621 | 0.2273 | 0.0055 |
| 2453760.7616 | 0.0903 | 0.8236 | 0.0231 | 2453760.7617 | 0.0903 | 0.1842 | 0.0174 |
| 2453765.7712 | -0.3188 | 0.6524 | 0.0238 | 2453765.7713 | -0.3187 | 0.2102 | 0.0384 |
| 2453766.7648 | -0.0032 | 0.8317 | 0.0216 | 2453766.7649 | -0.0032 | 0.1789 | 0.0322 |
| 2453767.7602 | 0.3129 | 0.6315 | 0.0167 | 2453767.7604 | 0.3130 | 0.2158 | 0.0181 |
| 2453768.7541 | -0.3715 | 0.6410 | 0.0137 | 2453768.7542 | -0.3714 | 0.2159 | 0.0138 |
| 2453782.7030 | 0.0584 | 0.8481 | 0.0250 | 2453782.7036 | 0.0586 | 0.1960 | 0.0382 |
| 2453808.6489 | 0.2982 | 0.6651 | 0.0145 | 2453808.6490 | 0.2982 | 0.2135 | 0.0216 |
| 2453812.6385 | -0.4348 | 0.6066 | 0.0179 | 2453812.6386 | -0.4348 | 0.2349 | 0.0232 |
| 2454396.9623 | 0.1331 | 0.7935 | 0.0152 | 2454396.9624 | 0.1331 | 0.1807 | 0.0156 |
| 2454400.0368 | 0.1095 | 0.8087 | 0.0090 | 2454400.0369 | 0.1095 | 0.1792 | 0.0082 |
| 2454400.9374 | 0.3955 | 0.6294 | 0.0089 | 2454400.9369 | 0.3954 | 0.2193 | 0.0087 |
| 2454401.9558 | -0.2811 | 0.7007 | 0.0066 | 2454401.9560 | -0.2810 | 0.1991 | 0.0076 |
| 2454403.9492 | 0.3520 | 0.6472 | 0.0089 | 2454403.9494 | 0.3521 | 0.2096 | 0.0060 |
| 2454405.9984 | 0.0028 | 0.8455 | 0.0139 | 2454405.9980 | 0.0026 | 0.1827 | 0.0129 |
| 2454407.9828 | -0.3670 | 0.6604 | 0.0061 | 2454407.9823 | -0.3672 | 0.2065 | 0.0073 |
| 2454409.9749 | 0.2656 | 0.6912 | 0.0079 | 2454409.9744 | 0.2655 | 0.2046 | 0.0081 |
| 2454412.9406 | 0.2075 | 0.7296 | 0.0124 | 2454412.9408 | 0.2075 | 0.1924 | 0.0092 |
| 2454413.9906 | -0.4591 | 0.6173 | 0.0116 | 2454413.9907 | -0.4591 | 0.2141 | 0.0082 |
| 2454414.8302 | -0.1925 | 0.7508 | 0.0114 | 2454414.8293 | -0.1927 | 0.1882 | 0.0078 |
| 2454415.9431 | 0.1610 | 0.7795 | 0.0139 | 2454415.9432 | 0.1610 | 0.1954 | 0.0142 |
| 2454417.9846 | -0.1907 | 0.7672 | 0.0058 | 2454417.9841 | -0.1908 | 0.1903 | 0.0095 |
| 2454418.9809 | 0.1257 | 0.8036 | 0.0208 | 2454418.9804 | 0.1256 | 0.1885 | 0.0260 |
| 2454422.9779 | 0.3951 | 0.6292 | 0.0120 | 2454422.9780 | 0.3951 | 0.2188 | 0.0164 |
| 2454423.9737 | -0.2887 | 0.7120 | 0.0167 | 2454423.9738 | -0.2887 | 0.1856 | 0.0203 |
| 2454425.8893 | 0.3197 | 0.6629 | 0.0135 | 2454425.8895 | 0.3197 | 0.2207 | 0.0097 |
| 2454456.8775 | 0.1608 | 0.7557 | 0.0133 | 2454456.8776 | 0.1608 | 0.1969 | 0.0127 |
| 2454458.8528 | -0.2119 | 0.7337 | 0.0184 | 2454458.8530 | -0.2118 | 0.2001 | 0.0212 |
| 2454459.8454 | 0.1033 | 0.8093 | 0.0286 | 2454459.8456 | 0.1034 | 0.1840 | 0.0302 |
| 2454460.8366 | 0.4181 | 0.6199 | 0.0087 | 2454460.8368 | 0.4182 | 0.2219 | 0.0137 |
| 2454461.8641 | -0.2556 | 0.7050 | 0.0121 | 2454461.8642 | -0.2555 | 0.2082 | 0.0147 |
| 2454464.8619 | -0.3035 | 0.6781 | 0.0227 | 2454464.8620 | -0.3035 | 0.2031 | 0.0285 |
| 2454465.8609 | 0.0137 | 0.8459 | 0.0264 | 2454465.8610 | 0.0138 | 0.1826 | 0.0361 |
| 2454466.6590 | 0.2672 | 0.6900 | 0.0180 | 2454466.6585 | 0.2670 | 0.2130 | 0.0306 |
| 2454475.7870 | 0.1660 | 0.7603 | 0.0140 | 2454475.7865 | 0.1659 | 0.1936 | 0.0129 |
| 2454476.7771 | 0.4805 | 0.6228 | 0.0129 | 2454476.7772 | 0.4805 | 0.2225 | 0.0133 |
| 2454479.7748 | 0.4325 | 0.6391 | 0.0123 | 2454479.7749 | 0.4325 | 0.2192 | 0.0130 |
| 2454480.7638 | -0.2535 | 0.7282 | 0.0252 | 2454480.7633 | -0.2536 | 0.1908 | 0.0217 |
| 2454481.7659 | 0.0648 | 0.8264 | 0.0377 | 2454481.7660 | 0.0648 | 0.1936 | 0.0322 |
| 2454482.7607 | 0.3807 | 0.6433 | 0.0110 | 2454482.7603 | 0.3806 | 0.2133 | 0.0160 |
| 2454483.7641 | -0.3006 | 0.7143 | 0.0246 | 2454483.7643 | -0.3006 | 0.1968 | 0.0233 |
| 2454484.7532 | 0.0135 | 0.8402 | 0.0196 | 2454484.7533 | 0.0135 | 0.1803 | 0.0180 |
| 2454486.7466 | -0.3534 | 0.6688 | 0.0234 | 2454486.7467 | -0.3534 | 0.2035 | 0.0235 |
| 2454494.7364 | 0.1839 | 0.7285 | 0.0217 | 2454494.7365 | 0.1840 | 0.1858 | 0.0238 |
| 2454495.7620 | -0.4904 | 0.6368 | 0.0124 | 2454495.7621 | -0.4903 | 0.2210 | 0.0163 |
| 2454496.7731 | -0.1693 | 0.7771 | 0.0133 | 2454496.7732 | -0.1692 | 0.1950 | 0.0186 |
| 2454498.7371 | 0.4545 | 0.6187 | 0.0104 | 2454498.7372 | 0.4545 | 0.2247 | 0.0136 |
| 2454499.7302 | -0.2302 | 0.7152 | 0.0287 | 2454499.7293 | -0.2304 | 0.1997 | 0.0295 |
| 2454503.7205 | 0.0371 | 0.8339 | 0.0347 | 2454503.7217 | 0.0375 | 0.1894 | 0.0379 |
| 2454504.7029 | 0.3491 | 0.6500 | 0.0190 | 2454504.7030 | 0.3491 | 0.2119 | 0.0225 |
| 2454505.7262 | -0.3260 | 0.6809 | 0.0112 | 2454505.7263 | -0.3259 | 0.2073 | 0.0129 |
| 2454508.7050 | -0.3800 | 0.6329 | 0.0118 | 2454508.7051 | -0.3799 | 0.2363 | 0.0158 |
| 2454509.6376 | -0.0838 | 0.8295 | 0.0215 | 2454509.6378 | -0.0837 | 0.1755 | 0.0262 |
| 2454821.7264 | 0.0285 | 0.8385 | 0.0102 | 2454821.7259 | 0.0283 | 0.1854 | 0.0114 |
| 2454821.8612 | 0.0713 | 0.8112 | 0.0246 | 2454821.8607 | 0.0711 | 0.1903 | 0.0181 |
| 2454822.7090 | 0.3405 | 0.6549 | 0.0043 | 2454822.7079 | 0.3402 | 0.2178 | 0.0075 |
| 2454829.7088 | -0.4365 | 0.6028 | 0.0075 | 2454829.7077 | -0.4368 | 0.2278 | 0.0062 |
| 2454829.8497 | -0.3917 | 0.6569 | 0.0090 | 2454829.8498 | -0.3917 | 0.2096 | 0.0082 |
| 2454830.6980 | -0.1223 | 0.8107 | 0.0083 | 2454830.6981 | -0.1223 | 0.1872 | 0.0110 |
| 2454831.7019 | 0.1965 | 0.7382 | 0.0158 | 2454831.7020 | 0.1965 | 0.2007 | 0.0162 |
| 2454832.8250 | -0.4469 | 0.6132 | 0.0082 | 2454832.8251 | -0.4468 | 0.2101 | 0.0058 |
| 2454838.6987 | 0.4185 | 0.6274 | 0.0120 | 2454838.6989 | 0.4186 | 0.2166 | 0.0219 |
| 2454838.8023 | 0.4514 | 0.6127 | 0.0104 | 2454838.8024 | 0.4514 | 0.2315 | 0.0088 |
| 2454839.6758 | -0.2712 | 0.7130 | 0.0060 | 2454839.6747 | -0.2715 | 0.1916 | 0.0062 |
| 2454839.7993 | -0.2320 | 0.7195 | 0.0044 | 2454839.7988 | -0.2321 | 0.1976 | 0.0073 |
| 2454840.6757 | 0.0464 | 0.8424 | 0.0096 | 2454840.6758 | 0.0464 | 0.1837 | 0.0105 |
| 2454842.6768 | -0.3181 | 0.6823 | 0.0062 | 2454842.6769 | -0.3181 | 0.1967 | 0.0088 |
| 2454844.6761 | 0.3168 | 0.6720 | 0.0129 | 2454844.6762 | 0.3168 | 0.2056 | 0.0102 |
| 2454844.7859 | 0.3517 | 0.6134 | 0.0165 | 2454844.7860 | 0.3517 | 0.2149 | 0.0142 |
| 2454846.7917 | -0.0113 | 0.8436 | 0.0204 | 2454846.7919 | -0.0113 | 0.1829 | 0.0215 |
| 2454847.7777 | 0.3018 | 0.6681 | 0.0129 | 2454847.7778 | 0.3018 | 0.2075 | 0.0125 |
| 2454856.7778 | 0.1600 | 0.7740 | 0.0128 | 2454856.7779 | 0.1600 | 0.1897 | 0.0122 |
| 2454865.7520 | 0.0100 | 0.8613 | 0.0144 | 2454865.7521 | 0.0100 | 0.1827 | 0.0140 |
| 2454867.7469 | -0.3565 | 0.6619 | 0.0060 | 2454867.7470 | -0.3564 | 0.2160 | 0.0045 |
| 2454873.7340 | -0.4551 | 0.6220 | 0.0061 | 2454873.7353 | -0.4547 | 0.2157 | 0.0053 |
| 2454875.7289 | 0.1784 | 0.7517 | 0.0163 | 2454875.7290 | 0.1785 | 0.1967 | 0.0128 |
| 2455213.7668 | -0.4684 | 0.6248 | 0.0068 | 2455213.7662 | -0.4686 | 0.2241 | 0.0061 |
| 2455221.6147 | 0.0239 | 0.8631 | 0.0079 | 2455221.6136 | 0.0235 | 0.1842 | 0.0087 |
| 2455221.7426 | 0.0645 | 0.8405 | 0.0159 | 2455221.7434 | 0.0648 | 0.1628 | 0.0200 |
| 2455222.6170 | 0.3422 | 0.6458 | 0.0069 | 2455222.6171 | 0.3422 | 0.2077 | 0.0081 |
| 2455236.5994 | -0.2173 | 0.7349 | 0.0151 | 2455236.5995 | -0.2173 | 0.1917 | 0.0067 |
| 2455236.7091 | -0.1825 | 0.7592 | 0.0141 | 2455236.7092 | -0.1825 | 0.1950 | 0.0110 |
| 2455240.7005 | 0.0851 | 0.8339 | 0.0161 | 2455240.7000 | 0.0849 | 0.1793 | 0.0183 |
| 2455241.6894 | 0.3991 | 0.6282 | 0.0051 | 2455241.6902 | 0.3994 | 0.2308 | 0.0080 |
| 2455243.7195 | 0.0439 | 0.8480 | 0.0147 | 2455243.7202 | 0.0441 | 0.1869 | 0.0116 |
| 2455253.6769 | 0.2061 | 0.7251 | 0.0177 | 2455253.6764 | 0.2059 | 0.1985 | 0.0183 |
| 2455258.6930 | -0.2009 | 0.7461 | 0.0219 | 2455258.6932 | -0.2008 | 0.1861 | 0.0198 |
| 2455259.6723 | 0.1101 | 0.8018 | 0.0136 | 2455259.6730 | 0.1103 | 0.1785 | 0.0145 |
| 2455268.6757 | -0.0306 | 0.8641 | 0.0212 | 2455268.6764 | -0.0304 | 0.1669 | 0.0191 |
| 2455270.6608 | -0.4002 | 0.6392 | 0.0058 | 2455270.6590 | -0.4008 | 0.2161 | 0.0093 |
| 2455271.6680 | -0.0803 | 0.8386 | 0.0121 | 2455271.6687 | -0.0801 | 0.1776 | 0.0179 |
| 2455273.6598 | -0.4478 | 0.6156 | 0.0069 | 2455273.6606 | -0.4475 | 0.2252 | 0.0084 |
| 2455274.6524 | -0.1326 | 0.8118 | 0.0171 | 2455274.6519 | -0.1327 | 0.1872 | 0.0138 |
| 2455275.6519 | 0.1849 | 0.7527 | 0.0189 | 2455275.6520 | 0.1849 | 0.1853 | 0.0188 |
| 2455580.5979 | 0.0287 | 0.8522 | 0.0082 | 2455580.5974 | 0.0286 | 0.1820 | 0.0075 |
| 2455580.7680 | 0.0828 | 0.8371 | 0.0133 | 2455580.7675 | 0.0826 | 0.1831 | 0.0149 |
| 2455581.5903 | 0.3439 | 0.6609 | 0.0074 | 2455581.5911 | 0.3442 | 0.2000 | 0.0065 |
| 2455581.7673 | 0.4001 | 0.6399 | 0.0082 | 2455581.7674 | 0.4002 | 0.2212 | 0.0088 |
| 2455582.5918 | -0.3380 | 0.6715 | 0.0068 | 2455582.5907 | -0.3384 | 0.2034 | 0.0077 |
| 2455582.7602 | -0.2846 | 0.6991 | 0.0105 | 2455582.7597 | -0.2847 | 0.1987 | 0.0094 |
| 2455583.5912 | -0.0207 | 0.8579 | 0.0160 | 2455583.5907 | -0.0208 | 0.1721 | 0.0117 |
| 2455583.7588 | 0.0326 | 0.8570 | 0.0105 | 2455583.7589 | 0.0326 | 0.1696 | 0.0107 |
| 2455584.5918 | 0.2971 | 0.6954 | 0.0165 | 2455584.5913 | 0.2970 | 0.2075 | 0.0112 |
| 2455584.7601 | 0.3506 | 0.6546 | 0.0085 | 2455584.7596 | 0.3504 | 0.2177 | 0.0061 |
| 2455585.7546 | -0.3336 | 0.6857 | 0.0075 | 2455585.7534 | -0.3340 | 0.1973 | 0.0068 |
| 2455588.5929 | -0.4322 | 0.6103 | 0.0090 | 2455588.5930 | -0.4322 | 0.2223 | 0.0096 |
| 2455588.7462 | -0.3835 | 0.6568 | 0.0080 | 2455588.7451 | -0.3839 | 0.2068 | 0.0084 |
| 2455589.7524 | -0.0640 | 0.8405 | 0.0120 | 2455589.7526 | -0.0639 | 0.1797 | 0.0128 |
| 2455591.7481 | -0.4302 | 0.6427 | 0.0085 | 2455591.7470 | -0.4306 | 0.2140 | 0.0078 |
| 2455592.5957 | -0.1610 | 0.7910 | 0.0137 | 2455592.5952 | -0.1612 | 0.1885 | 0.0116 |
| 2455595.7598 | -0.1562 | 0.7906 | 0.0142 | 2455595.7599 | -0.1561 | 0.1856 | 0.0132 |
| 2455596.7522 | 0.1590 | 0.7805 | 0.0079 | 2455596.7516 | 0.1588 | 0.1920 | 0.0100 |
| 2455597.7481 | 0.4753 | 0.6177 | 0.0107 | 2455597.7489 | 0.4755 | 0.2228 | 0.0089 |
| 2455599.6911 | 0.0923 | 0.8175 | 0.0114 | 2455599.6912 | 0.0923 | 0.1764 | 0.0101 |
| 2455600.7466 | 0.4275 | 0.6278 | 0.0050 | 2455600.7455 | 0.4272 | 0.2287 | 0.0057 |
| 2455601.7420 | -0.2564 | 0.7168 | 0.0100 | 2455601.7415 | -0.2565 | 0.1912 | 0.0130 |
| 2455602.7343 | 0.0588 | 0.8476 | 0.0114 | 2455602.7338 | 0.0586 | 0.1725 | 0.0139 |
| 2455603.6895 | 0.3621 | 0.6378 | 0.0080 | 2455603.6896 | 0.3621 | 0.2218 | 0.0064 |
| 2455604.7296 | -0.3076 | 0.6861 | 0.0166 | 2455604.7297 | -0.3075 | 0.2009 | 0.0130 |
| 2455605.7060 | 0.0025 | 0.8407 | 0.0163 | 2455605.7061 | 0.0025 | 0.1831 | 0.0123 |
| 2455607.7032 | -0.3632 | 0.6490 | 0.0052 | 2455607.7039 | -0.3630 | 0.2063 | 0.0082 |
| 2455608.7169 | -0.0413 | 0.8576 | 0.0116 | 2455608.7164 | -0.0415 | 0.1781 | 0.0060 |
| 2455614.7083 | -0.1386 | 0.8108 | 0.0185 | 2455614.7078 | -0.1387 | 0.1881 | 0.0182 |
| 2455616.6948 | 0.4923 | 0.6336 | 0.0066 | 2455615.6929 | 0.1741 | 0.1891 | 0.0086 |
| 2455617.6748 | -0.1965 | 0.7509 | 0.0081 | 2455616.6943 | 0.4921 | 0.2148 | 0.0056 |
| 2455621.6876 | 0.0779 | 0.8219 | 0.0093 | 2455617.6749 | -0.1964 | 0.1829 | 0.0111 |
| 2455944.7760 | -0.3166 | 0.6831 | 0.0063 | 2455621.6877 | 0.0779 | 0.1897 | 0.0127 |
| | | | | 2455944.7762 | -0.3165 | 0.2031 | 0.0075 |

## Photometry of VY Cyg

| HJD | Phase | V-mag | V-err | HJD | Phase | U-B | U-B err | HJD | Phase | B-V | B-V err |
|---|---|---|---|---|---|---|---|---|---|---|---|
| 2453665.7581 | -0.3147 | 9.0349 | 0.0040 | 2453665.7579 | -0.3147 | 1.2161 | 0.0067 | 2453665.7580 | -0.3147 | 1.3171 | 0.0043 |
| 2453667.7527 | -0.0608 | 8.3856 | 0.0053 | 2453667.7525 | -0.0609 | 0.7865 | 0.0102 | 2453667.7526 | -0.0608 | 0.9654 | 0.0073 |
| 2453669.6124 | 0.1759 | 8.4436 | 0.0043 | 2453669.6122 | 0.1758 | 0.8798 | 0.0046 | 2453669.6123 | 0.1758 | 1.0473 | 0.0047 |
| 2453670.6157 | 0.3036 | 8.4310 | 0.0028 | 2453670.6155 | 0.3035 | 0.9022 | 0.0071 | 2453670.6156 | 0.3035 | 1.0621 | 0.0039 |
| 2453671.6152 | 0.4308 | 8.7382 | 0.0057 | 2453671.6150 | 0.4307 | 1.0431 | 0.0051 | 2453671.6151 | 0.4307 | 1.2244 | 0.0064 |
| 2453672.6156 | -0.4419 | 8.8815 | 0.0036 | 2453672.6153 | -0.4420 | 1.1652 | 0.0063 | 2453672.6155 | -0.4419 | 1.2832 | 0.0037 |
| 2453673.6149 | -0.3147 | 9.0400 | 0.0041 | 2453673.6146 | -0.3148 | 1.2366 | 0.0033 | 2453673.6148 | -0.3147 | 1.3256 | 0.0046 |
| 2453674.6144 | -0.1875 | 8.9191 | 0.0055 | 2453674.6142 | -0.1875 | 1.0074 | 0.0068 | 2453674.6143 | -0.1875 | 1.2011 | 0.0062 |
| 2453675.7322 | -0.0453 | 8.3157 | 0.0055 | 2453675.7320 | -0.0453 | 0.7541 | 0.0120 | 2453675.7321 | -0.0453 | 0.9332 | 0.0086 |
| 2453677.7265 | 0.2086 | 8.4201 | 0.0064 | 2453677.7263 | 0.2085 | 0.8658 | 0.0090 | 2453677.7264 | 0.2086 | 1.0343 | 0.0074 |
| 2453678.7234 | 0.3354 | 8.5118 | 0.0042 | 2453678.7232 | 0.3354 | 0.9331 | 0.0117 | 2453678.7233 | 0.3354 | 1.1046 | 0.0054 |
| 2453679.6115 | 0.4485 | 8.7545 | 0.0031 | 2453679.6112 | 0.4484 | 1.0565 | 0.0080 | 2453679.6114 | 0.4485 | 1.2332 | 0.0033 |
| 2453680.6110 | -0.4243 | 8.9037 | 0.0031 | 2453680.6107 | -0.4244 | 1.1868 | 0.0043 | 2453680.6109 | -0.4243 | 1.2931 | 0.0040 |
| 2453688.6074 | -0.4066 | 8.9305 | 0.0069 | 2453688.6072 | -0.4066 | 1.2188 | 0.0109 | 2453688.6073 | -0.4066 | 1.3083 | 0.0081 |
| 2453689.6111 | -0.2788 | 9.0504 | 0.0092 | 2453689.6108 | -0.2789 | 1.1559 | 0.0129 | 2453689.6110 | -0.2789 | 1.3055 | 0.0096 |
| 2453690.6150 | -0.1511 | 8.8853 | 0.0090 | 2453690.6147 | -0.1511 | 0.8947 | 0.0283 | 2453690.6149 | -0.1511 | 1.1298 | 0.0117 |
| 2453691.6292 | -0.0220 | 8.2454 | 0.0086 | 2453691.6289 | -0.0220 | 0.7678 | 0.0185 | 2453691.6291 | -0.0220 | 0.9039 | 0.0103 |
| 2453693.6061 | 0.2296 | 8.3976 | 0.0047 | 2453693.6059 | 0.2296 | 0.8786 | 0.0091 | 2453693.6060 | 0.2296 | 1.0355 | 0.0063 |
| 2453694.6051 | 0.3568 | 8.5927 | 0.0109 | 2453694.6049 | 0.3567 | 0.9711 | 0.0058 | 2453694.6050 | 0.3567 | 1.1543 | 0.0116 |
| 2453695.6048 | 0.4840 | 8.7878 | 0.0050 | 2453695.6046 | 0.4840 | 1.0968 | 0.0058 | 2453695.6047 | 0.4840 | 1.2524 | 0.0055 |
| 2453699.5750 | -0.0107 | 8.2381 | 0.0059 | 2453699.5776 | -0.0104 | 0.7810 | 0.0058 | 2453699.5763 | -0.0105 | 0.9102 | 0.0068 |
| 2453701.5973 | 0.2467 | 8.3812 | 0.0038 | 2453701.5971 | 0.2466 | 0.8730 | 0.0108 | 2453701.5972 | 0.2467 | 1.0417 | 0.0051 |
| 2453702.6427 | 0.3797 | 8.6731 | 0.0026 | 2453702.6411 | 0.3795 | 0.9959 | 0.0036 | 2453702.6412 | 0.3795 | 1.1862 | 0.0033 |
| 2453703.6286 | -0.4948 | 8.8128 | 0.0044 | 2453703.6284 | -0.4948 | 1.1291 | 0.0065 | 2453703.6285 | -0.4948 | 1.2503 | 0.0058 |
| 2453705.5970 | -0.2443 | 9.0340 | 0.0027 | 2453705.5968 | -0.2443 | 1.1247 | 0.0099 | 2453705.5969 | -0.2443 | 1.2794 | 0.0030 |
| 2453706.5970 | -0.1170 | 8.6684 | 0.0030 | 2453706.5968 | -0.1170 | 0.8414 | 0.0106 | 2453706.5969 | -0.1170 | 1.0825 | 0.0046 |
| 2453708.6057 | 0.1387 | 8.4403 | 0.0325 | 2453708.6055 | 0.1386 | 0.8588 | 0.0131 | 2453708.6056 | 0.1386 | 1.0213 | 0.0328 |
| 2453709.5968 | 0.2648 | 8.3920 | 0.0035 | 2453709.5966 | 0.2648 | 0.8673 | 0.0069 | 2453709.5967 | 0.2648 | 1.0335 | 0.0038 |
| 2453710.5969 | 0.3921 | 8.6998 | 0.0044 | 2453710.5967 | 0.3921 | 1.0183 | 0.0050 | 2453710.5968 | 0.3921 | 1.1931 | 0.0062 |
| 2453711.5970 | -0.4806 | 8.8332 | 0.0040 | 2453711.5968 | -0.4807 | 1.1382 | 0.0115 | 2453711.5969 | -0.4806 | 1.2598 | 0.0048 |
| 2453712.5970 | -0.3534 | 9.0077 | 0.0043 | 2453712.5968 | -0.3534 | 1.2122 | 0.0097 | 2453712.5969 | -0.3534 | 1.3196 | 0.0046 |
| 2453714.5973 | -0.0988 | 8.5967 | 0.0036 | 2453714.5971 | -0.0988 | 0.8380 | 0.0095 | 2453714.5972 | -0.0988 | 1.0500 | 0.0048 |
| 2453724.5784 | 0.1715 | 8.4290 | 0.0280 | 2453724.5782 | 0.1715 | 0.8784 | 0.0309 | 2453724.5783 | 0.1715 | 1.0398 | 0.0317 |
| 2453725.5757 | 0.2985 | 8.4221 | 0.0119 | 2453725.5755 | 0.2985 | 0.8799 | 0.0123 | 2453725.5756 | 0.2985 | 1.0681 | 0.0128 |
| 2453726.5760 | 0.4258 | 8.7461 | 0.0005 | 2453726.5758 | 0.4258 | 1.0951 | 0.0234 | 2453726.5759 | 0.4258 | 1.2335 | 0.0048 |
| 2454001.8317 | 0.4584 | 8.7462 | 0.0042 | 2454001.8315 | 0.4584 | 1.0988 | 0.0118 | 2454001.8316 | 0.4584 | 1.2263 | 0.0046 |
| 2454003.7980 | -0.2913 | 9.0344 | 0.0031 | 2454003.7978 | -0.2914 | 1.1681 | 0.0023 | 2454003.7979 | -0.2913 | 1.3158 | 0.0036 |
| 2454004.7433 | -0.1710 | 8.8569 | 0.0066 | 2454004.7430 | -0.1711 | 0.9872 | 0.0042 | 2454004.7432 | -0.1710 | 1.1849 | 0.0069 |
| 2454005.7962 | -0.0370 | 8.2667 | 0.0169 | 2454005.7960 | -0.0370 | 0.7692 | 0.0170 | 2454005.7961 | -0.0370 | 0.9311 | 0.0193 |
| 2454008.8188 | 0.3477 | 8.5368 | 0.0075 | 2454008.8186 | 0.3477 | 0.9415 | 0.0072 | 2454008.8187 | 0.3477 | 1.1393 | 0.0085 |
| 2454010.7937 | -0.4010 | 8.9117 | 0.0102 | 2454010.7934 | -0.4010 | 1.1941 | 0.0036 | 2454010.7936 | -0.4010 | 1.3130 | 0.0106 |
| 2454012.7546 | -0.1514 | 8.7879 | 0.0177 | 2454012.7543 | -0.1514 | 0.9014 | 0.0118 | 2454012.7545 | -0.1514 | 1.1662 | 0.0182 |
| 2454017.6820 | 0.4757 | 8.7531 | 0.0102 | 2454017.6801 | 0.4755 | 1.2150 | 0.0090 | 2454017.6803 | 0.4755 | 1.2479 | 0.0131 |
| 2454018.7569 | -0.3875 | 8.9400 | 0.0052 | 2454018.7567 | -0.3875 | 1.2318 | 0.0155 | 2454018.7568 | -0.3875 | 1.3088 | 0.0080 |
| 2454023.7503 | 0.2481 | 8.3621 | 0.0078 | 2454023.7501 | 0.2480 | 0.8626 | 0.0080 | 2454023.7502 | 0.2480 | 1.0345 | 0.0079 |
| 2454025.7572 | -0.4965 | 8.7889 | 0.0124 | 2454025.7570 | -0.4965 | 1.1217 | 0.0109 | 2454025.7571 | -0.4965 | 1.2672 | 0.0129 |
| 2454030.7496 | 0.1389 | 8.4078 | 0.0128 | 2454030.7494 | 0.1389 | 0.8645 | 0.0113 | 2454030.7495 | 0.1389 | 1.0397 | 0.0156 |
| 2454031.7488 | 0.2661 | 8.3693 | 0.0120 | 2454031.7486 | 0.2660 | 0.8405 | 0.0073 | 2454031.7487 | 0.2660 | 1.0401 | 0.0127 |
| 2454033.7530 | -0.4789 | 8.8117 | 0.0173 | 2454033.7528 | -0.4789 | 1.1195 | 0.0097 | 2454033.7529 | -0.4789 | 1.2718 | 0.0178 |
| 2454037.7401 | 0.0286 | 8.2348 | 0.0040 | 2454037.7399 | 0.0286 | 0.7695 | 0.0050 | 2454037.7400 | 0.0286 | 0.9210 | 0.0061 |
| 2454039.7309 | 0.2820 | 8.3749 | 0.0118 | 2454039.7307 | 0.2819 | 0.8583 | 0.0158 | 2454039.7308 | 0.2819 | 1.0495 | 0.0147 |
| 2454040.7306 | 0.4092 | 8.7185 | 0.0008 | 2454040.7304 | 0.4092 | 1.0103 | 0.0371 | 2454040.7305 | 0.4092 | 1.2011 | 0.0012 |
| 2454047.7040 | 0.2967 | 8.3866 | 0.0226 | 2454047.7037 | 0.2967 | 0.9073 | 0.0179 | 2454047.7039 | 0.2967 | 1.0533 | 0.0228 |
| 2454049.7078 | -0.4483 | 8.8400 | 0.0052 | 2454049.7076 | -0.4483 | 1.1597 | 0.0098 | 2454049.7077 | -0.4483 | 1.2806 | 0.0072 |
| 2454050.7003 | -0.3219 | 9.0165 | 0.0109 | 2454050.7001 | -0.3220 | 1.2085 | 0.0063 | 2454050.7002 | -0.3219 | 1.3205 | 0.0123 |
| 2454057.6800 | -0.4336 | 8.8615 | 0.0069 | 2454057.6797 | -0.4336 | 1.1323 | 0.0076 | 2454057.6799 | -0.4336 | 1.3097 | 0.0073 |
| 2454058.6732 | -0.3072 | 9.0249 | 0.0054 | 2454058.6730 | -0.3072 | 1.2150 | 0.0016 | 2454058.6731 | -0.3072 | 1.3212 | 0.0055 |
| 2454059.6774 | -0.1794 | 8.8830 | 0.0067 | 2454059.6772 | -0.1794 | 0.9727 | 0.0081 | 2454059.6773 | -0.1794 | 1.1926 | 0.0070 |
| 2454060.6633 | -0.0539 | 8.3520 | 0.0039 | 2454060.6631 | -0.0539 | 0.7850 | 0.0120 | 2454060.6632 | -0.0539 | 0.9568 | 0.0049 |
| 2454061.6676 | 0.0739 | 8.2960 | 0.0137 | 2454061.6674 | 0.0739 | 0.7984 | 0.0086 | 2454061.6675 | 0.0739 | 0.9702 | 0.0139 |
| 2454062.6622 | 0.2005 | 8.4096 | 0.0038 | 2454062.6619 | 0.2005 | 0.8653 | 0.0085 | 2454062.6621 | 0.2005 | 1.0483 | 0.0056 |
| 2454064.6536 | 0.4539 | 8.7472 | 0.0044 | 2454064.6533 | 0.4539 | 1.0771 | 0.0080 | 2454064.6535 | 0.4539 | 1.2328 | 0.0080 |
| 2454067.6516 | -0.1645 | 8.8400 | 0.0113 | 2454067.6514 | -0.1645 | 0.9629 | 0.0146 | 2454067.6515 | -0.1645 | 1.1755 | 0.0115 |
| 2454068.6509 | -0.0373 | 8.2791 | 0.0097 | 2454068.6506 | -0.0373 | 0.7712 | 0.0126 | 2454068.6508 | -0.0373 | 0.9245 | 0.0098 |
| 2454069.6283 | 0.0871 | 8.3332 | 0.0060 | 2454069.6281 | 0.0871 | 0.8398 | 0.0030 | 2454069.6282 | 0.0871 | 0.9783 | 0.0066 |
| 2454070.6395 | 0.2158 | 8.4078 | 0.0069 | 2454070.6393 | 0.2158 | 0.8616 | 0.0190 | 2454070.6394 | 0.2158 | 1.0362 | 0.0071 |
| 2454071.6280 | 0.3416 | 8.5140 | 0.0109 | 2454071.6278 | 0.3416 | 0.9223 | 0.0256 | 2454071.6279 | 0.3416 | 1.1331 | 0.0134 |
| 2454072.6315 | 0.4693 | 8.7639 | 0.0055 | 2454072.6312 | 0.4693 | 1.1057 | 0.0150 | 2454072.6314 | 0.4693 | 1.2389 | 0.0074 |
| 2454075.6094 | -0.1517 | 8.7850 | 0.0017 | 2454075.6092 | -0.1517 | 0.9188 | 0.0052 | 2454075.6093 | -0.1517 | 1.1613 | 0.0044 |
| 2454076.6077 | -0.0246 | 8.2612 | 0.0002 | 2454076.6075 | -0.0246 | 0.7524 | 0.0295 | 2454076.6076 | -0.0246 | 0.9125 | 0.0051 |
| 2454077.6214 | 0.1044 | 8.3543 | 0.0078 | 2454077.6212 | 0.1044 | 0.8387 | 0.0092 | 2454077.6213 | 0.1044 | 1.0113 | 0.0099 |
| 2454078.6189 | 0.2314 | 8.3791 | 0.0066 | 2454078.6187 | 0.2313 | 0.8464 | 0.0187 | 2454078.6188 | 0.2313 | 1.0367 | 0.0076 |
| 2454079.6163 | 0.3583 | 8.5799 | 0.0070 | 2454079.6160 | 0.3583 | 0.9681 | 0.0125 | 2454079.6162 | 0.3583 | 1.1643 | 0.0074 |
| 2454080.6081 | 0.4845 | 8.7784 | 0.0137 | 2454080.6079 | 0.4845 | 1.1145 | 0.0060 | 2454080.6080 | 0.4845 | 1.2484 | 0.0141 |
| 2454081.6106 | -0.3879 | 8.9314 | 0.0142 | 2454081.6104 | -0.3879 | 1.1980 | 0.0059 | 2454081.6105 | -0.3879 | 1.3189 | 0.0152 |
| 2454085.5991 | 0.1197 | 8.3835 | 0.0048 | 2454085.5988 | 0.1197 | 0.8641 | 0.0162 | 2454085.5990 | 0.1197 | 1.0237 | 0.0062 |
| 2454086.5972 | 0.2468 | 8.3673 | 0.0091 | 2454086.5969 | 0.2467 | 0.8598 | 0.0120 | 2454086.5971 | 0.2468 | 1.0439 | 0.0096 |
| 2454087.5849 | 0.3725 | 8.6297 | 0.0079 | 2454087.5847 | 0.3725 | 1.0108 | 0.0063 | 2454087.5848 | 0.3725 | 1.1851 | 0.0084 |
| 2454090.5807 | -0.2462 | 9.0304 | 0.0017 | 2454090.5805 | -0.2463 | 1.1613 | 0.0023 | 2454090.5806 | -0.2462 | 1.2874 | 0.0020 |
| 2454733.7288 | -0.3908 | 8.9564 | 0.0099 | 2454733.7286 | -0.3909 | 1.1773 | 0.0065 | 2454733.7287 | -0.3909 | 1.3155 | 0.0099 |
| 2454734.6776 | -0.2701 | 9.0480 | 0.0051 | 2454734.6774 | -0.2701 | 1.1487 | 0.0060 | 2454734.6775 | -0.2701 | 1.3130 | 0.0058 |
| 2454737.7546 | 0.1215 | 8.4073 | 0.0060 | 2454737.7543 | 0.1215 | 0.8545 | 0.0139 | 2454737.7545 | 0.1215 | 1.0230 | 0.0085 |
| 2454738.6920 | 0.2408 | 8.3937 | 0.0002 | 2454738.6906 | 0.2407 | 0.8767 | 0.0055 | 2454738.6907 | 0.2407 | 1.0325 | 0.0020 |
| 2454741.7279 | -0.3728 | 8.9800 | 0.0025 | 2454741.7276 | -0.3728 | 1.2055 | 0.0054 | 2454741.7278 | -0.3728 | 1.3184 | 0.0032 |
| 2454746.6942 | 0.2593 | 8.3781 | 0.0020 | 2454746.6914 | 0.2590 | 0.8518 | 0.0095 | 2454746.6928 | 0.2591 | 1.0509 | 0.0081 |
| 2454766.6114 | -0.2058 | 8.9737 | 0.0051 | 2454766.6112 | -0.2058 | 1.0533 | 0.0078 | 2454766.6113 | -0.2058 | 1.2386 | 0.0055 |
| 2454767.6107 | -0.0786 | 8.5047 | 0.0077 | 2454767.6105 | -0.0786 | 0.8148 | 0.0099 | 2454767.6106 | -0.0786 | 1.0111 | 0.0098 |
| 2454772.6076 | -0.4426 | 8.8809 | 0.0038 | 2454772.6074 | -0.4426 | 1.1818 | 0.0026 | 2454772.6075 | -0.4426 | 1.2909 | 0.0045 |
| 2454776.6055 | 0.0662 | 8.3108 | 0.0046 | 2454776.6053 | 0.0662 | 0.8071 | 0.0052 | 2454776.6054 | 0.0662 | 0.9536 | 0.0048 |
| 2454777.6197 | 0.1953 | 8.4338 | 0.0022 | 2454777.6172 | 0.1950 | 0.8533 | 0.0097 | 2454777.6185 | 0.1951 | 1.0415 | 0.0026 |
| 2454785.6270 | 0.2144 | 8.4086 | 0.0023 | 2454785.6271 | 0.2144 | 0.8350 | 0.0161 | 2454785.6271 | 0.2144 | 1.0473 | 0.0044 |
| 2454786.6069 | 0.3391 | 8.5191 | 0.0038 | 2454786.6071 | 0.3391 | 0.9111 | 0.0093 | 2454786.6070 | 0.3391 | 1.1267 | 0.0044 |
| 2454788.6163 | -0.4051 | 8.9357 | 0.0032 | 2454788.6165 | -0.4051 | 1.1974 | 0.0027 | 2454788.6164 | -0.4051 | 1.2986 | 0.0035 |
| 2454791.5992 | -0.0255 | 8.2630 | 0.0069 | 2454791.5993 | -0.0255 | 0.7514 | 0.0216 | 2454791.5993 | -0.0255 | 0.9190 | 0.0145 |
| 2454801.5931 | 0.2465 | 8.3888 | 0.0038 | 2454801.5932 | 0.2465 | 0.8625 | 0.0050 | 2454801.5932 | 0.2465 | 1.0400 | 0.0051 |
| 2454803.5928 | -0.4990 | 8.8079 | 0.0031 | 2454803.5930 | -0.4990 | 1.1176 | 0.0038 | 2454803.5929 | -0.4990 | 1.2571 | 0.0040 |
| 2454811.5898 | -0.4812 | 8.8523 | 0.0041 | 2454811.5899 | -0.4812 | 1.1297 | 0.0259 | 2454811.5899 | -0.4812 | 1.2583 | 0.0050 |
| 2455097.7930 | -0.0553 | 8.3704 | 0.0107 | 2455097.7932 | -0.0553 | 0.7798 | 0.0125 | 2455097.7931 | -0.0553 | 0.9406 | 0.0149 |
| 2455098.6114 | 0.0489 | 8.2856 | 0.0130 | 2455098.6116 | 0.0489 | 0.7868 | 0.0115 | 2455098.6115 | 0.0489 | 0.9360 | 0.0160 |
| 2455098.7894 | 0.0715 | 8.3120 | 0.0082 | 2455098.7895 | 0.0715 | 0.8033 | 0.0194 | 2455098.7895 | 0.0715 | 0.9579 | 0.0164 |
| 2455099.6251 | 0.1779 | 8.4382 | 0.0023 | 2455099.6253 | 0.1779 | 0.8508 | 0.0126 | 2455099.6252 | 0.1779 | 1.0438 | 0.0039 |
| 2455099.7845 | 0.1982 | 8.4219 | 0.0150 | 2455099.7858 | 0.1983 | 0.8726 | 0.0090 | 2455099.7857 | 0.1983 | 1.0384 | 0.0154 |
| 2455100.7620 | 0.3226 | 8.4626 | 0.0154 | 2455100.7579 | 0.3221 | 0.9224 | 0.0119 | 2455100.7600 | 0.3223 | 1.0811 | 0.0156 |
| 2455101.8296 | 0.4585 | 8.7618 | 0.0021 | 2455101.8298 | 0.4585 | 1.0934 | 0.0086 | 2455101.8297 | 0.4585 | 1.2185 | 0.0028 |
| 2455102.6223 | -0.4406 | 8.8803 | 0.0008 | 2455102.6224 | -0.4406 | 1.1680 | 0.0063 | 2455102.6224 | -0.4406 | 1.2768 | 0.0027 |
| 2455102.7818 | -0.4203 | 8.9069 | 0.0047 | 2455102.7831 | -0.4202 | 1.1885 | 0.0041 | 2455102.7819 | -0.4203 | 1.2886 | 0.0050 |
| 2455106.7693 | 0.0872 | 8.3519 | 0.0055 | 2455106.7694 | 0.0872 | 0.7841 | 0.0228 | 2455106.7694 | 0.0872 | 0.9742 | 0.0169 |
| 2455131.7009 | 0.2603 | 8.3918 | 0.0079 | 2455131.7011 | 0.2603 | 0.8607 | 0.0138 | 2455131.7010 | 0.2603 | 1.0416 | 0.0101 |
| 2455133.6933 | -0.4861 | 8.8384 | 0.0062 | 2455133.6934 | -0.4861 | 1.1202 | 0.0079 | 2455133.6934 | -0.4861 | 1.2528 | 0.0079 |
| 2455134.6857 | -0.3598 | 8.9968 | 0.0050 | 2455134.6859 | -0.3598 | 1.2829 | 0.0096 | 2455134.6858 | -0.3598 | 1.3284 | 0.0089 |
| 2455135.6865 | -0.2325 | 9.0243 | 0.0038 | 2455135.6889 | -0.2322 | 1.0857 | 0.0049 | 2455135.6866 | -0.2325 | 1.2775 | 0.0053 |
| 2455136.6860 | -0.1053 | 8.6280 | 0.0054 | 2455136.6862 | -0.1052 | 0.8306 | 0.0099 | 2455136.6861 | -0.1052 | 1.0632 | 0.0072 |
| 2455137.6782 | 0.0210 | 8.2365 | 0.0046 | 2455137.6783 | 0.0210 | 0.8070 | 0.0164 | 2455137.6783 | 0.0210 | 0.9047 | 0.0059 |
| 2455138.6779 | 0.1483 | 8.4347 | 0.0122 | 2455138.6781 | 0.1483 | 0.8636 | 0.0072 | 2455138.6780 | 0.1483 | 1.0343 | 0.0126 |
| 2455139.6769 | 0.2754 | 8.3883 | 0.0033 | 2455139.6771 | 0.2754 | 0.8763 | 0.0307 | 2455139.6770 | 0.2754 | 1.0295 | 0.0043 |
| 2455141.6675 | -0.4712 | 8.8322 | 0.0051 | 2455141.6677 | -0.4712 | 1.1669 | 0.0131 | 2455141.6676 | -0.4712 | 1.2580 | 0.0060 |
| 2455143.6627 | -0.2173 | 8.9838 | 0.0009 | 2455143.6628 | -0.2173 | 1.0935 | 0.0039 | 2455143.6628 | -0.2173 | 1.2517 | 0.0014 |
| 2455144.6583 | -0.0906 | 8.5676 | 0.0047 | 2455144.6585 | -0.0906 | 0.8538 | 0.0168 | 2455144.6584 | -0.0906 | 1.0224 | 0.0051 |
| 2455145.6580 | 0.0366 | 8.2536 | 0.0076 | 2455145.6582 | 0.0367 | 0.7800 | 0.0067 | 2455145.6581 | 0.0366 | 0.9172 | 0.0085 |
| 2455146.6573 | 0.1638 | 8.4383 | 0.0028 | 2455146.6574 | 0.1638 | 0.8697 | 0.0091 | 2455146.6574 | 0.1638 | 1.0320 | 0.0068 |
| 2455151.6431 | -0.2016 | 8.9625 | 0.0011 | 2455151.6433 | -0.2016 | 1.0242 | 0.0083 | 2455151.6432 | -0.2016 | 1.2270 | 0.0021 |
| 2455468.7871 | 0.1623 | 8.4399 | 0.0068 | 2455468.7884 | 0.1624 | 0.8492 | 0.0112 | 2455468.7883 | 0.1624 | 1.0371 | 0.0071 |
| 2455469.8041 | 0.2917 | 8.4152 | 0.0059 | 2455469.8042 | 0.2917 | 0.8779 | 0.0053 | 2455469.8042 | 0.2917 | 1.0392 | 0.0065 |
| 2455470.7818 | 0.4161 | 8.7331 | 0.0027 | 2455470.7820 | 0.4161 | 1.0137 | 0.0038 | 2455470.7819 | 0.4161 | 1.1982 | 0.0046 |
| 2455471.8069 | -0.4534 | 8.8614 | 0.0052 | 2455471.8081 | -0.4533 | 1.1797 | 0.0053 | 2455471.8070 | -0.4534 | 1.2561 | 0.0055 |
| 2455476.7651 | 0.1776 | 8.4432 | 0.0104 | 2455476.7664 | 0.1778 | 0.8352 | 0.0051 | 2455476.7663 | 0.1778 | 1.0439 | 0.0115 |
| 2455477.7904 | 0.3081 | 8.4299 | 0.0064 | 2455477.7917 | 0.3083 | 0.8938 | 0.0096 | 2455477.7905 | 0.3081 | 1.0640 | 0.0079 |
| 2455478.7598 | 0.4315 | 8.7489 | 0.0050 | 2455478.7600 | 0.4315 | 1.0465 | 0.0037 | 2455478.7599 | 0.4315 | 1.2143 | 0.0056 |
| 2455479.7848 | -0.4380 | 8.8840 | 0.0045 | 2455479.7849 | -0.4380 | 1.1297 | 0.0038 | 2455479.7849 | -0.4380 | 1.2771 | 0.0052 |
| 2455480.7544 | -0.3146 | 9.0432 | 0.0081 | 2455480.7546 | -0.3146 | 1.2402 | 0.0080 | 2455480.7545 | -0.3146 | 1.3192 | 0.0084 |
| 2455481.7794 | -0.1842 | 8.9229 | 0.0022 | 2455481.7806 | -0.1840 | 1.0490 | 0.0109 | 2455481.7795 | -0.1842 | 1.1988 | 0.0047 |
| 2455488.7530 | -0.2966 | 9.0580 | 0.0029 | 2455488.7531 | -0.2966 | 1.2033 | 0.0038 | 2455488.7531 | -0.2966 | 1.2944 | 0.0032 |
| 2455490.7481 | -0.0427 | 8.3241 | 0.0050 | 2455490.7482 | -0.0427 | 0.8027 | 0.0077 | 2455490.7482 | -0.0427 | 0.9246 | 0.0065 |
| 2455492.7596 | 0.2133 | 8.4137 | 0.0091 | 2455492.7598 | 0.2133 | 0.8787 | 0.0092 | 2455492.7597 | 0.2133 | 1.0338 | 0.0097 |
| 2455493.7463 | 0.3389 | 8.5277 | 0.0073 | 2455493.7464 | 0.3389 | 0.9689 | 0.0070 | 2455493.7464 | 0.3389 | 1.0915 | 0.0079 |
| 2455495.7363 | -0.4078 | 8.9338 | 0.0037 | 2455495.7343 | -0.4081 | 1.1349 | 0.0101 | 2455495.7353 | -0.4080 | 1.2891 | 0.0042 |
| 2455498.7286 | -0.0270 | 8.2673 | 0.0117 | 2455498.7287 | -0.0270 | 0.7700 | 0.0185 | 2455498.7287 | -0.0270 | 0.9034 | 0.0153 |
| 2455499.7234 | 0.0996 | 8.3669 | 0.0084 | 2455499.7236 | 0.0996 | 0.8157 | 0.0080 | 2455499.7235 | 0.0996 | 0.9900 | 0.0088 |
| 2455501.7168 | 0.3533 | 8.5835 | 0.0059 | 2455504.7085 | -0.2659 | 1.1416 | 0.0074 | 2455501.7169 | 0.3533 | 1.1350 | 0.0059 |
| 2455502.7147 | 0.4803 | 8.7858 | 0.0056 | 2455508.6827 | 0.2399 | 0.8601 | 0.0039 | 2455502.7148 | 0.4803 | 1.2310 | 0.0064 |
| 2455503.7114 | -0.3928 | 8.9643 | 0.0022 | 2455510.6783 | 0.4939 | 1.1412 | 0.0056 | 2455503.7115 | -0.3928 | 1.2848 | 0.0030 |
| 2455504.7083 | -0.2660 | 9.0627 | 0.0017 | 2455511.6751 | -0.3793 | 1.2125 | 0.0053 | 2455504.7084 | -0.2659 | 1.2889 | 0.0033 |
| 2455508.6825 | 0.2399 | 8.3988 | 0.0023 | 2455512.6942 | -0.2496 | 1.1505 | 0.0028 | 2455508.6826 | 0.2399 | 1.0296 | 0.0030 |
| 2455510.6770 | 0.4937 | 8.8003 | 0.0037 | 2455513.6718 | -0.1251 | 0.8612 | 0.0093 | 2455510.6771 | 0.4937 | 1.2529 | 0.0062 |
| 2455511.6749 | -0.3793 | 8.9773 | 0.0023 | 2455515.6897 | 0.1317 | 0.8408 | 0.0119 | 2455511.6750 | -0.3793 | 1.3124 | 0.0033 |
| 2455512.6930 | -0.2497 | 9.0487 | 0.0038 | 2455516.6644 | 0.2557 | 0.8609 | 0.0194 | 2455512.6942 | -0.2496 | 1.2878 | 0.0040 |
| 2455513.6717 | -0.1252 | 8.7130 | 0.0036 | 2455517.6597 | 0.3824 | 1.0315 | 0.0098 | 2455513.6718 | -0.1251 | 1.0986 | 0.0041 |
| 2455515.6896 | 0.1317 | 8.4250 | 0.0050 | 2455518.6570 | -0.4907 | 1.1460 | 0.0080 | 2455515.6897 | 0.1317 | 1.0193 | 0.0104 |
| 2455516.6653 | 0.2559 | 8.3958 | 0.0041 | 2455519.6566 | -0.3634 | 1.1221 | 0.0069 | 2455516.6654 | 0.2559 | 1.0293 | 0.0058 |
| 2455517.6596 | 0.3824 | 8.6805 | 0.0054 | 2455522.6657 | 0.0195 | 0.7352 | 0.0113 | 2455517.6597 | 0.3824 | 1.1782 | 0.0060 |
| 2455518.6569 | -0.4907 | 8.8282 | 0.0044 | 2455523.6439 | 0.1440 | 0.8382 | 0.0119 | 2455518.6570 | -0.4907 | 1.2487 | 0.0048 |
| 2455519.6565 | -0.3634 | 9.0013 | 0.0040 | 2455525.6472 | 0.3990 | 0.9750 | 0.0084 | 2455519.6566 | -0.3634 | 1.3134 | 0.0050 |
| 2455522.6655 | 0.0195 | 8.2489 | 0.0041 | 2455526.6469 | -0.4738 | 1.1075 | 0.0045 | 2455522.6656 | 0.0195 | 0.9101 | 0.0050 |
| 2455523.6426 | 0.1439 | 8.4334 | 0.0051 | 2455527.6317 | -0.3484 | 1.1878 | 0.0127 | 2455523.6438 | 0.1440 | 1.0367 | 0.0056 |
| 2455525.6470 | 0.3990 | 8.7180 | 0.0041 | 2455528.6316 | -0.2212 | 1.0913 | 0.0027 | 2455525.6471 | 0.3990 | 1.1960 | 0.0049 |
| 2455526.6467 | -0.4738 | 8.8438 | 0.0041 | 2455530.6313 | 0.0333 | 0.7527 | 0.0053 | 2455526.6468 | -0.4738 | 1.2630 | 0.0046 |
| 2455527.6316 | -0.3484 | 9.0189 | 0.0041 | 2455531.6317 | 0.1607 | 0.8429 | 0.0124 | 2455527.6317 | -0.3484 | 1.3220 | 0.0042 |
| 2455528.6315 | -0.2212 | 9.0097 | 0.0042 | 2455543.5920 | -0.3171 | 1.1072 | 0.0108 | 2455528.6316 | -0.2212 | 1.2578 | 0.0044 |
| 2455530.6311 | 0.0333 | 8.2654 | 0.0085 | 2455544.5849 | -0.1907 | 0.9921 | 0.0049 | 2455530.6312 | 0.0333 | 0.9213 | 0.0089 |
| 2455531.6315 | 0.1606 | 8.4484 | 0.0058 | 2455545.5853 | -0.0634 | 0.7878 | 0.0100 | 2455531.6316 | 0.1607 | 1.0400 | 0.0060 |
| 2455543.5918 | -0.3171 | 9.0498 | 0.0054 | 2455546.6130 | 0.0674 | 0.7844 | 0.0178 | 2455543.5919 | -0.3171 | 1.3280 | 0.0059 |
| 2455544.5847 | -0.1908 | 8.9500 | 0.0041 | 2455554.5735 | 0.0805 | 0.7074 | 0.1245 | 2455544.5848 | -0.1907 | 1.2096 | 0.0044 |
| 2455545.5852 | -0.0634 | 8.4296 | 0.0045 | 2455823.8578 | 0.3532 | 0.9439 | 0.0044 | 2455545.5853 | -0.0634 | 0.9698 | 0.0066 |
| 2455546.6128 | 0.0674 | 8.3148 | 0.0044 | 2455824.8462 | 0.4790 | 1.0926 | 0.0079 | 2455546.6129 | 0.0674 | 0.9598 | 0.0070 |
| 2455554.5734 | 0.0805 | 8.3451 | 0.0052 | 2455826.8492 | -0.2661 | 1.1092 | 0.0092 | 2455554.5735 | 0.0805 | 0.9623 | 0.0055 |
| 2455823.8576 | 0.3531 | 8.5807 | 0.0039 | 2455827.8369 | -0.1404 | 0.8966 | 0.0107 | 2455823.8577 | 0.3532 | 1.1346 | 0.0054 |
| 2455824.8461 | 0.4790 | 8.7817 | 0.0069 | 2455828.8316 | -0.0138 | 0.7311 | 0.0045 | 2455824.8462 | 0.4790 | 1.2332 | 0.0079 |
| 2455826.8491 | -0.2661 | 9.0552 | 0.0043 | 2455830.8252 | 0.2399 | 0.8368 | 0.0189 | 2455826.8492 | -0.2661 | 1.2970 | 0.0055 |
| 2455827.8367 | -0.1404 | 8.7832 | 0.0042 | 2455833.8183 | -0.3791 | 1.1998 | 0.0013 | 2455827.8368 | -0.1404 | 1.1336 | 0.0055 |
| 2455828.8315 | -0.0138 | 8.2545 | 0.0025 | 2455834.8204 | -0.2516 | 1.1387 | 0.0024 | 2455828.8316 | -0.0138 | 0.9051 | 0.0045 |
| 2455830.8261 | 0.2400 | 8.3938 | 0.0019 | 2455839.8077 | 0.3832 | 0.9895 | 0.0122 | 2455830.8262 | 0.2401 | 1.0297 | 0.0045 |
| 2455833.8192 | -0.3790 | 8.9732 | 0.0057 | 2455840.7978 | -0.4908 | 1.1160 | 0.0116 | 2455833.8182 | -0.3791 | 1.3049 | 0.0057 |
| 2455834.8203 | -0.2516 | 9.0558 | 0.0048 | 2455841.7960 | -0.3638 | 1.0096 | 0.0141 | 2455834.8204 | -0.2516 | 1.2786 | 0.0052 |
| 2455839.8075 | 0.3831 | 8.6858 | 0.0030 | 2455842.7911 | -0.2371 | 1.1590 | 0.0072 | 2455839.8076 | 0.3831 | 1.1848 | 0.0045 |
| 2455840.7965 | -0.4910 | 8.8249 | 0.0040 | 2455843.7970 | -0.1091 | 0.8191 | 0.0137 | 2455840.7977 | -0.4908 | 1.2525 | 0.0052 |
| 2455841.7959 | -0.3638 | 8.9727 | 0.0100 | 2455844.5869 | -0.0086 | 0.7609 | 0.0183 | 2455841.7960 | -0.3638 | 1.3286 | 0.0141 |
| 2455842.7909 | -0.2372 | 9.0299 | 0.0028 | 2455844.6032 | -0.0065 | 0.7716 | 0.0151 | 2455842.7910 | -0.2371 | 1.2722 | 0.0028 |
| 2455843.7957 | -0.1093 | 8.6523 | 0.0100 | 2455844.6196 | -0.0044 | 0.7679 | 0.0279 | 2455843.7958 | -0.1093 | 1.0784 | 0.0122 |
| 2455844.5849 | -0.0088 | 8.2569 | 0.0008 | 2455844.6360 | -0.0023 | 0.7405 | 0.0106 | 2455844.5860 | -0.0087 | 0.9063 | 0.0040 |
| 2455844.6013 | -0.0067 | 8.2503 | 0.0013 | 2455844.6523 | -0.0003 | 0.7724 | 0.0116 | 2455844.6023 | -0.0066 | 0.9152 | 0.0072 |
| 2455844.6176 | -0.0047 | 8.2494 | 0.0016 | 2455844.6699 | 0.0020 | 0.7683 | 0.0056 | 2455844.6187 | -0.0045 | 0.9128 | 0.0023 |
| 2455844.6340 | -0.0026 | 8.2456 | 0.0017 | 2455844.6852 | 0.0039 | 0.7685 | 0.0386 | 2455844.6351 | -0.0024 | 0.9170 | 0.0053 |
| 2455844.6504 | -0.0005 | 8.2456 | 0.0019 | 2455844.7019 | 0.0061 | 0.7705 | 0.0116 | 2455844.6514 | -0.0004 | 0.9081 | 0.0069 |
| 2455844.6668 | 0.0016 | 8.2458 | 0.0036 | 2455844.7183 | 0.0081 | 0.7457 | 0.0092 | 2455844.6679 | 0.0017 | 0.9092 | 0.0066 |
| 2455844.6833 | 0.0037 | 8.2481 | 0.0030 | 2455844.7835 | 0.0164 | 0.7502 | 0.0332 | 2455844.6843 | 0.0038 | 0.9073 | 0.0042 |
| 2455844.6999 | 0.0058 | 8.2475 | 0.0049 | 2455845.7831 | 0.1437 | 0.8402 | 0.0153 | 2455844.7010 | 0.0059 | 0.9104 | 0.0060 |
| 2455844.7164 | 0.0079 | 8.2423 | 0.0012 | 2455846.7802 | 0.2706 | 0.8920 | 0.0132 | 2455844.7174 | 0.0080 | 0.9154 | 0.0054 |
| 2455844.7816 | 0.0162 | 8.2539 | 0.0048 | 2455847.7849 | 0.3984 | 1.0290 | 0.0150 | 2455844.7826 | 0.0163 | 0.9053 | 0.0058 |
| 2455845.7829 | 0.1436 | 8.4380 | 0.0054 | 2455848.7822 | -0.4746 | 1.0221 | 0.0140 | 2455845.7830 | 0.1437 | 1.0269 | 0.0092 |
| 2455846.7822 | 0.2708 | 8.3986 | 0.0057 | 2455849.7725 | -0.3486 | 1.1215 | 0.0030 | 2455846.7812 | 0.2707 | 1.0270 | 0.0062 |
| 2455847.7847 | 0.3984 | 8.7109 | 0.0045 | 2455850.7694 | -0.2217 | 1.0633 | 0.0134 | 2455847.7848 | 0.3984 | 1.2041 | 0.0062 |
| 2455848.7821 | -0.4746 | 8.8393 | 0.0060 | 2455851.7743 | -0.0938 | 0.8719 | 0.0140 | 2455848.7822 | -0.4746 | 1.2553 | 0.0076 |
| 2455849.7712 | -0.3488 | 9.0170 | 0.0035 | 2455852.7681 | 0.0327 | 0.7576 | 0.0202 | 2455849.7713 | -0.3487 | 1.3172 | 0.0045 |
| 2455850.7692 | -0.2217 | 9.0154 | 0.0049 | 2455853.7687 | 0.1600 | 0.8567 | 0.0173 | 2455850.7693 | -0.2217 | 1.2561 | 0.0056 |
| 2455851.7741 | -0.0938 | 8.5855 | 0.0011 | 2455854.7631 | 0.2866 | 0.8346 | 0.0047 | 2455851.7742 | -0.0938 | 1.0387 | 0.0014 |
| 2455852.7679 | 0.0326 | 8.2567 | 0.0171 | 2455855.7512 | 0.4123 | 1.0387 | 0.0101 | 2455852.7680 | 0.0327 | 0.9177 | 0.0177 |
| 2455853.7685 | 0.1600 | 8.4402 | 0.0068 | 2455856.7463 | -0.4610 | 1.1494 | 0.0087 | 2455853.7686 | 0.1600 | 1.0394 | 0.0073 |
| 2455854.7618 | 0.2864 | 8.4121 | 0.0045 | 2455857.7444 | -0.3340 | 1.1728 | 0.0105 | 2455854.7619 | 0.2864 | 1.0435 | 0.0052 |
| 2455855.7499 | 0.4122 | 8.7234 | 0.0101 | 2455858.7411 | -0.2071 | 1.0373 | 0.0100 | 2455855.7511 | 0.4123 | 1.2155 | 0.0106 |
| 2455856.7462 | -0.4610 | 8.8573 | 0.0025 | 2455860.7411 | 0.0474 | 0.7716 | 0.0125 | 2455856.7463 | -0.4610 | 1.2698 | 0.0029 |
| 2455857.7442 | -0.3340 | 9.0276 | 0.0007 | 2455861.7384 | 0.1743 | 0.8758 | 0.0130 | 2455857.7443 | -0.3340 | 1.3159 | 0.0017 |
| 2455858.7410 | -0.2071 | 8.9794 | 0.0019 | 2455862.7490 | 0.3030 | 0.8907 | 0.0053 | 2455858.7411 | -0.2071 | 1.2418 | 0.0035 |
| 2455860.7410 | 0.0474 | 8.2747 | 0.0082 | 2455864.7332 | -0.4445 | 1.1640 | 0.0071 | 2455860.7411 | 0.0474 | 0.9321 | 0.0117 |
| 2455861.7382 | 0.1743 | 8.4448 | 0.0031 | 2455865.7387 | -0.3165 | 1.2457 | 0.0159 | 2455861.7383 | 0.1743 | 1.0359 | 0.0049 |
| 2455862.7477 | 0.3028 | 8.4363 | 0.0040 | 2455866.7258 | -0.1909 | 1.0404 | 0.0039 | 2455862.7478 | 0.3028 | 1.0513 | 0.0046 |
| 2455864.7331 | -0.4445 | 8.8860 | 0.0039 | | | | | 2455864.7332 | -0.4445 | 1.2844 | 0.0047 |
| 2455865.7386 | -0.3165 | 9.0443 | 0.0043 | | | | | 2455865.7387 | -0.3165 | 1.3088 | 0.0062 |
| 2455866.7257 | -0.1909 | 8.9416 | 0.0060 | | | | | 2455866.7258 | -0.1909 | 1.2105 | 0.0067 |

**Photometry of VY Cyg, cont…**

| HJD | Phase | V-R | V-R err | HJD | Phase | R-I | R-I err |
|---|---|---|---|---|---|---|---|
| 2453665.7581 | -0.3147 | 0.9045 | 0.0044 | 2453665.7582 | -0.3147 | 0.8174 | 0.0024 |
| 2453667.7527 | -0.0608 | 0.7058 | 0.0068 | 2453667.7528 | -0.0608 | 0.6857 | 0.0057 |
| 2453669.6125 | 0.1759 | 0.7582 | 0.0048 | 2453669.6126 | 0.1759 | 0.7179 | 0.0036 |
| 2453670.6158 | 0.3036 | 0.7621 | 0.0032 | 2453670.6159 | 0.3036 | 0.7278 | 0.0023 |
| 2453671.6153 | 0.4308 | 0.8523 | 0.0061 | 2453671.6154 | 0.4308 | 0.7838 | 0.0027 |
| 2453672.6156 | -0.4419 | 0.8922 | 0.0037 | 2453672.6157 | -0.4419 | 0.8015 | 0.0027 |
| 2453673.6149 | -0.3147 | 0.9049 | 0.0043 | 2453673.6150 | -0.3147 | 0.8139 | 0.0023 |
| 2453674.6145 | -0.1875 | 0.8546 | 0.0057 | 2453674.6146 | -0.1875 | 0.7749 | 0.0019 |
| 2453675.7322 | -0.0453 | 0.6835 | 0.0064 | 2453675.7323 | -0.0452 | 0.6741 | 0.0053 |
| 2453677.7266 | 0.2086 | 0.7668 | 0.0069 | 2453677.7267 | 0.2086 | 0.7159 | 0.0039 |
| 2453678.7234 | 0.3354 | 0.7960 | 0.0052 | 2453678.7236 | 0.3355 | 0.7401 | 0.0063 |
| 2453679.6115 | 0.4485 | 0.8615 | 0.0035 | 2453679.6116 | 0.4485 | 0.7861 | 0.0027 |
| 2453680.6110 | -0.4243 | 0.8896 | 0.0032 | 2453680.6111 | -0.4243 | 0.8063 | 0.0018 |
| 2453688.6074 | -0.4066 | 0.8939 | 0.0071 | 2453688.6076 | -0.4066 | 0.8075 | 0.0023 |
| 2453689.6111 | -0.2788 | 0.8978 | 0.0099 | 2453689.6112 | -0.2788 | 0.8161 | 0.0043 |
| 2453690.6150 | -0.1511 | 0.7874 | 0.0109 | 2453690.6151 | -0.1511 | 0.7297 | 0.0068 |
| 2453691.6292 | -0.0220 | 0.6664 | 0.0089 | 2453691.6293 | -0.0220 | 0.6664 | 0.0059 |
| 2453693.6061 | 0.2296 | 0.7472 | 0.0052 | 2453693.6063 | 0.2296 | 0.7176 | 0.0028 |
| 2453694.6051 | 0.3568 | 0.8169 | 0.0117 | 2453694.6053 | 0.3568 | 0.7552 | 0.0053 |
| 2453695.6048 | 0.4840 | 0.8687 | 0.0057 | 2453695.6050 | 0.4840 | 0.7879 | 0.0048 |
| 2453699.5750 | -0.0107 | 0.6692 | 0.0060 | 2453699.5766 | -0.0105 | 0.6547 | 0.0025 |
| 2453701.5974 | 0.2467 | 0.7377 | 0.0039 | 2453701.5975 | 0.2467 | 0.7271 | 0.0056 |
| 2453702.6413 | 0.3795 | 0.8397 | 0.0040 | 2453702.6400 | 0.3794 | 0.7704 | 0.0066 |
| 2453703.6286 | -0.4948 | 0.8802 | 0.0059 | 2453703.6287 | -0.4948 | 0.7877 | 0.0043 |
| 2453705.5970 | -0.2443 | 0.8908 | 0.0032 | 2453705.5972 | -0.2442 | 0.8047 | 0.0039 |
| 2453706.5970 | -0.1170 | 0.7866 | 0.0037 | 2453706.5971 | -0.1170 | 0.7337 | 0.0023 |
| 2453708.6057 | 0.1387 | 0.7628 | 0.0329 | 2453708.6059 | 0.1387 | 0.7127 | 0.0054 |
| 2453709.5968 | 0.2648 | 0.7525 | 0.0045 | 2453709.5969 | 0.2648 | 0.7157 | 0.0047 |
| 2453710.5969 | 0.3921 | 0.8509 | 0.0049 | 2453710.5971 | 0.3921 | 0.7695 | 0.0051 |
| 2453711.5971 | -0.4806 | 0.8807 | 0.0041 | 2453711.5972 | -0.4806 | 0.7924 | 0.0032 |
| 2453712.5970 | -0.3534 | 0.9088 | 0.0052 | 2453712.5971 | -0.3533 | 0.8146 | 0.0033 |
| 2453714.5973 | -0.0988 | 0.7597 | 0.0046 | 2453714.5974 | -0.0988 | 0.7186 | 0.0044 |
| 2453724.5785 | 0.1716 | 0.7615 | 0.0295 | 2453724.5786 | 0.1716 | 0.7181 | 0.0100 |
| 2453725.5757 | 0.2985 | 0.7694 | 0.0127 | 2453725.5758 | 0.2985 | 0.7233 | 0.0047 |
| 2453726.5761 | 0.4258 | 0.8621 | 0.0009 | 2453726.5762 | 0.4258 | 0.7810 | 0.0007 |
| 2454001.8317 | 0.4584 | 0.8774 | 0.0056 | 2454001.8318 | 0.4584 | 0.7908 | 0.0063 |
| 2454003.7980 | -0.2913 | 0.9046 | 0.0037 | 2454003.7982 | -0.2913 | 0.8223 | 0.0037 |
| 2454004.7433 | -0.1710 | 0.8451 | 0.0067 | 2454004.7434 | -0.1710 | 0.7787 | 0.0066 |
| 2454005.7962 | -0.0370 | 0.6826 | 0.0177 | 2454005.7964 | -0.0370 | 0.6733 | 0.0108 |
| 2454008.8205 | 0.3479 | 0.8116 | 0.0075 | 2454008.8206 | 0.3479 | 0.7669 | 0.0014 |
| 2454010.7937 | -0.4010 | 0.8880 | 0.0105 | 2454010.7955 | -0.4007 | 0.8350 | 0.0041 |
| 2454012.7563 | -0.1512 | 0.8352 | 0.0178 | 2454012.7580 | -0.1510 | 0.7716 | 0.0016 |
| 2454017.6804 | 0.4755 | 0.8749 | 0.0105 | 2454017.6789 | 0.4753 | 0.8058 | 0.0033 |
| 2454018.7570 | -0.3875 | 0.9070 | 0.0062 | 2454018.7571 | -0.3874 | 0.8142 | 0.0057 |
| 2454023.7503 | 0.2481 | 0.7453 | 0.0088 | 2454023.7504 | 0.2481 | 0.7372 | 0.0045 |
| 2454025.7572 | -0.4965 | 0.8915 | 0.0126 | 2454025.7573 | -0.4965 | 0.7929 | 0.0085 |
| 2454030.7496 | 0.1389 | 0.7558 | 0.0135 | 2454030.7498 | 0.1389 | 0.7209 | 0.0045 |
| 2454031.7488 | 0.2661 | 0.7577 | 0.0132 | 2454031.7489 | 0.2661 | 0.7282 | 0.0092 |
| 2454033.7531 | -0.4789 | 0.8785 | 0.0207 | 2454033.7532 | -0.4788 | 0.8135 | 0.0143 |
| 2454037.7401 | 0.0286 | 0.6775 | 0.0046 | 2454037.7402 | 0.0286 | 0.6797 | 0.0030 |
| 2454039.7309 | 0.2820 | 0.7639 | 0.0125 | 2454039.7311 | 0.2820 | 0.7298 | 0.0085 |
| 2454040.7306 | 0.4092 | 0.8714 | 0.0013 | 2454040.7307 | 0.4092 | 0.7876 | 0.0043 |
| 2454047.7040 | 0.2967 | 0.7751 | 0.0234 | 2454047.7041 | 0.2967 | 0.7339 | 0.0067 |
| 2454049.7079 | -0.4482 | 0.8830 | 0.0071 | 2454049.7080 | -0.4482 | 0.8030 | 0.0114 |
| 2454050.7003 | -0.3219 | 0.9277 | 0.0122 | 2454050.7005 | -0.3219 | 0.8157 | 0.0140 |
| 2454057.6800 | -0.4336 | 0.8966 | 0.0086 | 2454057.6801 | -0.4336 | 0.8154 | 0.0053 |
| 2454058.6732 | -0.3072 | 0.9182 | 0.0056 | 2454058.6733 | -0.3072 | 0.8264 | 0.0045 |
| 2454059.6774 | -0.1794 | 0.8585 | 0.0069 | 2454059.6775 | -0.1794 | 0.7856 | 0.0042 |
| 2454060.6634 | -0.0539 | 0.7009 | 0.0053 | 2454060.6635 | -0.0539 | 0.6900 | 0.0039 |
| 2454061.6676 | 0.0739 | 0.7122 | 0.0157 | 2454061.6678 | 0.0739 | 0.6976 | 0.0122 |
| 2454062.6622 | 0.2005 | 0.7697 | 0.0061 | 2454062.6623 | 0.2005 | 0.7345 | 0.0054 |
| 2454064.6536 | 0.4539 | 0.8769 | 0.0045 | 2454064.6537 | 0.4540 | 0.7922 | 0.0016 |
| 2454067.6517 | -0.1645 | 0.8482 | 0.0115 | 2454067.6518 | -0.1645 | 0.7786 | 0.0055 |
| 2454068.6509 | -0.0373 | 0.6878 | 0.0116 | 2454068.6510 | -0.0373 | 0.6813 | 0.0077 |
| 2454069.6283 | 0.0871 | 0.7147 | 0.0079 | 2454069.6285 | 0.0871 | 0.6982 | 0.0083 |
| 2454070.6396 | 0.2158 | 0.7579 | 0.0077 | 2454070.6397 | 0.2158 | 0.7207 | 0.0059 |
| 2454071.6281 | 0.3416 | 0.8069 | 0.0115 | 2454071.6282 | 0.3416 | 0.7613 | 0.0048 |
| 2454072.6315 | 0.4693 | 0.8748 | 0.0072 | 2454072.6316 | 0.4693 | 0.7931 | 0.0051 |
| 2454075.6094 | -0.1517 | 0.8282 | 0.0038 | 2454075.6095 | -0.1517 | 0.7701 | 0.0038 |
| 2454076.6077 | -0.0246 | 0.6779 | 0.0033 | 2454076.6078 | -0.0246 | 0.6744 | 0.0087 |
| 2454077.6215 | 0.1044 | 0.7369 | 0.0099 | 2454077.6216 | 0.1044 | 0.7087 | 0.0093 |
| 2454078.6189 | 0.2313 | 0.7671 | 0.0070 | 2454078.6190 | 0.2314 | 0.7332 | 0.0033 |
| 2454079.6163 | 0.3583 | 0.8270 | 0.0070 | 2454079.6164 | 0.3583 | 0.7691 | 0.0017 |
| 2454080.6098 | 0.4847 | 0.8806 | 0.0137 | 2454080.6099 | 0.4848 | 0.8016 | 0.0068 |
| 2454081.6106 | -0.3879 | 0.9089 | 0.0144 | 2454081.6107 | -0.3879 | 0.8195 | 0.0066 |
| 2454085.5991 | 0.1197 | 0.7480 | 0.0077 | 2454085.5992 | 0.1198 | 0.7206 | 0.0069 |
| 2454086.5972 | 0.2468 | 0.7553 | 0.0091 | 2454086.5973 | 0.2468 | 0.7269 | 0.0056 |
| 2454087.5850 | 0.3725 | 0.8448 | 0.0088 | 2454087.5851 | 0.3725 | 0.7700 | 0.0064 |
| 2454090.5807 | -0.2462 | 0.9026 | 0.0029 | 2454090.5808 | -0.2462 | 0.8140 | 0.0024 |
| 2454733.7289 | -0.3908 | 0.8989 | 0.0103 | 2454733.7290 | -0.3908 | 0.8188 | 0.0043 |
| 2454734.6777 | -0.2701 | 0.9023 | 0.0065 | 2454734.6778 | -0.2701 | 0.8220 | 0.0069 |
| 2454737.7546 | 0.1215 | 0.7374 | 0.0079 | 2454737.7547 | 0.1216 | 0.7202 | 0.0074 |
| 2454738.6908 | 0.2407 | 0.7714 | 0.0030 | 2454738.6897 | 0.2406 | 0.7153 | 0.0068 |
| 2454741.7279 | -0.3728 | 0.9037 | 0.0026 | 2454741.7280 | -0.3728 | 0.8202 | 0.0016 |
| 2454746.6930 | 0.2592 | 0.7384 | 0.0089 | 2454746.6918 | 0.2590 | 0.7227 | 0.0143 |
| 2454766.6114 | -0.2058 | 0.8731 | 0.0051 | 2454766.6115 | -0.2058 | 0.7900 | 0.0025 |
| 2454767.6107 | -0.0786 | 0.7351 | 0.0082 | 2454767.6109 | -0.0786 | 0.7051 | 0.0069 |
| 2454772.6076 | -0.4426 | 0.8925 | 0.0040 | 2454772.6077 | -0.4426 | 0.8106 | 0.0026 |
| 2454776.6056 | 0.0662 | 0.6964 | 0.0048 | 2454776.6057 | 0.0662 | 0.6893 | 0.0019 |
| 2454777.6199 | 0.1953 | 0.7583 | 0.0022 | 2454777.6202 | 0.1954 | 0.7210 | 0.0009 |
| 2454785.6282 | 0.2146 | 0.7503 | 0.0030 | 2454785.6285 | 0.2146 | 0.7303 | 0.0035 |
| 2454786.6071 | 0.3391 | 0.7832 | 0.0038 | 2454786.6073 | 0.3392 | 0.7462 | 0.0029 |
| 2454788.6165 | -0.4051 | 0.9004 | 0.0039 | 2454788.6167 | -0.4051 | 0.8171 | 0.0044 |
| 2454791.5993 | -0.0255 | 0.6650 | 0.0070 | 2454791.5996 | -0.0254 | 0.6767 | 0.0013 |
| 2454801.5933 | 0.2465 | 0.7491 | 0.0047 | 2454801.5935 | 0.2465 | 0.7198 | 0.0034 |
| 2454803.5930 | -0.4990 | 0.8761 | 0.0034 | 2454803.5932 | -0.4990 | 0.7974 | 0.0024 |
| 2454811.5900 | -0.4812 | 0.8738 | 0.0041 | 2454811.5902 | -0.4812 | 0.7913 | 0.0017 |
| 2455097.7932 | -0.0553 | 0.7215 | 0.0134 | 2455097.7934 | -0.0552 | 0.6906 | 0.0095 |
| 2455098.6116 | 0.0489 | 0.6913 | 0.0144 | 2455098.6129 | 0.0491 | 0.6934 | 0.0071 |
| 2455098.7895 | 0.0715 | 0.7294 | 0.0108 | 2455098.7898 | 0.0716 | 0.6917 | 0.0085 |
| 2455099.6263 | 0.1780 | 0.7656 | 0.0035 | 2455099.6266 | 0.1781 | 0.7289 | 0.0081 |
| 2455099.7847 | 0.1982 | 0.7695 | 0.0154 | 2455099.7850 | 0.1982 | 0.7373 | 0.0052 |
| 2455100.7622 | 0.3226 | 0.8012 | 0.0154 | 2455100.7613 | 0.3225 | 0.7417 | 0.0008 |
| 2455101.8298 | 0.4585 | 0.8982 | 0.0027 | 2455101.8311 | 0.4587 | 0.8069 | 0.0025 |
| 2455102.6224 | -0.4406 | 0.9095 | 0.0039 | 2455102.6227 | -0.4406 | 0.8268 | 0.0040 |
| 2455102.7830 | -0.4202 | 0.9122 | 0.0053 | 2455102.7833 | -0.4202 | 0.8241 | 0.0045 |
| 2455106.7695 | 0.0872 | 0.7161 | 0.0073 | 2455106.7697 | 0.0872 | 0.7090 | 0.0058 |
| 2455131.7011 | 0.2603 | 0.7531 | 0.0093 | 2455131.7013 | 0.2603 | 0.7300 | 0.0075 |
| 2455133.6935 | -0.4861 | 0.8929 | 0.0063 | 2455133.6937 | -0.4861 | 0.8085 | 0.0028 |
| 2455134.6859 | -0.3598 | 0.9077 | 0.0062 | 2455134.6861 | -0.3598 | 0.8238 | 0.0036 |
| 2455135.6867 | -0.2324 | 0.8963 | 0.0057 | 2455135.6869 | -0.2324 | 0.8187 | 0.0047 |
| 2455136.6862 | -0.1052 | 0.7763 | 0.0058 | 2455136.6864 | -0.1052 | 0.7499 | 0.0034 |
| 2455137.6795 | 0.0212 | 0.6956 | 0.0046 | 2455137.6797 | 0.0212 | 0.6776 | 0.0060 |
| 2455138.6781 | 0.1483 | 0.7752 | 0.0129 | 2455138.6783 | 0.1483 | 0.7326 | 0.0070 |
| 2455139.6771 | 0.2754 | 0.7700 | 0.0069 | 2455139.6773 | 0.2755 | 0.7403 | 0.0085 |
| 2455141.6677 | -0.4712 | 0.9018 | 0.0052 | 2455141.6691 | -0.4710 | 0.8174 | 0.0040 |
| 2455143.6628 | -0.2173 | 0.8869 | 0.0014 | 2455143.6631 | -0.2173 | 0.8132 | 0.0013 |
| 2455144.6585 | -0.0906 | 0.7872 | 0.0060 | 2455144.6587 | -0.0906 | 0.7276 | 0.0042 |
| 2455145.6582 | 0.0367 | 0.7013 | 0.0092 | 2455145.6585 | 0.0367 | 0.6884 | 0.0101 |
| 2455146.6574 | 0.1638 | 0.7879 | 0.0043 | 2455146.6577 | 0.1639 | 0.7355 | 0.0036 |
| 2455151.6433 | -0.2016 | 0.8789 | 0.0029 | 2455151.6435 | -0.2016 | 0.7936 | 0.0030 |
| 2455468.7873 | 0.1623 | 0.7747 | 0.0071 | 2455468.7876 | 0.1623 | 0.7365 | 0.0067 |
| 2455469.8043 | 0.2917 | 0.7893 | 0.0072 | 2455469.8045 | 0.2917 | 0.7510 | 0.0055 |
| 2455470.7820 | 0.4161 | 0.8717 | 0.0031 | 2455470.7822 | 0.4162 | 0.8069 | 0.0029 |
| 2455471.8071 | -0.4534 | 0.9174 | 0.0081 | 2455471.8073 | -0.4534 | 0.8251 | 0.0088 |
| 2455476.7664 | 0.1778 | 0.7748 | 0.0118 | 2455476.7666 | 0.1778 | 0.7345 | 0.0097 |
| 2455477.7906 | 0.3082 | 0.7896 | 0.0079 | 2455477.7908 | 0.3082 | 0.7446 | 0.0088 |
| 2455478.7600 | 0.4315 | 0.8857 | 0.0052 | 2455478.7602 | 0.4316 | 0.8025 | 0.0018 |
| 2455479.7850 | -0.4380 | 0.9131 | 0.0062 | 2455479.7852 | -0.4380 | 0.8230 | 0.0052 |
| 2455480.7546 | -0.3146 | 0.9381 | 0.0082 | 2455480.7548 | -0.3146 | 0.8360 | 0.0032 |
| 2455481.7796 | -0.1842 | 0.8669 | 0.0030 | 2455481.7798 | -0.1841 | 0.8007 | 0.0026 |
| 2455488.7531 | -0.2966 | 0.9395 | 0.0038 | 2455488.7534 | -0.2966 | 0.8435 | 0.0040 |
| 2455490.7494 | -0.0425 | 0.7195 | 0.0065 | 2455490.7496 | -0.0425 | 0.6908 | 0.0066 |
| 2455492.7598 | 0.2133 | 0.7735 | 0.0114 | 2455492.7600 | 0.2134 | 0.7307 | 0.0071 |
| 2455493.7475 | 0.3390 | 0.8236 | 0.0077 | 2455493.7478 | 0.3391 | 0.7600 | 0.0034 |
| 2455495.7354 | -0.4080 | 0.9261 | 0.0041 | 2455495.7345 | -0.4081 | 0.8291 | 0.0027 |
| 2455498.7287 | -0.0270 | 0.7031 | 0.0149 | 2455498.7289 | -0.0270 | 0.6885 | 0.0126 |
| 2455499.7236 | 0.0996 | 0.7411 | 0.0103 | 2455499.7238 | 0.0997 | 0.7263 | 0.0070 |
| 2455501.7170 | 0.3533 | 0.8372 | 0.0066 | 2455501.7173 | 0.3534 | 0.7744 | 0.0046 |
| 2455502.7148 | 0.4803 | 0.8915 | 0.0084 | 2455502.7151 | 0.4804 | 0.8169 | 0.0077 |
| 2455503.7116 | -0.3928 | 0.9314 | 0.0042 | 2455503.7118 | -0.3928 | 0.8388 | 0.0048 |
| 2455504.7085 | -0.2659 | 0.9291 | 0.0023 | 2455504.7087 | -0.2659 | 0.8340 | 0.0017 |
| 2455508.6838 | 0.2400 | 0.7732 | 0.0027 | 2455508.6840 | 0.2400 | 0.7349 | 0.0044 |
| 2455510.6772 | 0.4937 | 0.8791 | 0.0048 | 2455510.6775 | 0.4938 | 0.8111 | 0.0057 |
| 2455511.6751 | -0.3793 | 0.9178 | 0.0040 | 2455511.6753 | -0.3792 | 0.8334 | 0.0039 |
| 2455512.6931 | -0.2497 | 0.8924 | 0.0039 | 2455512.6934 | -0.2497 | 0.8402 | 0.0047 |
| 2455513.6719 | -0.1251 | 0.8032 | 0.0040 | 2455513.6721 | -0.1251 | 0.7613 | 0.0030 |
| 2455515.6898 | 0.1317 | 0.7439 | 0.0053 | 2455515.6900 | 0.1317 | 0.7304 | 0.0069 |
| 2455516.6644 | 0.2557 | 0.7725 | 0.0046 | 2455516.6635 | 0.2556 | 0.7344 | 0.0034 |
| 2455517.6598 | 0.3824 | 0.8531 | 0.0054 | 2455517.6600 | 0.3825 | 0.7942 | 0.0025 |
| 2455518.6570 | -0.4907 | 0.8887 | 0.0044 | 2455518.6573 | -0.4906 | 0.8190 | 0.0021 |
| 2455519.6566 | -0.3634 | 0.9089 | 0.0044 | 2455519.6569 | -0.3634 | 0.8403 | 0.0033 |
| 2455522.6657 | 0.0195 | 0.6872 | 0.0046 | 2455522.6659 | 0.0196 | 0.6846 | 0.0028 |
| 2455523.6428 | 0.1439 | 0.7618 | 0.0055 | 2455523.6442 | 0.1441 | 0.7306 | 0.0043 |
| 2455525.6472 | 0.3990 | 0.8626 | 0.0044 | 2455525.6474 | 0.3990 | 0.7954 | 0.0030 |
| 2455526.6480 | -0.4736 | 0.8939 | 0.0043 | 2455526.6493 | -0.4735 | 0.8183 | 0.0024 |
| 2455527.6317 | -0.3484 | 0.9140 | 0.0044 | 2455527.6320 | -0.3484 | 0.8304 | 0.0027 |
| 2455528.6317 | -0.2211 | 0.8921 | 0.0043 | 2455528.6330 | -0.2210 | 0.8144 | 0.0013 |
| 2455530.6313 | 0.0333 | 0.6816 | 0.0090 | 2455530.6315 | 0.0334 | 0.6890 | 0.0034 |
| 2455531.6328 | 0.1608 | 0.7645 | 0.0059 | 2455531.6330 | 0.1608 | 0.7317 | 0.0016 |
| 2455543.5920 | -0.3171 | 0.9080 | 0.0057 | 2455543.5923 | -0.3171 | 0.8304 | 0.0022 |
| 2455544.5860 | -0.1906 | 0.8665 | 0.0046 | 2455544.5863 | -0.1906 | 0.8010 | 0.0022 |
| 2455545.5853 | -0.0634 | 0.7161 | 0.0056 | 2455545.5856 | -0.0634 | 0.7011 | 0.0036 |
| 2455546.6130 | 0.0674 | 0.6992 | 0.0074 | 2455546.6132 | 0.0674 | 0.7153 | 0.0059 |
| 2455554.5735 | 0.0805 | 0.7233 | 0.0053 | 2455554.5738 | 0.0806 | 0.7067 | 0.0041 |
| 2455823.8578 | 0.3532 | 0.8276 | 0.0042 | 2455823.8580 | 0.3532 | 0.7800 | 0.0074 |
| 2455824.8462 | 0.4790 | 0.8873 | 0.0086 | 2455824.8465 | 0.4790 | 0.8120 | 0.0096 |
| 2455826.8504 | -0.2660 | 0.9429 | 0.0046 | 2455826.8517 | -0.2658 | 0.8309 | 0.0019 |
| 2455827.8369 | -0.1404 | 0.8193 | 0.0048 | 2455827.8372 | -0.1404 | 0.7710 | 0.0029 |
| 2455828.8317 | -0.0138 | 0.6759 | 0.0037 | 2455828.8319 | -0.0138 | 0.6838 | 0.0057 |
| 2455830.8252 | 0.2399 | 0.7636 | 0.0031 | 2455830.8243 | 0.2398 | 0.7438 | 0.0035 |
| 2455833.8194 | -0.3790 | 0.9262 | 0.0062 | 2455833.8196 | -0.3790 | 0.8326 | 0.0055 |
| 2455834.8205 | -0.2516 | 0.9145 | 0.0049 | 2455834.8207 | -0.2516 | 0.8387 | 0.0028 |
| 2455839.8077 | 0.3832 | 0.8684 | 0.0036 | 2455839.8079 | 0.3832 | 0.7809 | 0.0040 |
| 2455840.7967 | -0.4910 | 0.8971 | 0.0044 | 2455840.7969 | -0.4909 | 0.8223 | 0.0040 |
| 2455841.7960 | -0.3638 | 0.9094 | 0.0141 | 2455841.7963 | -0.3637 | 0.8544 | 0.0141 |
| 2455842.7911 | -0.2371 | 0.8957 | 0.0035 | 2455842.7913 | -0.2371 | 0.8254 | 0.0050 |
| 2455843.7959 | -0.1092 | 0.7866 | 0.0101 | 2455843.7961 | -0.1092 | 0.7397 | 0.0029 |
| 2455844.5860 | -0.0087 | 0.6712 | 0.0014 | 2455844.5871 | -0.0086 | 0.6754 | 0.0034 |
| 2455844.6024 | -0.0066 | 0.6663 | 0.0022 | 2455844.6035 | -0.0065 | 0.6804 | 0.0041 |
| 2455844.6187 | -0.0045 | 0.6689 | 0.0028 | 2455844.6199 | -0.0044 | 0.6722 | 0.0040 |
| 2455844.6351 | -0.0024 | 0.6638 | 0.0029 | 2455844.6362 | -0.0023 | 0.6684 | 0.0061 |
| 2455844.6515 | -0.0004 | 0.6698 | 0.0038 | 2455844.6526 | -0.0002 | 0.6744 | 0.0040 |
| 2455844.6679 | 0.0017 | 0.6657 | 0.0042 | 2455844.6690 | 0.0019 | 0.6783 | 0.0053 |
| 2455844.6844 | 0.0038 | 0.6707 | 0.0039 | 2455844.6855 | 0.0040 | 0.6783 | 0.0035 |
| 2455844.7010 | 0.0059 | 0.6697 | 0.0062 | 2455844.7022 | 0.0061 | 0.6780 | 0.0056 |
| 2455844.7175 | 0.0080 | 0.6646 | 0.0045 | 2455844.7186 | 0.0082 | 0.6795 | 0.0045 |
| 2455844.7827 | 0.0163 | 0.6877 | 0.0054 | 2455844.7838 | 0.0165 | 0.6727 | 0.0056 |
| 2455845.7831 | 0.1437 | 0.7562 | 0.0083 | 2455845.7833 | 0.1437 | 0.7264 | 0.0074 |
| 2455846.7813 | 0.2707 | 0.7739 | 0.0102 | 2455846.7815 | 0.2707 | 0.7428 | 0.0087 |
| 2455847.7849 | 0.3984 | 0.8582 | 0.0046 | 2455847.7851 | 0.3985 | 0.8006 | 0.0012 |
| 2455848.7822 | -0.4746 | 0.8936 | 0.0066 | 2455848.7825 | -0.4746 | 0.8086 | 0.0039 |
| 2455849.7714 | -0.3487 | 0.9218 | 0.0036 | 2455849.7727 | -0.3486 | 0.8395 | 0.0026 |
| 2455850.7694 | -0.2217 | 0.9066 | 0.0050 | 2455850.7696 | -0.2217 | 0.8188 | 0.0014 |
| 2455851.7743 | -0.0938 | 0.7799 | 0.0062 | 2455851.7746 | -0.0938 | 0.7348 | 0.0076 |
| 2455852.7681 | 0.0327 | 0.6894 | 0.0175 | 2455852.7683 | 0.0327 | 0.6900 | 0.0038 |
| 2455853.7687 | 0.1600 | 0.7628 | 0.0079 | 2455853.7689 | 0.1600 | 0.7380 | 0.0074 |
| 2455854.7620 | 0.2864 | 0.7712 | 0.0058 | 2455854.7622 | 0.2865 | 0.7442 | 0.0038 |
| 2455855.7501 | 0.4122 | 0.8713 | 0.0105 | 2455855.7503 | 0.4122 | 0.8066 | 0.0043 |
| 2455856.7464 | -0.4610 | 0.9055 | 0.0052 | 2455856.7466 | -0.4610 | 0.8205 | 0.0064 |
| 2455857.7444 | -0.3340 | 0.9380 | 0.0034 | 2455857.7446 | -0.3340 | 0.8416 | 0.0037 |
| 2455858.7411 | -0.2071 | 0.8929 | 0.0053 | 2455858.7414 | -0.2071 | 0.8084 | 0.0051 |
| 2455860.7412 | 0.0474 | 0.6978 | 0.0085 | 2455860.7414 | 0.0475 | 0.6977 | 0.0060 |
| 2455861.7384 | 0.1743 | 0.7680 | 0.0062 | 2455861.7387 | 0.1744 | 0.7387 | 0.0057 |
| 2455862.7479 | 0.3028 | 0.7770 | 0.0066 | 2455862.7481 | 0.3029 | 0.7463 | 0.0060 |
| 2455864.7333 | -0.4445 | 0.9037 | 0.0044 | 2455864.7335 | -0.4445 | 0.8269 | 0.0063 |
| 2455865.7399 | -0.3164 | 0.9178 | 0.0046 | 2455865.7412 | -0.3162 | 0.8319 | 0.0023 |
| 2455866.7259 | -0.1909 | 0.8552 | 0.0061 | 2455866.7272 | -0.1907 | 0.8052 | 0.0017 |